\documentclass[amssymb,amsmath,eqsecnum,showpacs,nofootinbib,twocolumn] {revtex4}
\usepackage{color}
\usepackage{footnote}
\usepackage{graphicx}
\usepackage{ulem}
\usepackage{physics}
\usepackage{verbatim}
\usepackage{bigints}
\usepackage{float} 
\usepackage[colorlinks]{hyperref}

\hypersetup{
	colorlinks,
	citecolor=blue,
	filecolor=gray,
	linkcolor=blue,
	urlcolor=blue
}

\begin{document}

\author{Jonathan Claros $^{1}$ and Emanuel Gallo$^1$
}

\newcommand{\red}[1]{\textcolor{red}{#1}}
\newcommand{\blue}[1]{\textcolor{blue}{#1}}

\affiliation{$^1$FaMAF, UNC; Instituto de Física Enrique Gaviola (IFEG), CONICET, \\
Ciudad Universitaria, (5000) C\'ordoba, Argentina. }

\title{
Accurate analytical modeling of light rays in spherically symmetric spacetimes: Applications in the study of black hole accretion disks and polarimetry}

\begin{abstract}

We present new, simple analytical formulas to accurately describe light rays in spherically symmetric static spacetimes. These formulas extend those introduced by Beloborodov and refined by Poutanen for the Schwarzschild metric.
Our enhanced formulas are designed to be applicable to a broader range of spacetimes, making them particularly valuable for describing phenomena around compact objects like neutron stars and black holes. As an illustration of their application, we present analytical studies of images of thin accretion disks surrounding black holes and explore their associated polarimetry.

\end{abstract}

\maketitle
\section{Introduction}

 In recent years, we have entered a fascinating era in observational studies of compact objects such as neutron stars and black holes.  The high-energy X-ray luminosity and polarization of light rays emanating from neutron stars have begun to be analyzed through various instruments, such as \cite{ixpe,XMM-Newton,NuSTAR,2017gravity}. Simultaneously, the first images of the supermassive black holes M87* and SgrA* have been achieved through the Event Horizon Telescope (EHT) consortium \cite{EventHorizonTelescope:2022wkp}. In the case of M87*, the study of the polarization of light rays emanating from the accretion disk surrounding it has also been initiated \cite{EventHorizonTelescope:2021bee}. In the case of SgrA*, using instruments from the gravity collaboration \cite{2017gravity} and information from The Very Large Telescope Interferometer (VLTI) \cite{vlti}, high-resolution imaging and polarization studies of hotspots orbiting very close to the event horizon have been conducted. These studies have allowed the characterization of the disk material and the magnetic fields of SgrA* \cite{GRAVITY:2023avo,GRAVITY:2020lpa,GRAVITY:2020hwn}, confirming the existence of material following relativistic orbits at a few Schwarzschild radii. It is anticipated that in the coming years, with the aid of future observatories, image resolution will improve, enabling us to study not only the dynamics of accretion disks in greater detail but also to test Einstein's theory of relativity in regions of intense gravitational fields (examples and analysis of different mission concepts can be found in \cite{Andrianov:2021,Johnson:2021,Kudri2021, Likha2022,Rudnit:2023,roelofs_ngeht_2023,Peter:2022,Trippe2023,Hudson_2023,Tamar:2024qxt})  . Hence, in addition to technological advancements and signal processing for such observations, models predicting the expected types of observations are necessary. Particularly, it would be desirable to understand how different gravitational theories can affect the type of images, studies of luminosity curves, polarization, etc., that are expected to be observed. Generally, to carry out this type of analysis, ray tracing models are needed, which must be numerically solved for each spacetime one wishes to study. This can be time-consuming, as it requires simulation for each possible choice of parameters describing the system. Therefore, in recent times, there has been interest in approximate analytical methods to perform these tasks much faster \cite{Beloborodov_2002,Sotani_2017,Semerak:2014kra,DeFalco:2016yox,LaPlaca:2019rjz,Poutanen:2019tcd}.

The essence of these approximate methods is to have analytical formulas describing light rays connecting emission points near compact objects with a distant observer. Among these formulas, the Beloborodov formula stands out (Eq. (1) of \cite{Beloborodov_2002}), which allows, for a Schwarzschild spacetime, to relate the emission angle of a light ray (measured with respect to the radial direction) with the angular (and radial) position at which the same point is observed in the asymptotic region. This formula has been successfully used to describe a variety of phenomena around neutron stars and black holes \cite{Poutanen:2003yd,Viironen:2004ze,Gierlinski:2004tu,Kulkarni:2005cs, Suleimanov:2006pk, vanAdelsberg:2006uu,Poutanen:2006hw, Bogdanov:2006zd,Morsink:2007tv,Ho:2008bq, Ibragimov:2009js, Dipanjan:2011ef, Poutanen:2013xaa, Potekhin:2014fla, Mohan:2015jsa, Watts:2016uzu,Perego:2017fho,Mushtukov:2017ubg,Ascenzi:2024osa,Saathoff:2024pzk,Cardenas-Avendano:2022csp,Loktev:2023cty,Markozov:2023zxq,Ahlberg:2023iza,EventHorizonTelescope:2021btj,Hu:2022ehk,Loktev:2020tin,BG_2023} (this list is not exhaustive). However, this formula has three limitations. It becomes less accurate when the emission angle is greater than $90^{\circ}$; it is limited to spherical symmetry and therefore, spin effects of the compact objects on the spacetime cannot be considered and it is only designed to approximate light rays in the Schwarzschild spacetime. The first limitation was recently overcome by a superior formula obtained by Poutanen \cite{Poutanen:2019tcd}  (also valid for a Schwarzschild spacetime), which has been successfully used in different situations. Regarding the second limitation, it has been observed that it is not always a severe constraint. In particular, in the study of images of black hole accretion disks and their polarization, comparisons have been made between ray-tracing studies in Kerr spacetime and their contrast with the expected results using the Beloborodov formula. It has been found that for moderate spins, the resulting direct images and polarization are similar \cite{Cardenas-Avendano:2022csp,Loktev:2023cty}. Specifically, as outlined in reference \cite{Loktev:2023cty}, the Poutanen formula is utilized to present ARTPOL, a rapid analytical method for tracing polarized light rays. This method serves as an efficient alternative to the traditionally time-consuming numerical ray-tracing computations. It demonstrates accuracy when applied to Kerr black holes with a dimensionless spin parameter of up to $0.94$. Furthermore, as discussed in \cite{Loktev:2023cty}, it achieves this accuracy while being more than four orders of magnitude quicker than direct ray-tracing calculations.  Regarding the last limitation, it is our intention to present in this work new approximate formulas that generalize those of Beloborodov and Poutanen for a class of more general spherically symmetric spacetimes. These new formulas then allow describing optical phenomena of electromagnetic processes originating near these compact objects considering different models for the spacetime around them, and therefore allow analytically obtaining observables whose dependence with the different parameters describing these spacetimes can also be given in terms of simple closed mathematical expressions.

In this work, all analytical expressions are applicable to scenarios where light rays complete less than half a revolution around the compact object before reaching the observer. These rays are commonly referred to as direct or primary rays in the literature. Rays completing half a revolution are termed secondary rays, while those completing $n$ half-turns are known as $n$-th order rays. Recent analytical inquiries into the shapes of higher-order images of equatorial emission rings for high-order rays have been conducted by other researchers, focusing on Schwarzschild spacetime. Bisnovatyi-Kogan and Tsupko \cite{Bisnovatyi-Kogan:2022ujt} investigated scenarios with the observer positioned on the disk's axis of symmetry, while Tsupko extended the analysis to include any inclination angle of the accretion disk relative to the observer \cite{Tsupko:2022kwi}. Expansion of these findings to more general spacetimes were recently presented in \cite{Aratore:2024bro}. All these analytical tools offer fresh insights for studying gravitational fields in the strong-field regime.

The organization of this work is outlined as follows: In Sec. \ref{sec:analyaprox}, we introduce the new approximations, with detailed derivations provided in Sec. \ref{sec:derivation}.
Sec. \ref{sec:test} demonstrates the accuracy of the new formulas by examining their fit to the exact relationship between $\alpha$ and $\psi$ as provided by equations \eqref{eq:belo-ap1} and \eqref{eq:psip}.
In Sec. \ref{sec:acr}, we explore various applications of the analytical approximations derived in Sec. \ref{sec:analyaprox} in the context of black hole image analysis. First, Sec. \ref{sec:gralframework} offers an overview of the general framework for the accretion disk model used in this study, accompanied by new simple analytical formulas for mapping emission points to their corresponding images in the observer's plane. Section \ref{subsec:isora} computes the images of elliptical orbits and isoradial equatorial curves of the accretion disks using these analytical formulas, comparing them with exact images. In Sec. \ref{c3s2}, we calculate the flux of the accretion disk based on the thin disk model proposed by Novikov, Page, and Thorne \cite{NoviThorne:1973,Page:1974he}. Section \ref{subs:pol} presents polarized images of synchrotron-emitting gas rings orbiting black holes, following the methodology outlined by Narayan et al. \cite{EventHorizonTelescope:2021btj}, while Section \ref{Sec:QU_loops} briefly discusses $QU$-diagrams. 
We conclude with some final remarks in Section \ref{Sec:final} and include two appendices.

\section{Analytical approximation for light bending}
\label{sec:analyaprox}
\begin{figure}[htbp]
    {\includegraphics[scale=0.63]{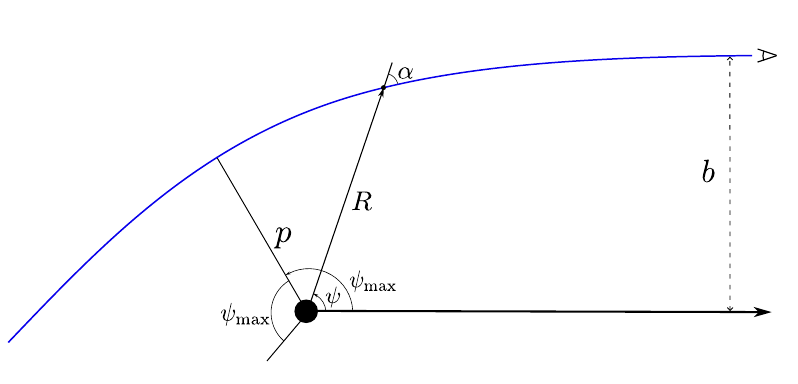}} 
        \caption{ Geometry associated with a light ray that, after passing through the vicinity of a compact object, reaches an observer situated in the asymptotic region $\psi=0$ with an impact parameter $b$. The closest approach of the ray to the compact object (the periastron) occurs at the radial coordinate $r=p$. $\alpha$ indicates the angle between the radial direction and the direction of the light ray's orbit at a radial coordinate $r=R$ from the origin.
        }
        \label{fig:0}
    \end{figure}

Let us consider a light ray passing close to a compact object and reaching an asymptotic observer with an impact parameter $b$. In a coordinate system as depicted in Fig. \ref{fig:0} , for a given point on the orbit with coordinates $(R,\psi)$, the light ray forms an emission angle $\alpha$ with the radial direction. The original Beloborodov formula \cite{Beloborodov_2002}, valid for a Schwarzschild spacetime with Schwarzschild radius $r_H$, is an approximation formula relating $x=1-\cos\alpha$ to $y=1-\cos\psi$ through the following simple relationship:    
\begin{equation} \label{eq:belo-original}
        x=(1-u)y,
    \end{equation}
   where $u=r_H/R$. 

Equation \eqref{eq:belo-original} was later generalized by Poutanen and Beloborodov \cite{poutanen2006pulse} to include corrections of order $y^3$: 
\begin{equation} \label{eq:belo-pou2006}
        x=(1-u)y(1+\frac{u^2}{112}y^2).
    \end{equation}

More recently, an empirical fitting obtained by Poutanen in \cite{Poutanen:2019tcd} substantially improves the accuracy of Eq.\eqref{eq:belo-original} in the regime where $\psi\to\pi$,
\begin{equation} \label{eq:Poutanen}
        x=(1-u)y\left\{1+\frac{u^2}{112}y^2-\frac{e}{100}uy\left[\ln(1-\frac{y}{2})+\frac{y}{2}\right]\right\}.
    \end{equation}

    As shown by Poutanen, the error remains below $0.06\%$ for $\psi < 120^\circ$ and radii exceeding $1.5 r_{\text{H}}$. Beyond these radii, the error exceeds $0.2\%$ only for $\psi > 162^\circ$, corresponding to emission points behind the compact object.

   Here, we present extensions of these formulas to cover a broader range of spherically symmetric spacetimes of the form
    \begin{equation} \label{eq:METRIC}
        ds^2 = -A(r)dt^{2}+\frac{dr^{2}}{A(r)}+r^2(d\theta^{2}+\sin^{2}{\theta}d\phi^{2}). 
    \end{equation}
    In particular, we will show that the generalization of Eq.\eqref{eq:belo-pou2006} preserving terms up to $y^3$ reads,
\begin{widetext}
\begin{equation}\label{eq:belo-altoorden}
x = A(R) y \left[1 + \left(\frac{I_{1}}{2} + \frac{1}{6}\right) y+\frac{\left(315I_{1}^{2} - 135I_{2} + 180A(R)^{2} + 120I_{1} + 32
\right)}{720} y^{2} \right].
\end{equation}
\end{widetext}
where 
\begin{eqnarray}
I_1&=&\int^1_0 P(u')du',\label{eq:i1}\\
I_2&=&\int^1_0 P^2(u')du',\label{eq:i2}
\end{eqnarray}
with 
\begin{equation}\label{eq:p}
    P(u')=-4A(u')u'^2 + A(R),
\end{equation}
and $A(u')$ obtained from  $A(r)$ after the change of variable $r=R/u'$. For a wide variety of metrics, integrals $I_1$ and $I_2$ have exact explicit forms.

Moreover, for the case where the metric component $A(r)$ takes the form: 
\begin{equation}\label{eq:ar}
A(r)=1+\sum^N_{n=1} \frac{a_n}{r^n},
\end{equation}
we propose a generalization of the Poutanen formula Eq.\eqref{eq:Poutanen} given by:
\begin{widetext}
\begin{equation}\label{eq:our}
 x = A(R) y \left\{1 + \left(\frac{I_{1}}{2} + \frac{1}{6}\right) y+\frac{\left(315I_{1}^{2} - 135I_{2} + 180A(R)^{2} + 120I_{1} + 32
\right)}{720}y^2 -\frac{e}{100}(1-A(R))y\left[\ln(1-\frac{y}{2})+\frac{y}{2}\right]\right\},   
\end{equation}
\end{widetext}
which, as we will show, improves the accuracy of  Eq.\eqref{eq:belo-altoorden} in the regime where $\psi\to\pi$. The explicit expression for the integrals $I_1$ and $I_2$ of Eq.\eqref{eq:our} will be given in the next section. For a consistency check, note that for the 
Schwarzschild metric ($A(r)=1-
{r_H}/{r}$), $I_1$ and $I_2$ reduce to: 
\begin{eqnarray}
I_1&=&-\frac{1}{3},\\
I_2&=&\frac{161R^2-280r_HR+135r^2_H}{105R^2}.
\end{eqnarray}
With these expressions at hand, it is easy to check that Eq.\eqref{eq:our} reduces to the 
Poutanen formula Eq.\eqref{eq:Poutanen}. 

Observe that if we 
only preserve the linear term in $y$ in Eq.\eqref{eq:our}, we 
obtain a simple generalization of the Beloborodov formula:
\begin{equation}\label{eq:Belo-general}
 x = A(R) y. 
\end{equation}
Even though in \cite{BG_2023}, we have 
found that the Beloborodov procedure 
yields an expression as in Eq. 
\eqref{eq:Belo-general} for two 
particular kinds of metrics, to our 
knowledge, this is the first time that is observed in the literature that it retains 
this simple form in the more general 
case. As we will show, this is a very 
good approximation for most 
astrophysical situations and for more 
general $A(R)$, not necessarily of the 
form given by Eq. \eqref{eq:ar}.

Equations \eqref{eq:belo-altoorden}, 
\eqref{eq:our}, and the generalized 
Beloborodov formula in Eq. 
\eqref{eq:Belo-general} stand out as 
key outcomes of our study.

Note that for metrics of type Eq.\eqref{eq:ar}, one generally expects that for sufficiently large radii compared to $r_H$, the contribution of terms $a_i$ with $i \geq 2$ is a higher-order correction compared to the one introduced by the dominant term $1+a_1/r$. Therefore, for radii sufficiently large compared to the event horizon's radius, it is expected that Poutanen's formula remains an excellent approximation even for metrics other than Schwarzschild. However, in the vicinity of a non-Schwarzschild spherically symmetric spacetime (including the vicinity of the photon sphere) originating from a more general gravitational theory, it is expected that this approximation becomes ineffective. In such situations, the newly introduced formulas provide simple analytical expressions that are useful for testing and comparing the strong-field regions of compact objects whose gravity is modeled by alternative theories.

In the next two sections, we will 
discuss and test these formulas.

\section{Derivation}\label{sec:derivation}

The Beloborodov formula  was introduced in \cite{Beloborodov_2002}, without derivation. One possible derivation is described by De Falco et.al. in \cite{DeFalco:2016yox}. In our work, we will follow an alternative approach.
Let $(\mathcal{M},g_{ab})$ be a static and spherically symmetric spacetime described by the metric \eqref{eq:METRIC}. In the optical geometric limit, photons are assumed to follow null geodesics, with their dynamics governed by the Lagrangian:
    \begin{equation} \label{eq:1_2}
        \mathcal{L} =\frac{1}{2}g_{\mu \nu} \frac{dx^{\mu}}{d\lambda}\frac{dx^{\nu}}{d\lambda},  
    \end{equation}
    together to the on-shell condition $g_{\mu \nu} k^{\mu}k^{\nu}=0$. Here, $\lambda$ is assumed to be an affine parameter and $\vb*{k}$ (with components $k^\mu$) is  the null vector tangent to the null geodesic curve.
    As the orbits of the null geodesics are planar in these geometries, and in order to use the same convention for the notation of the angles as in Figure \ref{fig:0}, we make (without loss of generality) a coordinate transformation on the angular coordinates of the metric, such as in the new coordinates $(\tilde{\theta},\psi)$, the orbit of the null geodesics is on the equatorial plane $\tilde\theta=\pi/2$, that is, the metric now takes the form:
    \begin{equation}
    ds^2=-A(r)dt^2+\frac{dr^2}{A(r)}+r^2d\tilde\theta^2+\sin^2\tilde\theta d\psi^2.
    \end{equation}
    This change of coordinates will only be applied in this section with the aim of deducing the final formulas. For the rest of the sections, we will revert to using the coordinate system as described in Sec. \ref{sec:analyaprox}.

The condition $k_\mu k^\mu=0$ on the equatorial plane, reads:
    
        \begin{equation} \label{eq:1_3_2}
             -A(r)\dot{t}^{2}+\frac{\dot{r}^{2}}{A(r)}+r^2 \dot{\psi}^{2}=0.
        \end{equation}
  
     From the existence of the time-translation Killing vectors $\partial_t$ and the axially symmetric Killing vector field $\partial_\psi$, we have two respective conserved quantities, the energy $E$ and the angular momentum $L$ of the photon, defined as  
    \begin{equation} \label{eq:1_5}
        \dot{t}=\frac{E}{A}, \hspace{4mm} \dot{\psi}=\frac{L}{r^{2}},
    \end{equation}
    with $\dot{f}=\frac{df}{d\lambda}$.
This allows to rewrite Eq.\eqref{eq:1_3_2} as
\begin{equation} \label{eq:1_7}
        \dot{r}^{2}+V(r)=E^2, 
    \end{equation}
    with 
    \begin{equation}
        V(r)=\frac{A(r)}{r^{2}}b^{2},
    \end{equation}
    and $b=L/E$ the impact parameter.
    
Assuming that the light ray reaches the asymptotic region coming from the vicinity of the compact object  as shown in Fig.\ref{fig:0}, it follows that for the part of the motion where the particle has already passed the periastron $p$ (i.e. considering the region where $\dot\psi<0$ and $\dot{r}>0$) we have,
  
    \begin{equation} \label{eq:1_8}
        \frac{d\psi}{dr} = \frac{\dot{\psi}}{\dot{r}}=-\frac{1}{r^{2}}\left[\frac{1}{b^{2}} - \frac{A(r)}{r^2} \right]^{-1/2} ,
    \end{equation}
whose integration gives
    \begin{equation} \label{eq:1_9}
    \psi(R,b)=\int_{\infty}^{R}\frac{d\psi}{dr}dr=\int_{R}^{\infty}\frac{dr}{r^{2}}\left[\frac{1}{b^{2}} - \frac{A(r)}{r^2} \right]^{-1/2}.
    \end{equation}

After doing the change of variable $r=R/u'$, and expressing $b$ in terms of $\alpha$ \cite{Pechenick_1983,Beloborodov_2002,Briozzo_2022}:
\begin{equation} \label{eq:impact_parameter}
    b=\frac{R}{\sqrt{A(R)}}\sin\alpha,
\end{equation} we can rewrite
Eq.\eqref{eq:1_9} as:
{
\begin{eqnarray} \label{eq:belo-ap1}
        \psi(R,\alpha)&=&\bigintss_{0}^{1}\frac{\sin\alpha}{\sqrt{A(R)-A(\frac{R}{u'})u'^2\sin^2\alpha}}du'\\
        &=&\bigintss_{0}^{1}\frac{\sqrt{1-\cos^2\alpha}}{\sqrt{A(R)-A(\frac{R}{u'})u'^2(1-\cos^2\alpha})}du'.\nonumber
    \end{eqnarray}
    }

Let us remark that the expression \eqref{eq:belo-ap1} is valid in the region $\alpha\in[0,\frac{\pi}{2}]$. For values of $\alpha>\frac{\pi}{2}$, the expression for $\psi$ must be modified to \cite{Poutanen:2019tcd}:
\begin{equation}\label{eq:psip}
    \psi(R,\alpha)=2\psi_{max}-\psi(R,\pi-\alpha),
\end{equation}
where $\psi(R,\pi-\alpha)$ is computed using  Eq.\eqref{eq:belo-ap1} and $\psi_{max}=\psi(p,\frac{\pi}{2})$ is the value of $\psi$ at the periastron $p$ obtained by solving $\frac{dr}{d\psi}\bigg.|_{r=p}=0$, which taking into account Eq.\eqref{eq:1_8} reduces to 
\begin{equation}\label{eq:perias}
    p^2-b^2A(p)=0.
\end{equation}    
In general, Eqs.\eqref{eq:belo-ap1} and \eqref{eq:perias} must be solved numerically.
Alternatively, an excellent analytical approximation can be found that circumvents the need to solve these equations numerically by following the steps outlined below.

We start  by performing a Taylor expansion of the integrand of Eq.\eqref{eq:belo-ap1} around $\cos\alpha=1$. Denoting $1-\cos\alpha=v^2$, we obtain: 

\begin{widetext}   
{
\begin{equation}\label{eq:psiapp}
    \psi(R,v)\approx \frac{\sqrt{2}}{\sqrt{A(R)}}v-\int^1_0\frac{P(u')}{2\sqrt{2A(R)}A(R)}du' v^{3}+
    \int^1_0\frac{-4A^2(R)+3P^2(u')}{16\sqrt{2A(R)}A^2(R)} v^{5}+\mathcal{O}(v^{7}),
\end{equation}
}
\end{widetext}
with $P(u')$ as defined in Eq.\eqref{eq:p}.

With this expression at hand, we can now compute $y=1-\cos\psi(R,v)$  by performing a Taylor expansion of $y$ in terms of $v$. 
{Hence, in terms of $x=v^2=1-\cos\alpha$ we obtain: 
\begin{equation}\label{eq:pogen}
y=A_1 x+A_2x^2+A_3x^3+\mathcal{O}(x^4),
\end{equation}}
with
\begin{eqnarray}
A_1&=&\frac{1}{A(R)},\\
A_2&=&-\frac{3I_1+1}{6A(R)^2},\\
A_3&=&\frac{45(I^2_1+3I_2-4A^2(R))+120I_1+8}{720A^3(R)}.
\end{eqnarray}

Equation \eqref{eq:pogen} is a generalization to arbitrary spherically symmetric spacetimes of Eq. (2) of \cite{Beloborodov_2002}. However, here we are interested in the inverse relation, i.e., an expression of $x$ in terms of $y$. To achieve this, we assume that $x$ also has an expansion in terms of powers of $y$, i.e., $x=\sum_{i=1}^{3} B_i y^i+\mathcal{O}(y^4)$. After replacing this expansion into Eq. \eqref{eq:pogen} and comparing both sides of the resulting expression, we obtain the relation between the coefficients $B_i$ and $A_i$, namely:
\begin{eqnarray}
B_1&=&\frac{1}{A_1},\\
B_2&=&-\frac{A_2}{A^3_1},\\
B_3&=&\frac{2A^2_2-A_1A_3 }{A^5_1}.
\end{eqnarray}
With these relations we recover Eq. \eqref{eq:belo-altoorden} as presented in the previous section.

Now, let us consider the scenario where $A(r)$ takes the form given by Eq. \eqref{eq:ar}. Specifically, we focus on the case where $N=6$, indicating the consideration of 6 parameters $a_i$ to describe the metric,
\begin{equation}\label{eq:ar6}
A(r)=1+\sum^6_{n=1} \frac{a_n}{r^n}.
\end{equation} Following a straightforward computation of the integrals $I_1$ and $I_2$ (their expressions are provided in Appendix \ref{app:A}), Eq. \eqref{eq:our} takes the explicit form:
\begin{widetext}
\begin{equation} \label{eq:our_o6}
\begin{aligned}
   x= & A(R)\left\{y  +  \left( a_2R^4 + \frac{5}{3}a_3R^3 + \frac{15}{7}a_4R^2 + \frac{5}{2}a_5R + \frac{25}{9}a_6 \right) \frac{y^2}{10R^{6}}\right. \\
    &\left.+ \frac{1}{112} \left[ \left( a_1^2 + \frac{24}{5}a_2 \right)R^{10} + \left( \frac{28}{5}a_1a_2 + \frac{28}{3}a_3 \right)R^9 \right.\right.\\
    &\left.\left. + \left( \frac{28}{3}a_1a_3 + \frac{392}{75}a_2^2 + \frac{40}{3}a_4 \right)R^8 + \left( \frac{64}{5}a_1a_4 + \frac{224}{15}a_2a_3 + \frac{84}{5}a_5 \right)R^7 \right.\right.\\
    &\left.\left. + \left( \frac{175}{11}a_1a_5 + \frac{208}{11}a_2a_4 + \frac{980}{99}a_3^2 + \frac{1960}{99}a_6 \right)R^6 + \left( \frac{56}{3}a_1a_6 + \frac{112}{5}a_2a_5 + 24a_3a_4 \right)R^5 \right.\right.\\
    &\left.\left. + \left( \frac{14896}{585}a_2a_6 + \frac{1078}{39}a_3a_5 + \frac{184}{13}a_4^2 \right)R^4 + \left( \frac{832}{27}a_3a_6 + 32a_4a_5 \right)R^3 \right.\right.\\
    &\left.\left. + \left( \frac{176}{5}a_4a_6 + \frac{357}{20}a_5^2 \right)R^2 + \frac{350Ra_5a_6}{9} + \frac{28952}{1377}a_6^2 \right]\frac{y^3}{R^{12}}\right. \\
    &\left.- \frac{e}{100}y^2 ( 1-A(R) )\left[\ln\left(1 - \frac{y}{2}\right) + \frac{y}{2} \right]\right\}.
\end{aligned}
\end{equation}
\end{widetext}
Concerning the last term, the inspiration for the structure of the logarithmic term stems from the expression proposed by Poutanen in \cite{poutanen2006pulse}. Essentially, we have adopted a contribution of the following kind: $\gamma y^2 ( 1-A(R) )A(R)[\ln\left(1 - \frac{y}{2}\right) + \frac{y}{2}]$, 
and numerically optimized the parameter $\gamma$ to enhance the fitting. We observed that by varying the choices of the parameters $a_i$ describing the metric, the value of $\gamma$ does not undergo significant changes, consistently remaining close to $e/100$. 

Before concluding this section, it is 
important to emphasize that the form of 
the metric Eq.\eqref{eq:ar} is shared 
by many well-known spacetimes from 
different gravitational theories, for 
example if only $a_1$ and $a_2$ are nonvanishing, the associated family of metrics includes Reissner-Nordstr\"{o}m (RN), black holes with a conformally coupled scalar field \cite{Astorino:2013sfa}, asymptotically flat solutions of the Horndeski theory \cite{Babichev:2017guv}, black holes in Modify Gravity (MOG) \cite{Moffat:2014aja}, braneworld (BW) gravity \cite{Aliev:2005bi} and nonconmutative gravity \cite{Anacleto:2019tdj}.  Metrics with \( a_i \neq 0 \) also include solutions derived from loop quantum cosmology (LQC) \cite{Lewandowski:2022zce} or Einstein-Bel-Robinson gravity \cite{Sajadi:2023bwe}, among others. This list is not exhaustive. For an analysis of the optical appearance of images from thin accretion disks and shadows generated by several of these metrics, as well as other metrics with non-zero \( a_i \) parameters, and their validation based on shadow observations, we refer to \cite{Vagnozzi:2022moj,daSilva:2023jxa,Wang:2022yvi,Khodadi:2020jij}.

On the other hand, even in many cases where the coefficient $A(r)$ does not have the exact form $A(r)=1+\sum^6_{n=1} \frac{a_n}{r^n}$, it can be used as a very good approximation of the exact form of $A(r)$ after performing a Taylor expansion of it in terms of powers of the parameters describing the metric. One such case is that of a particular Einstein-Maxwell-dilaton (EMD) black hole solution with $A(r)$ given by \cite{Garfinkle:1990qj}:
\begin{equation} \label{eq:emd_bh}
A(r)=1-\frac{2M}{r}\left(\sqrt{1+\frac{q^4}{4M^2r^2}}-\frac{q^2}{2Mr}\right),
\end{equation}
 as will be shown in Sec.\ref{sec:acr}.

\section{Testing the new formulas}\label{sec:test}

 \begin{figure*}[htbp]
        \centering
        \begin{tabular}{ccc}
            {\includegraphics[scale=0.4,trim=0 0 0 0]{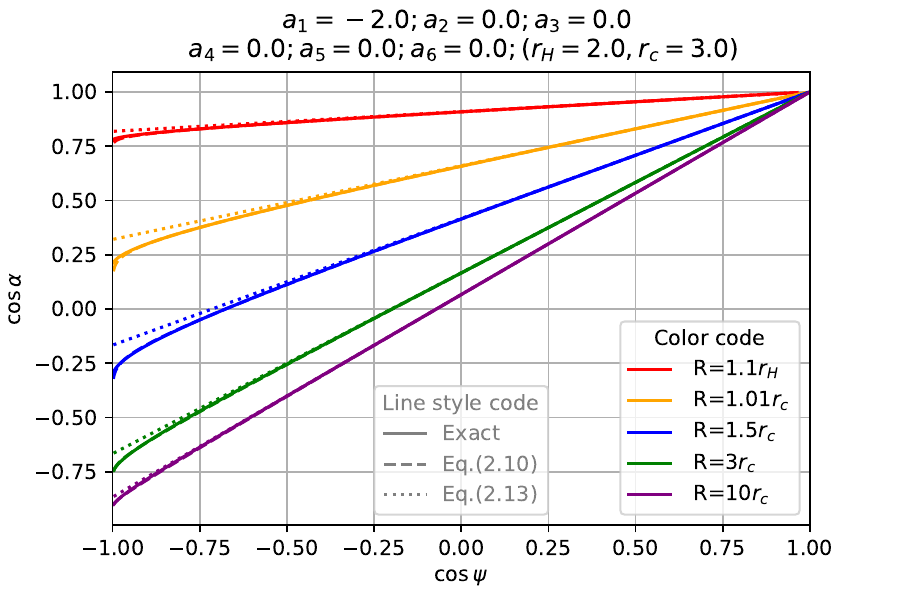}} &
            {\includegraphics[scale=0.4,trim=0 0 0 0]{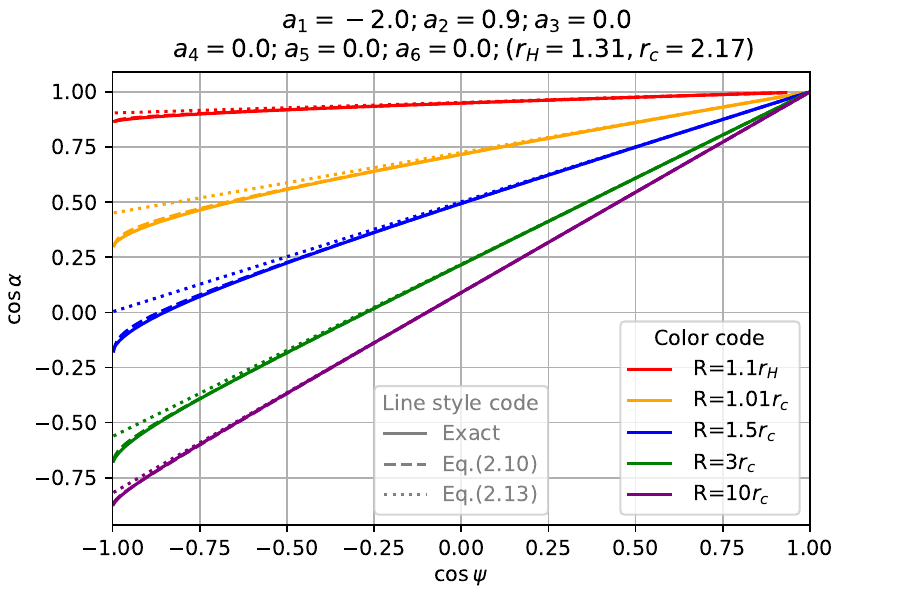}} &
            {\includegraphics[scale=0.4,trim=0 0 0 0]{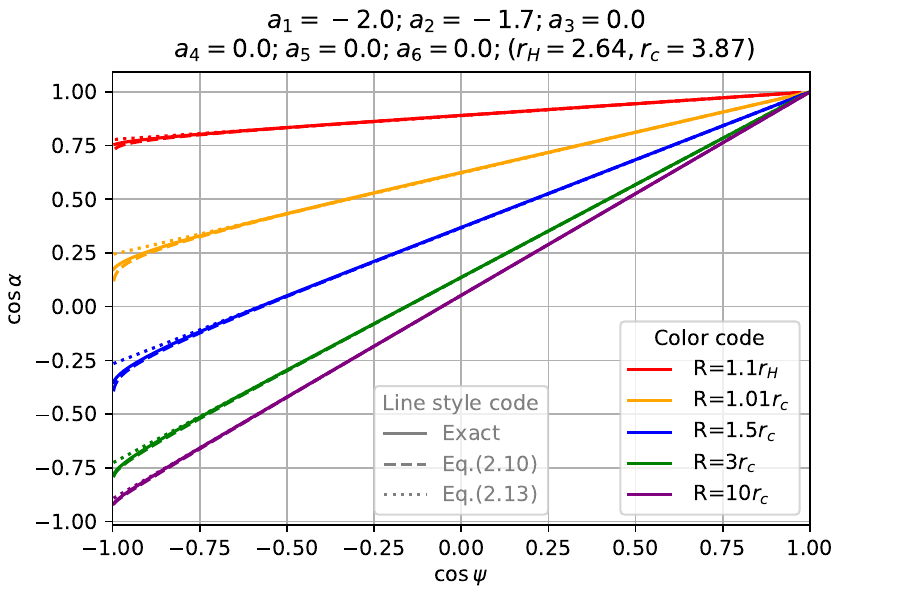}} \\
            {\includegraphics[scale=0.4,trim=0 0 0 0]{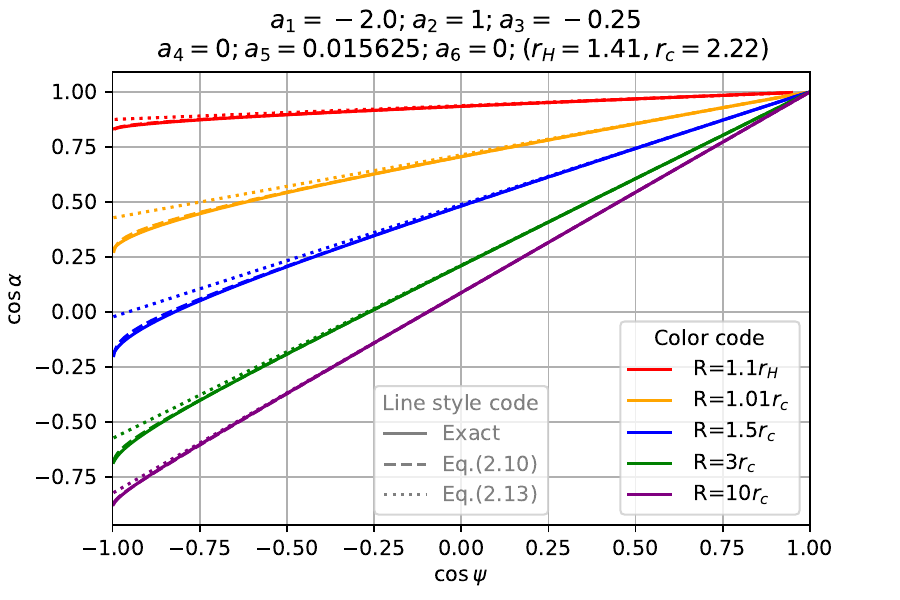}} &
            {\includegraphics[scale=0.4,trim=0 0 0 0]{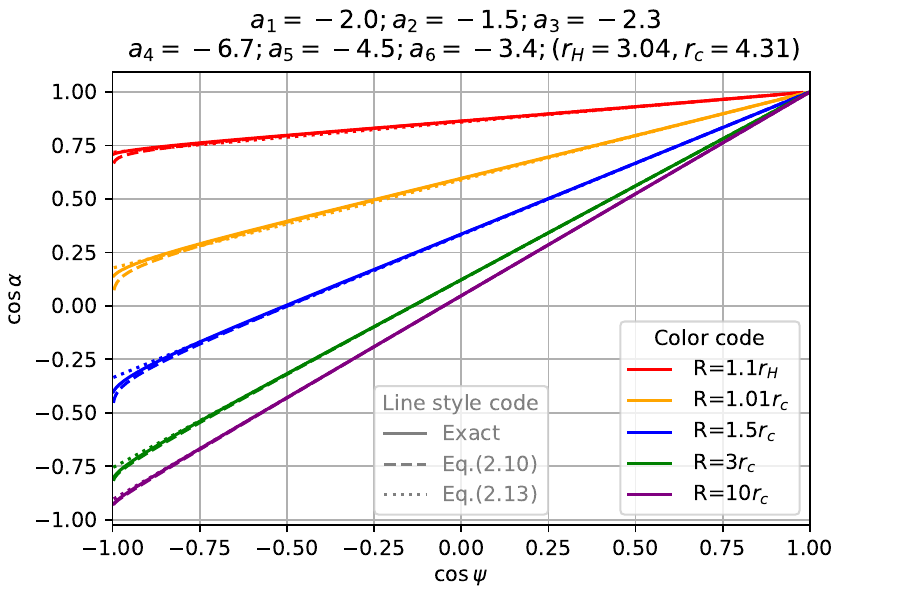}} &
            {\includegraphics[scale=0.4,trim=0 0 0 0]{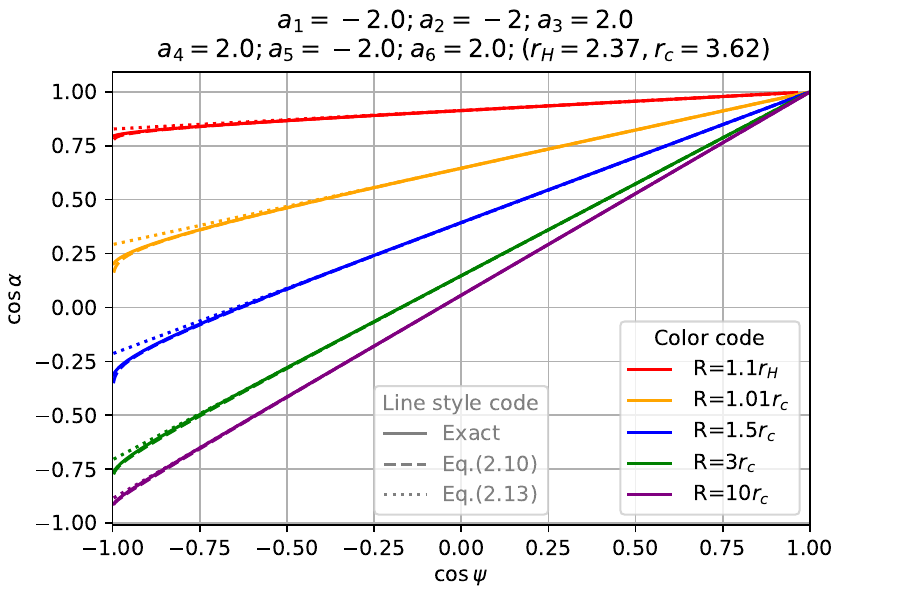}} \\
            {\includegraphics[scale=0.4,trim=0 0 0 0]{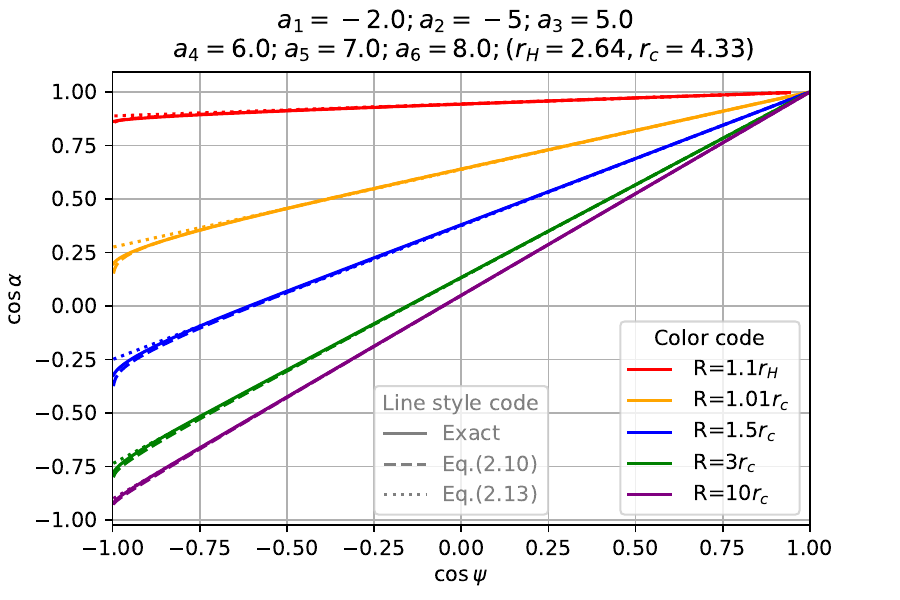}} &
            {\includegraphics[scale=0.4,trim=0 0 0 0]{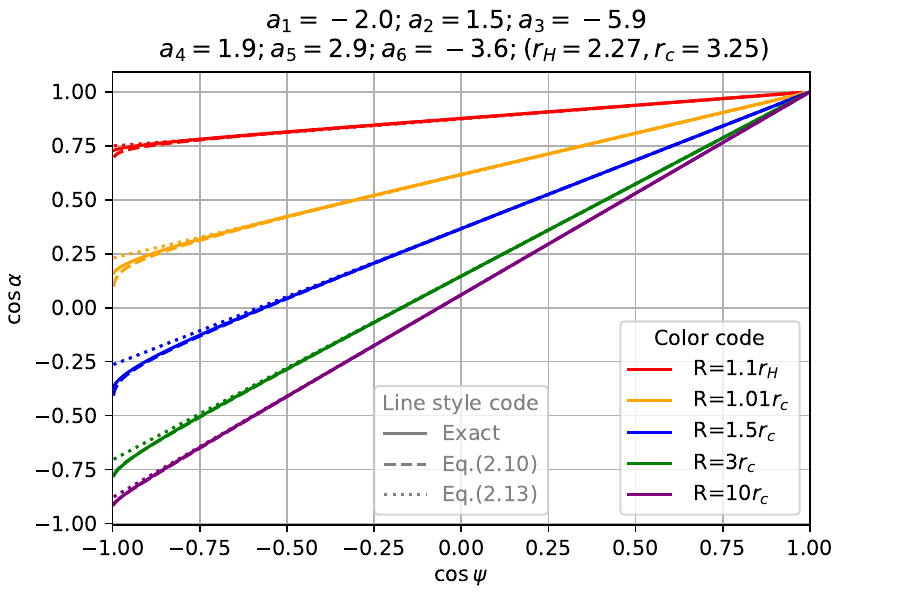}} &
            {\includegraphics[scale=0.4,trim=0 0 0 0]{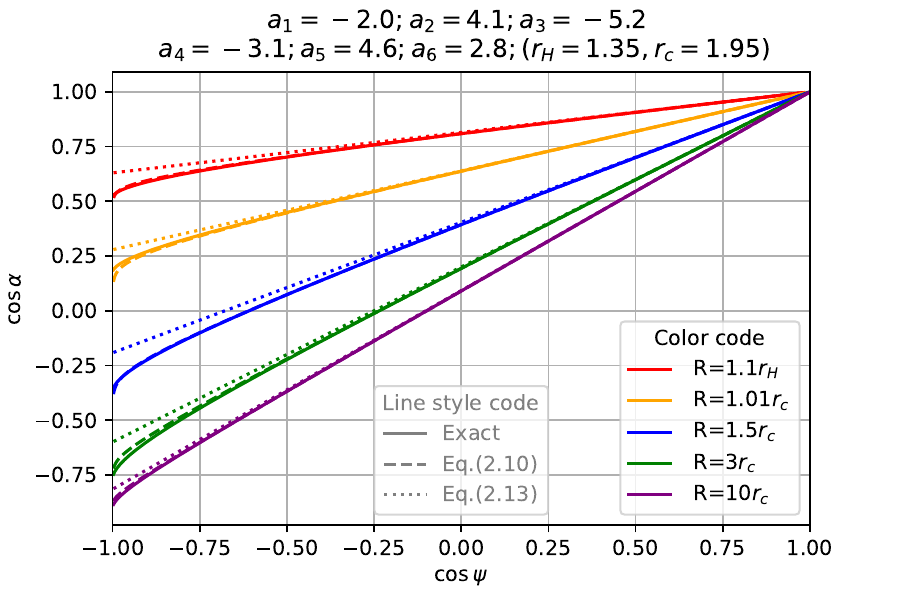}} \\
        \end{tabular}
        \caption{Comparison between the exact formula relating $\cos\alpha$ to $\cos\psi$ (obtained from numerical integration) and the approximations given by equations \eqref{eq:our} and \eqref{eq:Belo-general}. This comparison is carried out for different choices of the parameters $a_n$. $r_H$ represents the radial coordinate of the event horizon, and $r_c$ represents the radial coordinate of the photon sphere.  In all cases, an excellent approximation of the analytical formulas compared to the ``exact" solution obtained numerically is observed. Particularly, the equation \eqref{eq:our} significantly improves the accuracy of the equation as $\psi$ tends to $\pi$.
        }
        \label{fig:1}
    \end{figure*}

     \begin{figure*}[htbp]
        \centering
        \begin{tabular}{ccc}
            {\includegraphics[scale=0.4,trim=0 0 0 0]{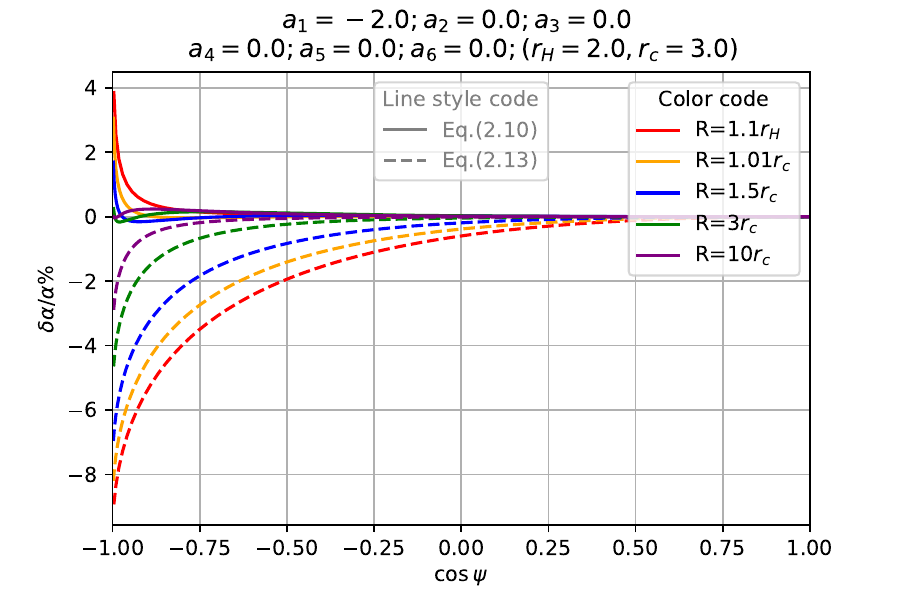}} &
            {\includegraphics[scale=0.4,trim=0 0 0 0]{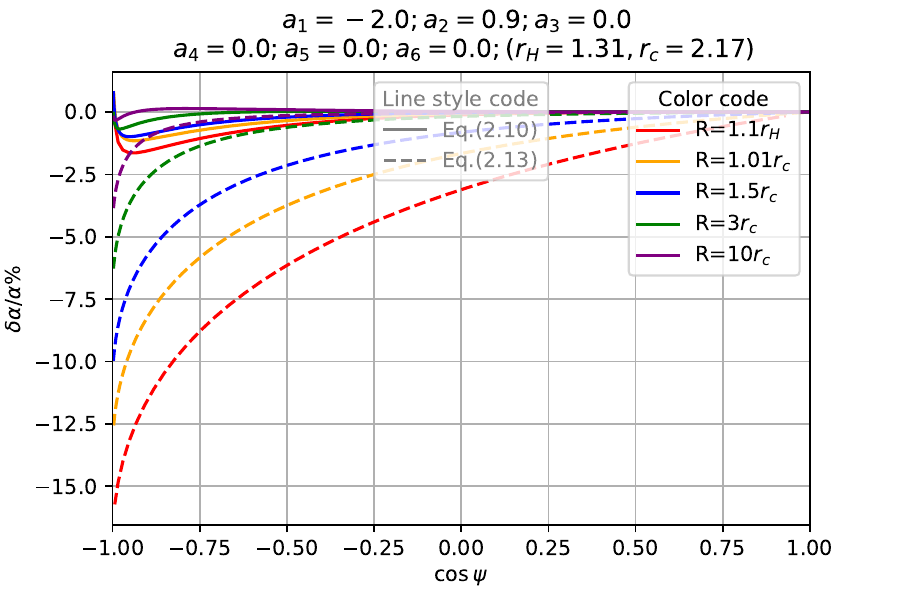}} &
            {\includegraphics[scale=0.4,trim=0 0 0 0]{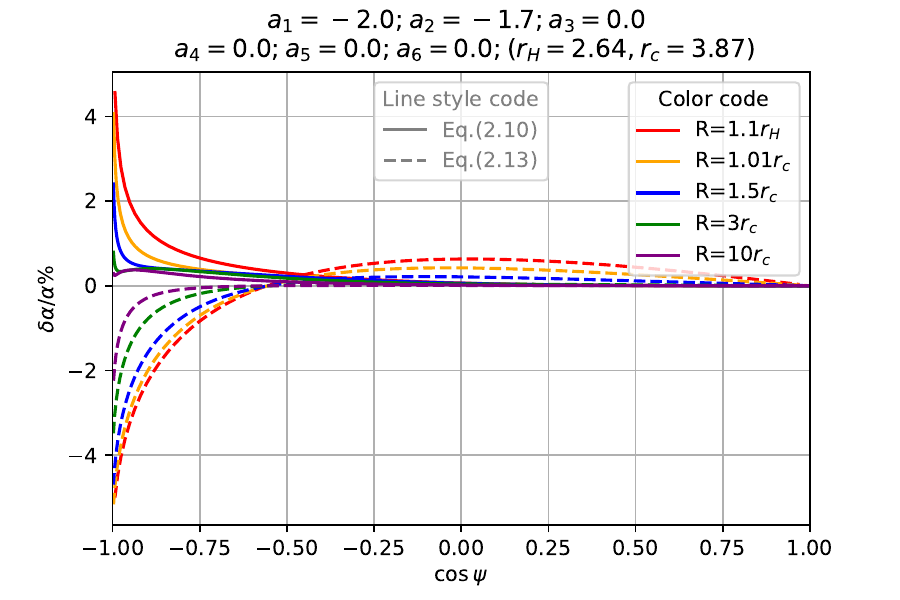}} \\
            {\includegraphics[scale=0.4,trim=0 0 0 0]{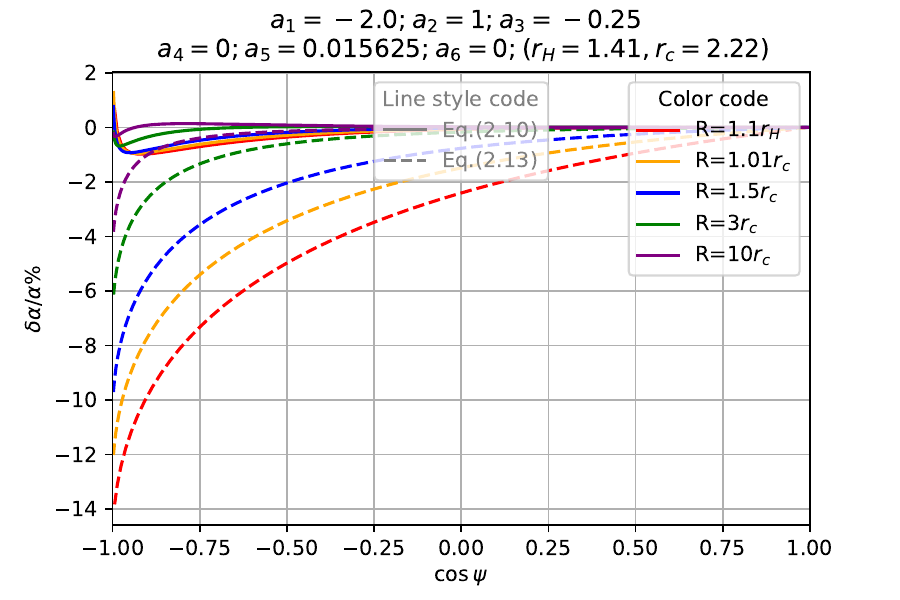}} &
            {\includegraphics[scale=0.4,trim=0 0 0 0]{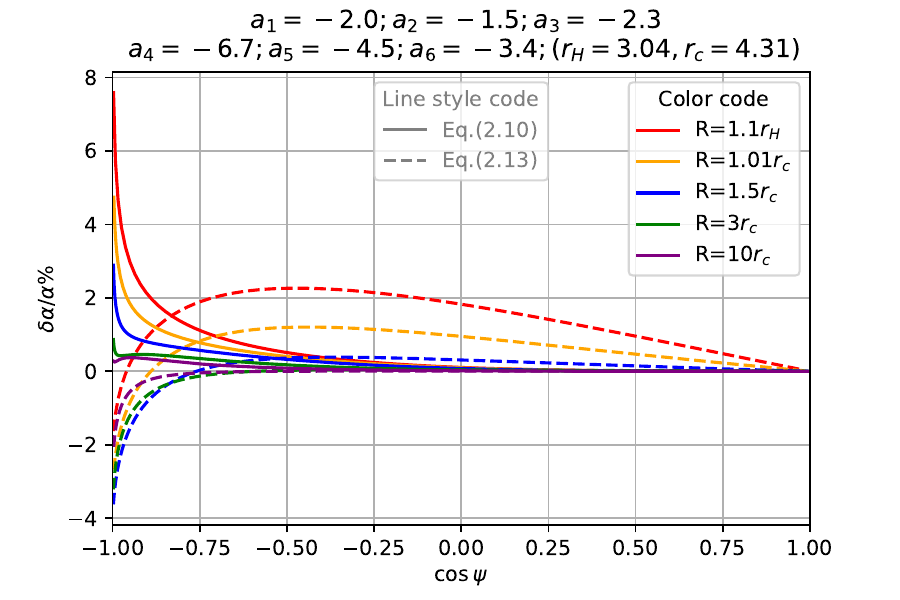}} &
            {\includegraphics[scale=0.4,trim=0 0 0 0]{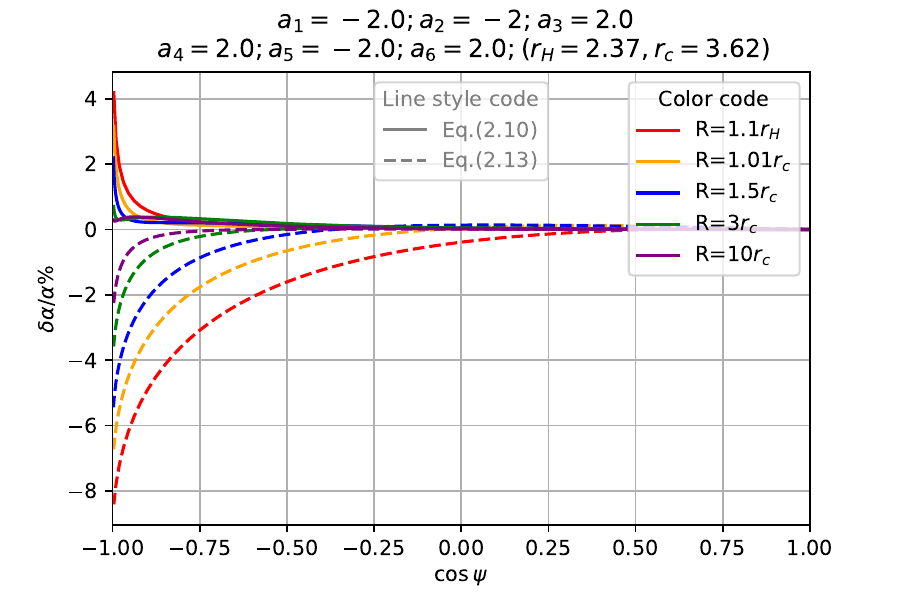}} \\
            {\includegraphics[scale=0.4,trim=0 0 0 0]{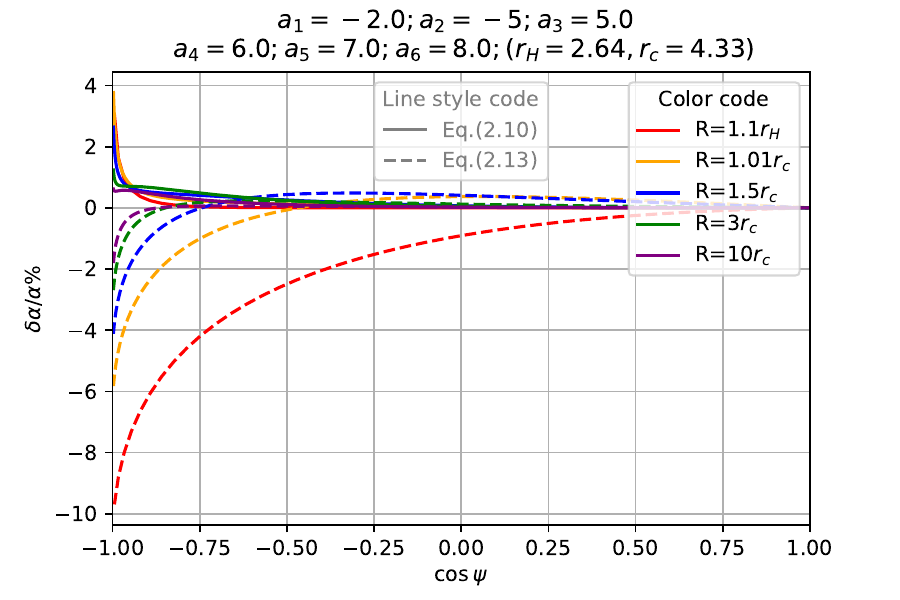}} &
            {\includegraphics[scale=0.4,trim=0 0 0 0]{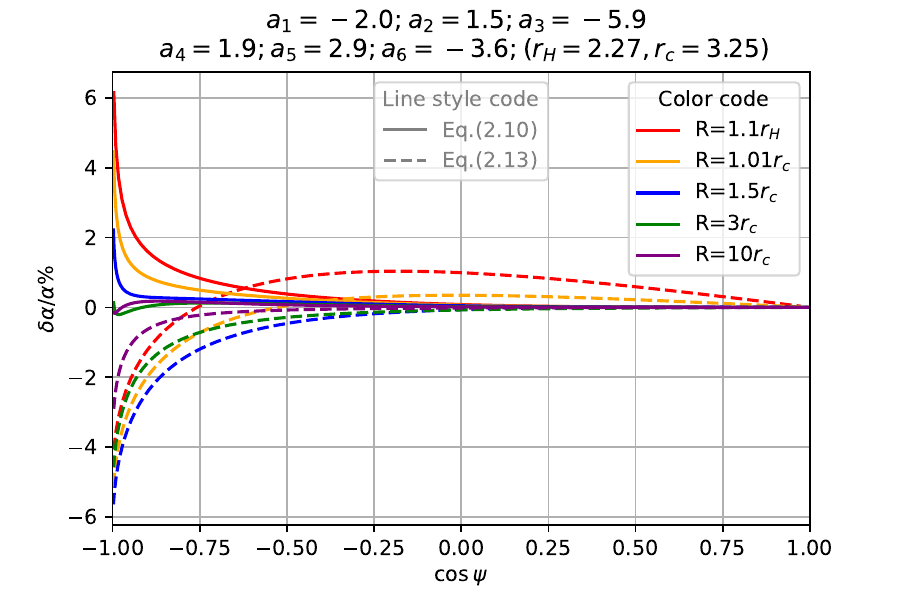}} &
            {\includegraphics[scale=0.4,trim=0 0 0 0]{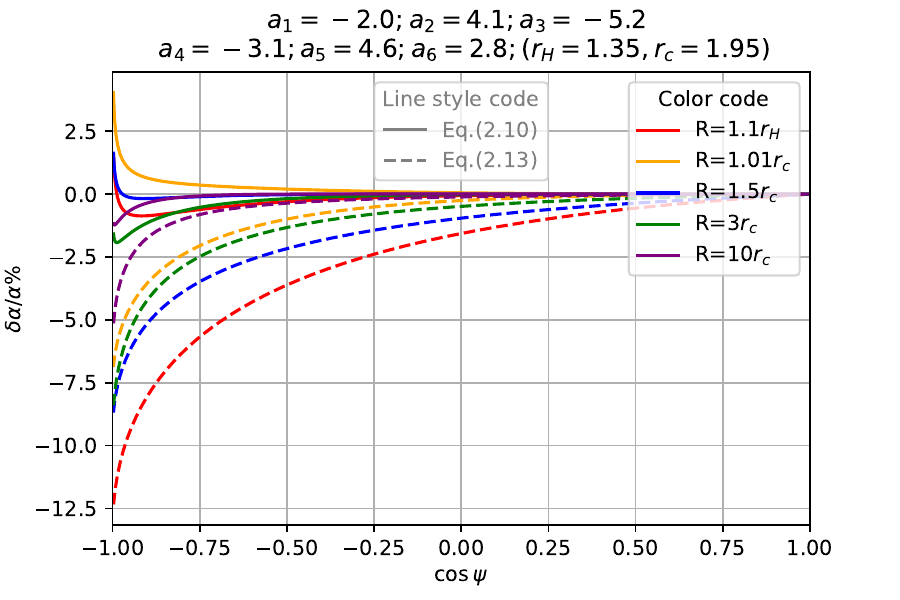}} \\
        \end{tabular}
        \caption{Relative error for the angle $\alpha$ for the same choice of parameters $a_n$ as in the referenced Figure \ref{fig:1}. As can be observed, the generalized Beloborodov approximation remains reasonable for values of $R>1.5r_c$, losing precision for angles $\psi$ close to $\pi$. However, the new formula \eqref{eq:our} can keep the error below $2-4\%$ even for values of $r$ close to $r_c$ and even the event horizon.
        }
        \label{fig:Ab_p7}
    \end{figure*}

    \begin{figure*}[htbp]
        \centering
        \begin{tabular}{ccc}
            {\includegraphics[scale=0.4,trim=0 0 0 0]{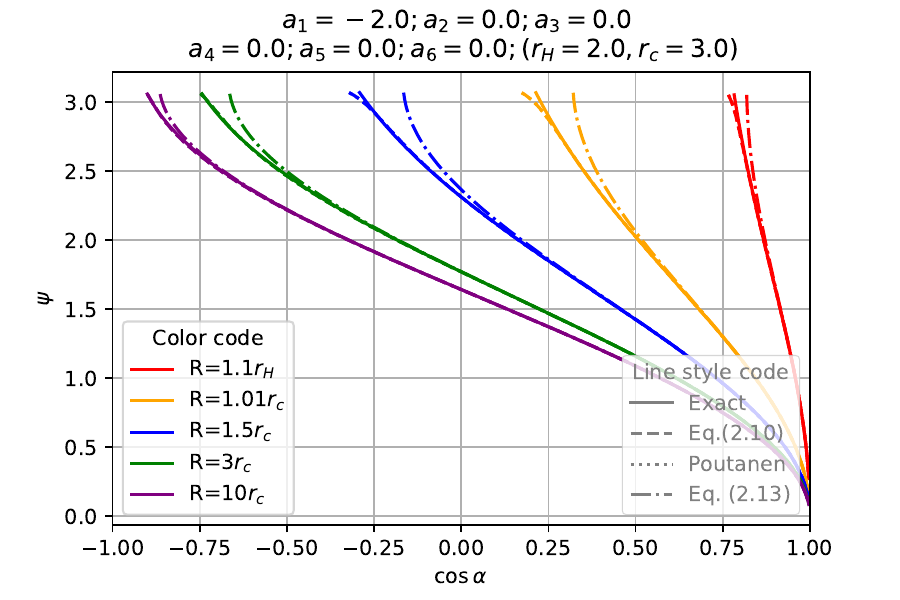}} &
            {\includegraphics[scale=0.4,trim=0 0 0 0]{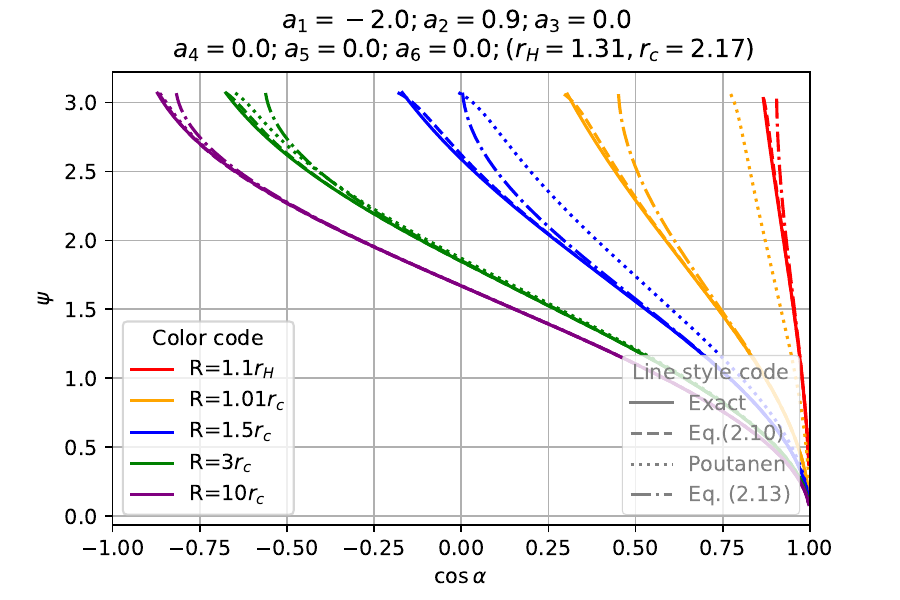}} &
            {\includegraphics[scale=0.4,trim=0 0 0 0]{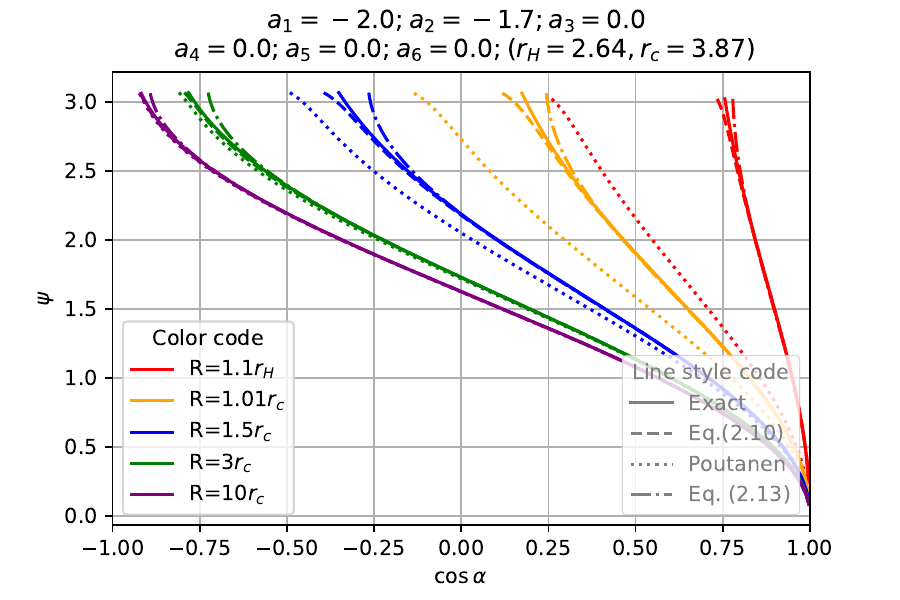}} \\
            {\includegraphics[scale=0.4,trim=0 0 0 0]{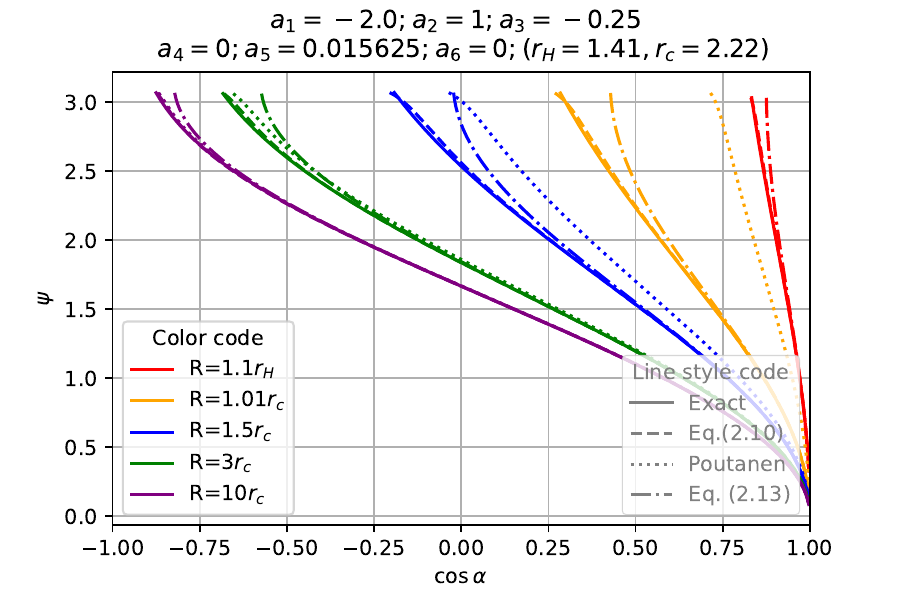}} &
            {\includegraphics[scale=0.4,trim=0 0 0 0]{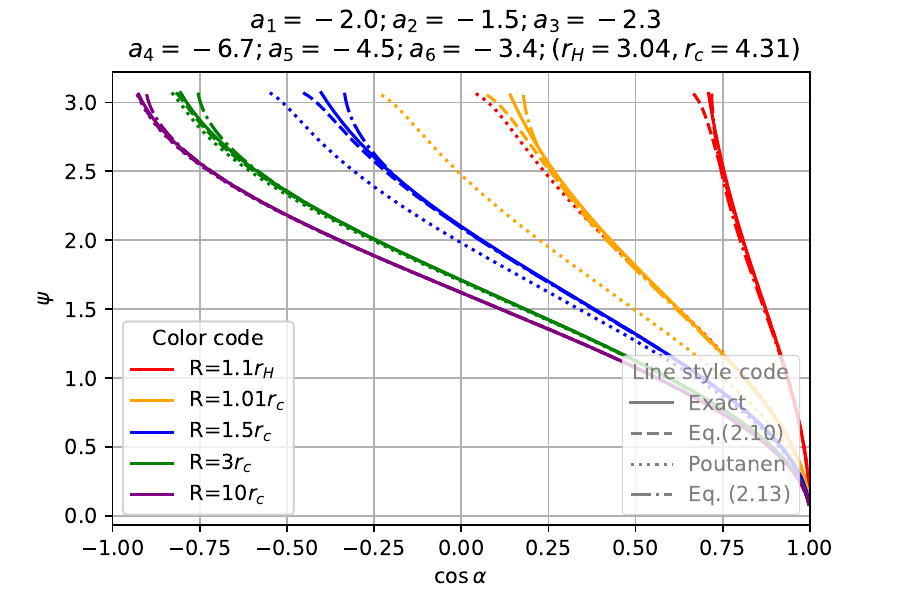}} &
            {\includegraphics[scale=0.4,trim=0 0 0 0]{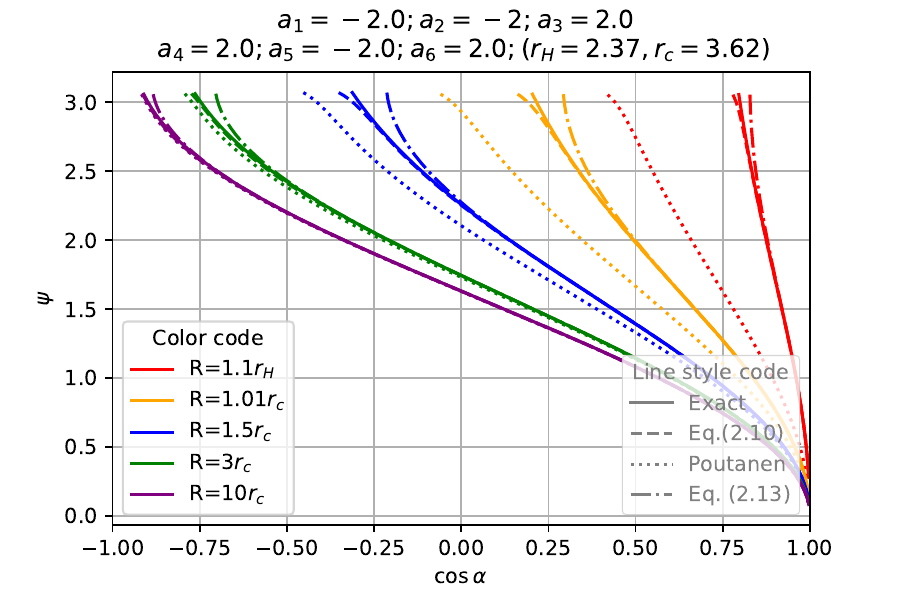}} \\
            {\includegraphics[scale=0.4,trim=0 0 0 0]{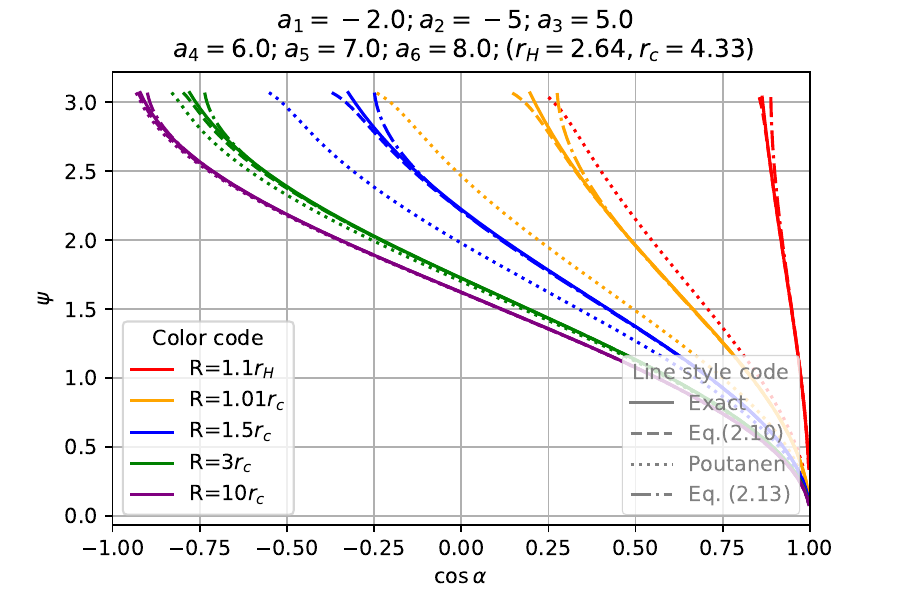}} &
            {\includegraphics[scale=0.4,trim=0 0 0 0]{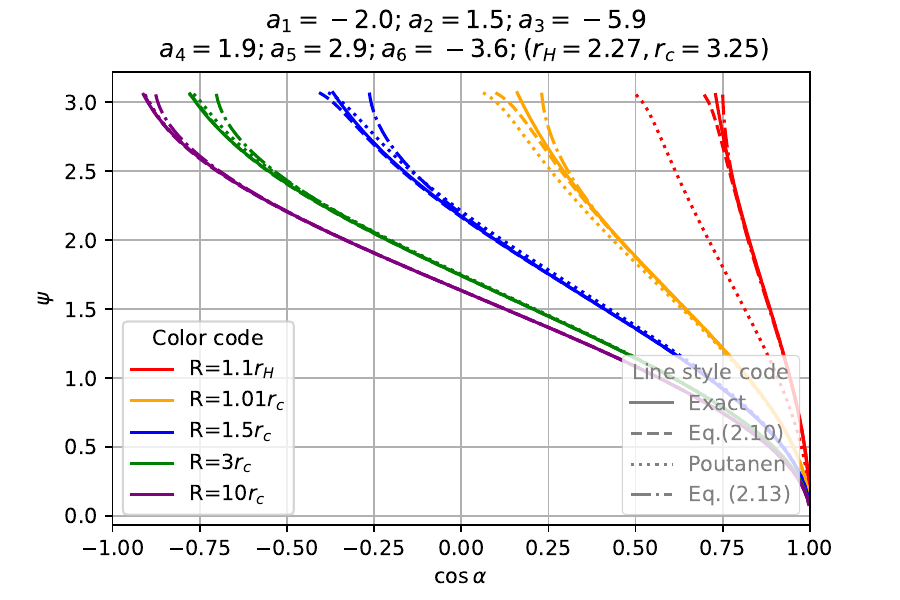}} &
            {\includegraphics[scale=0.4,trim=0 0 0 0]{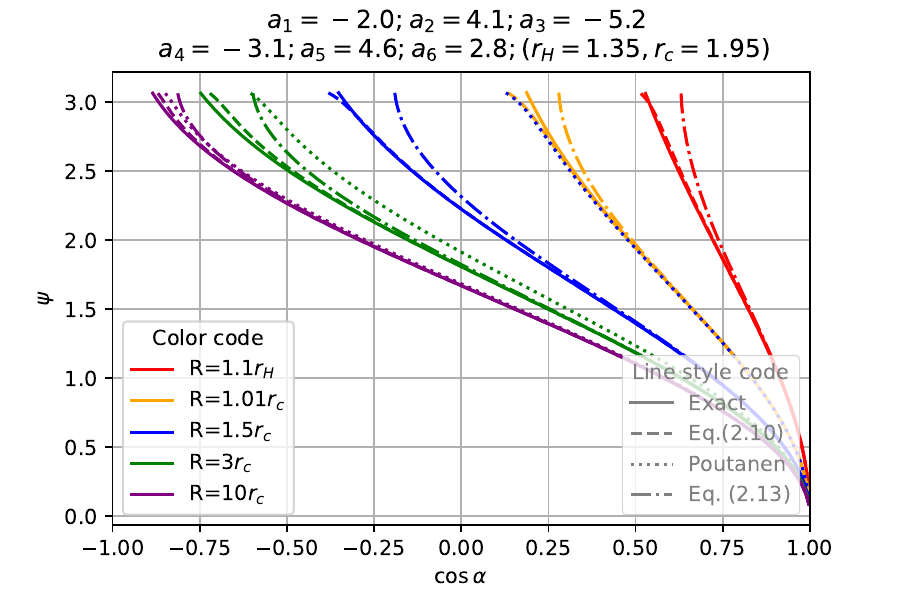}} \\
        \end{tabular}
        \caption{An alternative version to Figure \ref{fig:1}, adding the comparison with the Poutanen formula. Even though the Poutanen formula can be used as a good approximation for relatively large radii compared to the event horizon's, for radii of the order of $1.5r_c$ or smaller, the Poutanen formula loses accuracy for the other metrics compared to those given by \eqref{eq:our} and \eqref{eq:Belo-general}.
        }
        \label{fig:Ab_p8}
    \end{figure*}

    \begin{figure*}[htbp]
        \centering
        \begin{tabular}{ccc}
            {\includegraphics[scale=0.4,trim=0 0 0 0]{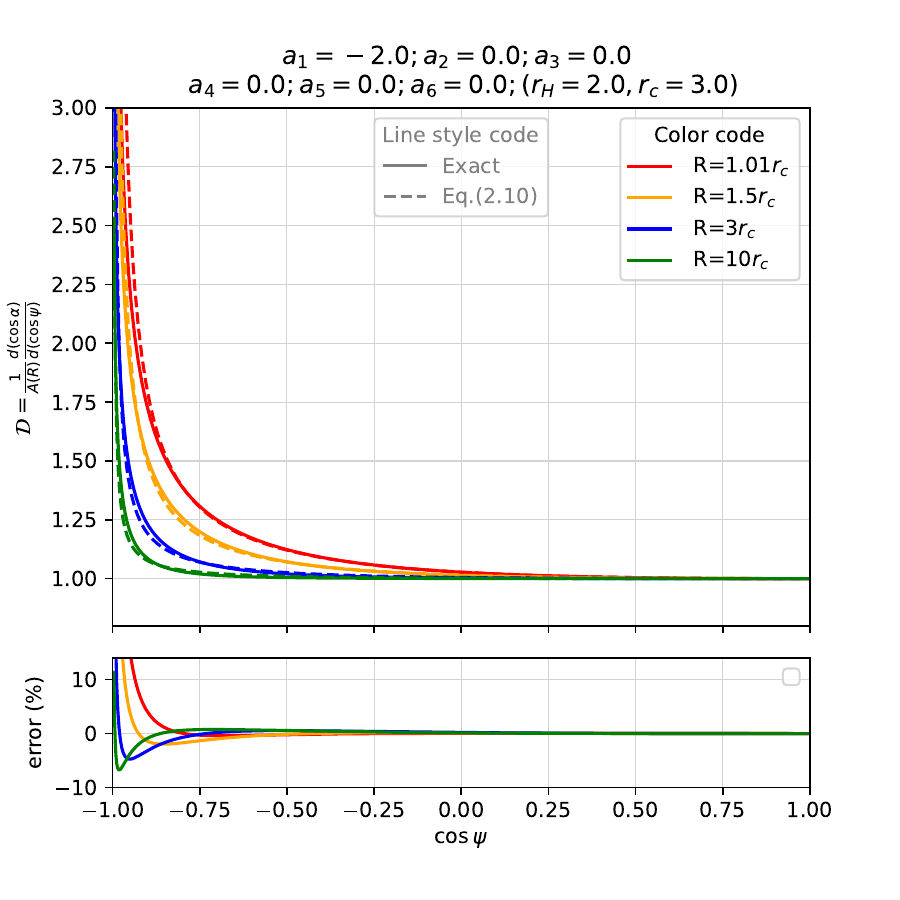}} &
            {\includegraphics[scale=0.4,trim=0 0 0 0]{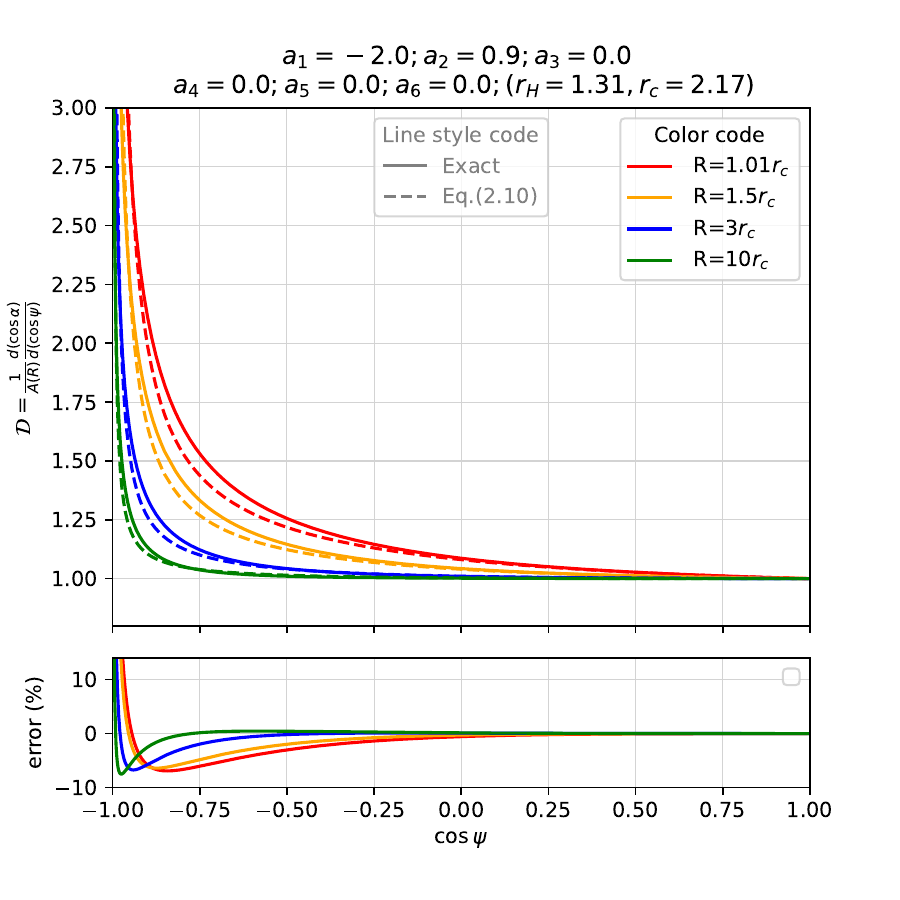}} &
            {\includegraphics[scale=0.4,trim=0 0 0 0]{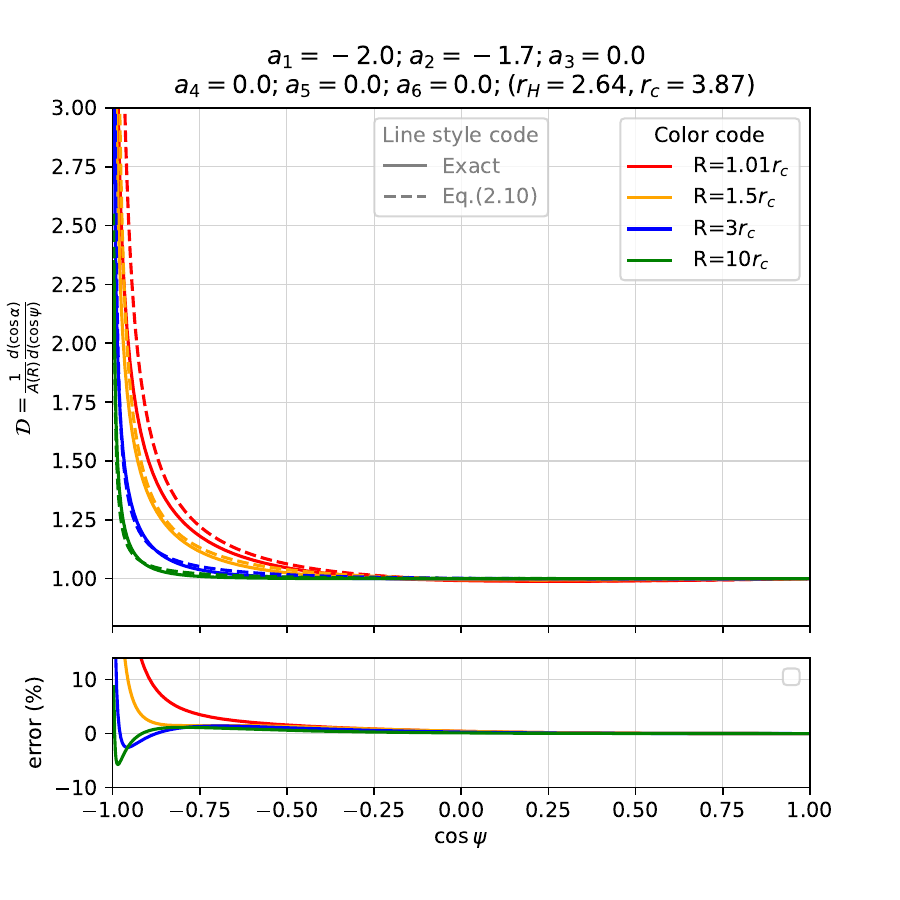}} \\
            {\includegraphics[scale=0.4,trim=0 0 0 0]{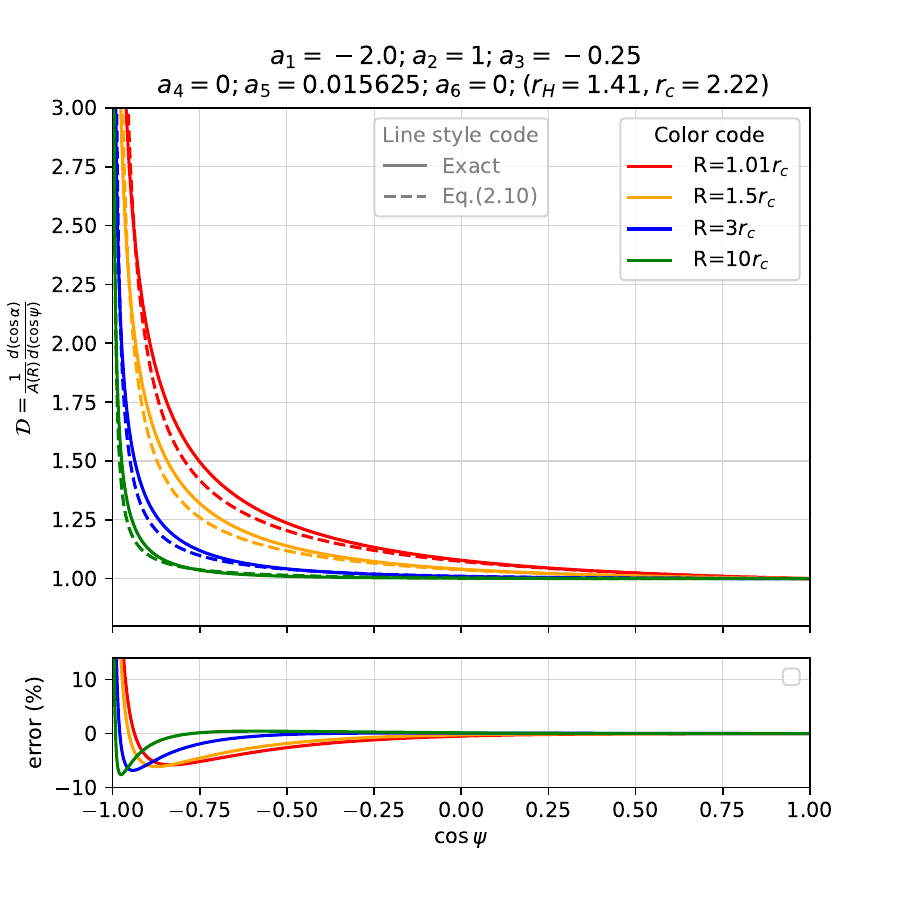}} &
            {\includegraphics[scale=0.4,trim=0 0 0 0]{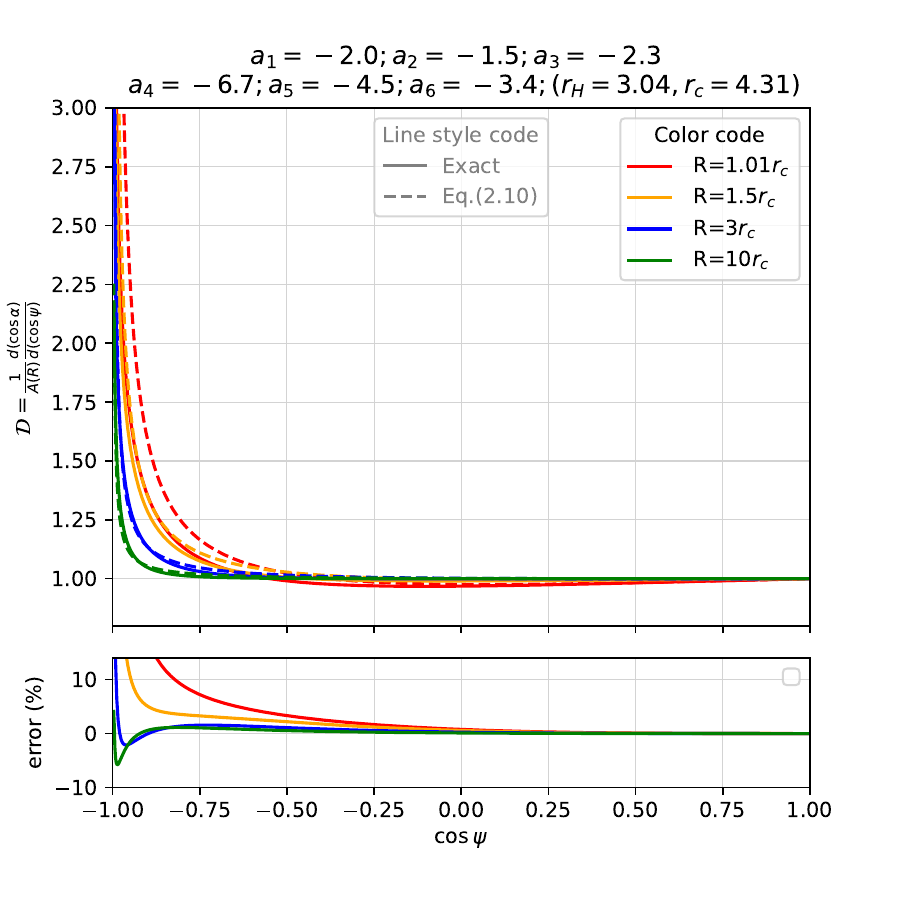}} &
            {\includegraphics[scale=0.4,trim=0 0 0 0]{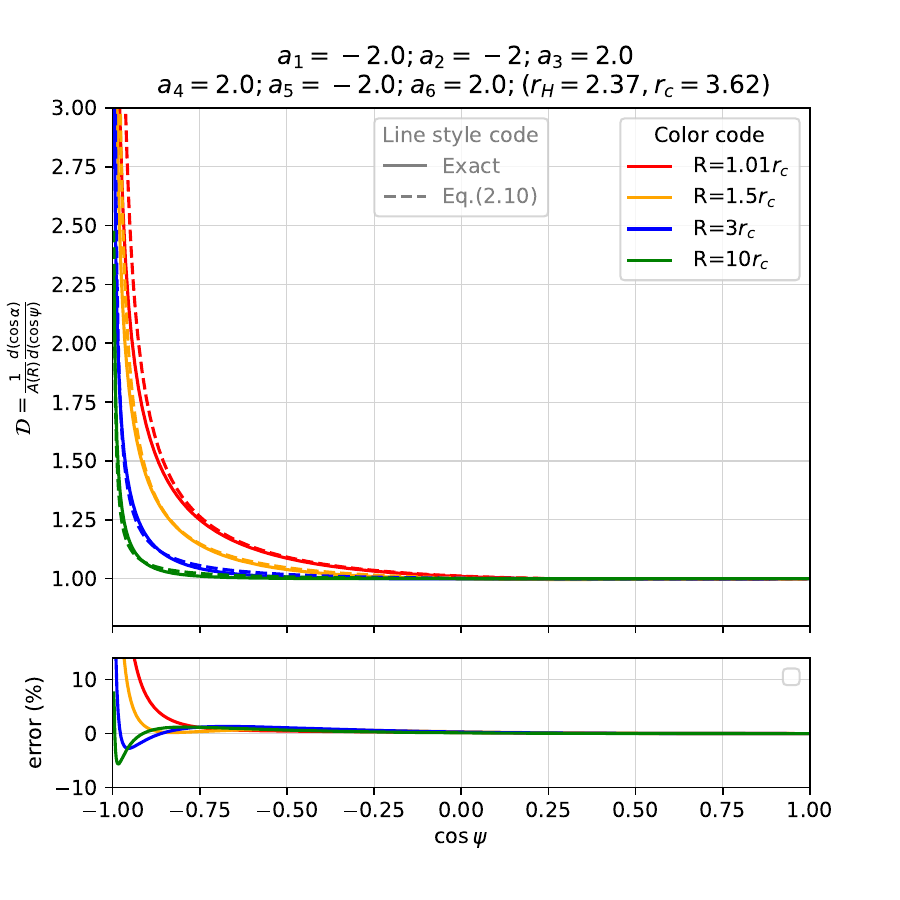}} \\
            {\includegraphics[scale=0.4,trim=0 0 0 0]{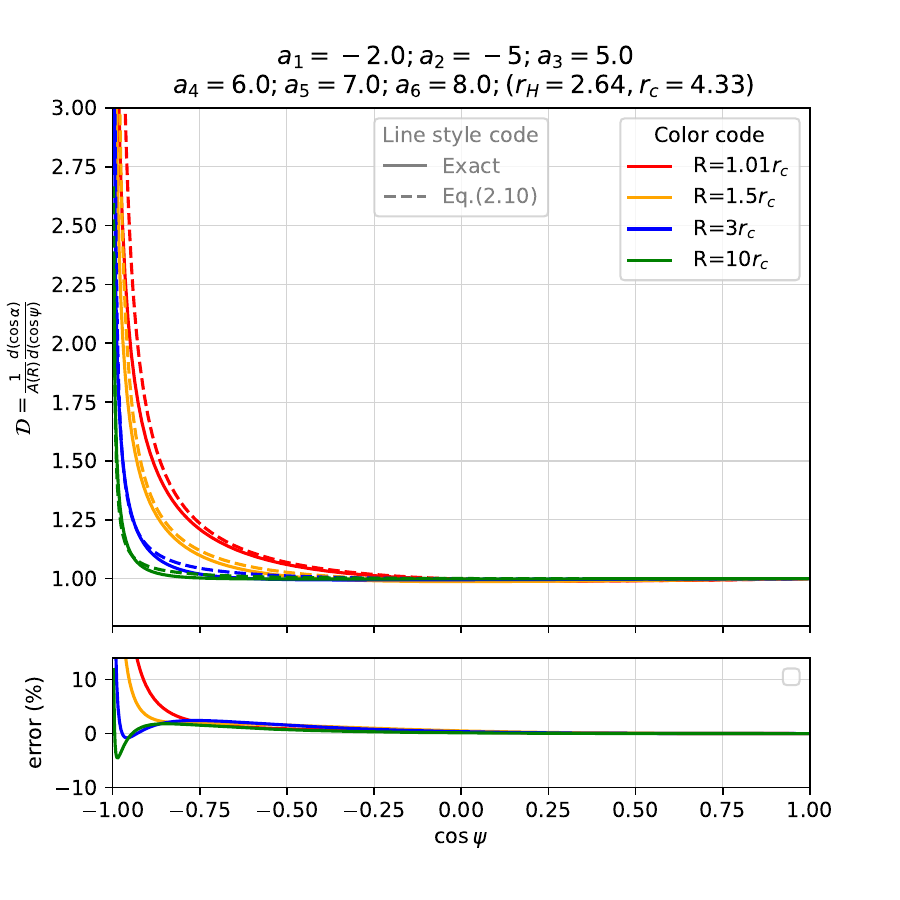}} &
            {\includegraphics[scale=0.4,trim=0 0 0 0]{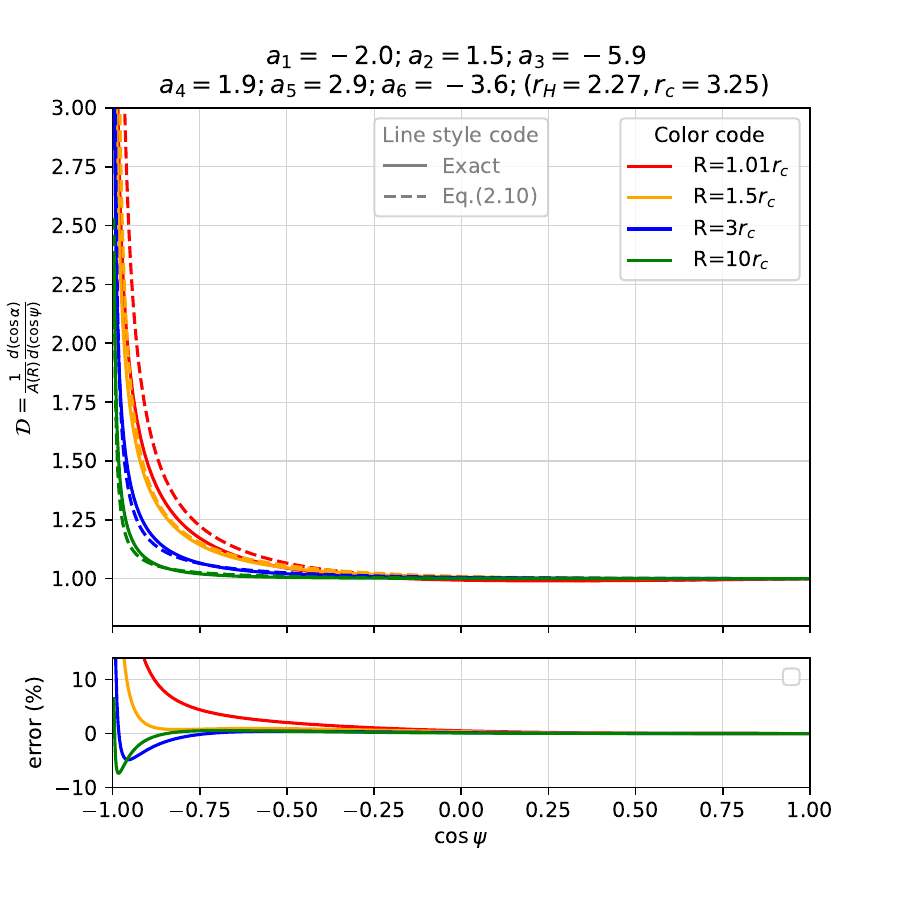}} &
            {\includegraphics[scale=0.4,trim=0 0 0 0]{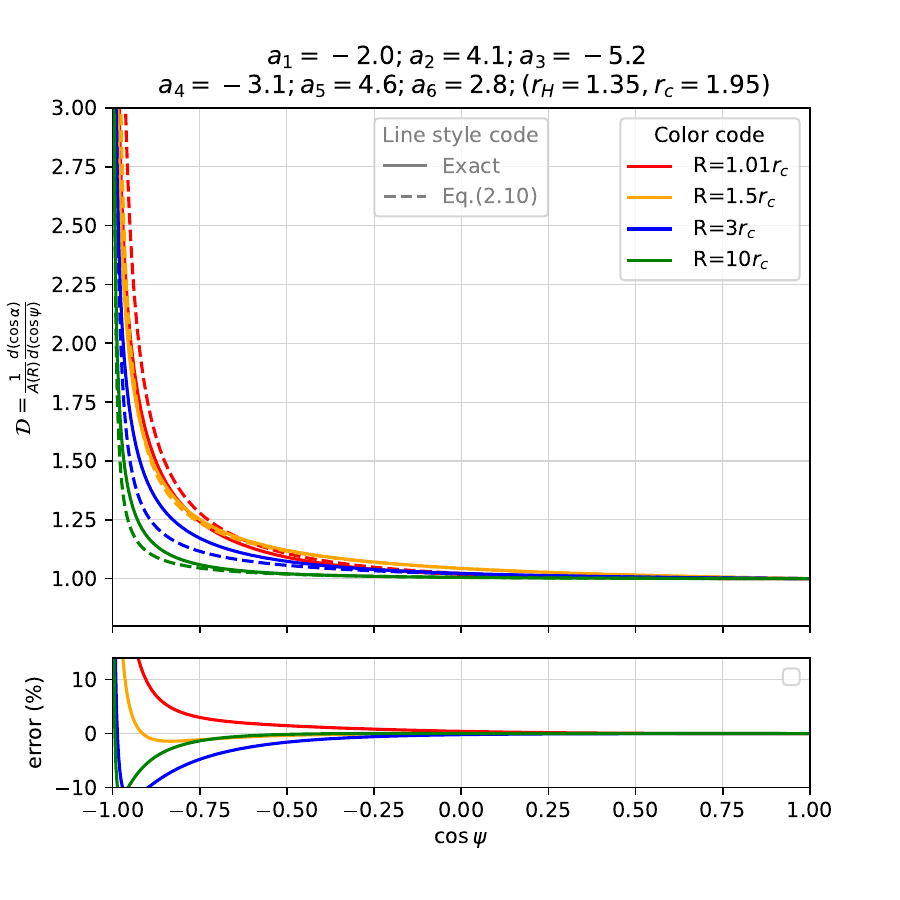}} \\
        \end{tabular}
        \caption{Comparison between the exact expression and the approximations for the lensing factor $\mathcal{D}$, along with the associated relative error. The parameters $a_n$ used are the same as in Figure \ref{fig:1}. Similar to the case with the Poutanen formula for the Schwarzschild metric, the new generalized formula Eq.\eqref{eq:our} yields errors not exceeding 5\% for $R\gtrsim1.5r_c$ and angles at $\cos\psi\lesssim -0.8$.
        }
        \label{fig:Ab_p9}
    \end{figure*}

   Hereafter, and throughout the remainder of this work, we employ geometrized units $(G = c = 1)$, with the radius $R$ measured in units of the Arnowitt-Deser-Misner (ADM) mass $M$. Similarly, the parameters $a_n$ are considered in units of $n$-powers of the mass $M^n$. Specifically, under these units, and considering that the metrics are modifications of the Schwarzschild solution, we set $a_1 = -2$ for the rest of the work.
 
    In Fig. \ref{fig:1}, we compare the plot of $\cos\alpha$ versus $\cos\psi$ (obtained from the numerical integration of Eq.\eqref{eq:belo-ap1} and Eq.\eqref{eq:perias}) with the alternative formulas proposed in our work, specifically \eqref{eq:our} and the generalized Beloborodov formula \eqref{eq:Belo-general}. 
    
    This comparison is conducted for a sample of various (and arbitrary) choices of the parameters $a_n$. As before, $r_H$ denotes the radial coordinate of the event horizon, and $r_c$ represents the radial coordinate of the photon sphere obtained by solving the equation $2A(r)-r\frac{dA(r)}{dr}=0$ \cite{Chandra:1983,Cardoso:2008bp}.  Both of them are in general numerically obtained and also expressed in units of $M$. The top-left figure corresponds to the Schwarzschild metric, where the introduced approximate formulas reduce to those of Beloborodov and Poutanen respectively. The parameter selections in the top-middle graph can be used to describe a Reissner-Nordström metric, with $a_2$ playing the role of the electric charge $q^2$. Conversely, negative values of $a_2$ arise in various theories, such as braneworld metrics (top-right graph). Across all scenarios, we observe an excellent approximation of the analytical formulas compared to the ``exact" solution obtained numerically. We observe that, for most values of $\psi$, the generalized Beloborodov formula provides an accurate fit, but as with the original Beloborodov formula for the Schwarzschild metric (Eq.\eqref{eq:belo-original}), it loses fidelity as $\psi\to\pi$.
 In such cases, the additional logarithmic term in \eqref{eq:our} introduces the necessary curvature to more accurately align with the exact curve.

{In Fig. \ref{fig:Ab_p7}, we depict the relative error for the angle $\alpha$ for the same selection of parameters $a_n$ as in the Figure \ref{fig:1}. Here $\delta\alpha/\alpha\%$ means $(\alpha_{aprox}-\alpha_{num})/\alpha_{num}\times 100$, with $\alpha_{num}$ numerically obtained and $\alpha_{aprox}$ computed using Eqs.\eqref{eq:our} or \eqref{eq:Belo-general}. As evident, the generalized Beloborodov approximation remains acceptable for $R>1.5r_c$, albeit losing precision for angles $\psi$ approaching $\pi$. Conversely, the new formula \eqref{eq:our} manages to maintain the error below $2-4\%$, even for values of $r$ near $r_c$ and the event horizon.}

    {Fig. \ref{fig:Ab_p8} provides an alternative presentation to Fig. \ref{fig:1}, incorporating the comparison with the Poutanen formula. As anticipated, the Poutanen formula serves as a reliable approximation for relatively large radii compared to the event horizon's, as the corrections introduced by the parameters $a_2,...,a_6$ in such cases are higher-order corrections to the Schwarzschild metric. However, for radii of the order of $3r_c$ or smaller, the accuracy of the Poutanen formula diminishes compared to those provided by \eqref{eq:our} and \eqref{eq:Belo-general}.}

   {
   Now let us study the lensing factor as defined in \cite{Poutanen:2019tcd}.
   Let us consider a surface element of area $dS$ at the neighborhood of the compact object emitting light with a radiation intensity $I$. The observed flux coming from this element is proportional to the solid angle $d\Omega$ occupied by the surface element in the observer's sky. This is given by
   \begin{equation}
       d\Omega=\frac{bdbd\varphi}{D^2},
   \end{equation}
   with $D$ the distance to the source and $\varphi$ the azimuthal angle in the observer plane (see Fig.\ref{fig:framework}). Taking into account that $dS=R^2d\cos\psi d\varphi$ this solid angle can be rewritten as
   \begin{equation}
       d\Omega=\frac{dS}{D^2}\frac{b}{R^2}\bigg|\frac{db}{d\cos\psi}\bigg|=\frac{dS\cos\alpha}{D^2}\frac{1}{A(R)}\frac{d\cos\alpha}{d\cos\psi}.
   \end{equation}
   The lensing factor is defined as 
   \begin{equation}
       \mathcal{D}=\frac{1}{A(R)}\frac{d\cos\alpha}{d\cos\psi}.
   \end{equation}

   As noted by Poutanen in \cite{Poutanen:2019tcd}, Beloborodov's basic relationship Eq.\eqref{eq:belo-original} demonstrates limited accuracy for large emission angles $\alpha$ and high compactness $r_H/R > 1/2$. While the error remains below $0.7\%$ for $r_H/R \leq 1/3$, it escalates to $10\%$ at $r_H/R = 1/2$. The discrepancy exacerbates for the lensing factor $\mathcal{D}$, where Eq. \eqref{eq:belo-original} implies $\mathcal{D} = 1$, yet the actual value notably increases at negative $\cos\psi$. As observed in \cite{Poutanen:2019tcd}, for $\cos\psi = -0.7$ ($\psi = 134^\circ$), the deviation exceeds $10\%$ for $r_H/R = 1/3$ and reaches $15\%$ for $r_H/R = 1/2$. Consequently, this approximation could significantly impact observed flux, particularly for the luminosity curves of neutron stars or inclined accretion disks around black holes. Recognizing this challenge motivated Poutanen to pursue a more accurate approximation given by his formula Eq.\eqref{eq:Poutanen}. Here, we repeat their analysis but with formula \eqref{eq:our} for the more general metrics Eq.\eqref{eq:ar}.

   Figure \ref{fig:Ab_p9} illustrates a comparison between the exact expression and the approximations for the lensing factor, accompanied by the corresponding relative error. 
   The choice for the parameters $a_n$ is identical to those in Figure \ref{fig:1}. Analogous to the scenario with the Poutanen formula for the Schwarzschild metric, the novel generalized formula Eq. \eqref{eq:our} yields errors of no more than $5\%$ for $R>1.5r_c$ and angles near $\cos\psi=-0.8$.}

\begin{figure}[htbp]
        \centering
        \begin{tabular}{lll}
            {\includegraphics[scale=0.6,trim=1cm 0 0  0]{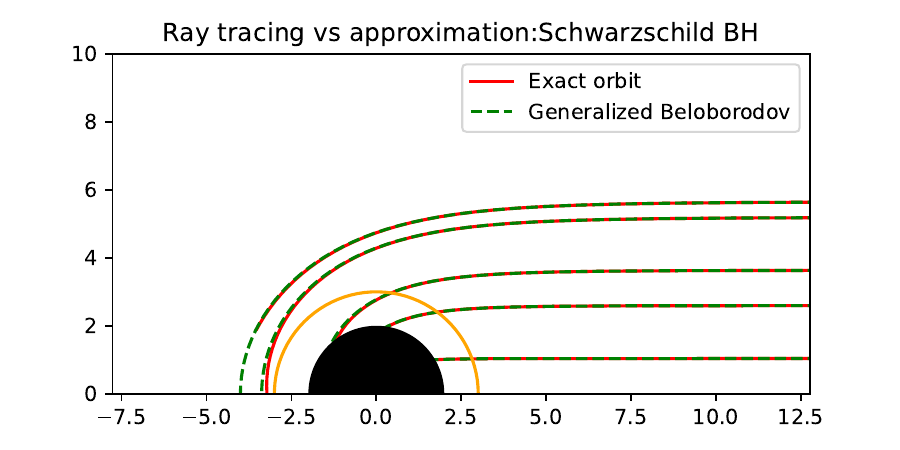}} &\\
            {\includegraphics[scale=0.6,trim=1cm 0 0 0]{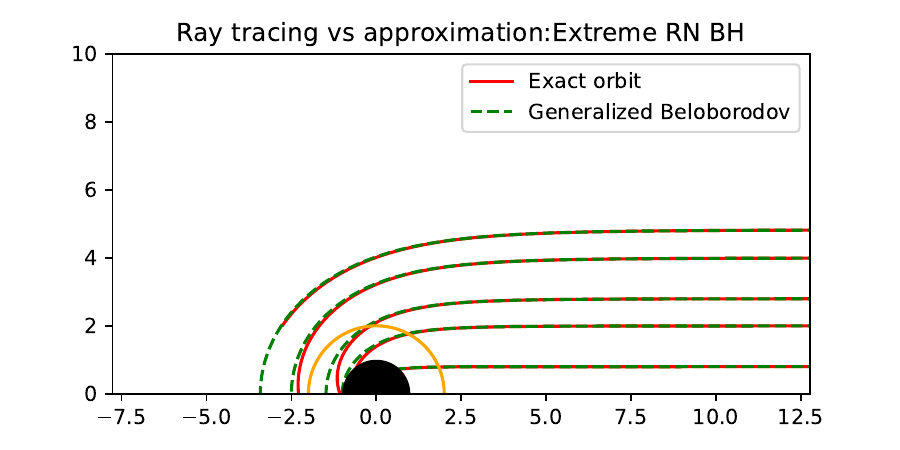}} &\\
            {\includegraphics[scale=0.6,trim=1cm 0 0 0]{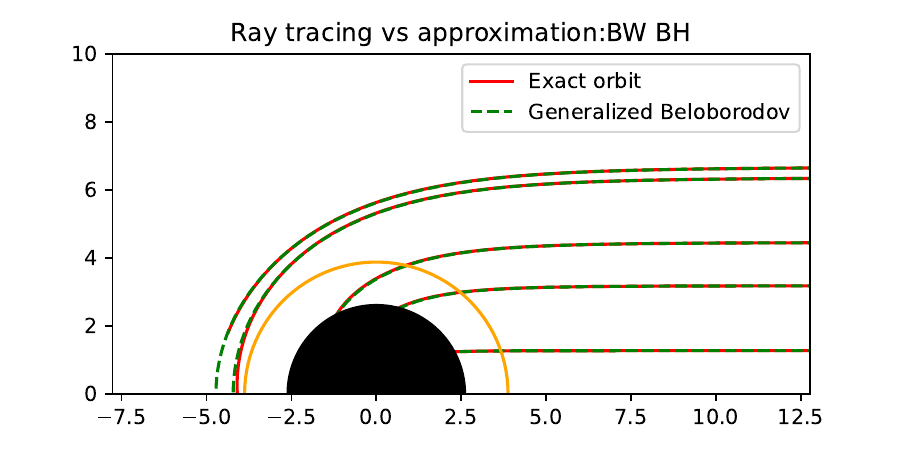}}
        \end{tabular}
        \caption{Comparison between the approximations to the orbits given by the equation \eqref{eq:orbit} and the one obtained numerically considering various values of the impact parameter. The black semicircle represents the event horizon of the black hole; the orange semicircle locates the photon sphere. The solid red line is the numerical solution of the geodesic equations and the dashed green one is obtained by our approximations. In the top figure, a Schwarzschild spacetime is considered, in the center one an extreme Reissner-Nordstr\"om black hole, and in the bottom figure a braneworld-type solution with $a_2=-1.7$. For comparison, the exact orbit is only shown up to the point of closest approach to the black hole (where $\alpha=\frac{\pi}{2}$).}
        \label{fig:raytracing}
    \end{figure}

   Finally, before concluding this section, note that given the generalized Beloborodov expression \eqref{eq:Belo-general}, one could use it to invert the equation \eqref{eq:impact_parameter} for $b$ and thus obtain the orbital equation for the radial coordinate $r$ in terms of $\psi$ and $b$. To this end, one should solve the following equation \begin{equation}\label{eq:orbit}
       b^2=-r^2(A(r)y^2-2y).
   \end{equation} 
   Unfortunately, this equation cannot be solved analytically for a general $A(r)$. However, in the case where $A(r)$ takes the form of Eq. \eqref{eq:ar6} with only $a_1$ and $a_2\neq 0$, such a solution can be found analytically. In such a case, if we denote $r_+$ and $r_-$  as the two roots of $A(r)=1+a_1/r+a_2/r^2$ (with $r_+\geq r_-$), the solution of Eq.\eqref{eq:orbit} for the orbit reads:
   
 \begin{widetext}
     \begin{equation} \label{eq:orbnue}  r(\psi)=\left[\frac{(r_{+} -r_{-})^2(1-\cos\psi)^2}{4(1+\cos\psi)^2}+\frac{8r_{+}r_{-}(1-\cos\psi)}{(1+\cos\psi)^2}+\frac{b^2}{\sin^2\psi}\right]^{1/2}-\frac{(r_{+}+r_{-})(1-\cos\psi)}{(1+\cos\psi)},
 \end{equation}
 \end{widetext}
 Eq.\eqref{eq:orbnue} generalizes the formula obtained by Beloborodov for the Schwarzschild spacetime \cite{Beloborodov_2002}. It is worth emphasizing that this equation maintains accuracy as long as we are in the region where $\alpha\leq\pi/2$, and its accuracy will be greater as $R$ becomes larger than the radius of the event horizon. However, even for values of the impact parameter $b$ close to the photon sphere, the accuracy of this equation proves to be high as long as the emission angle remains not very large. Three specific examples of the comparison between this formula and the numerical solution are shown in Fig. \ref{fig:raytracing}.

\section{Applications to black holes: Accretion disk images and polarimetry}\label{sec:acr}

\subsection{The general framework} \label{sec:gralframework}
\begin{widetext}
    \begin{figure}[htbp] 
        \centering        \includegraphics[scale=0.41]{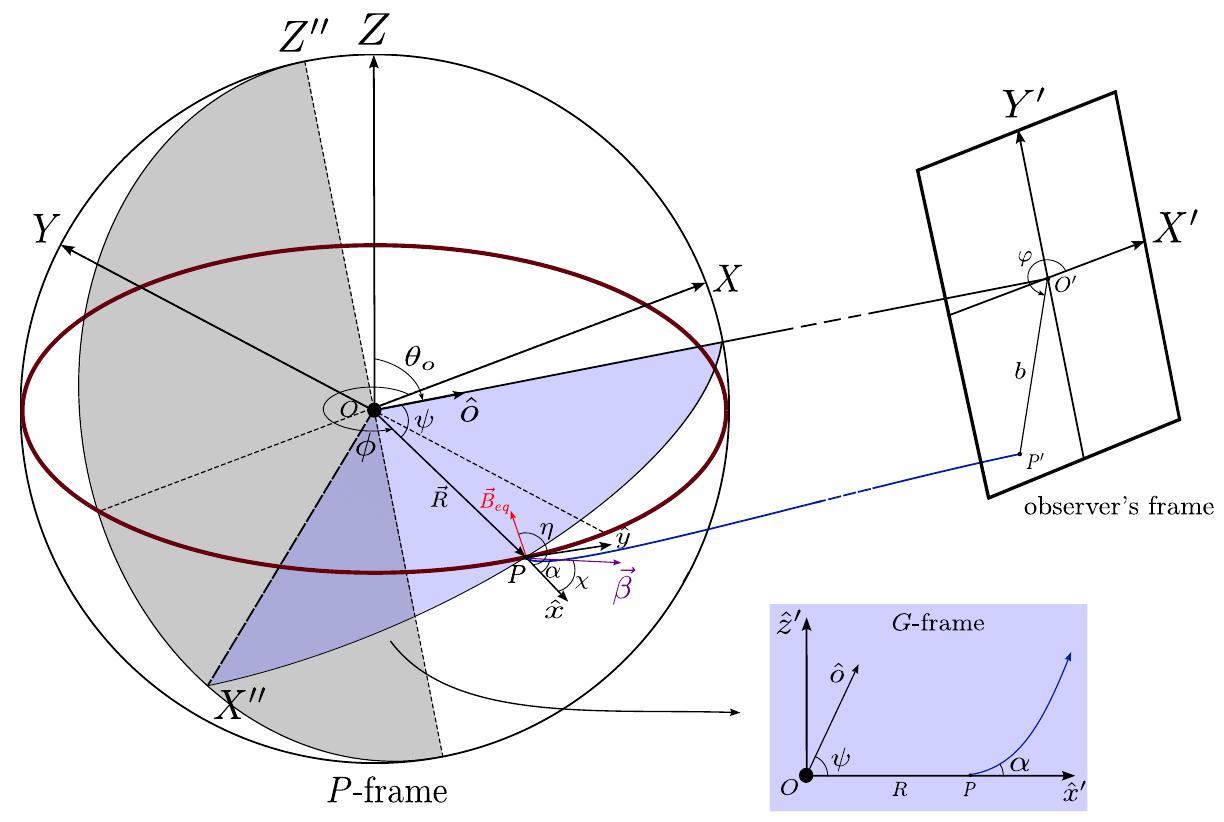}
        \caption{Geometry of the different coordinate frames. The orbit of the fluid element $P$ with $r=R$ (board), the plane where the geodesic trajectory is located (blue) and the plane of the distant observer with its respective projection (gray) are shown.}
        \label{fig:framework}
    \end{figure}
    \end{widetext}

        In order to study the images of accretion disks around black holes, we begin by outlining the coordinate and geometric conventions that we will use, as illustrated in Figure \ref{fig:framework}, following the design proposed in \cite{Lumi:1979}. Let us examine a black hole surrounded by a thin accretion disk. Consider a sphere of radius $R$ centered at the origin of coordinates $O$ of a coordinates system $(X,Y,Z)$. The black hole is centered at $O$ while its accretion disk is in the equatorial plane $XY$. We denote as $P$ the fluid element of the disk  (described by the vector position $\vb*{R}$), from which a light beam is emitted that reaches the plane $X'Y'$ of the asymptotic observer $O'$. The angle $\theta_{o}$ refers to the disk's inclination angle relative to the distant observer and $\phi$ is the azimuthal angle on the equatorial plane characterizing the emission point $P$.  To represent the complete geometry, the following relations are given: $\overline{OX}\parallel \overline{O'X'}$ and $\overline{OZ''}\parallel \overline{O'Y'}$ (the gray plane is parallel to the observer's plane), $\overline{OX''}||\overline{O'P'}$ and $\overline{OO'}$ is perpendicular to the observer's frame. The other components of the diagram will be detailed in Sec.\ref{subs:pol} when we study the polarization patterns. 

    The path that the light follows is in the blue plane, as shown in the subgraph at the bottom right. The coordinate point $P'$ of the observer's frame where the light arrives, is described through Cartesian coordinates $(X'_{P'},Y'_{P'})$ or alternatively by  the polar coordinates $(b,\varphi)$, 
    \begin{equation} \label{eq:polar_coord}
        X'_{P'}=b\cos \varphi, \hspace{4mm} Y'_{P'}=b\sin \varphi,
    \end{equation}
    with $b$ the impact parameter related to $R$ and $\alpha$ by Eq. \eqref{eq:impact_parameter}, or equivalently, in terms of $x=1-\cos\alpha$ as:
\begin{equation}\label{eq:bgenx}
    b(R,x)=\frac{R}{\sqrt{A(R)}}\sqrt{1-(1-x)^2}.
\end{equation}

     To relate the observation angle $\varphi$ with the angular coordinate $\phi$ of the emission point $P$,  note that the angle between $\overline{OZ''}$ and the ${Y}$ axis is $\theta_{o}$. Additionally, since the $X'$ and $X$ axes are parallel, it follows that:   
    \begin{equation} \label{eq:phis_relation}
        \tan\varphi=\tan\phi \cos \theta_{o}. 
    \end{equation}

    Alternatively, as in Eq.\eqref{eq:polar_coord} we only need to compute $\cos\varphi$ and $\sin\varphi$ we can use the following relationships:

    \begin{eqnarray}
    \cos\varphi&=&\frac{\cos\phi}{\sqrt{1-\sin^2\theta_o\sin^2\phi}},\label{cosvarp}\\
    \sin\varphi&=&\frac{\sin\phi\cos\theta_{o}}{\sqrt{1-\sin^2\theta_o\sin^2\phi}}.\label{sinvarp}
    \end{eqnarray}
    Equations \eqref{eq:phis_relation}, \eqref{cosvarp} and \eqref{sinvarp} are derived in Appendix A.
    
    To complete the analytic relation between the coordinates $(b, \varphi)$ of the image point $P'$ with those of the emission point $P$ (with coordinates $(R, \phi)$), we also need to determine the relationship between $x$ and $\phi$ in Eq.\eqref{eq:bgenx}. To do this, observe that $x$ is connected to $y=1-\cos\psi$ through Eqs. \eqref{eq:our} or \eqref{eq:Belo-general}. Thus, we only require a connection between $\cos\psi$ and $\phi$.
    This connection becomes evident when considering the unit vector $\vb*{o}$ originating from the black hole and extending toward the distant observer along the line $\overline{OO'}$ (with components $\vb*{o}=(0,-\sin\theta_{o},\cos\theta_{o})$ in the $XYZ$ frame), forming an angle $\psi$ with $\vb*{R}$ (with components $\vb*{R}=R(\cos\phi,\sin\phi,0)$ in the same frame). Consequently:
    \begin{equation} \label{eq:cospsi_eq}
        \cos\psi=\vb*{o}\cdot \frac{\vb*{R}}{|\vb*{R}|}=-\sin\theta_{o}\sin\phi.     
    \end{equation}
    
    Hence, to analytically obtain the coordinates $(b(\varphi),\varphi)$ of the observation point $P'$ associated with the coordinates $(R,\phi)$ of the emission point $P$, we proceed as follows: $i)$ to determine the angular position, we use Eq.\eqref{eq:phis_relation} (or directly \eqref{cosvarp} and \eqref{sinvarp}); and $ii)$ to calculate $b(\varphi)$ in terms of $(R,\phi)$, we first use the relationship between $\cos\psi$ and $\phi$ given by Eq.\eqref{eq:cospsi_eq} to compute $y=1-\cos\psi=1+\sin\theta_{o}\sin\phi$ , then we use Eq.\eqref{eq:our} or \eqref{eq:Belo-general} according to the desired approximation to calculate $x=1-\cos\alpha$, and finally we substitute into equation \eqref{eq:bgenx} to find $b$.
    In contrast to this analytical approach, the exact expression for $b(R,\phi)$ can only be obtained after a numerical integration of \eqref{eq:psip}.
    
    In summary, with the use of \eqref{eq:our} we have established an explicit analytical mapping between the emission point $P$ (with local coordinates $(R,\phi)$) and its corresponding image point $P'$ (with local coordinates $(X'_{P'},Y'_{P'})$). In the case where our focus is exclusively on the use of the generalized Beloborodov approximation, these relationships are reduced to the following explicit and simple forms:
    
    \begin{eqnarray}\label{eq:Xapp1alt}
    \begin{bmatrix}
     X'_{P'}(R,\phi)\\
     Y'_{P'}(R,\phi)\\
     \end{bmatrix}
= b(R,\phi)
 \begin{bmatrix}   
   \frac{\cos\phi}{\sqrt{1-\sin^2\theta_o\sin^2\phi}}\\
 {\frac{\sin\phi\cos\theta_{o}}{\sqrt{1-\sin^2\theta_o\sin^2\phi}}}
    \end{bmatrix},
\end{eqnarray}
    with 
    \begin{equation}
        b(R,\phi)=\frac{R}{\sqrt{A(R)}} \left\{ 1-\left[1 - A(R)\left(1 + \sin\theta_{o}\sin\phi\right)\right]^2 \right\}^{1/2}.
    \end{equation}
Despite the simplicity of these equations, we are not aware of their prior presentation in the literature.

    Alternatively, in the case of only being interested to find expressions for the shape of the image of the isoradial curve $R=$const that are explicitly parameterized by $\varphi$ (i.e., without going through the calculation of $\phi$), we can proceed as follows: First, we relate $\cos\psi$ and $\varphi$ through the  relation:
\begin{equation}\label{eq:cospsilum}
        \cos\psi=-\frac{\sin\varphi}{\sqrt{\sin^2\varphi+\cot^2\theta_{o}}}.
    \end{equation}
   This relation was derived by Luminet using spherical trigonometry \cite{Lumi:1979}. An alternative and more direct derivation is presented in Appendix \ref{app:A}.
   
    Second, from Eq.\eqref{eq:cospsilum}, we compute $y=1-\cos\psi$ which yields,
    \begin{equation}\label{eq:yvarphi}
        y=1+\frac{\sin\varphi}{\sqrt{\sin^2\varphi+\cot^2\theta_{o}}}.
    \end{equation}
After that, $b$ is determined by Eq. \eqref{eq:bgenx}. That is, the images of the isoradial equatorial curves $R$=const are approximated (and as we will see accurately described) by the following formally compact analytical expressions:
    \begin{eqnarray}
\begin{bmatrix}\label{eq:x'casi}
        X'_{P'}(\varphi)\\
        Y'_{P'}(\varphi)\\
    \end{bmatrix}
    =      \frac{R}{\sqrt{A(R)}}\sqrt{1-(1-x)^2}
    \begin{bmatrix}
    \cos\varphi\\
\sin\varphi\\
\end{bmatrix},
    \end{eqnarray}
    with $x$ given by Eq.\eqref{eq:our}  and $y$ by \eqref{eq:yvarphi}. 
    In the generalized Beloborodov approximation, Eq.\eqref{eq:x'casi} reduces to:
    \begin{widetext}
    \begin{eqnarray}
\begin{bmatrix}\label{eq:Xapp1}
    X'_{P'}(\varphi)\\
    Y'_{P'}(\varphi)\\
    \end{bmatrix}
    =\frac{R}{\sqrt{A(R)}} \left\{ 1-\left[1 - A(R)\left(1 + \frac{\sin\varphi}{\sqrt{\sin^2\varphi + \cot^2\theta_{o}}}\right)\right]^2 \right\}^{1/2} 
    \begin{bmatrix}
    {\cos\varphi}\\
    {\sin\varphi}\\
\end{bmatrix}.
\end{eqnarray}
    \end{widetext}

Before concluding this subsection, it is worth noting that the mapping provided between $P$ and $P'$ by Eqs.\eqref{eq:bgenx}, \eqref{cosvarp} and \eqref{sinvarp}, or alternatively in the generalized Beloborodov approximation by Eq.\eqref{eq:Xapp1alt}, also allows for the analytical computation of images of any curve on the equatorial plane. As an illustration of this concept, in the subsequent subsection, we will not only compute isoradial curves but also generate images of elliptical orbits, which will be further compared with exact solutions.

\begin{figure*}[htbp]
        \centering
        \begin{tabular}{ccc}
            \hspace{-5.5mm}
            {\includegraphics[scale=0.34,trim=1.32cm 1.65cm 1.24cm 1.47cm]{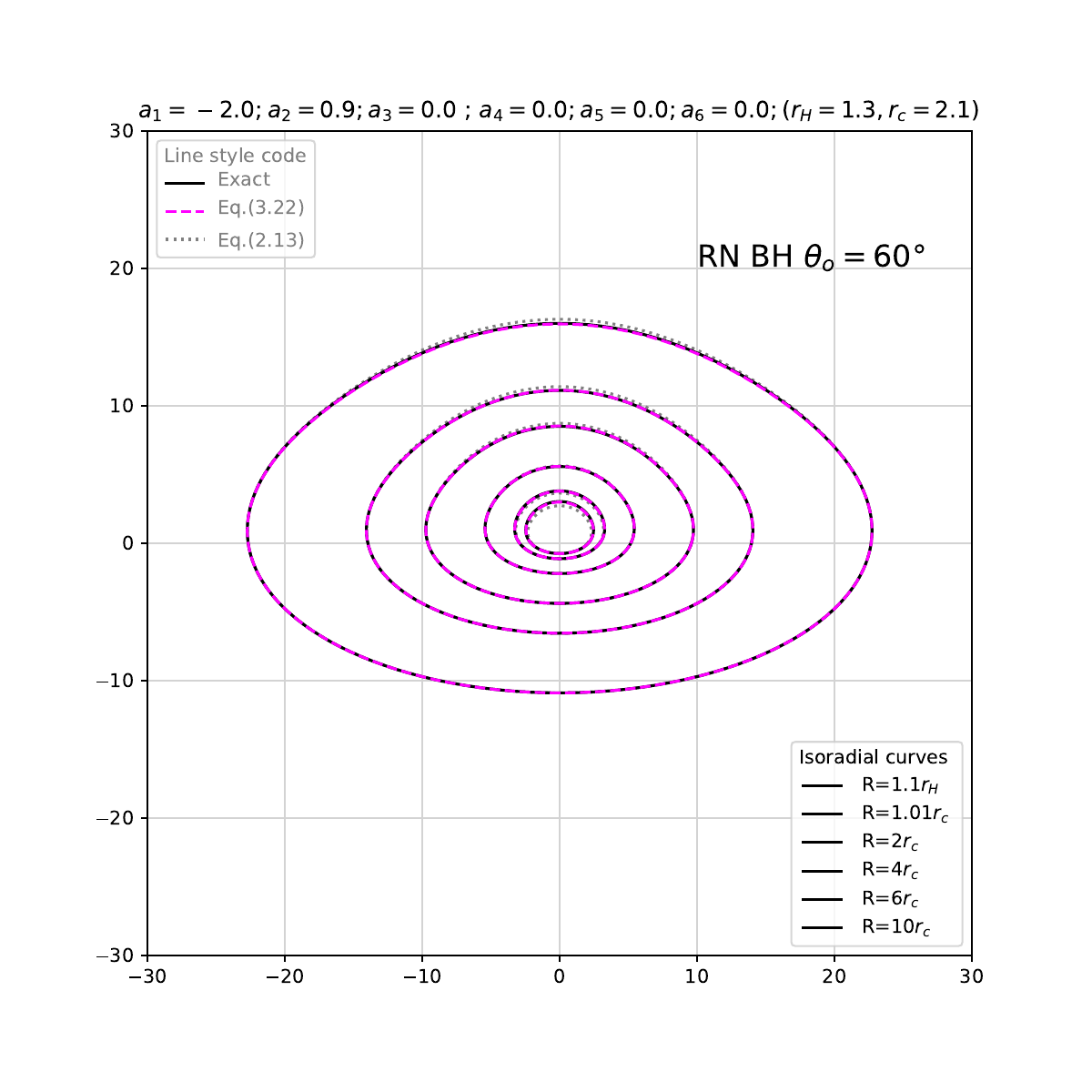}} &
            {\includegraphics[scale=0.34,trim=1.32cm 1.65cm 1.24cm 1.47cm]{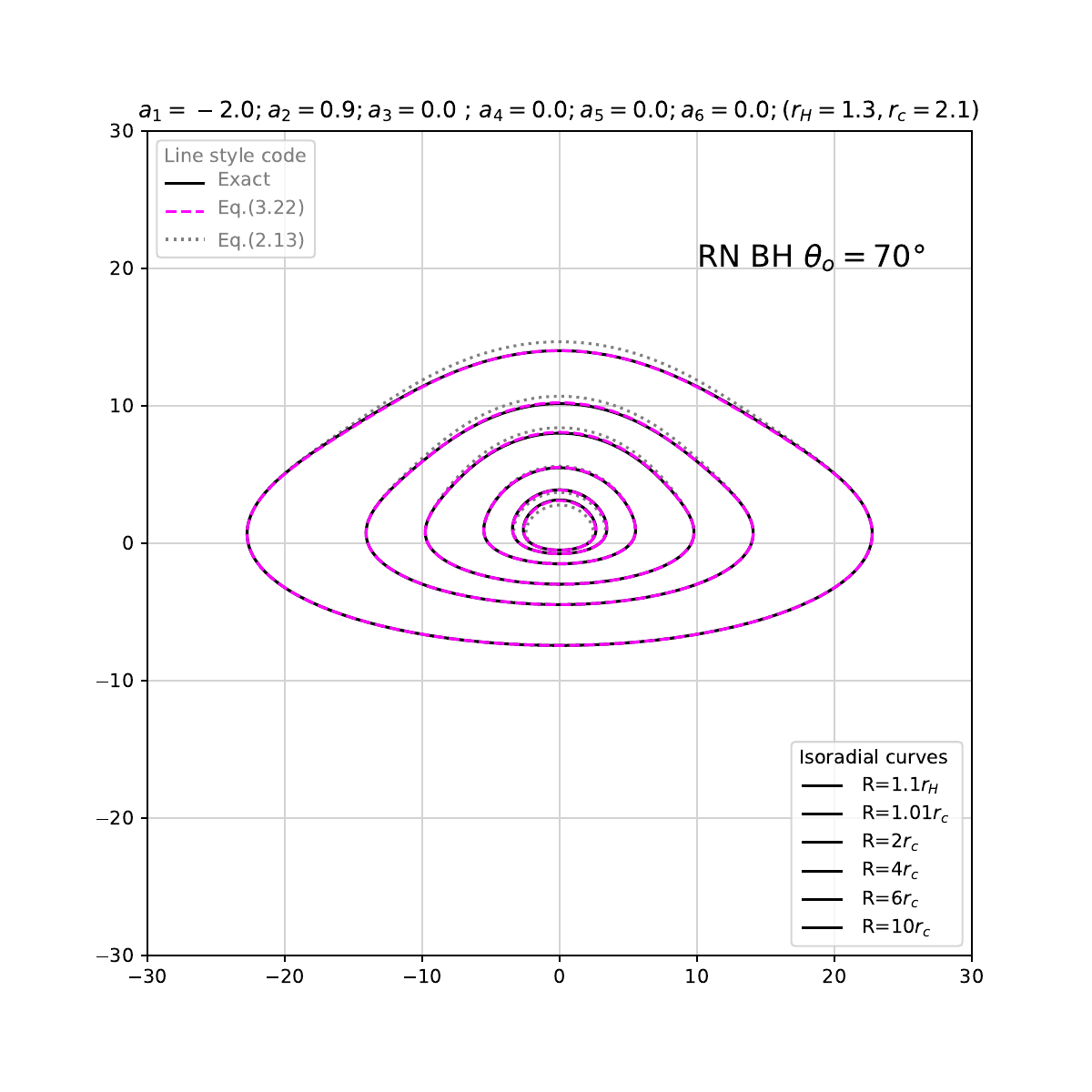}} &
            {\includegraphics[scale=0.34,trim=1.32cm 1.65cm 1.24cm 1.47cm]{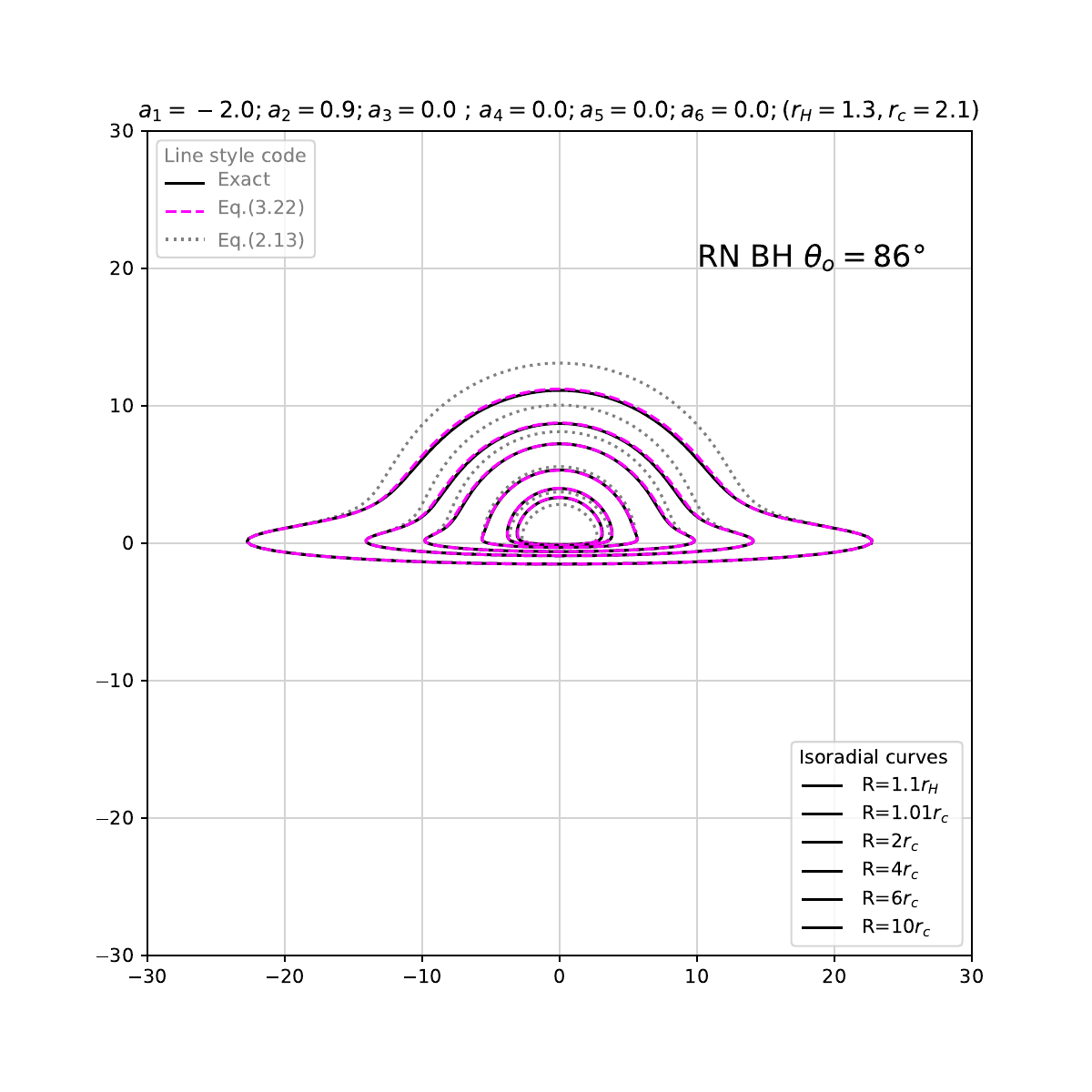}} \\
            \hspace{-5.5mm}
            {\includegraphics[scale=0.34,trim=1.32cm 1.65cm 1.24cm 1.47cm]{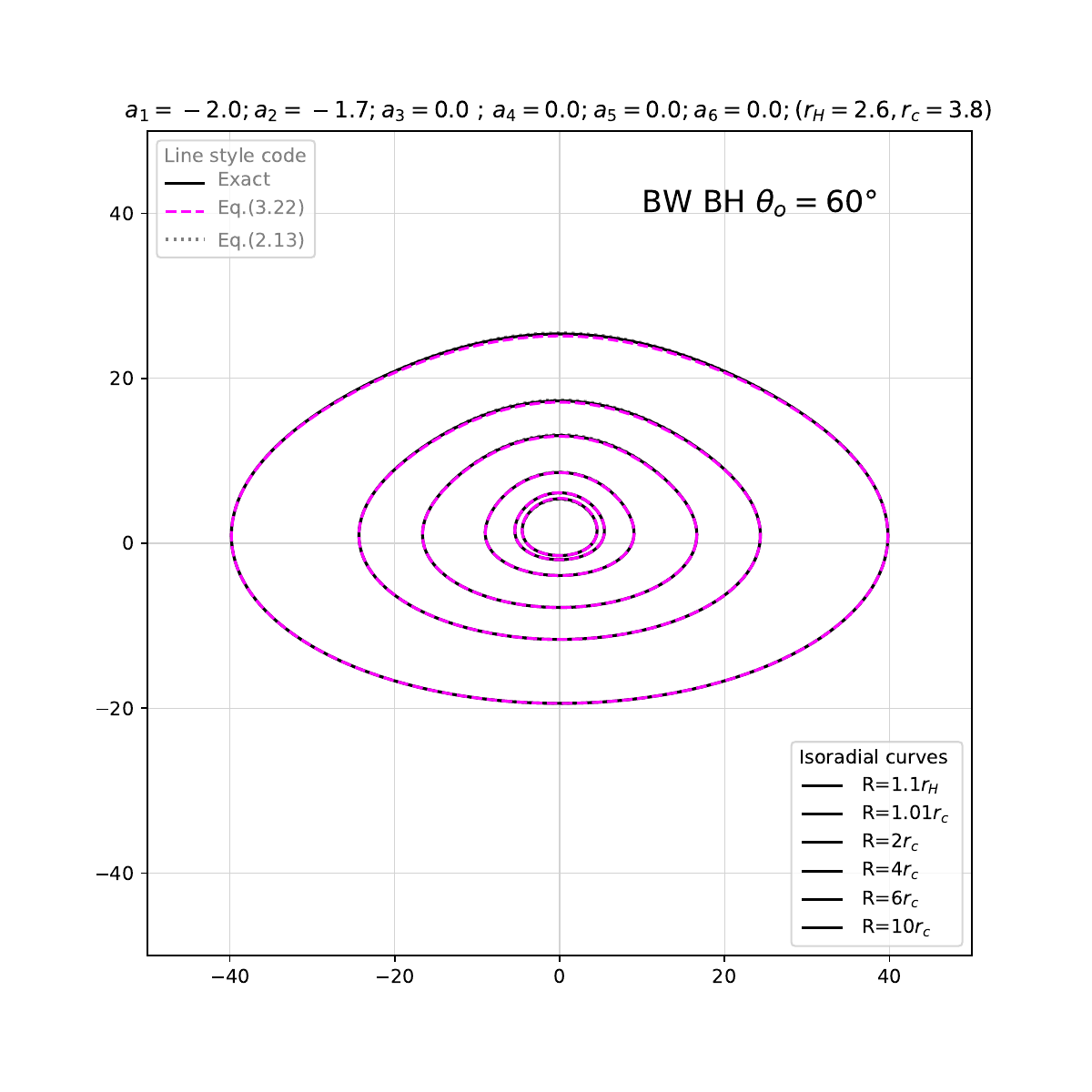}} &
            {\includegraphics[scale=0.34,trim=1.32cm 1.65cm 1.24cm 1.47cm]{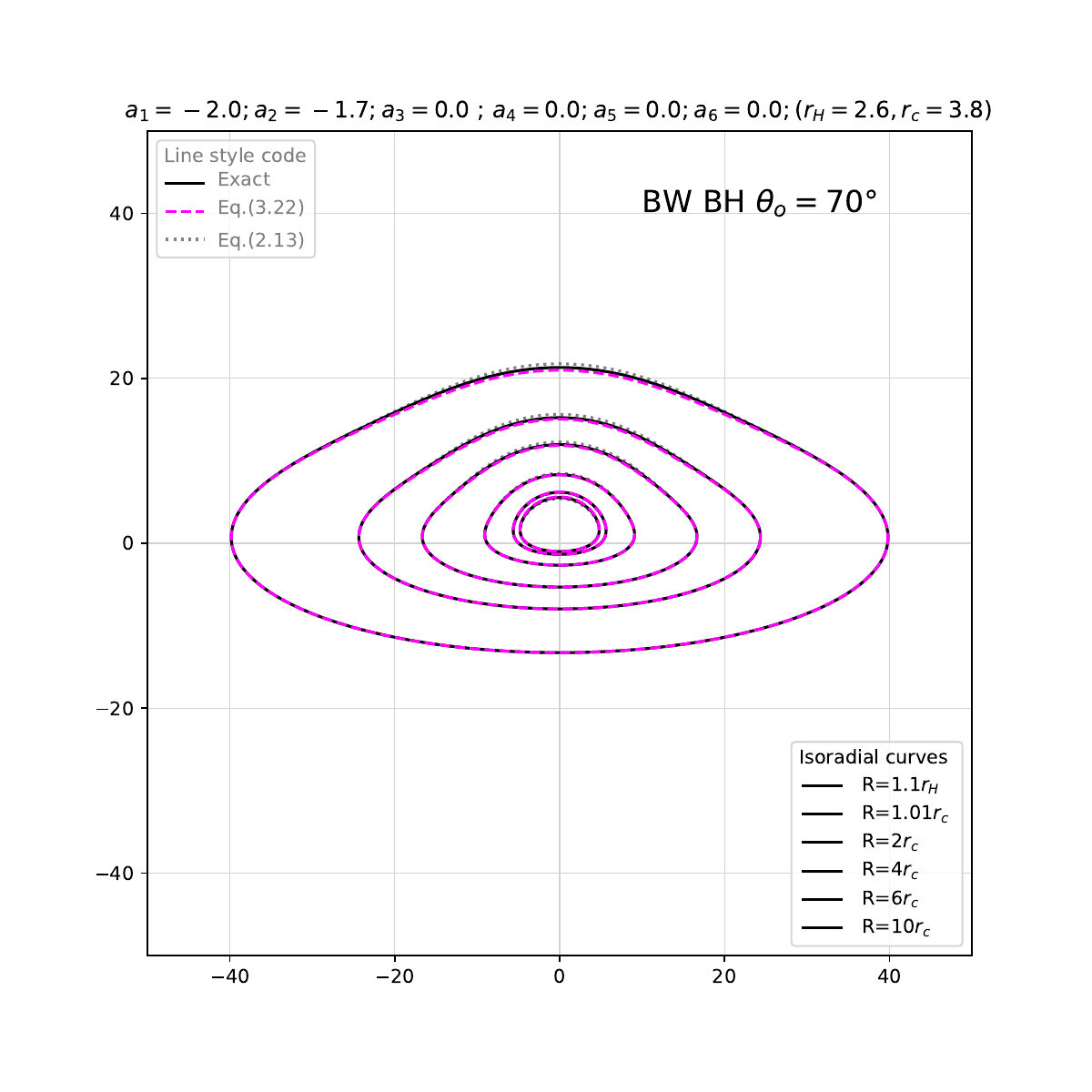}} &
            {\includegraphics[scale=0.34,trim=1.32cm 1.65cm 1.24cm 1.47cm]{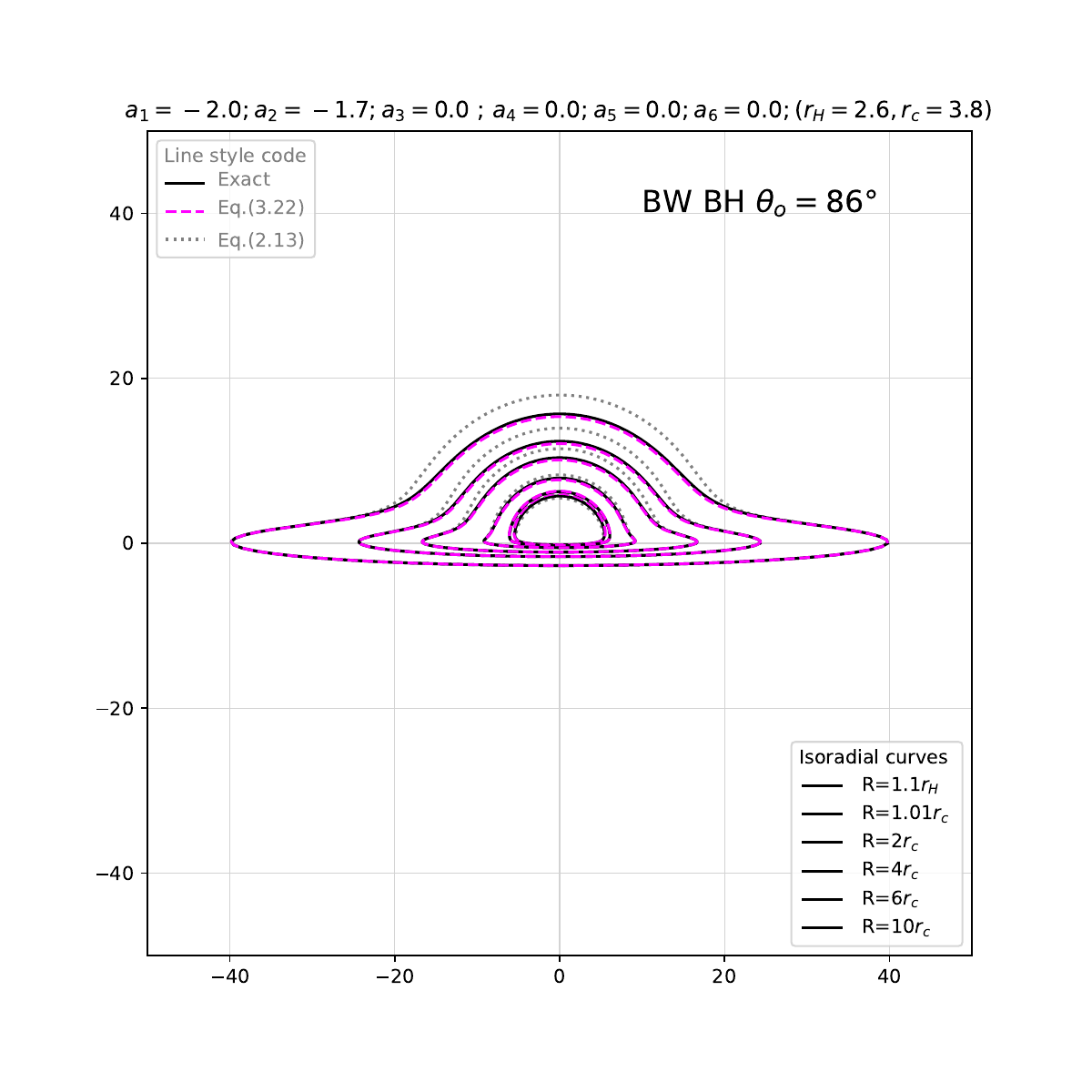}} 
        \end{tabular}
        \caption{Images of isoradial curves for two different spacetimes are depicted. In the top panel, we examine a Reissner-Nordström black hole with $a_2=0.9$, while in the bottom panel, we repeat the analysis, now considering a braneworld metric with $a_2=-1.7$. Here, $r_H$ represents the event horizon radius, and $r_c$ denotes the radius of the photon sphere.}
        \label{fig:Ab_p10}
    \end{figure*}

    \begin{figure*}[htbp]
        \centering
        \begin{tabular}{ccc}
            \hspace{-5.5mm}
            {\includegraphics[scale=0.34,trim=1.32cm 1.65cm 1.24cm 1.47cm]{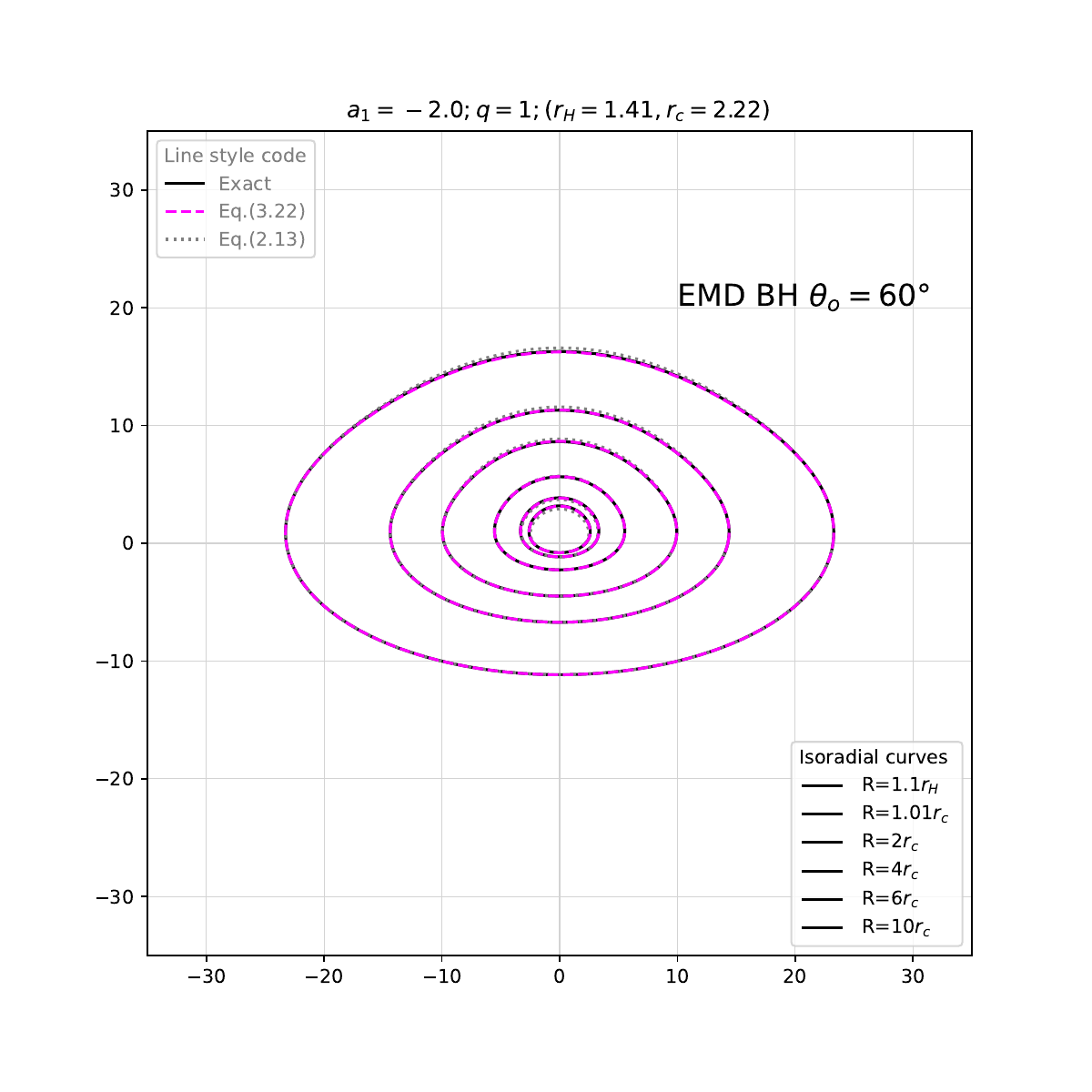}} &
            {\includegraphics[scale=0.34,trim=1.32cm 1.65cm 1.24cm 1.47cm]{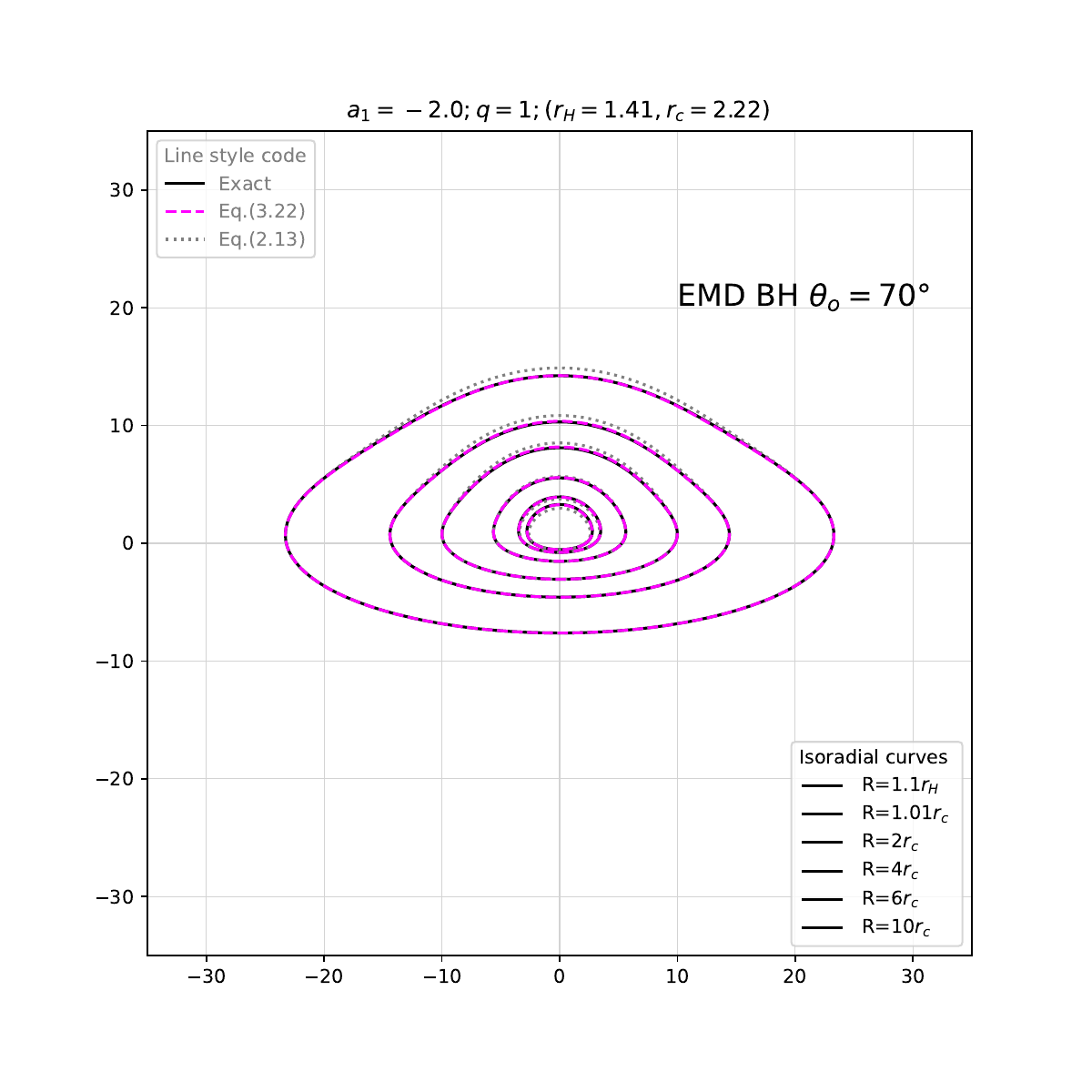}} &
            {\includegraphics[scale=0.34,trim=1.32cm 1.65cm 1.24cm 1.47cm]{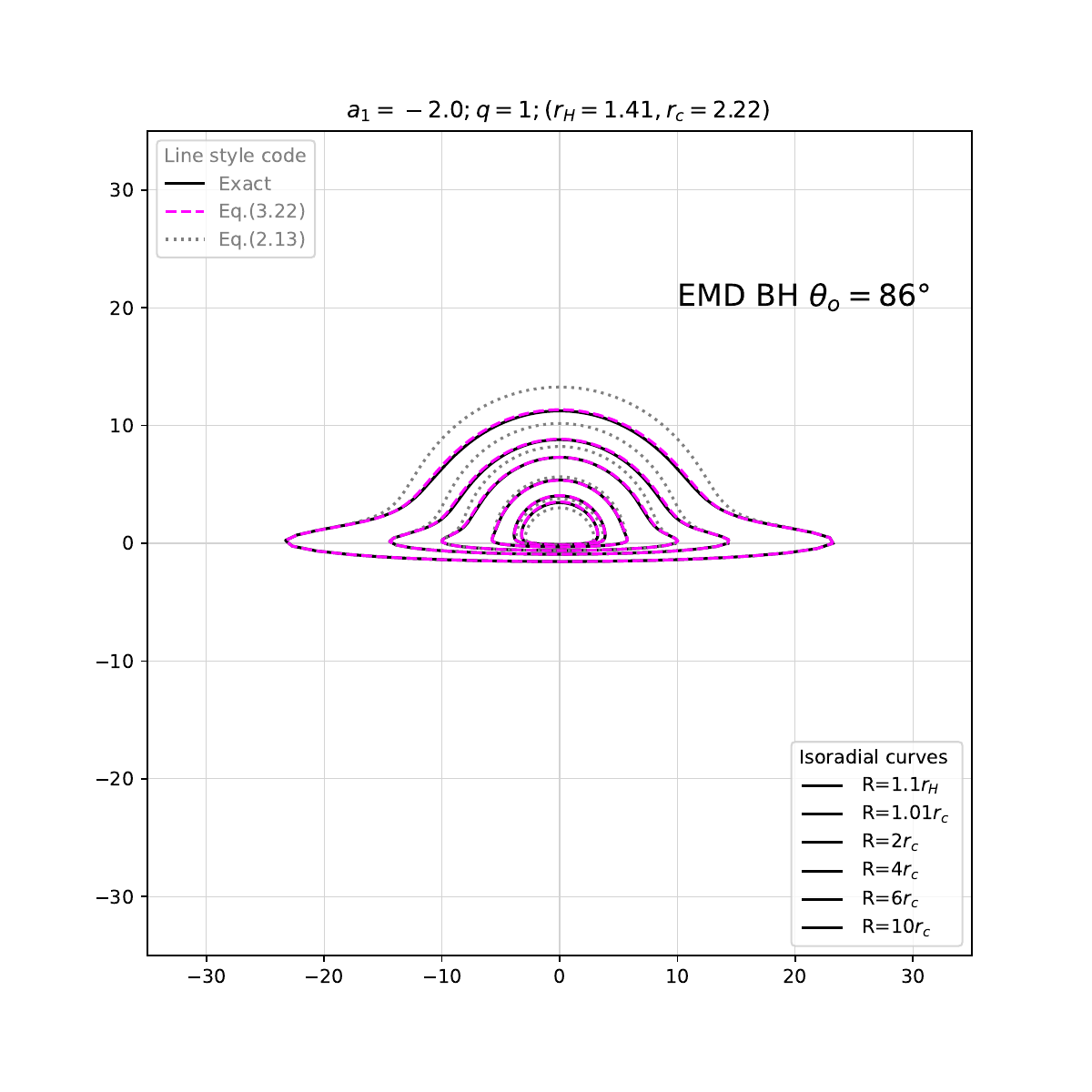}} \\
            \hspace{-5.5mm}
            {\includegraphics[scale=0.34,trim=1.32cm 1.65cm 1.24cm 1.47cm]{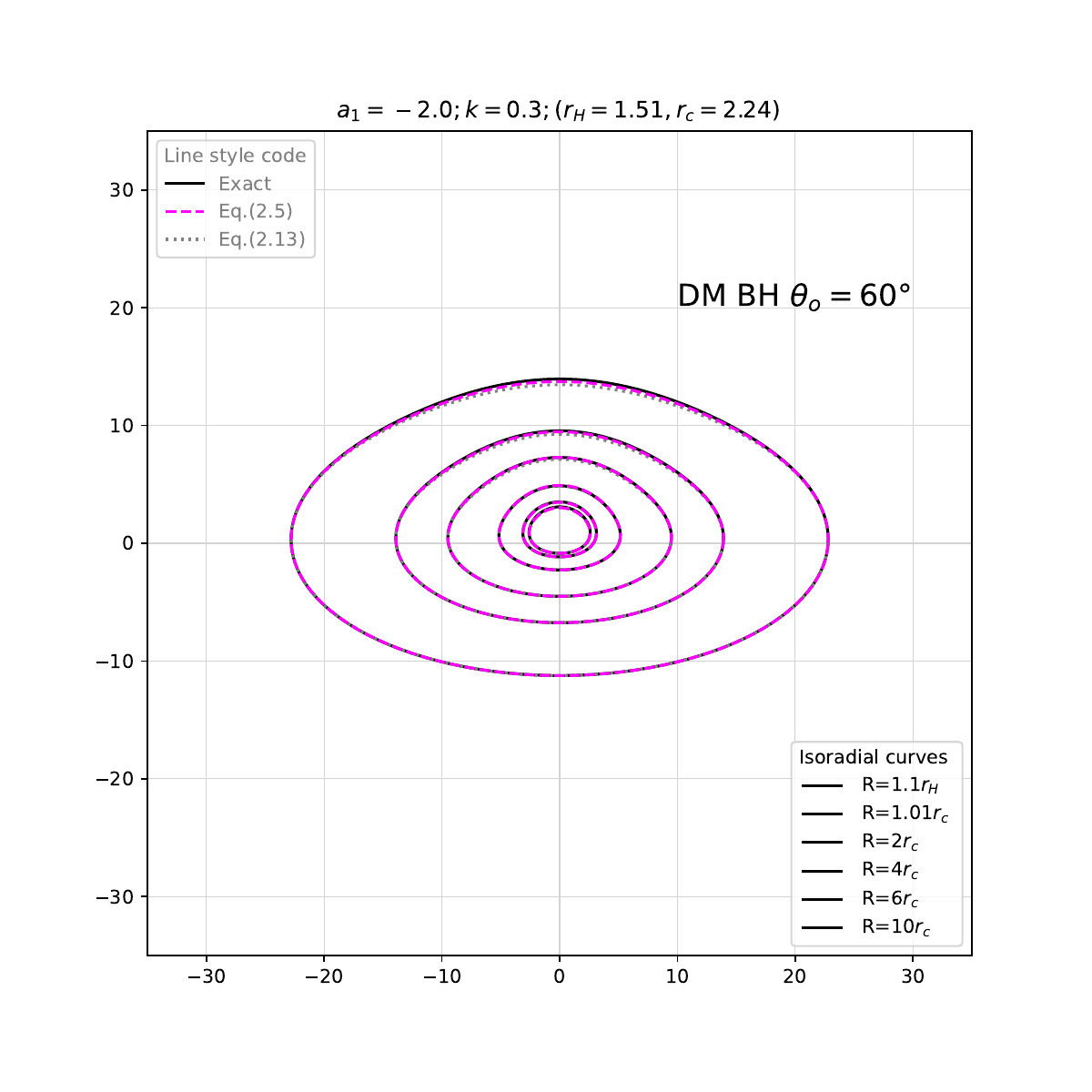}} &
            {\includegraphics[scale=0.34,trim=1.32cm 1.65cm 1.24cm 1.47cm]{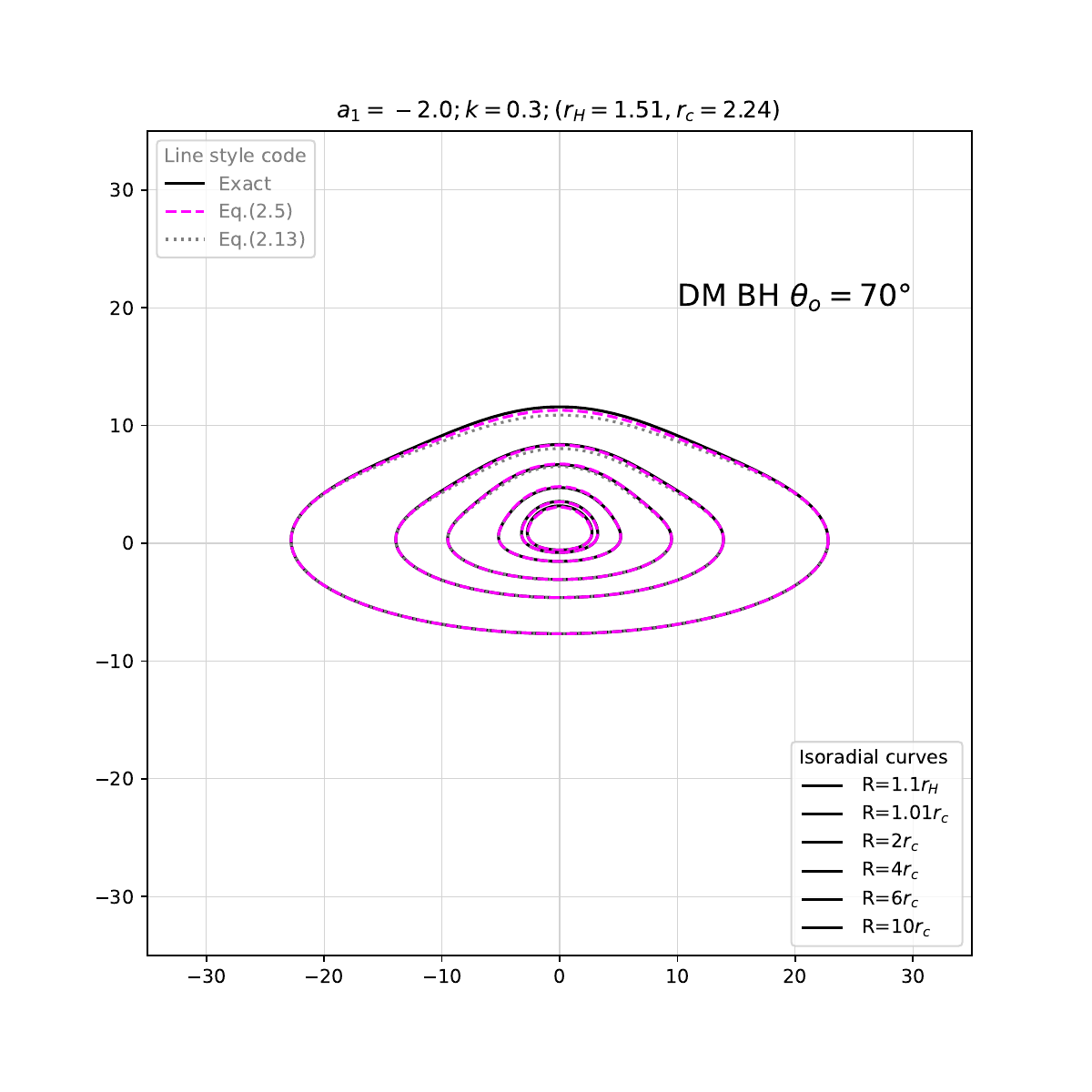}} &
            {\includegraphics[scale=0.34,trim=1.32cm 1.65cm 1.24cm 1.47cm]{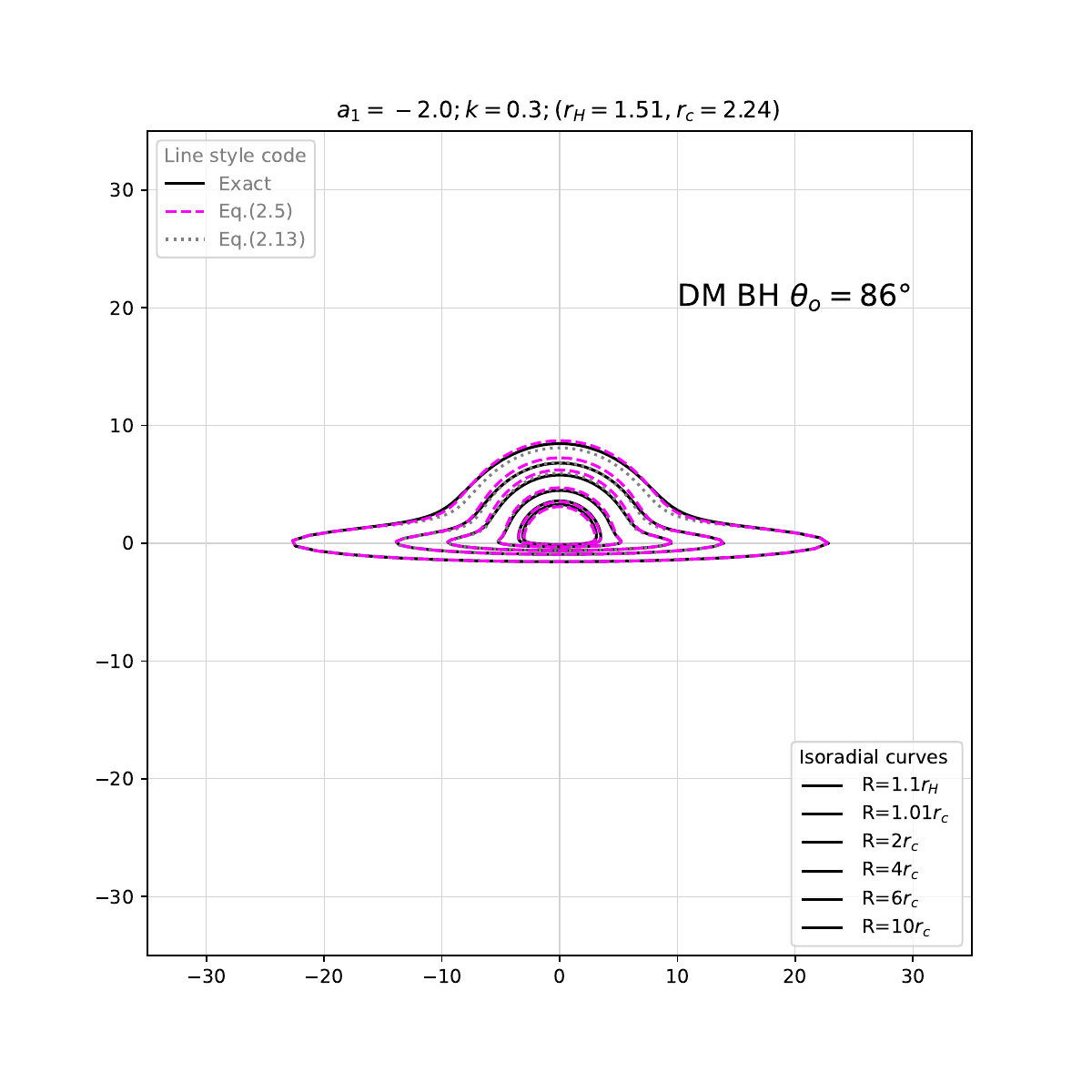}} 
        \end{tabular}
        \caption{Similar to Fig. \ref{fig:Ab_p10}, but now the top panel considers an Einstein-Maxwell-dilaton metric described in Eq. \eqref{eq:emd_bh}, while the bottom panel features a metric representing a black hole surrounded by dark matter, as described by Eq. \eqref{eq:metricDM}.}
        \label{fig:Ab_p11}
    \end{figure*}

    \begin{figure}[htbp] 
        \centering
        \includegraphics[scale=0.7
]{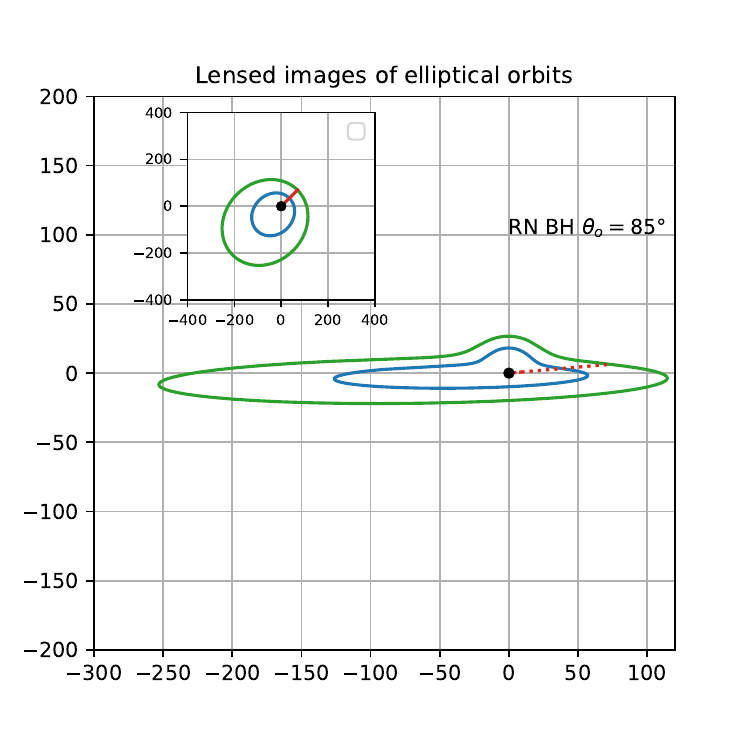}
       \caption{Image of two elliptical orbits in the equatorial plane of a black hole (black circle) as seen by a distant observer at $\theta_o=85^{\circ}$. The metric $A(r)$ of the black hole  is of the form \eqref{eq:ar6} with $a_2=0.6$.}
        \label{fig:ellip}
    \end{figure}

\subsection{Images of accretion disks: Isoradial and elliptical curves}\label{subsec:isora}
In order to test our formulas for image generation, in Figs. \ref{fig:Ab_p10} and \ref{fig:Ab_p11}, we show how they approximate the isoradial curves $R=$const for different metrics and inclinations. In the top panel of Fig. \ref{fig:Ab_p10}, we display isoradial curves for a Reissner-Nordstr\"om black hole with parameter $a_2=0.9$, for three inclinations of the equatorial disk with respect to the observer: $\theta_o=60^{\circ}, 70^{\circ}$, and $86^{\circ}$. In the bottom panel, we conduct a similar analysis for a braneworld-type metric with $a_2=-1.7$. The isoradial curves are chosen at multiples of the respective photon sphere locations, with the exception of the innermost curve, which is an image of the isoradial curve located at $1.1r_H$. In all cases, we compare them with numerically obtained images. As can be seen, even for inclinations of $\theta_o=70^{\circ}$, the generalized Beloborodov formula (or directly Eq.\eqref{eq:Xapp1alt}) provides an excellent approximation, although it loses accuracy for large radii. However, if we use the equation \eqref{eq:our} to relate $\psi$ to $\alpha$, we observe that the approximate analytical curves fit very well to the isoradial curves, even for inclination angles close to $90^{\circ}$ and large radii.

In Fig. \ref{fig:Ab_p11}, we repeat the comparison but for two metrics that do not have the exact form given by Eq. \eqref{eq:ar6}. In the top panel, we consider the case of the Einstein-Maxwell-dilaton metric presented in Eq. \eqref{eq:emd_bh} of Sec. \ref{sec:derivation}. The ``exact" isoradial curves are again obtained through numerical integration of Eq. \eqref{eq:belo-ap1} (with $A(r)$ given by Eq.\eqref{eq:emd_bh}) and its inversion to relate $\alpha$ to $\psi$. However, to calculate the approximate images using Eq.\eqref{eq:our}, instead of performing the exact integrals $I_1$ and $I_2$ with the exact expression of $A(r)$, it is more convenient to perform a Taylor expansion in powers of $q$ of $A(r)$ retaining terms up to order $q^8$. After performing such a procedure, it can be seen that the relation between $q$ and the $a_i$ up to the considered order is (with $M=1$) $a_1=-2$, $a_2=q^2$, $a_3=-q^4/4$, $a_5=q^8/64$, and $a_4=a_6=0$. The fit of the curves using the generalized Beloborodov formula, i.e. \eqref{eq:Xapp1} is again performed with the exact $A(R)$. We can see again that in all the cases considered, Beloborodov's approximation remains very good for inclination angles $\theta_o\leq 70^{\circ}$, and that the approximate formula using Eq.\eqref{eq:our_o6} continues to provide excellent agreement even for inclination angles of the order of $86^{\circ}$.
Finally, we consider a model of a spherically symmetric black hole surrounded by dark matter and a phantom field (describing dark energy) as studied in \cite{Li:2012zx}. The  metric is given by:
\begin{equation}\label{eq:metricDM}
A(r)=1+\frac{a_1}{r}+k\ln\left(\frac{r}{|k|}\right).
\end{equation}
This metric is not in the form of Eq. \eqref{eq:ar6}, nor can it be approximated by an expansion of that form in the limit of $k\ll 1$. However, we can still use the approximations given by \eqref{eq:belo-altoorden} and \eqref{eq:Belo-general}.
In the bottom panel of Fig. \ref{fig:Ab_p11}, we can see the comparison between the images of the isoradial curves obtained numerically and their comparison with the approximate formulas for a value of $k=0.3$. We can observe that even for large inclination angles, in this case, the generalized Beloborodov formula \eqref{eq:Belo-general} provides a very good approximation of the isoradial curves. If instead of using Eq. \eqref{eq:belo-altoorden} , we had used Eq. \eqref{eq:our}, the approximate curves do not improve the numerical approximation (they even worsen it). Neither Eq. \eqref{eq:belo-altoorden}  nor Eq. \eqref{eq:Belo-general} provide a good approximation if we consider values of $k$ such that $|k|>0.3$.

Finally, in Fig. \ref{fig:ellip}, we illustrate how the images of two elliptical orbits situated in the equatorial plane of a Reissner-Nordstr\"om black hole with the set $a_2=0.6$. These orbits have their periastron at $R=45M$ and $90M$ respectively, with the major axis rotated by $\phi=45^{\circ}$ with respect to the X-axis of Fig. \ref{fig:framework}. Both ellipses have an eccentricity $e=0.5$. The observer views these orbits at an inclination $\theta_o=85^{\circ}$ minimally affected by gravitational lensing effects, except for that portion of the orbit located just behind the black hole along the line of sight. For completeness, the segment parallel to the major axis is depicted in red, both in the plane of the source's equatorial disk and in its image in the observer's image plane. Analytical studies of images depicting elliptical orbits can significantly contribute to the analysis of various processes occurring around black holes. These processes include orbiting hotspots resulting from the disruption of stars as they plunge into the black hole \cite{Piran:2015gha}, the modeling of elliptical accretion disks \cite{1995ApJ...438..610E,Liu:2020ljq,Cao:2018coa}, and the exploration of luminosity curves of binary systems, where one component is a neutron star orbiting along elliptical paths, among others. Further examination of these studies will be conducted in future works.

\subsection{Redshift and Bolometric Flux} \label{c3s2}
    Let us focus now on the study of images of accretion disks around black holes.
    We adopt the conventional thin accretion disk model proposed by Novikov, Page and Thorne (NPT) \cite{Page:1974he,NoviThorne:1973}, where a black hole accompanied by a disk is assumed to be optically thick and geometrically thin. In this subsection, we use the framework developed by \cite{Hu:2022lek} to explore the dynamics. Additionally, we confine the analysis to motion of massive particles with 4-velocity $u^\alpha$ conforming the thin disk in the equatorial plane.  Consequently, from $g_{\mu \nu} u^{\mu}u^{\nu}=-1$ we have:

    \begin{equation} \label{eq:L_timelike}
         -A(r)\dot{t}^{2}+\frac{\dot{r}^{2}}{A(r)}+r^2 \dot{\phi}^{2}=-1,
     \end{equation}
    with $\dot{f}$ meaning the derivative of $f$ with respect to proper time $\tau$.
    Using the Euler-Lagrange equations and considering the independence of the metric in the coordinates $t$ and $\phi$, it follows that: 
    
    \begin{equation} \label{eq:cte_quant}
        \dot{t}=\frac{\mathcal{E}}{A(r)}, \hspace{4mm} \dot{\phi}=\frac{\mathcal{J}}{r^{2}},
    \end{equation}
    where $\mathcal{E}$ and $\mathcal{J}$ represent the specific energy and specific angular momentum, respectively. By substituting Eq. \eqref{eq:cte_quant} into Eq. \eqref{eq:L_timelike}, we obtain:
    \begin{equation} \label{eq:effect_pot}
        \dot{r}^{2}+V_{\text{eff}}(r)=\mathcal{E}^{2}, \hspace{4mm} V_{\text{eff}}(r)=A(r)\left(1+\frac{\mathcal{J}^{2}}{r^{2}}\right),
    \end{equation}
    where we have introduced the effective potential $V_{\text{eff}}(r)$. Considering the local minimum, we can determine the radius of the innermost stable orbit circular (ISCO). The ISCO radius, denoted as $r_{\text{isco}}$, satisfies the condition $V'_{\text{eff}}(r_{\text{isco}})=0=V''_{\text{eff}}(r_{\text{isco}})$, where the prime indicates the derivative with respect to $r$. These conditions yield: 
    \begin{equation} \label{eq:r_isco}           r_{\text{isco}}=\frac{3A(r_{\text{isco}})A'(r_{\text{isco}})}{2A'(r_{\text{isco}})^{2}-A(r_{\text{isco}})A''(r_{\text{isco}})}.
    \end{equation}
    
    The ISCO location is necessary for the calculation of the accretion disk flux since, in the NPT model, it is assumed to be the innermost orbit from which the accreted disk emits light \cite{Page:1974he,NoviThorne:1973}.

    For particles following circular orbits of radius $r$, the energy $\mathcal{E}$, orbital angular momentum $\mathcal{J}$  and the angular velocity $\Omega$ (which follow from the conditions $V_{\text{eff}}(r)=\mathcal{E}$ and  $V'_{\text{eff}}(r)=0$), are

    \begin{subequations}
        \begin{equation} \label{eq:energy_part}
            \mathcal{E}(r)=\frac{\sqrt{2}A(r)}{\sqrt{2A(r)-rA'(r)}},
        \end{equation}
        \begin{equation} \label{eq:momang_part}
            \mathcal{J}(r)= \frac{r\sqrt{rA'(r)}}{\sqrt{2A(r)-rA'(r)}},
        \end{equation}
        \begin{equation} \label{eq:velang_part}
            \Omega(r)=\frac{d\phi}{dt} = \frac{A'(r)}{\sqrt{2rA'(r)}}.
        \end{equation}
    \end{subequations}
   These expressions hold true for values of $r$ such that $r\geq r_{c}$ with $r_c$ the radius of the photon sphere, and they are stable if $r\geq r_{\text{isco}}$.
   
    Finally, taking into account the NPT model \cite{NoviThorne:1973,Page:1974he}, the bolometric flux of radiation emitted by the accretion disk is expressed as: 
        \begin{equation} \label{eq:flux_emit}
        F_{e}(r)=-\frac{dM/dt}{4\pi r}\mathcal{G}(r)\int^{r}_{r_{\text{isco}}} \left[\mathcal{E}(\tilde{r})-\Omega(\tilde{r})\mathcal{J}(\tilde{r})\right]\mathcal{J}'(\tilde{r}) d\tilde{r},
    \end{equation}
    where $\frac{dM}{dt}$ represents the accretion rate, which we assume to be constant and 
    \begin{equation}
\mathcal{G}(r)= \frac{\Omega'(r)}{\left[\mathcal{E}(r)-\Omega(r)\mathcal{J}(r)\right]^{2}}. 
    \end{equation}.

    Since our objective is to calculate the bolometric observed flux $F_{o}$, distinct from the bolometric emitted flux $F_{e}$ for the accretion disk, we need to consider the redshift factor $g$ \cite{Ellis:1971pg,Lumi:1979}. This factor is defined as:
    \begin{equation} \label{eq:rshift_g}
        g=1+z=\frac{E_e}{E_o},
    \end{equation}
    where $z$ represents the redshift, $E_{e}$ denotes the emitted energy for a photon from the accretion disk, and $E_{o}$ is the observed energy measured by a distant observer. As before, $k^{\mu}$ denotes the 4-momentum of photons. In the rest frame of the emitting disk element with 4-velocity $u_{\text{\tiny S}}^{\mu}=(\dot{t},0,\dot{\phi},0)=\dot{t}(1,0,\Omega,0)$, the photon energy is given by \cite{Lumi:1979}:
    \begin{equation} \label{eq:energy_emit}
        E_{e}=-k_{\mu}u_{\text{\tiny S}}^{\mu}=-k_{t}\dot{t}\left(1+\Omega \frac{k_{\phi}}{k_{t}}\right).
    \end{equation}
    Note that both $k_{\phi}$ and $k_{t}$ are conserved quantities along the photon path. From the geometry illustrated in the Figure \ref{fig:framework}, it follows that $k_{\phi}/k_{t}=b\sin\theta_{o}\cos\varphi$ , so:
    \begin{equation} \label{eq:energy_emit2}
        E_{e}=-k_{t}\dot{t}\left(1+\Omega \hspace{0.5mm} b\sin\theta_{o}\cos\varphi\right).
    \end{equation}
    
    On the other hand,  the 4-velocity of a static distant observer is $u_{\text{\tiny o}}^{\mu}=(1,0,0,0)$. Therefore, the projection of the photon's 4-momentum onto this observer is:
    \begin{equation} \label{eq:energy_obs}
        E_{o}=-k_{\mu}u_{\text{\tiny o}}^{\mu}=-k_{t}.
    \end{equation}
    
    Combining Eqs. \eqref{eq:energy_emit2} and \eqref{eq:energy_obs} (together with \eqref{eq:cte_quant}) , the redshift factor Eq. \eqref{eq:rshift_g} is
    \begin{equation} \label{eq:rshift_g2}
        g=1+z=\left(1+\Omega \hspace{0.4mm} b\sin\theta_{o}\cos\varphi\right)\frac{\mathcal{E}(r)}{A(r)},
    \end{equation}

     Therefore, to analytically calculate the redshift factor $g$ of the emission point $P$, we employ the connection between the observed point $P'$ and $P$ via the mapping $(R,\phi)\to(b(\varphi),\varphi)$ as described in Sec. \ref{sec:gralframework}. Specifically, we compute $b$ using Eq. \eqref{eq:bgenx}, with $x$ expressed in terms of $y=1+\sin\theta_o\sin\phi$ through Eq. \eqref{eq:our_o6}, and $\cos\varphi$ given by Eq. \eqref{cosvarp}.

\begin{figure*}[htbp]
        \centering
        \begin{tabular}{ccc}
        \hspace{-6mm}
        {\includegraphics[scale=0.095,trim=0 0 0 0]{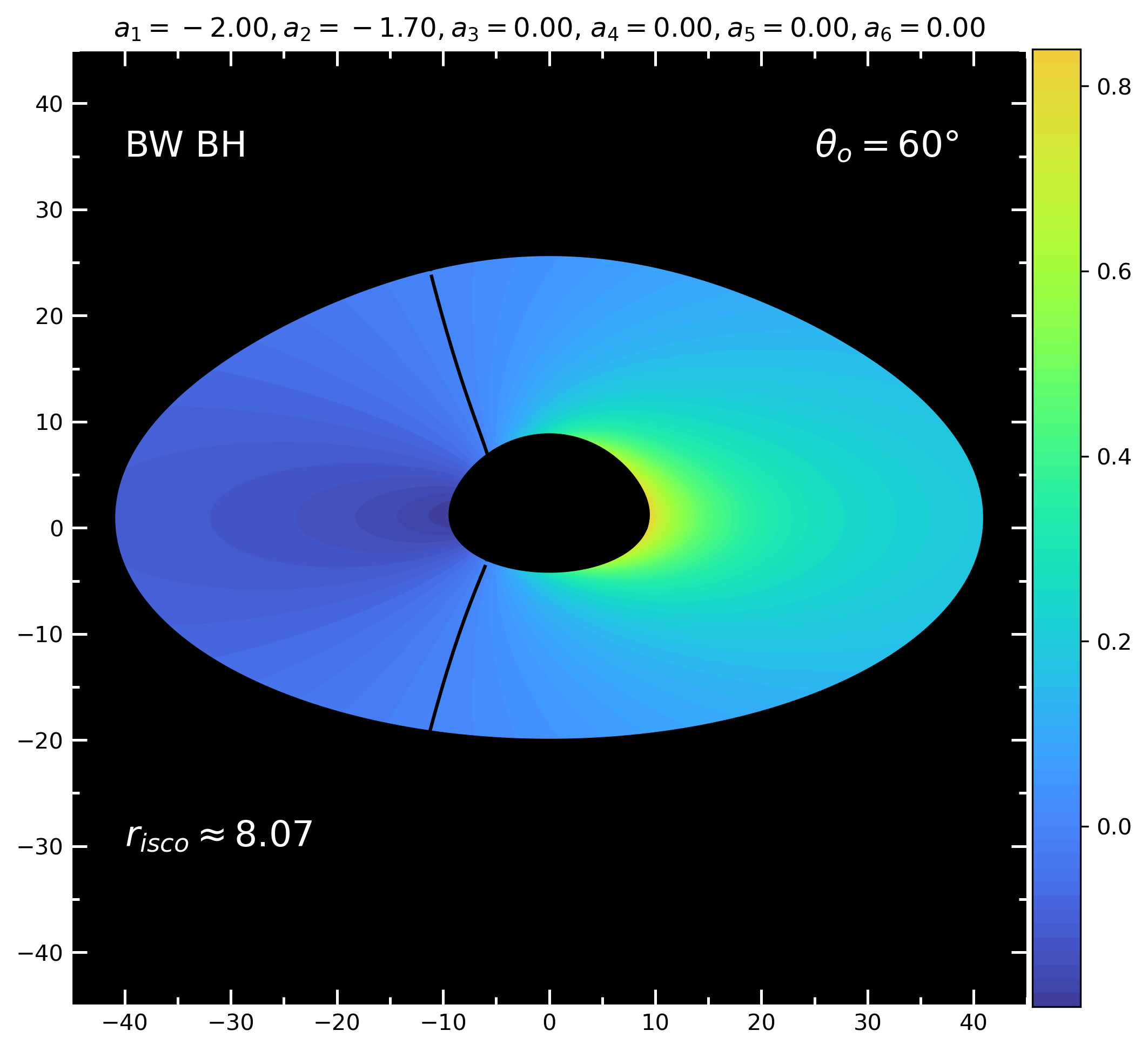}} &
        \hspace{3mm} 
        {\includegraphics[scale=0.095,trim=0 0 0 0]{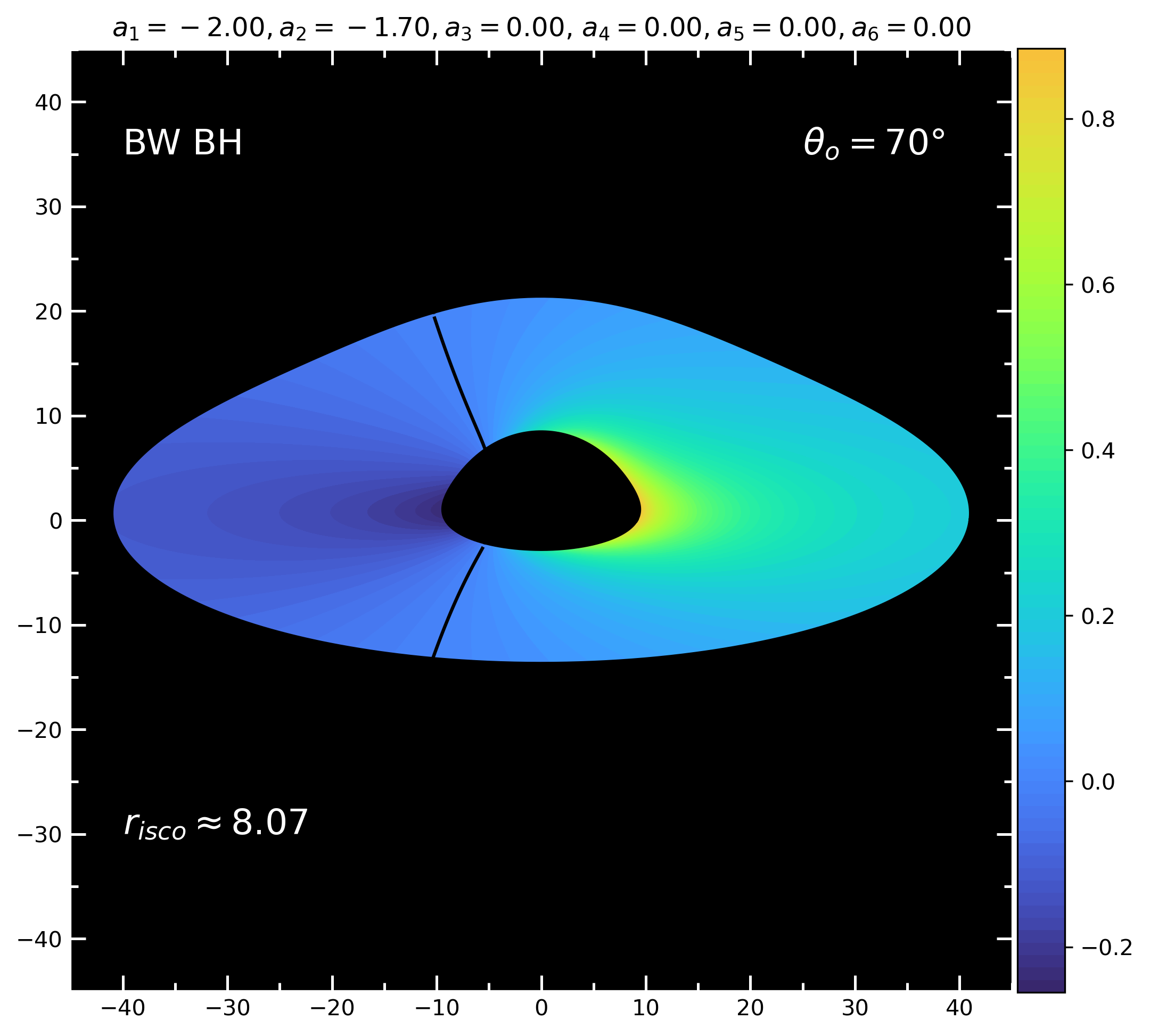}} &
        \hspace{3mm} 
        {\includegraphics[scale=0.095,trim=0 0 0 0]{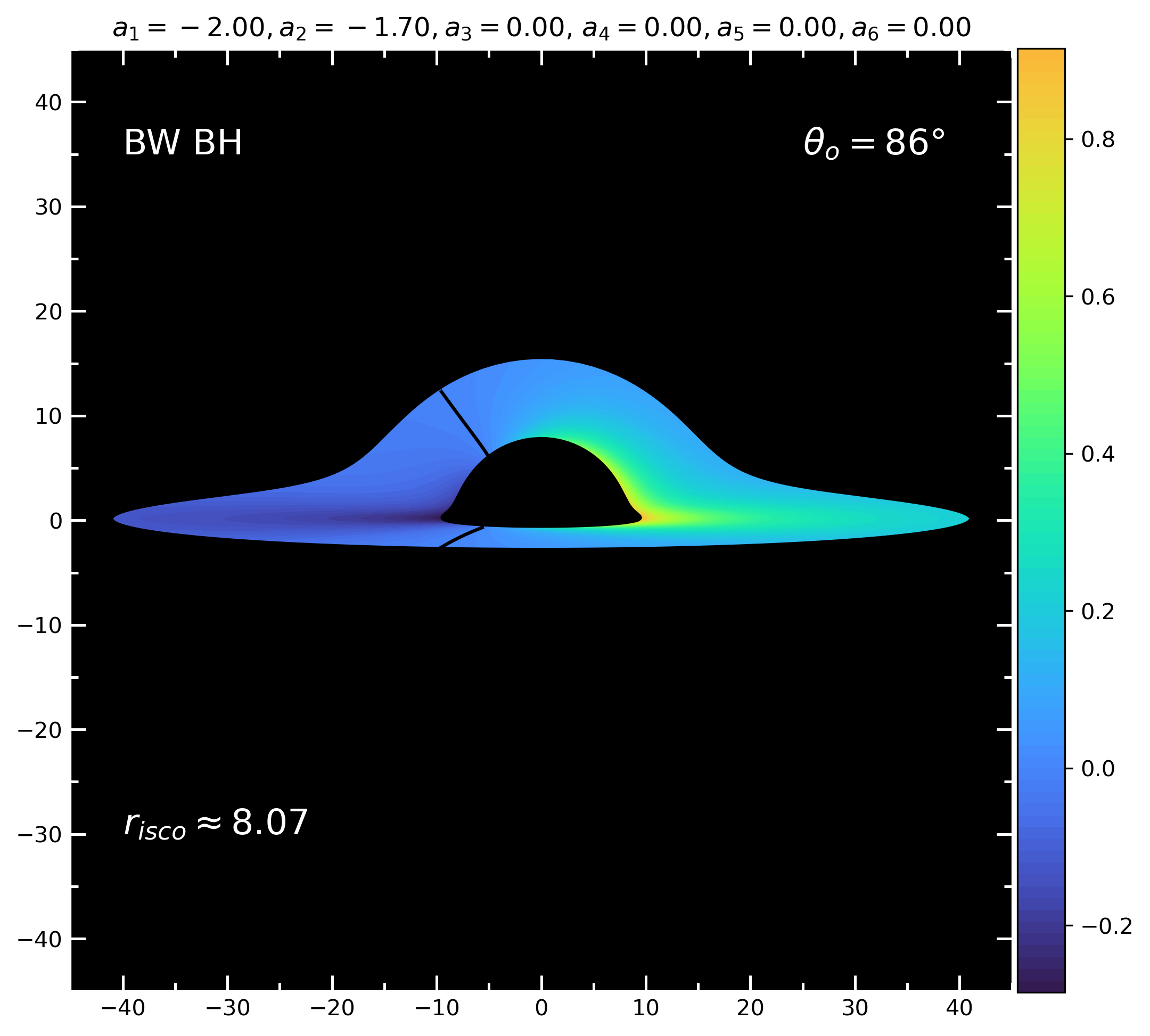}} \\
        \hspace{-6mm}
        {\includegraphics[scale=0.095,trim=0 0 0 0]{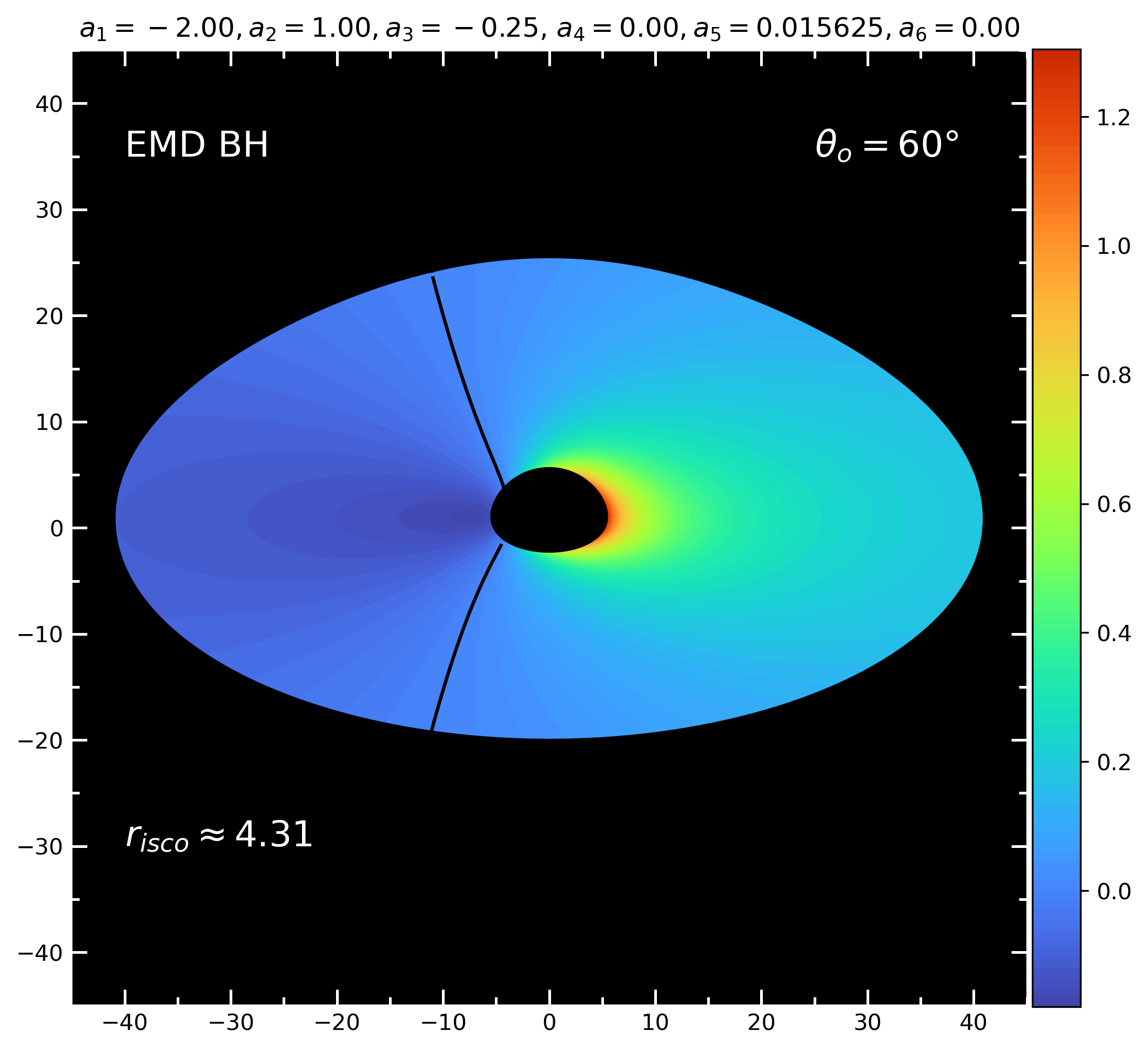}} &
        \hspace{3mm} 
        {\includegraphics[scale=0.095,trim=0 0 0 0]{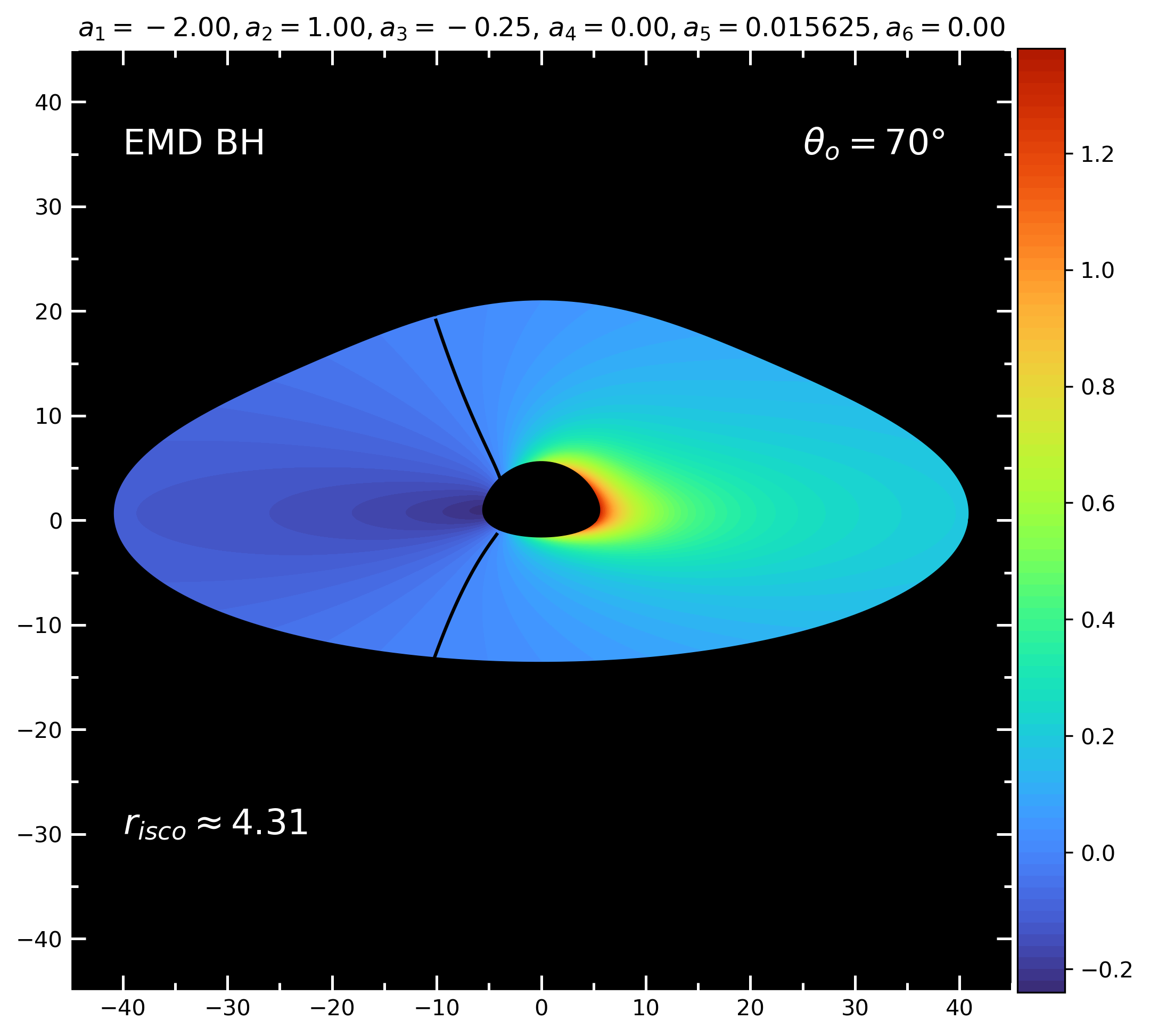}} &
        \hspace{3mm} 
        {\includegraphics[scale=0.095,trim=0 0 0 0]{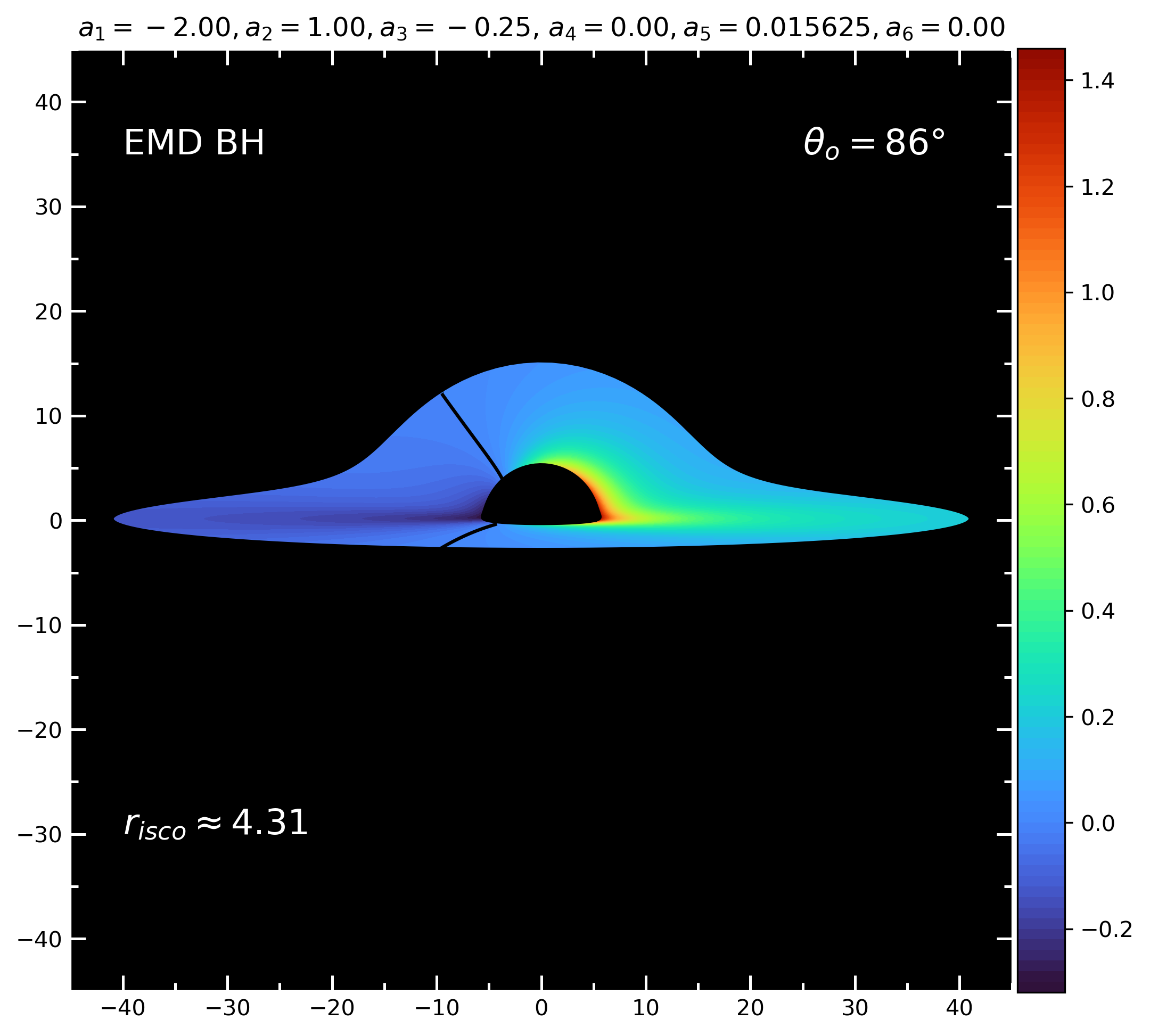}} \\
        \end{tabular}
        \caption{Redshift distributions of the braneworld and Einstein-Maxwell-dilaton black holes taking into account different inclinations. The maximum values of the redshift tend to red while the minimum values tend to dark blue. It is also possible to observe the contour for $z=0$ corresponding to the black line in the figures. These images are generated with $10^{6}$ points $(R,\phi)$ from the inner orbit corresponding to $r=r_{\text{isco}}$ to the outermost orbit $r=40$.}
        \label{fig:redshift1}
    \end{figure*}

\begin{figure*}[htbp]
        \centering
        \begin{tabular}{ccc}
        \hspace{-6mm}
        {\includegraphics[scale=0.095,trim=0 0 0 0]{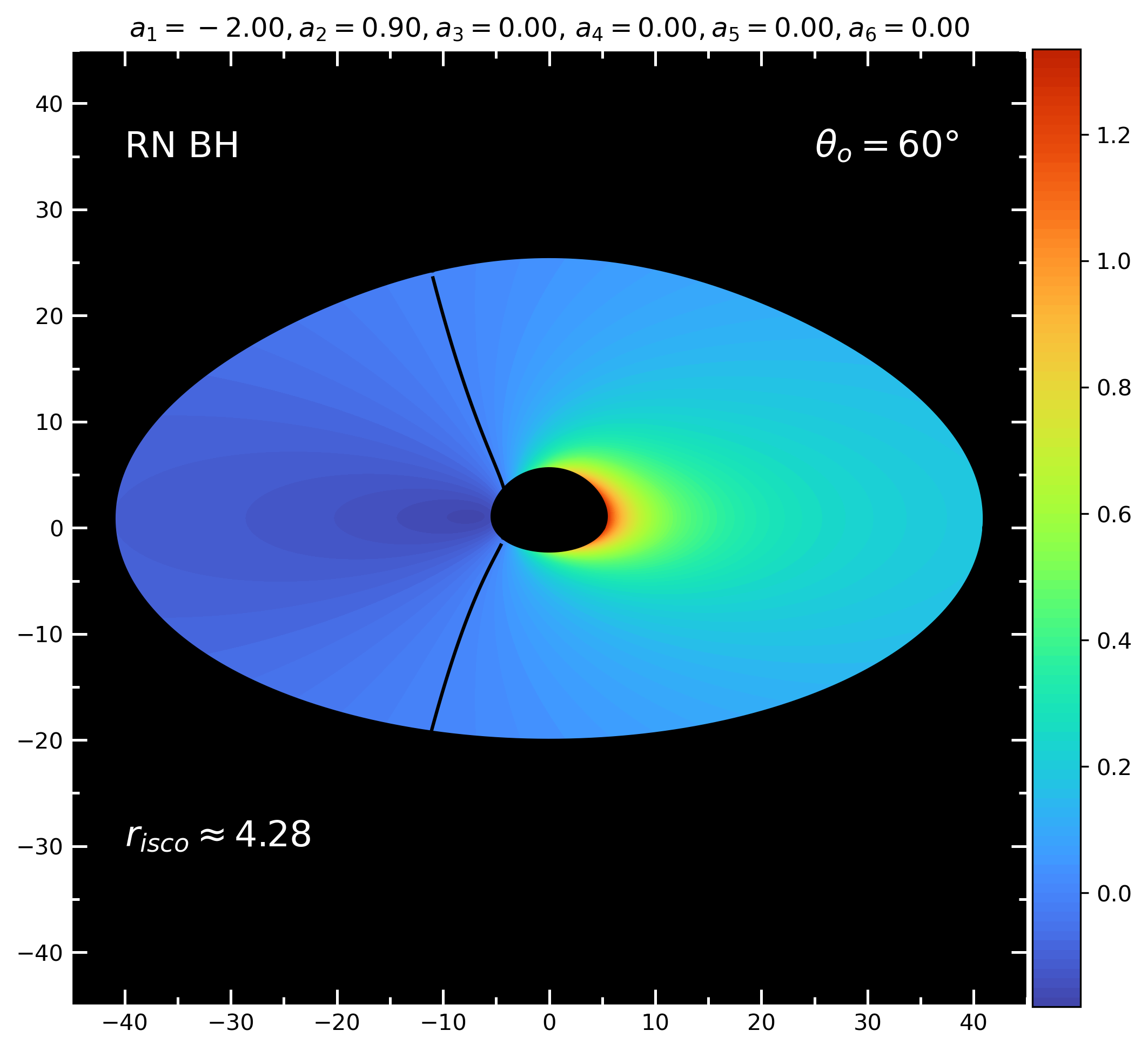}} &
        \hspace{3mm} 
        {\includegraphics[scale=0.095,trim=0 0 0 0]{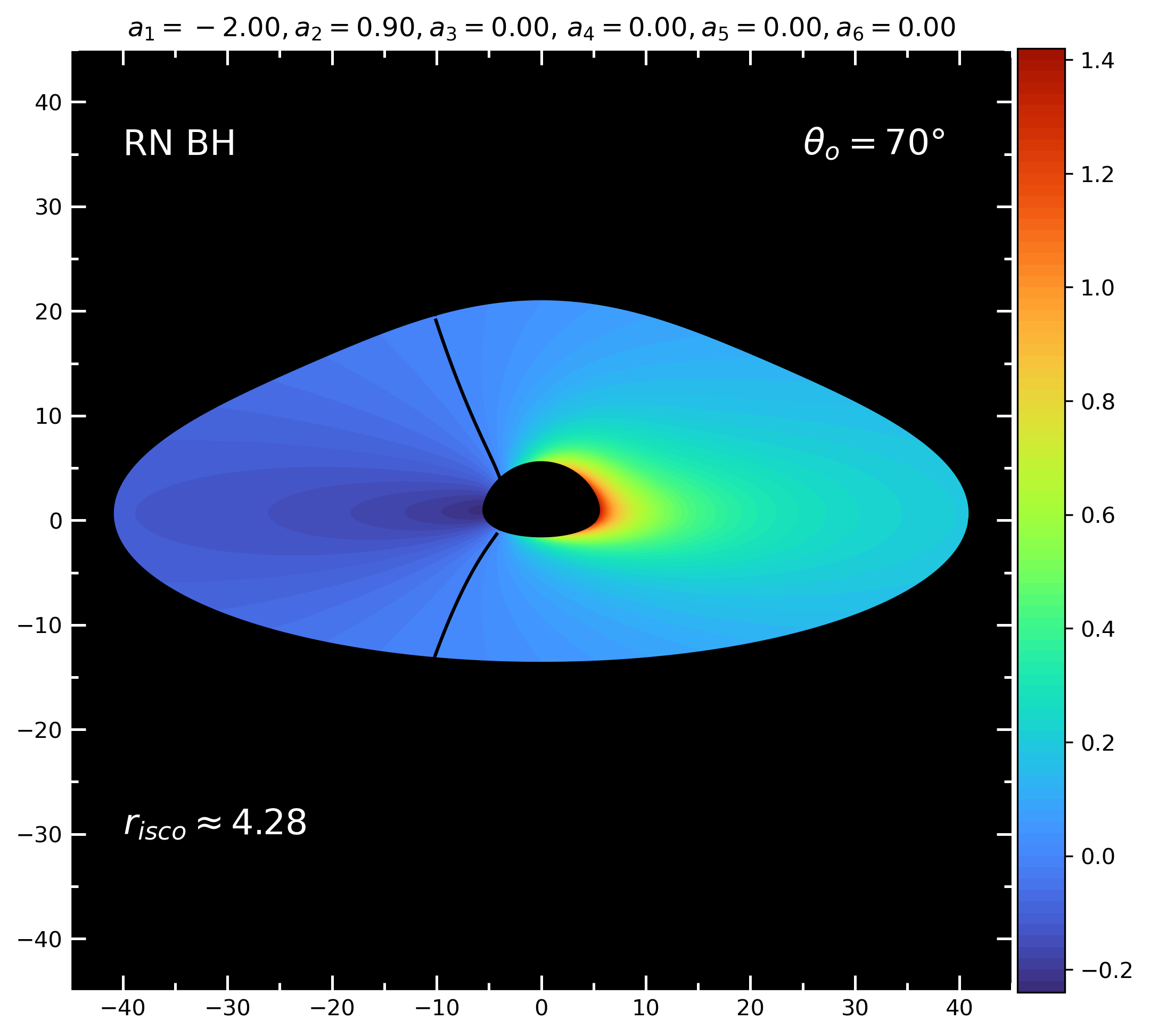}} &
        \hspace{3mm} 
        {\includegraphics[scale=0.095,trim=0 0 0 0]{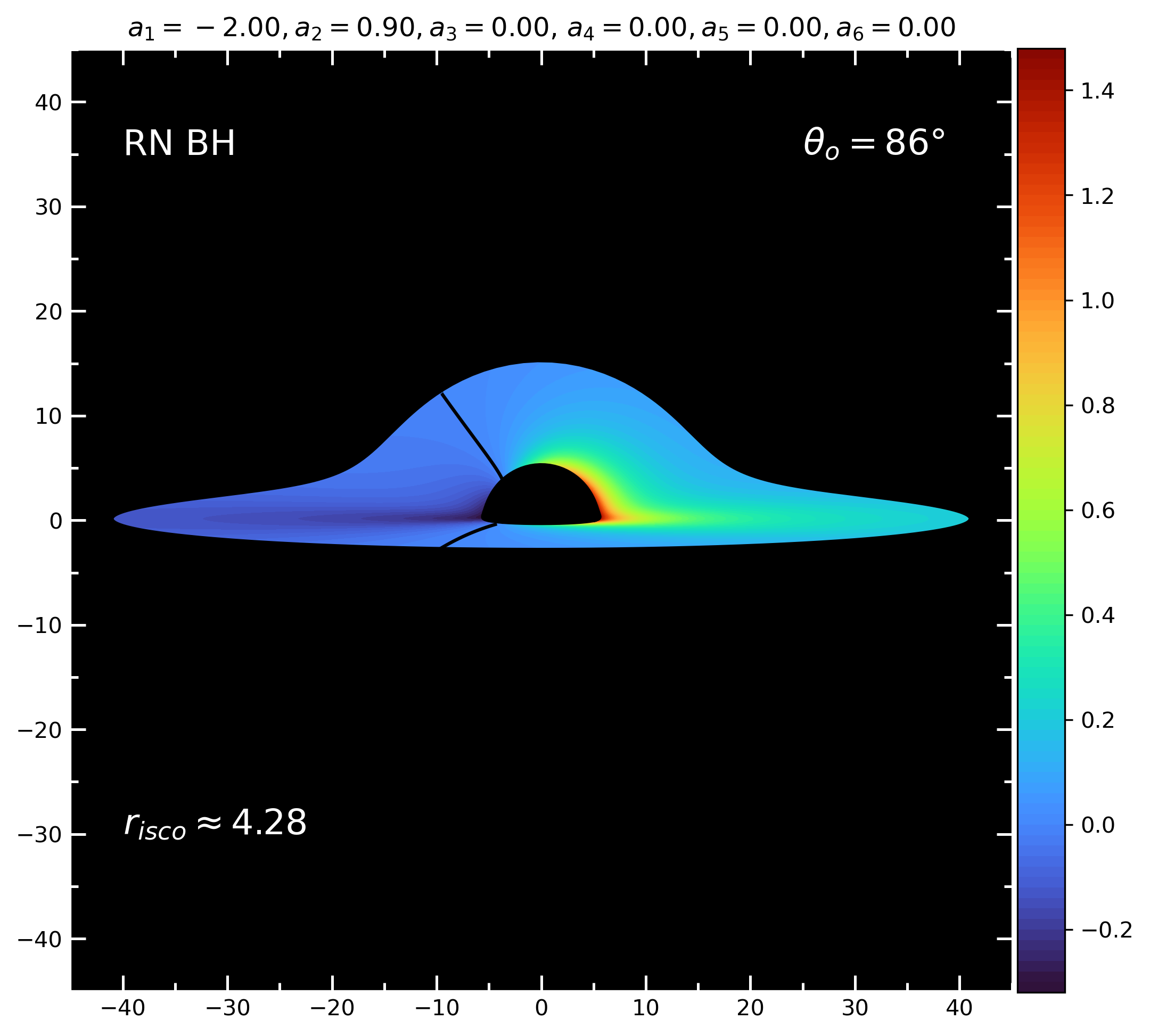}} \\
        \hspace{-6mm}
        {\includegraphics[scale=0.095,trim=0 0 0 0]{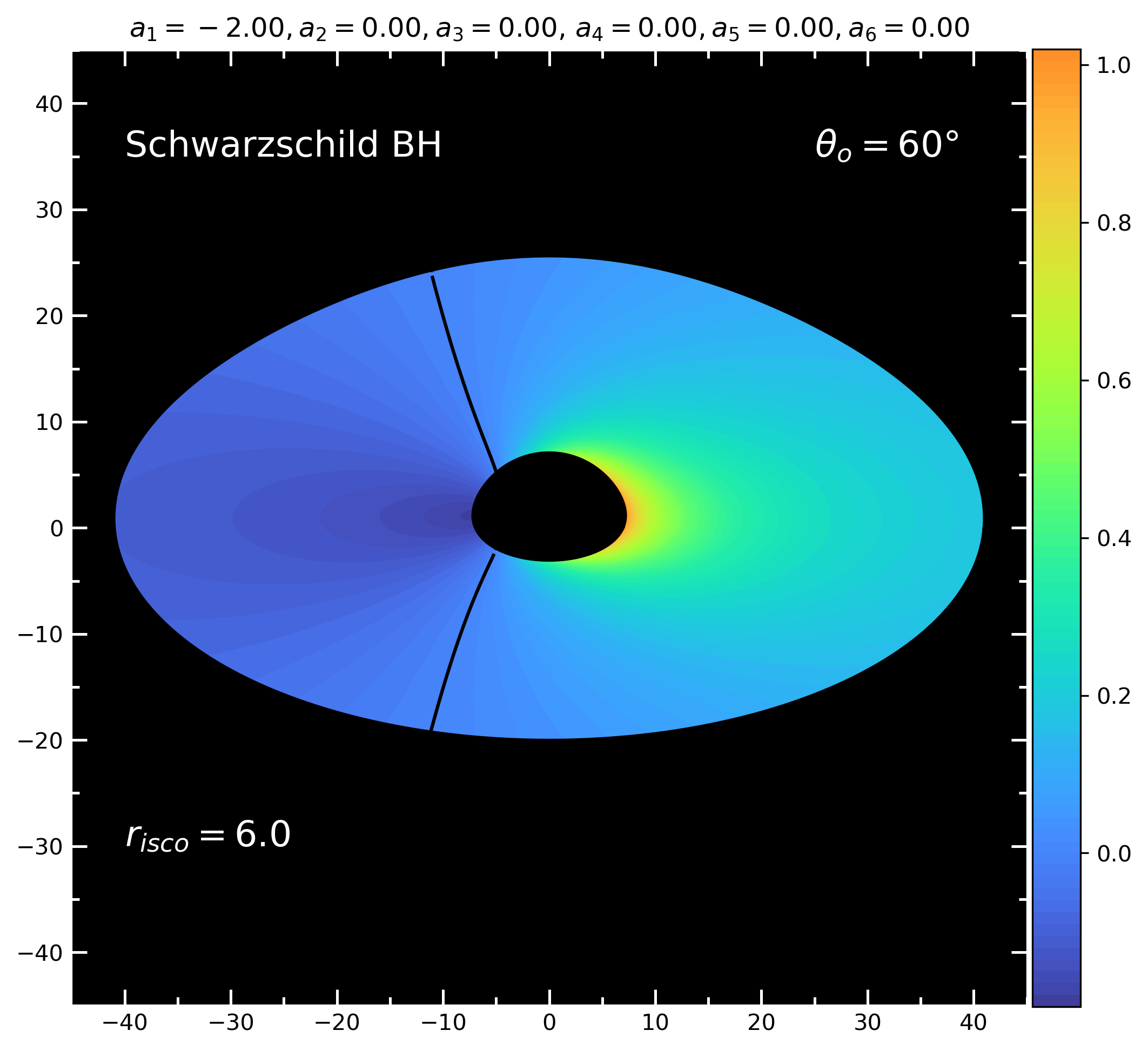}} &
        \hspace{3mm} 
        {\includegraphics[scale=0.095,trim=0 0 0 0]{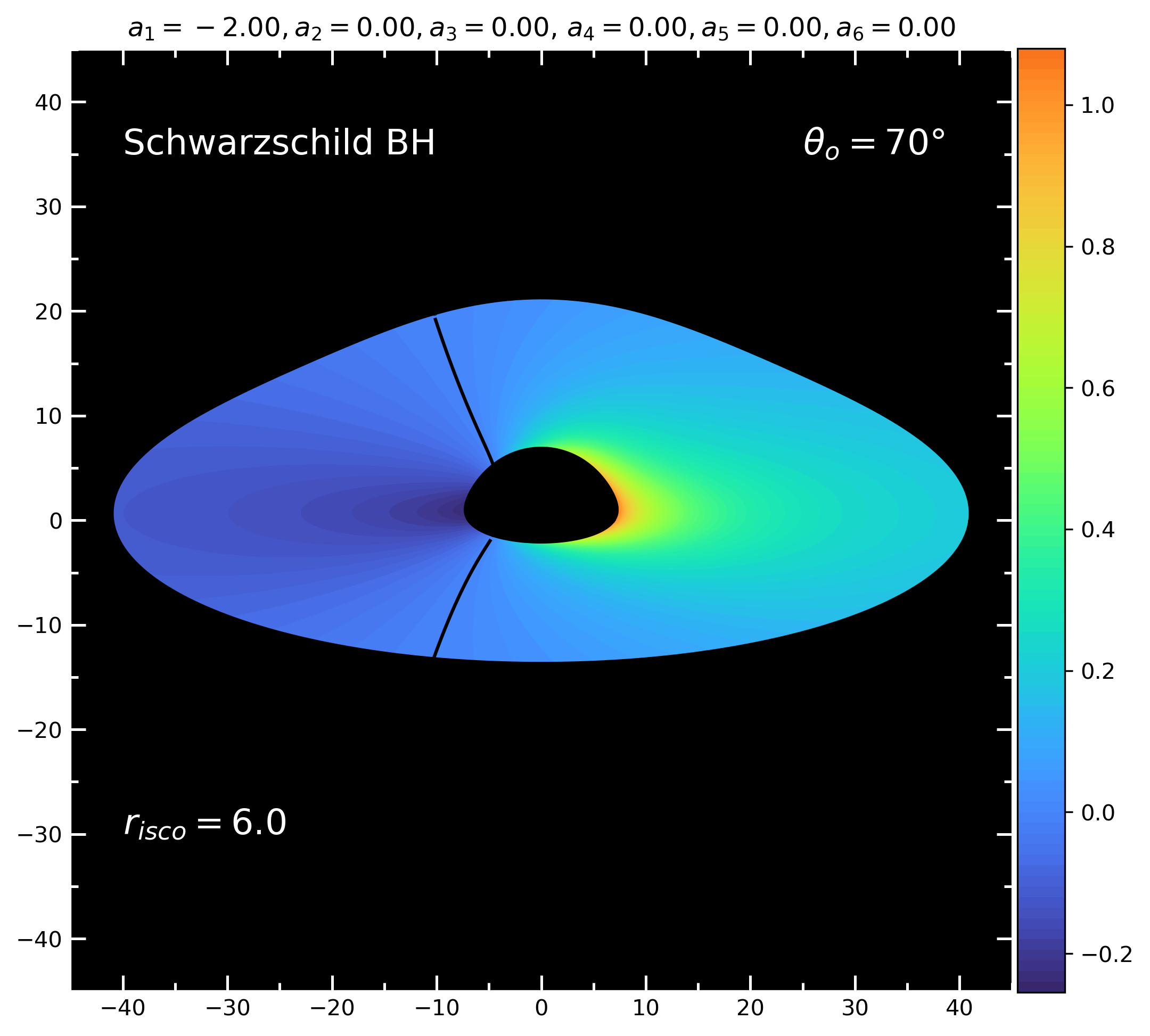}} &
        \hspace{3mm} 
        {\includegraphics[scale=0.095,trim=0 0 0 0]{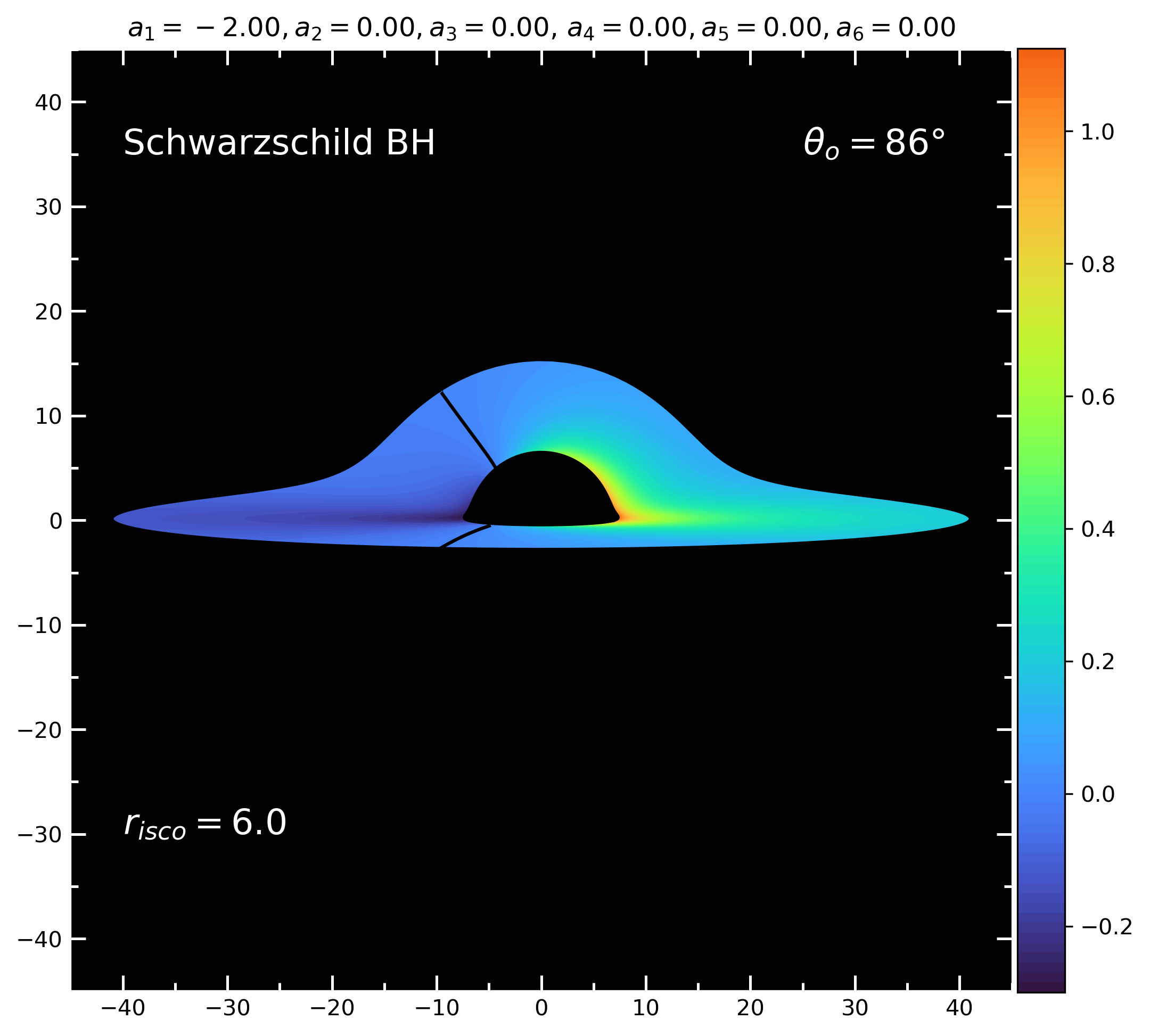}} \\
        \end{tabular}
        \caption{Similar to Figure \ref{fig:redshift1} for Reissner-Nordstr\"om and Schwarzschild black holes.}
        \label{fig:redshift2}
    \end{figure*}

    In Figs. \ref{fig:redshift1} and \ref{fig:redshift2}, we present the redshift distribution $z=g-1$ for different spacetime metrics employing the analytic approximation \eqref{eq:our_o6}. Specifically, in Fig. \ref{fig:redshift1}, we depict the redshift distributions of braneworld and Einstein-Maxwell-dilaton black holes, considering various inclinations of the accretion disk. Redshift values tend toward red at their maximum, while they tend toward dark blue at their minimum. Additionally, the contour for $z=0$, represented by the black line in the figures, is shown. Circular orbits vary from $r=r_{\text{isco}}$ to $r=40$. Similar plots are displayed in Fig. \ref{fig:redshift2} for Schwarzschild and Reissner-Nordström spacetimes.
     
    Recognizing that the outflow from the accretion disk differs with respect to the observed flux by a factor $g^{-4}$ due to the gravitational and relativistic Doppler effect \cite{Lumi:1979,Ellis:1971pg}, $F_{o}$ can be expressed as:
    
    \begin{equation} \label{eq:flux_obs}
        F_{o}(P')=\frac{F_{e}(P)}{g^{4}}.
    \end{equation}

    In summary, the bolometric flux observed by a distant observer is calculated using Eqs. \eqref{eq:flux_emit},\eqref{eq:rshift_g2} and \eqref{eq:flux_obs}. 
    
\begin{figure*}[htbp]
        \centering
        \begin{tabular}{ccc}
            \hspace{-6mm}
            {\includegraphics[scale=0.08,trim=0 0 0 0]{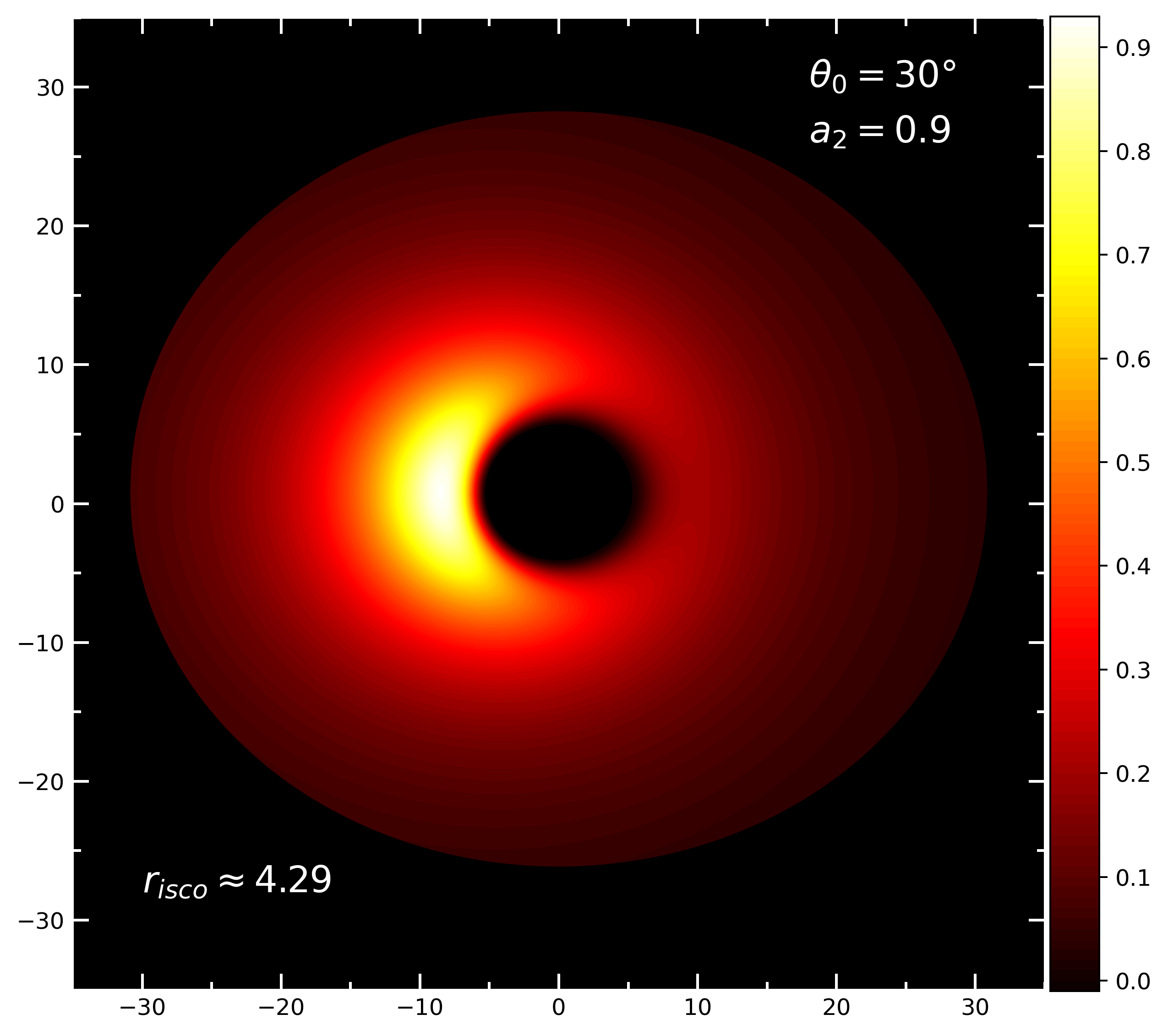}} &
            \hspace{3mm}  
            {\includegraphics[scale=0.08,trim=0 0 0 0]{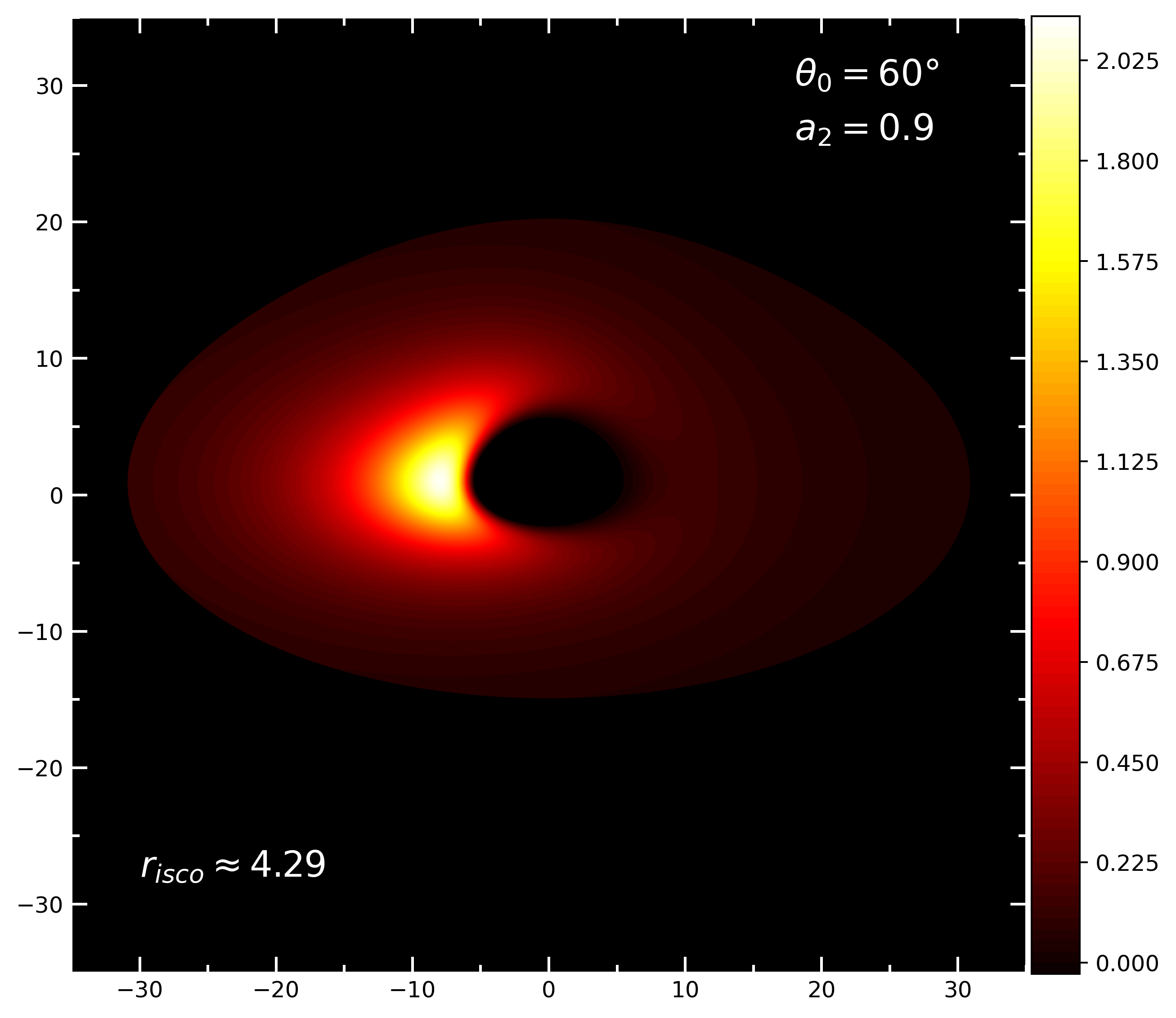}} & 
            \hspace{3mm}
            {\includegraphics[scale=0.08,trim=0 0 0 0]{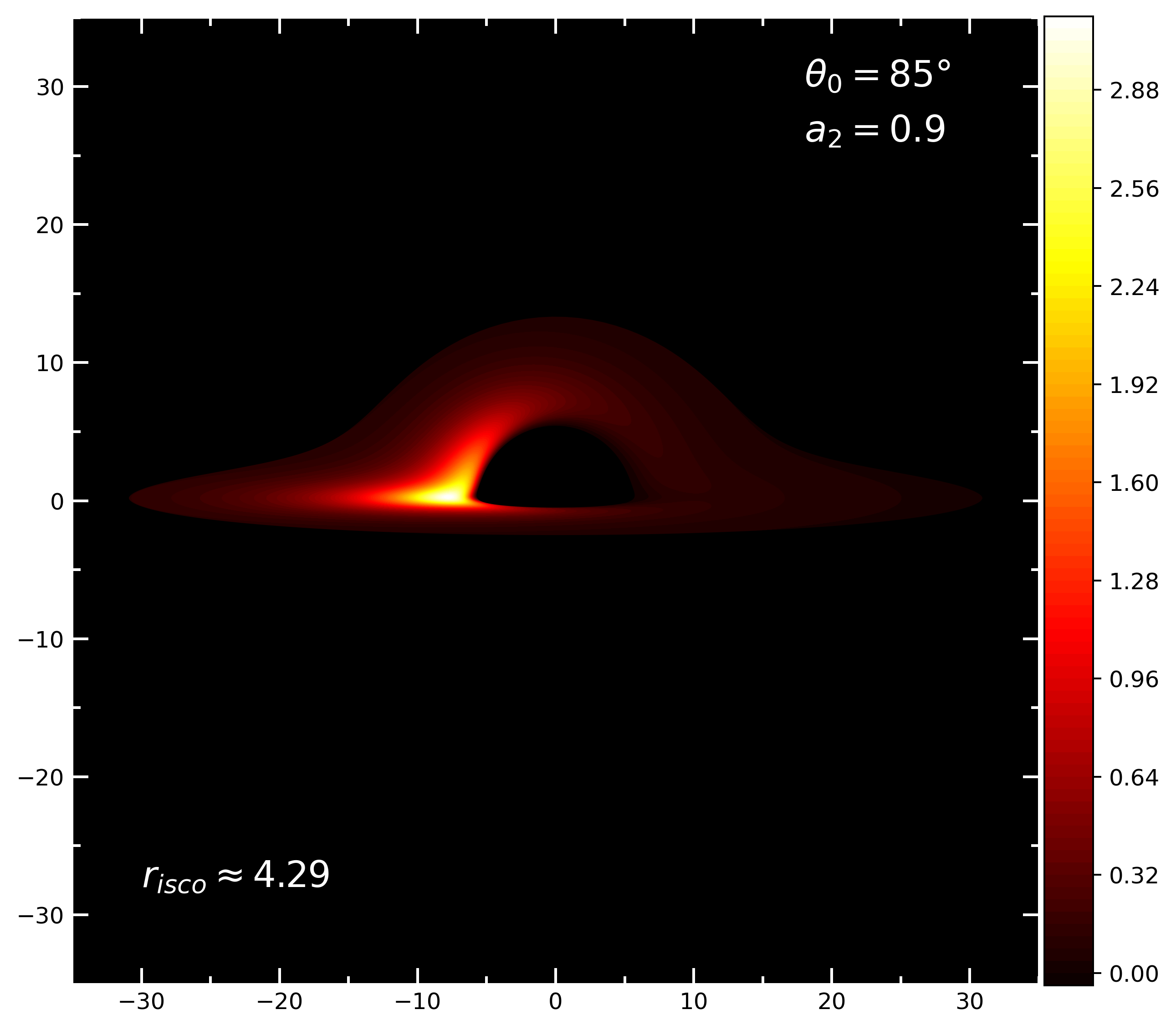}}\\
            \hspace{-6mm}
            {\includegraphics[scale=0.08,trim=0 0 0 0]{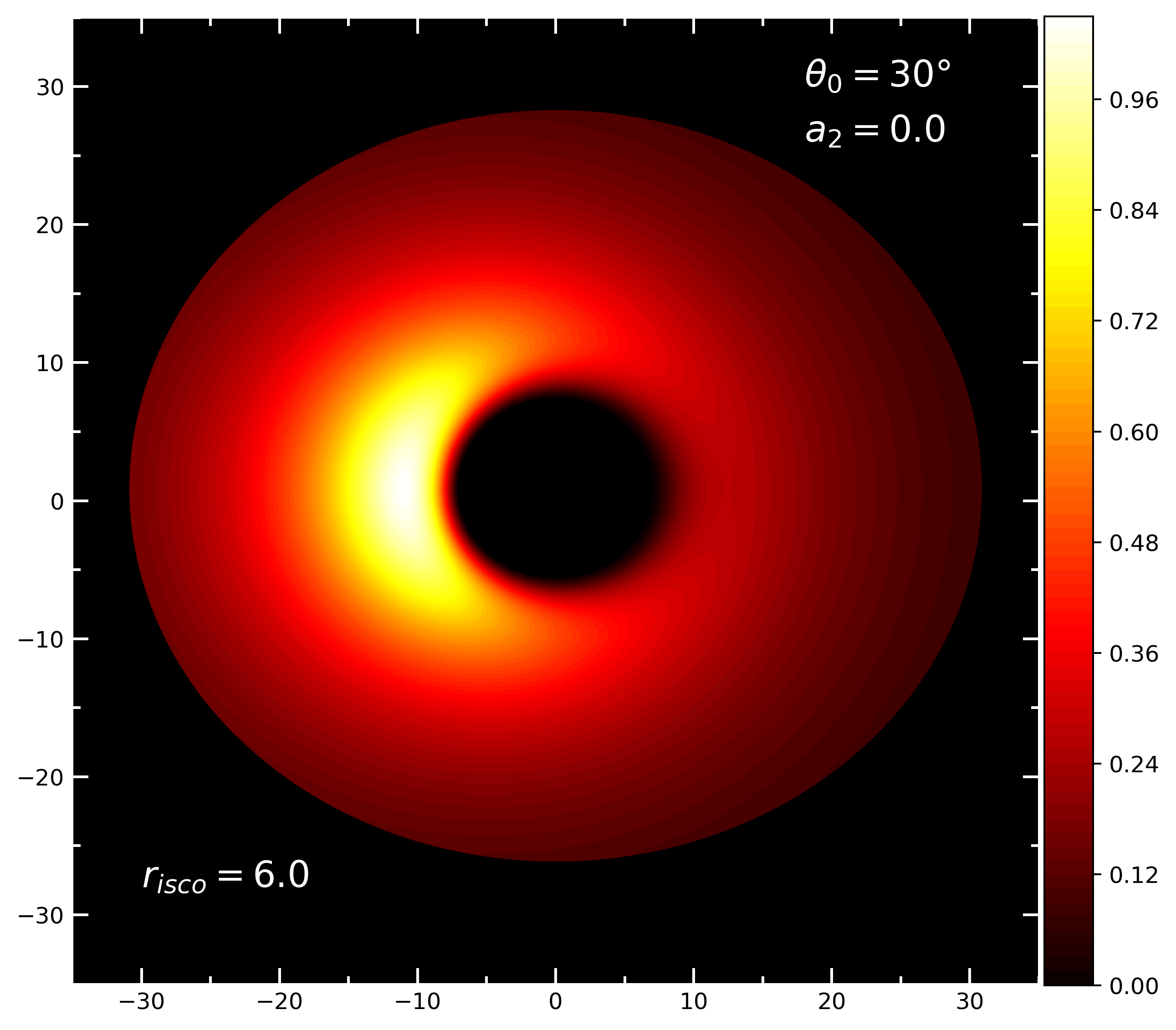}} &
            \hspace{3mm}
            {\includegraphics[scale=0.08,trim=0 0 0 0]{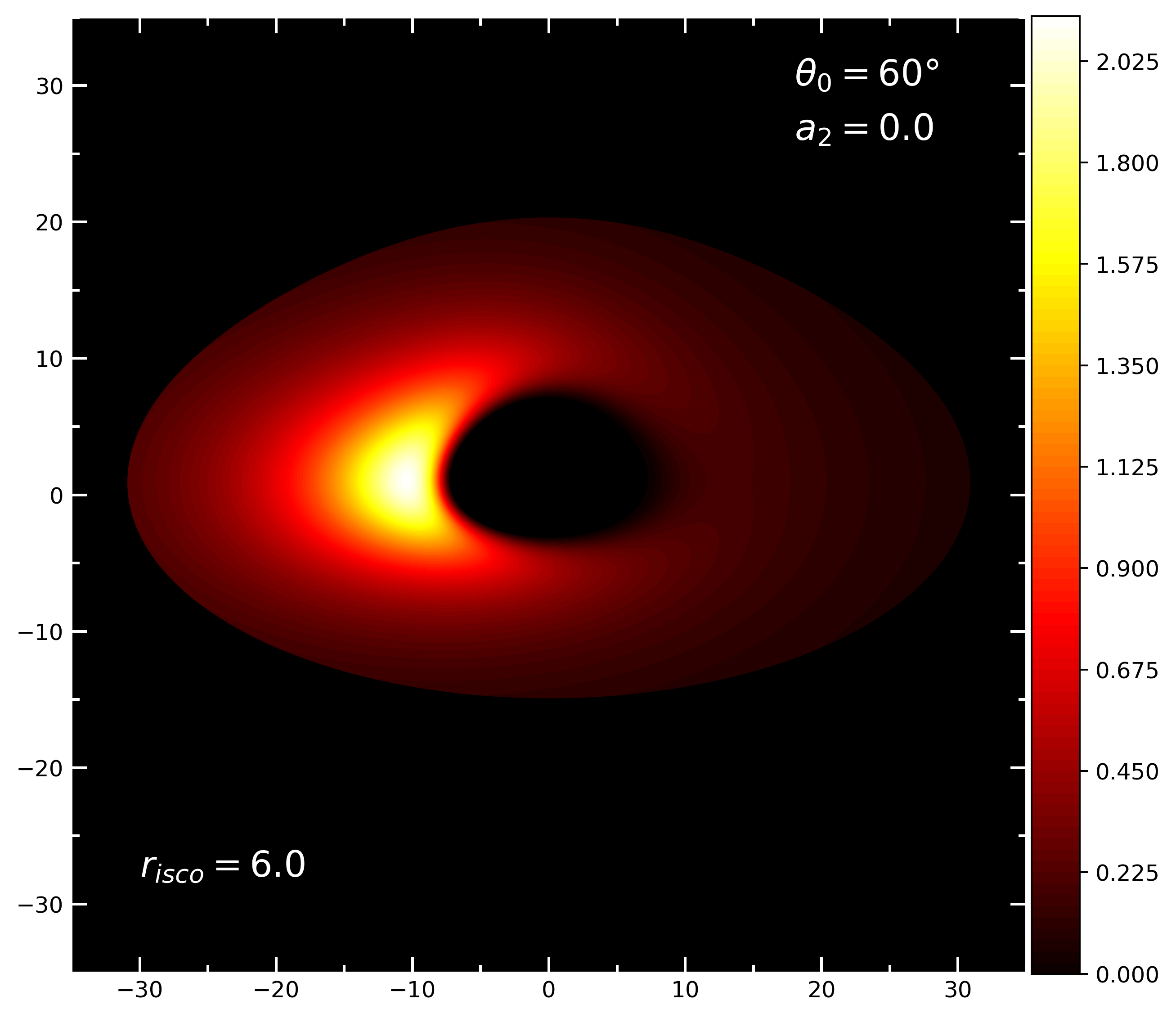}} & 
            \hspace{3mm}
            {\includegraphics[scale=0.08,trim=0 0 0 0]{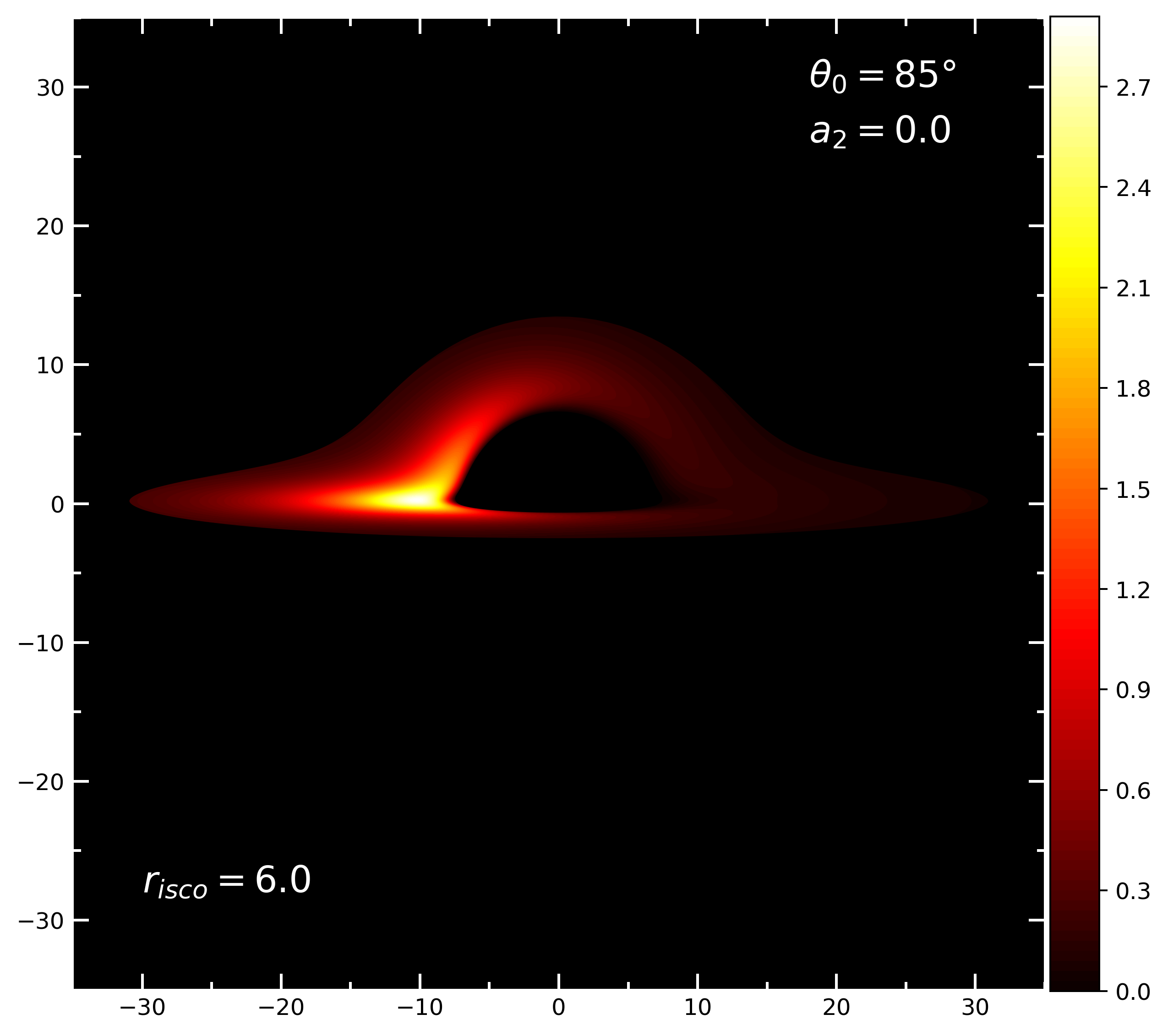}}\\
            \hspace{-6mm}
            {\includegraphics[scale=0.08,trim=0 0 0 0]{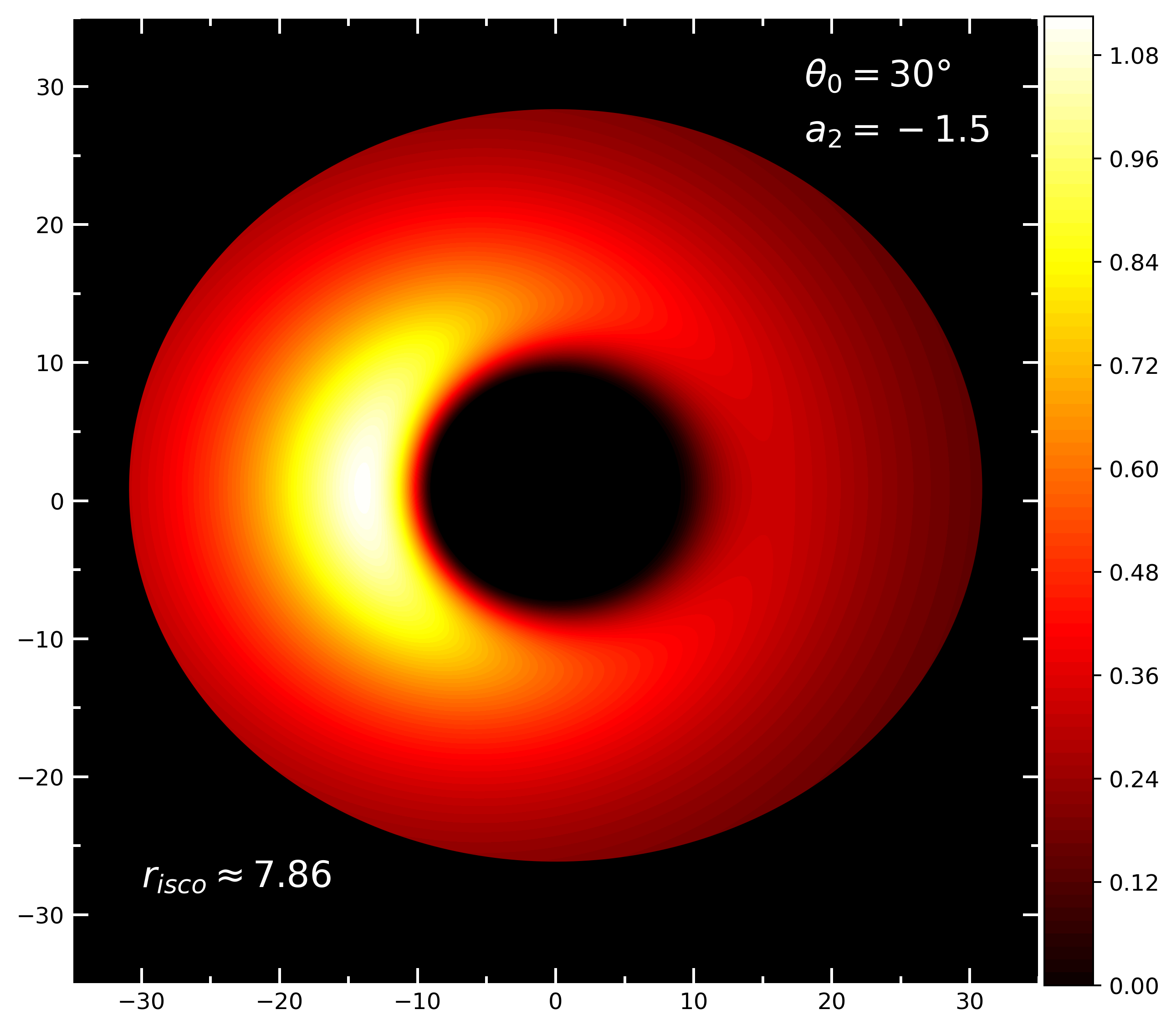}} &
            \hspace{3mm}
            {\includegraphics[scale=0.08,trim=0 0 0 0]{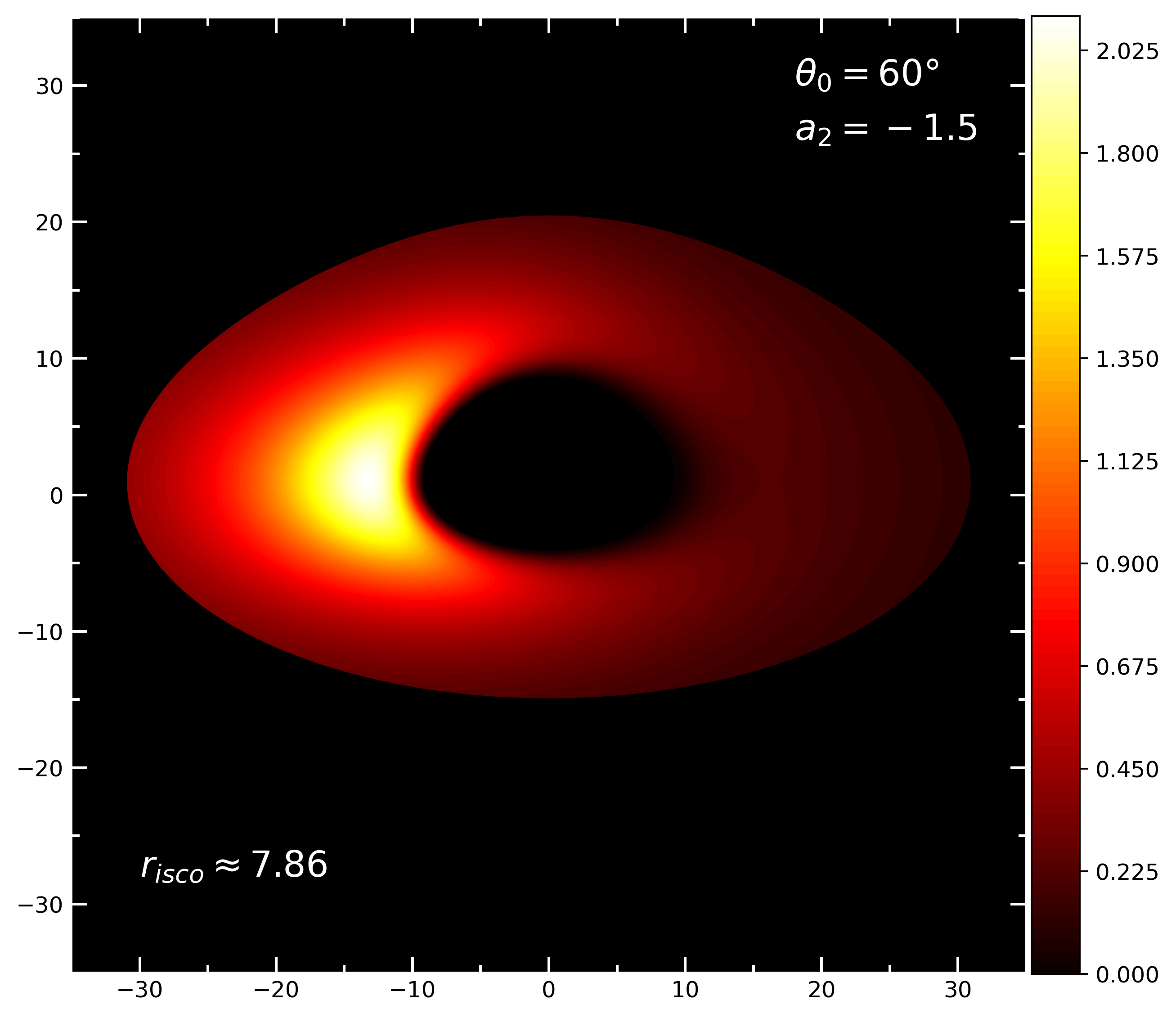}} &
            \hspace{3mm}
            {\includegraphics[scale=0.08,trim=0 0 0 0]{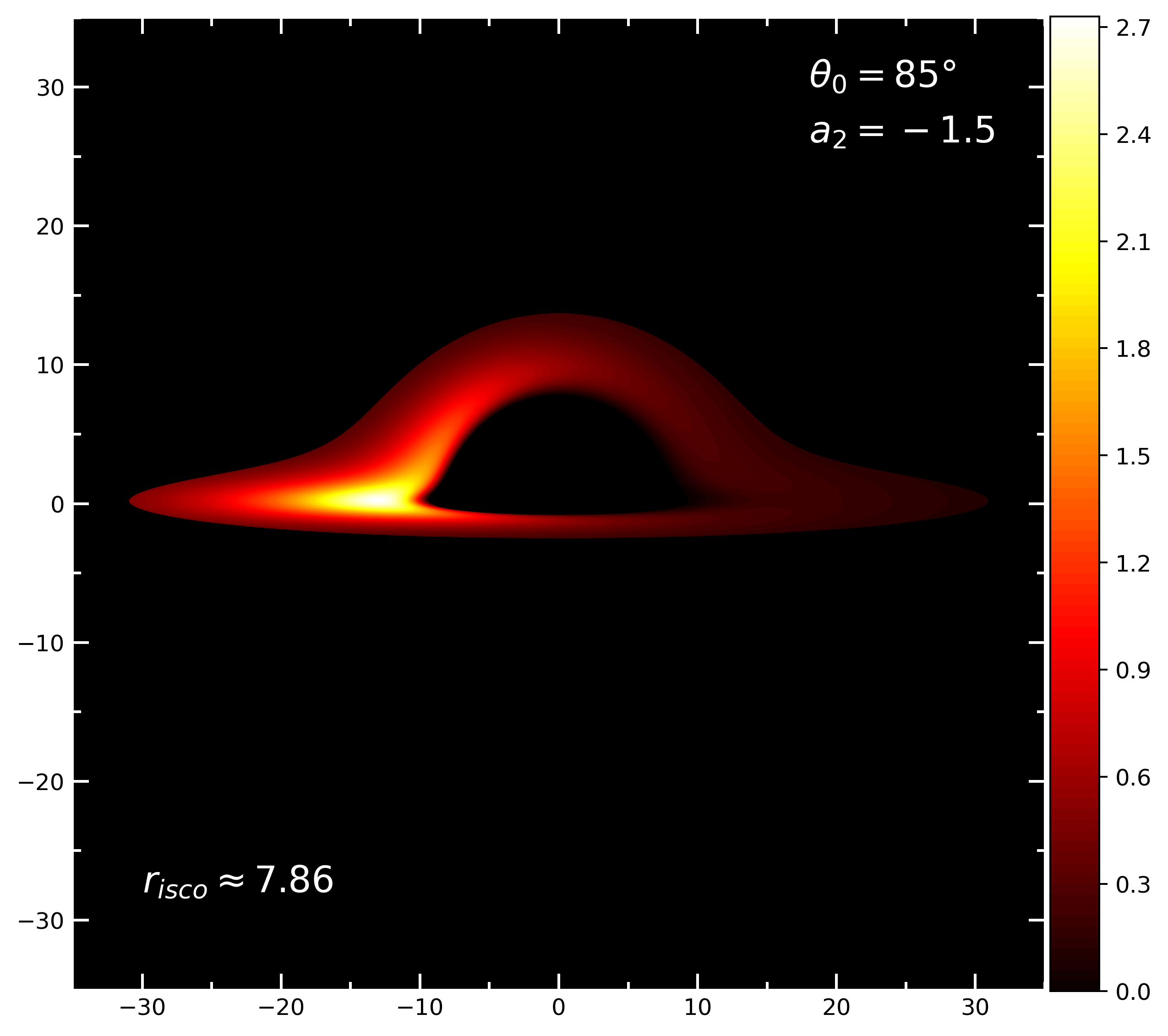}}\\
        \end{tabular}
        \caption{Bolometric flow profiles obtained by fixing the parameter $a_{1}=-2$, with varying values of $a_{2}$ and for different inclinations of the disk. These images were generated analytically using $10^{6}$ points $(R,\phi)$ from the inner orbit corresponding to $r=r_{\text{ISCO}}$ to the outermost orbit $r=30$.} 
        \label{fig:flux_o2}
    \end{figure*}

    Let us apply this formalism to the construction of analytic images of thin accretion disks whose flux is described by the Novikov-Page-Thorne model. For the metric model, let us consider a spacetime as Eq.\eqref{eq:METRIC} with \begin{equation}
        A(r)=1-\frac{2M}{r}+\frac{a_2}{r^2}.
    \end{equation}
    The analytical expression for the emitted flux is provided in Appendix \ref{app:B}. Varying the parameter $a_2$ (in units of $M^2$) leads to the analysis of different flow profiles corresponding to various disk inclinations, as illustrated in Figure \ref{fig:flux_o2}, where each flux profile has been normalized to the maximum value of $\tfrac{F_{e}}{dM/dt/4\pi}$. It is evident that, for a fixed value of $a_2$, the intensity increases with the inclination angle $\theta_{o}$. Furthermore, it is noticeable that the brightness predominantly accumulates on the left side of the image, attributed to the blueshift resulting from the Doppler effect induced by the rotating particles of the accretion disk, as shown in Figs. \ref{fig:redshift1} and \ref{fig:redshift2}. Additionally, it is worth noting that the innermost stable circular orbit described by Eq. \eqref{eq:r_isco} increases as  $a_2$ decreases, reflecting the gravitational effect exerted by the parameter $a_2$; a larger value of $r$ would be necessary to maintain stable circular orbits. Consequently, for this same reason, the maximum emitted value decreases. Specifically, for $a_2$ equal to $0.9$, $0$, and $-1.5$, the maximum value of $\tfrac{F_{e}}{dM/dt/4\pi}$ is $4.20 \times 10^{-4}$, $1.72 \times 10^{-4}$, and $7.84 \times 10^{-5}$, respectively, consistent with \cite{Lumi:1979} for the special case of $a_2=0$, corresponding to a Schwarzschild black hole.

    Before concluding this section, it is worth noting that the same analytical techniques can also be applied to consider other models of the accretion disk temperature profile. For instance, in \cite{Boero:2021afh} and \cite{Boero2}, by employing a two-temperature model and a thin disk, and through the numerical resolution of geodesic deviation equations of null congruences in a Kerr spacetime, images of disks have been obtained that are consistent with the images obtained by the Event Horizon Telescope for M87* and SgrA*.
    Also note that in that reference, numerical integration is performed from the observer toward the black hole. In particular, the authors observe numerically that for inclinations $\theta_0$ very close to $\pi/2$, those light ray trajectories that reach the disc are almost aligned along the line in the equatorial plane that is opposite to the line of sight. This result is consistent with what is directly obtained from Eqs. \eqref{cosvarp} or \eqref{sinvarp}.

\subsection{Polarization}\label{subs:pol}
    Let us now focus on the analytical study of polarization for a synchrotron emission model due to the charged gas orbiting (and falling) into the black hole with different distributions of magnetic field lines. To accomplish this, we will closely follow the pioneering framework developed in \cite{EventHorizonTelescope:2021btj} for a Schwarzschild black hole, where they used the Beloborodov approximation.
    
    Again, we consider the geometry depicted in Fig. \ref{fig:framework}. Specifically, we direct our attention to a discrete fluid element $P$ of the accretion disk. As mentioned earlier, this element has coordinates $(R,\phi)$. Let us examine the trajectory of a null geodesic originating from point $P$ and extending toward the asymptotic observer. The trajectory of the photon is confined to the $\overline{POO'}$ plane, as illustrated by the blue plane in Fig. \ref{fig:framework}, where Cartesian coordinates $(x',y',z')$ are established. Within this framework, the $x'$-axis aligns parallel to the line $\overline{OP}$, while the distant observer $O'$ resides within the $x'z'$-plane. We complement this coordinate system with the $t'$-axis aligned parallel to the $t$-axis. Following the convention outlined in \cite{EventHorizonTelescope:2021btj}, we denote this established coordinate system, along with its associated orthogonal tetrad $\{\vb*{\hat{t}}',\vb*{\hat{x}}',\vb*{\hat{y}}',\vb*{\hat{z}}'\}$ at $P$, as the $G$-frame.

    In this frame, the 4-momentum $\vb*{k}$ of the photon forms an emission angle $\alpha$ with the $\vb*{\hat{x}}'$-axis. We normalize the photon energy measured by the distant observer as $k_{t'}=-1$, hence the 4-momentum tetrad components of the emitted photon become:
    \begin{equation} \label{eq:gframe_photon}
        \begin{aligned}
            &k^{\hat{t}'}_{(G)}=\frac{1}{\sqrt{A(R)}}, \hspace{4mm} &k^{\hat{x}'}_{(G)}=k^{\hat{t'}}_{(G)}\cos\alpha, \\
            &k^{\hat{y}'}_{(G)}=0, \hspace{4mm} &k^{\hat{z}'}_{(G)}=k^{\hat{t'}}_{(G)}\sin\alpha,
        \end{aligned}   
    \end{equation}
   Note that the $y'$-component is zero since $\vb*{k}$ lies on the $x'z'$-plane. In the calculation of $\alpha$, we shall incorporate the approximations previously mentioned in Sec.\ref{sec:analyaprox}, in particular the Eq. \eqref{eq:our_o6}.
    
    We now consider a new Cartesian frame whose origin is also $O$, where the unit vector $\vb*{\hat{x}}$ is aligned along the line $\overline{OP}$, and $\vb*{\hat{y}}$ is tangent to the equatorial circle of the radius $R$ containing $P$. Additionally, $\vb*{\hat{z}}$ is orthogonal to the plane of the disk, so that this unit vector represents the direction of the coordinate axis $Z$. We complement this coordinate system with the $t$-axis and associated base vector $\vb*{t}$ at $P$ which agrees with $\vb*{t}'$. This frame is denoted as the $P$-frame, and it is explicitly associated with the rotational dynamics of the fluid element within the accretion disk. It is noteworthy that the $x'$-axis of the $G$-frame is aligned with the $x$-axis of the $P$-frame. The relationship between these frames is given by an angle $\xi$, representing a rotation about the $x$-axis given by \cite{EventHorizonTelescope:2021btj}:

    \begin{equation} \label{eq:xi_eqs}
          \cos\xi=\frac{\cos\theta_{o}}{\sin\psi}, \hspace{4mm} \sin\xi=\frac{\sin\theta_{o}\cos\phi}{\sin\psi}.    
    \end{equation}

    Using this rotation operation, the tetrad components of $\vb*{k}$ in the $P$-frame can be expressed as:

    \begin{equation} \label{eq:pframe_photon}
        \begin{aligned}
            &k^{\hat{t}}_{(P)}=\frac{1}{\sqrt{A(R)}}, \hspace{3mm} &k^{\hat{x}}_{(P)}=\frac{\cos\alpha}{\sqrt{A(R)}}, \\
            &k^{\hat{y}}_{(P)}=-\frac{\sin\xi\sin\alpha}{\sqrt{A(R)}}, \hspace{4mm} &k^{\hat{z}}_{(P)}=\frac{\cos\xi\sin\alpha}{\sqrt{A(R)}}.
        \end{aligned}   
    \end{equation}

    On the other hand, the velocity $\boldsymbol{\beta}$ of the fluid element situated at point $P$, as observed in the local $P$-frame, can be expressed as:
    
    \begin{equation} \label{eq:beta}
      \boldsymbol{\beta}=\beta(\cos\chi \ \vb*{\hat{x}}+\sin\chi \ \vb*{\hat{y}}),
    \end{equation}
    where the angle $\chi$ is measured with respect to the unit vector $\vb*{\hat{x}}$, increasing in a counter-clockwise direction, as illustrated in the Figure  \ref{fig:framework}.
        
   To calculate the radiation fields associated with the motion of charged particles, it is convenient to adopt a reference frame moving with the fluid, known as the fluid frame or $F$-frame, with the associated tetrad $(\vb*{\tilde{t}},\vb*{\tilde{x}},\vb*{\tilde{y}},\vb*{\tilde{z}})$.
To express the 4-momentum of the photon within the $F$-frame, a Lorentz boost between the $P$-frame and the $F$-frame must be employed. The resulting tetrad components of $\vb*{k}$ in the $F$-frame are expressed as follows \cite{EventHorizonTelescope:2021btj}:
    
    \begin{equation} \label{eq:fframe_photon}
        \begin{aligned}
            k^{\hat{\tilde{t}}}_{(F)}=&\gamma k^{\hat{t}}_{(P)}-\gamma\beta\cos\chi k^{\hat{x}}_{(P)}-\gamma\beta\sin\chi k^{\hat{y}}_{(P)}, \\
            k^{\hat{\tilde{x}}}_{(F)}=&-\gamma\beta\cos\chi k^{\hat{t}}_{(P)}+\left(1+(\gamma-1)\cos^{2}\chi \right)k^{\hat{x}}_{(P)} \\ &+ (\gamma-1)\cos\chi\sin\chi k^{\hat{y}}_{(P)},\\
            k^{\hat{\tilde{y}}}_{(F)}=&-\gamma\beta\sin\chi k^{\hat{t}}_{(P)}+ (\gamma-1)\cos\chi\sin\chi k^{\hat{x}}_{(P)} \\ &+\left(1+(\gamma-1)\sin^{2}\chi \right)k^{\hat{y}}_{(P)}, \\ k^{\hat{\tilde{z}}}_{(F)}=\hspace{0.5mm}&k^{\hat{z}}_{(P)},
        \end{aligned}
    \end{equation}
    with $\gamma=(1-\beta^2)^{-1/2}$ the Lorentz factor.

    Additionally, influenced by both the motion of the source and the gravitational field, the energy of the light observed by an observer undergoes a shift due to gravitational redshift and relativistic Doppler effect. This additional correction factor, denoted as $\delta$, is given by:
\begin{equation} \label{eq:delta}
\delta = \frac{E_{o}}{E_{(F)}} = \frac{1}{k^{\hat{t}}_{(F)}},
\end{equation}
where $E_{o}$ represents the photon energy as measured by the distant observer, previously set to 1, while $E_{(F)}$ denotes the energy emitted within the $F$-frame.

    Let us consider the magnetic field $\vb*{B}$ in the $F$-frame. Its components are:
    
    \begin{equation} \label{eq:field_b}
        \begin{aligned}
            &\vb*{B}=B_{r}\vb*{\tilde{x}} +B_{\phi}\vb*{\tilde{y}}+B_{z}\vb*{\tilde{z}}=\vb*{{B}}_{eq}+B_{z}\vb*{\tilde{z}},
        \end{aligned}  
    \end{equation}
    with \begin{equation}
        \vb*{{B}}_{eq}\equiv B_{r}\vb*{\tilde{x}} +B_{\phi}\vb*{\tilde{y}}=B_{eq}(\cos\eta \ \vb*{\tilde{x}} + \sin\eta \ \vb*{\tilde{y}}).
    \end{equation}
    
   Here, $\vb*{B}_{eq}$ represents the projection of $\vb*{B}$ onto the equatorial plane, and $\eta$ is measured with respect to the $\vb*{\tilde{x}}$ direction (See our Fig.\ref{fig:framework} or Fig. 2 of \cite{EventHorizonTelescope:2021btj}) . The explicit relation between these components of the magnetic field in the $F$-frame and the corresponding components in the $P$-frame are given in \cite{EventHorizonTelescope:2021btj}. The angle between the 3-vector $\vb*{k}_{(F)}$ and the magnetic field $\vb*{B}$, can  be expressed as:
    
    \begin{equation} \label{eq:sinzeta}
        \sin\zeta=\frac{|\vb*{k}_{(F)}\times\vb*{B}|}{|\vb*{k}_{(F)}||\vb*{B}|}.
    \end{equation}
    This factor becomes relevant in the computation of intensity. In the pursuit of determining the electric field $\vb*{E}$, calculating the polarization vector is essential, given its orthogonality to the plane defined by $\vb*{k}_{(F)}$ and $\vb*{B}$. Consequently, the 4-vector of polarization, denoted as $\vb*{f}$ is expressed in the $F$-frame  as:
     
    \begin{equation} \label{eq:fframe_pol}
        \begin{aligned}
            &f^{\hat{t}}_{(F)}=0, \\       &f^{i}_{(F)}=\frac{(\vb*{k}_{(F)}\times\vb*{B})^{i}}{|\vb*{k}_{(F)}|}, \hspace{4mm} i=\hat{x},\hat{y},\hat{z}
        \end{aligned}  
    \end{equation}
    and satisfies
    
    \begin{equation} \label{eq:pol_cond}
        f^{\mu}k_{\mu}=0, \hspace{6mm} f^{\mu}f_{\mu}=\sin^{2}\zeta |\vb*{B}|^{2}.
    \end{equation}
    
   To return to the $P$-frame, we apply the inverse Lorentz transformations, which yield: 
    \begin{equation} \label{eq:pframe_pol}
        \begin{aligned}
            f^{\hat{t}}_{(P)}=&\gamma f^{\hat{\tilde{t}}}_{(F)} +\gamma\beta\cos\chi f^{\hat{\tilde{x}}}_{(F)} +\gamma\beta\sin\chi f^{\hat{\tilde{y}}}_{(F)}, \\
            f^{\hat{x}}_{(P)}=&\gamma\beta\cos\chi f^{\hat{\tilde{t}}}_{(F)}+\left(1+(\gamma-1)\cos^{2}\chi \right)f^{\hat{\tilde{x}}}_{(F)} \\ &+ (\gamma-1)\cos\chi\sin\chi f^{\hat{\tilde{y}}}_{(F)},\\
            f^{\hat{y}}_{(P)}=&\gamma\beta\sin\chi f^{\hat{\tilde{t}}}_{(F)}+ (\gamma-1)\cos\chi\sin\chi f^{\hat{\tilde{x}}}_{(F)} \\ &+\left(1+(\gamma-1)\sin^{2}\chi \right)f^{\hat{\tilde{y}}}_{(F)}, \\ f^{\hat{z}}_{(P)}=\hspace{0.5mm}&f^{\hat{\tilde{z}}}_{(F)}.
        \end{aligned}
    \end{equation}
   
The polarization vector $\vb*{f}$ not only satisfies Eqs. \eqref{eq:pol_cond}, but also must undergo parallel transport along the null geodesic, i.e., $k^{\mu}\nabla_{\mu}f^{\nu}=0$. In the case of a Schwarzschild metric, one can take advantage of the existence of a conserved quantity $\kappa_{\text{WP}}$, known as the Walker-Penrose constant \cite{Walker:1970un}, along the null geodesic, i.e., $k^\mu\nabla_\mu \kappa_{\text{WP}}=0$. This constant generally takes the form \cite{Chandra:1983}:
\begin{equation}\label{eq:walpengen}
    \kappa_{\text{WP}} = 2[(\vb*{k}\cdot\vb*{l})(\vb*{f}\cdot\vb*{n}) - (\vb*{k}\cdot\vb*{m})(\vb*{f}\cdot\vb*{\bar{m}})]\Psi^{-1/3}_2,
\end{equation}
where $\Psi_2:=C_{\alpha\beta\gamma\delta}l^{\alpha}m^{\beta}\bar{m}^{\gamma}n^{\delta}$, with $C_{\alpha\beta\gamma\delta}$ being the Weyl tensor, and $\{\vb*{l},\vb*{n},\vb*{m},\vb*{\bar{m}}\}$ being the principal null tetrad associated with these type D spacetimes.

For a Schwarzschild metric, in terms of the Schwarzschild coordinate components of $\vb*{k}$ and $\vb*{f}$, and omitting the global constant factor $M^{-1/3}$, it can be rewritten in the simple form:
\begin{equation} \label{eq:walkpenrose_const_or}
    \begin{aligned}
        &\kappa = \kappa_{1}+i\kappa_{2}, \\
        &\kappa_{1} = r(k^{t}f^{r}-k^{r}f^{t}), \\ 
        &\kappa_{2} = -r^{3}\sin\theta(k^{\phi}f^{\theta}-k^{\theta}f^{\phi}),
    \end{aligned}  
\end{equation}
which, when evaluated at $P$ and in terms of the tetrad components of $\vb*{k}$ and $\vb*{f}$ in the $P$-frame, reduces to:
    \begin{equation} \label{eq:walkpenrose_const}
        \begin{aligned}
            &\kappa_{1}=R\left(k^{\hat{t}}_{(P)}f^{\hat{x}}_{(P)}-k^{\hat{x}}_{(P)}f^{\hat{t}}_{(P)}\right), \\ &\kappa_{2}=R\left(k^{\hat{y}}_{(P)}f^{\hat{z}}_{(P)}-k^{\hat{z}}_{(P)}f^{\hat{y}}_{(P)}\right).
        \end{aligned}  
    \end{equation}

   Unfortunately, while every static and spherically symmetric spacetime is automatically of type D, in general, the Walker-Penrose constant, as defined by Eq. \eqref{eq:walpengen}, is not necessarily a conserved quantity, as such spacetimes will not be solutions to the vacuum Einstein equations. However, even in these more general spacetimes, the quantities in Eqs. \eqref{eq:walkpenrose_const_or} are constants of motion, as demonstrated by the existence of Killing-Yano tensors \cite{Howarth2000,PhysRevD.106.104024} or can be derived from an argument inspired by Pineault's work \cite{1977MNRAS.179..691P}, as discussed in \cite{Gallo24}.

Our goal is to determine the polarized electric field in the distant observer's frame $X'Y'$ by applying the conservation of $\kappa$. This involves computing $\kappa$ at the emission point $P$ using Eqs. \eqref{eq:walkpenrose_const_or}, followed by utilizing a known relation between the electric fields (proportional to $\vb*{f}$) and $\kappa$ at the observation point $P'$. Specifically, adhering to the original normalization as described in Eq. \eqref{eq:pol_cond}, the electric field $\vb*{E}$ at point $P'$ in the $X'Y'$-frame is expressed as follows \cite{Li:2008zr,Himwich:2020msm,Aimar:2023vcs}: 
    \begin{equation} \label{eq:efield}
        \begin{aligned}
            E_{X'}&=\frac{Y'\kappa_{2}+X'\kappa_{1}}{X'^{2}+Y'^{2}}, \\
            E_{Y'}&=\frac{Y'\kappa_{1}-X'\kappa_{2}}{X'^{2}+Y'^{2}}, \\
            E_{X'}^{2}+E_{Y'}^{2}&=\sin^{2}\zeta|\vb*{B}|^{2}.
        \end{aligned}  
    \end{equation}
    An alternative and simple geometrical interpretation of these relations will be presented elsewhere \cite{Gallo24}.
    
If we assume that the observed polarized intensity $\mathrm{P}$ arises from synchrotron radiation emitted from point $P$, and consider a model of an optically and geometrically thin accretion disk, it can be approximated as \cite{EventHorizonTelescope:2021btj, Gelles:2021kti}:
\begin{equation} \label{eq:i_aprox}
    \mathrm{P}=\delta^{3+\alpha_{\nu}}l_{p}|\vb*{B}|^{1+\alpha_{\nu}}\sin^{1+\alpha_{\nu}}\zeta,
\end{equation}
where $\delta$ represents the Doppler factor \eqref{eq:delta}, $\alpha_\nu$ denotes the spectral index, and $l_p$ stands for the geodesic path length through the emitted region. The spectral index depends on the disk properties and the ratio of emitted energy to the electron temperature. In this study, we set it to $1$ \cite{EventHorizonTelescope:2021btj, Gelles:2021kti}. Additionally, $l_p$ can be expressed as:
\begin{equation} \label{eq:pathlength}
    l_{p}=\frac{k^{\hat{t}}_{(F)}}{k^{\hat{z}}_{(F)}}H,
\end{equation}
where $H$ denotes the thickness of the disk, which for simplicity can be assumed as a constant equal to $1$.    Therefore, the components of the observed polarization vector in the $X'Y'$-frame are
    \begin{equation} \label{eq:efield_obs}
        \begin{aligned}
            E_{X',obs}&=\delta^{2}l_{p}^{1/2}E_{X'}, \\
            E_{Y',obs}&=\delta^{2}l_{p}^{1/2}E_{Y'},
        \end{aligned}  
    \end{equation}
    so, the total polarization intensity and the electric vector position angle (EVPA) are expressed:     

    \begin{equation} \label{eq:stokes_1}
        \begin{aligned}
            \mathrm{P}&=E_{X',\text{obs}}^{2}+E_{Y',\text{obs}}^{2},\\
            \text{EVPA}&=\arctan\frac{E_{Y',\text{obs}}}{E_{X',\text{obs}}}=\frac{1}{2}\arctan\frac{U}{Q},
        \end{aligned}  
    \end{equation}
    where the Stokes parameters $Q$ and $U$ are given by
    
    \begin{equation} \label{eq:stokes_2}
        \begin{aligned}
            Q&= E_{X',\text{obs}}^{2}-E_{Y',\text{obs}}^{2}, \\
            U&=2E_{X',\text{obs}}E_{Y',\text{obs}}. \\
        \end{aligned}  
    \end{equation}

    \begin{figure*}[htbp]
        \centering
        \begin{tabular}{ccc}
            \hspace{-6mm}
            {\includegraphics[scale=0.304,trim=0 0 0 0]{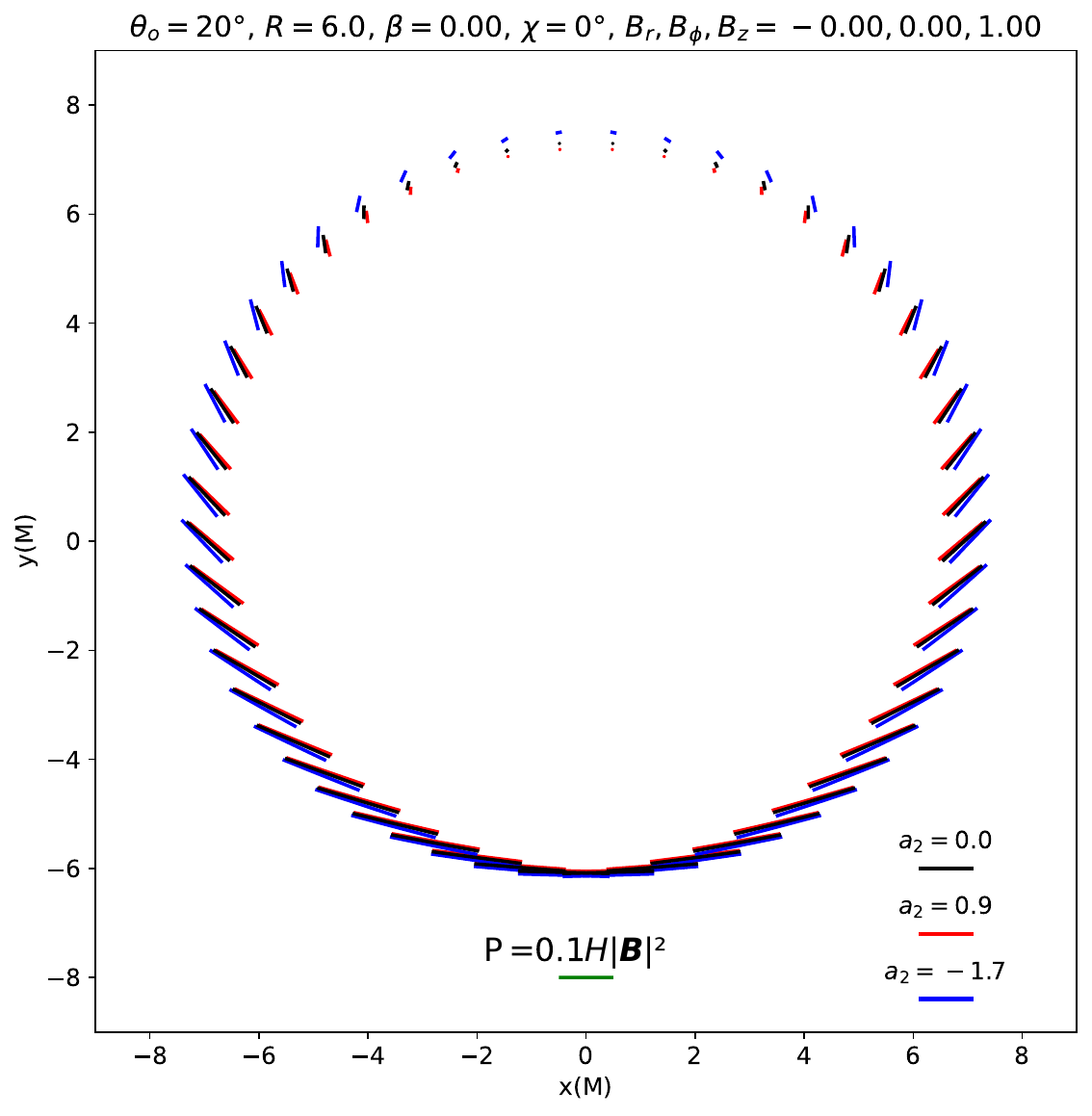}} &
            \hspace{2mm} 
            {\includegraphics[scale=0.304,trim=0 0 0 0]{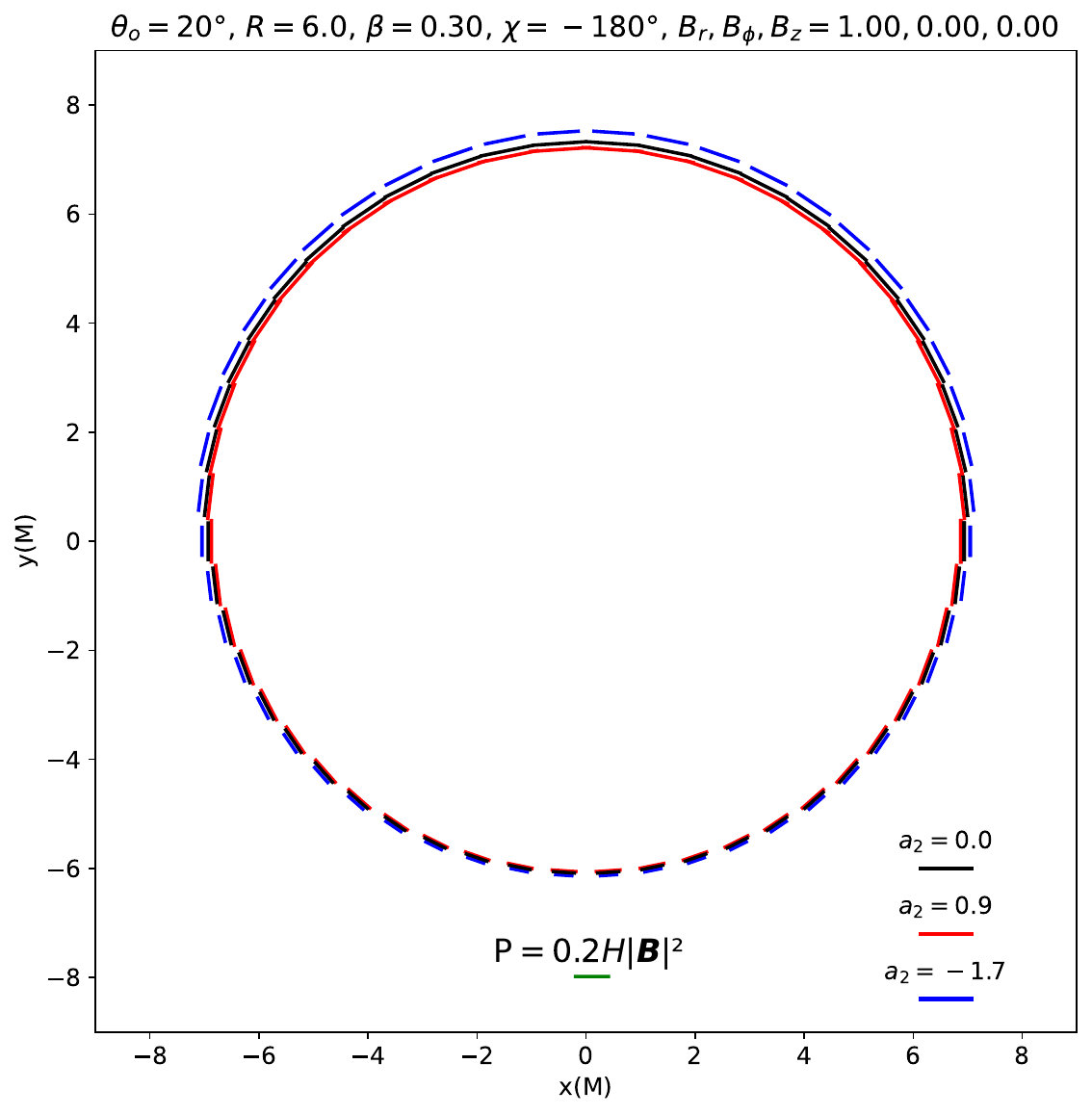}} &
            \hspace{2mm}  
            {\includegraphics[scale=0.304,trim=0 0 0 0]{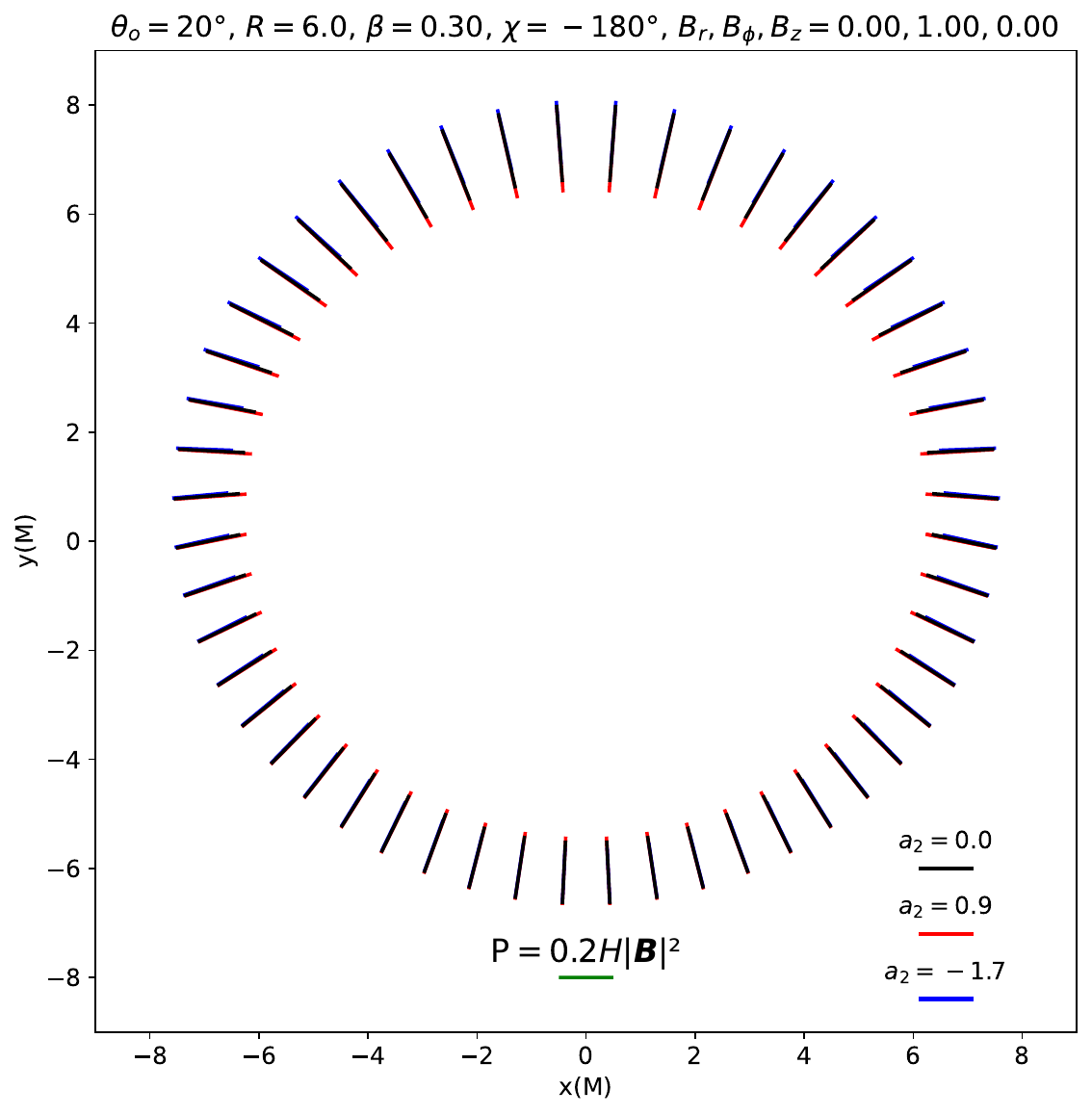}} 
            \hspace{2mm}\\
            \hspace{-6mm}
            {\includegraphics[scale=0.304,trim=0 0 0 0]{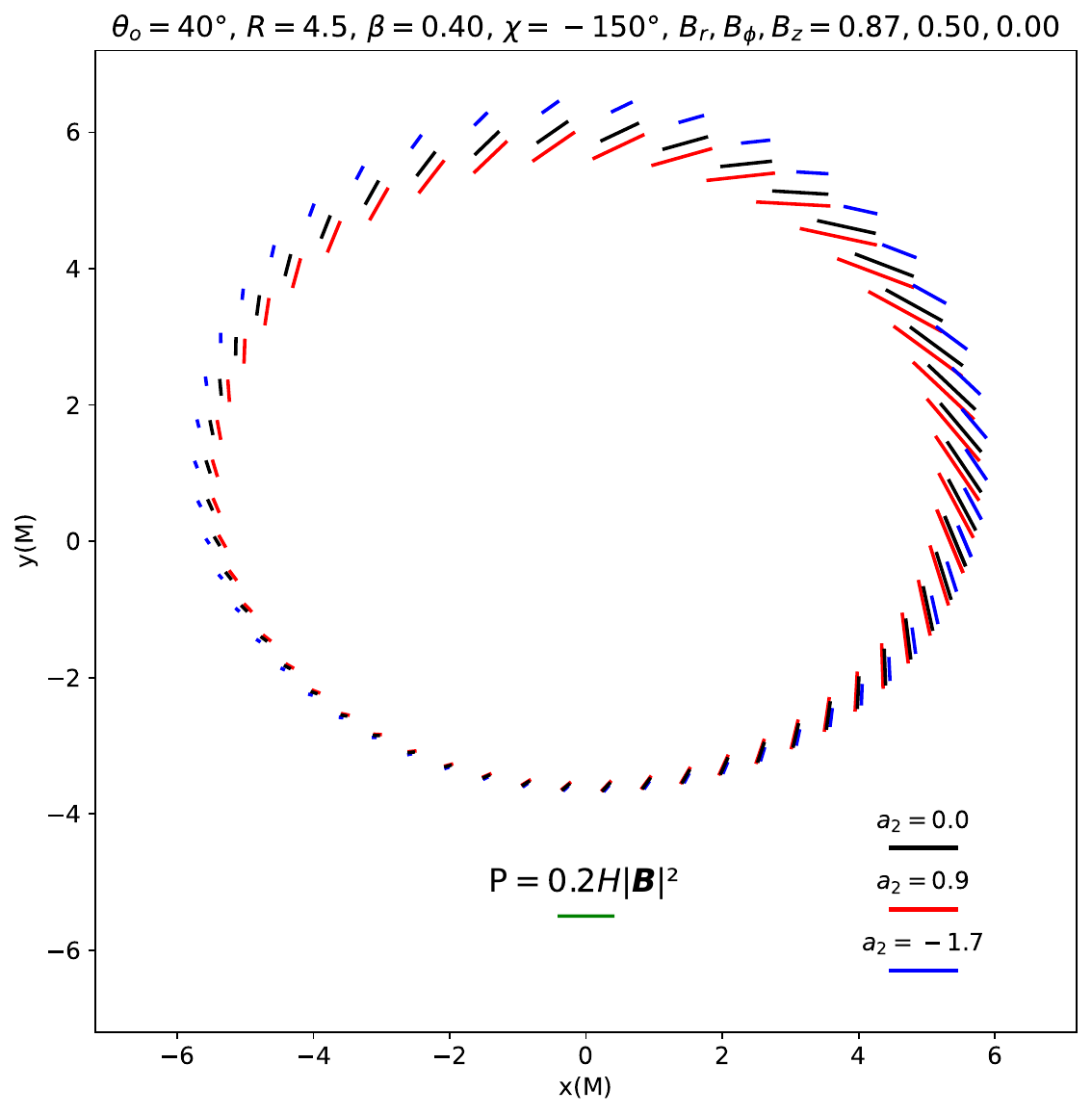}} &
            \hspace{2mm}
            {\includegraphics[scale=0.304,trim=0 0 0 0]{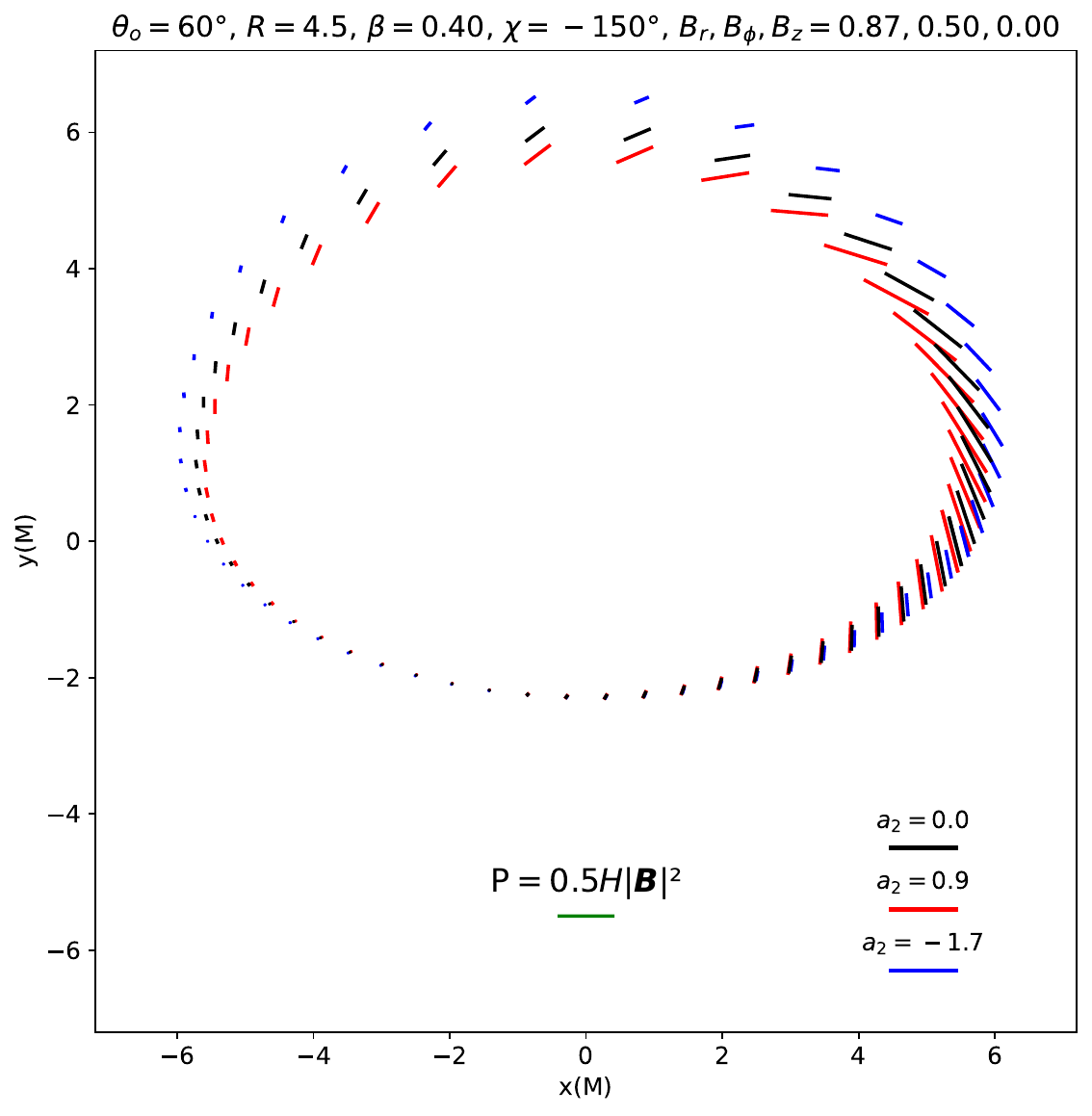}} &
            \hspace{2mm}
            {\includegraphics[scale=0.304,trim=0 0 0 0]{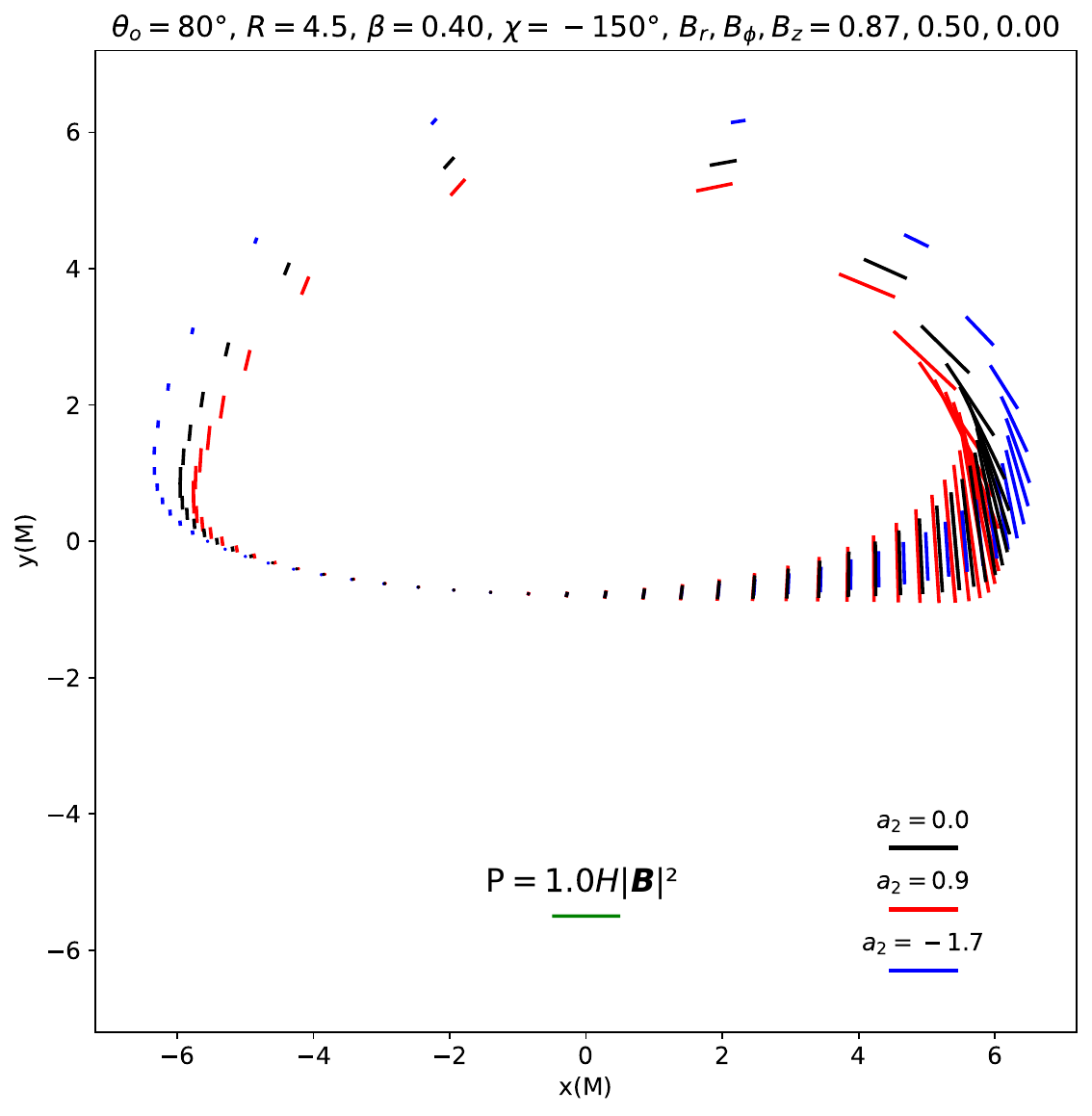}}
            \hspace{2mm}\\
            \hspace{-6mm}
            {\includegraphics[scale=0.304,trim=0 0 0 0]{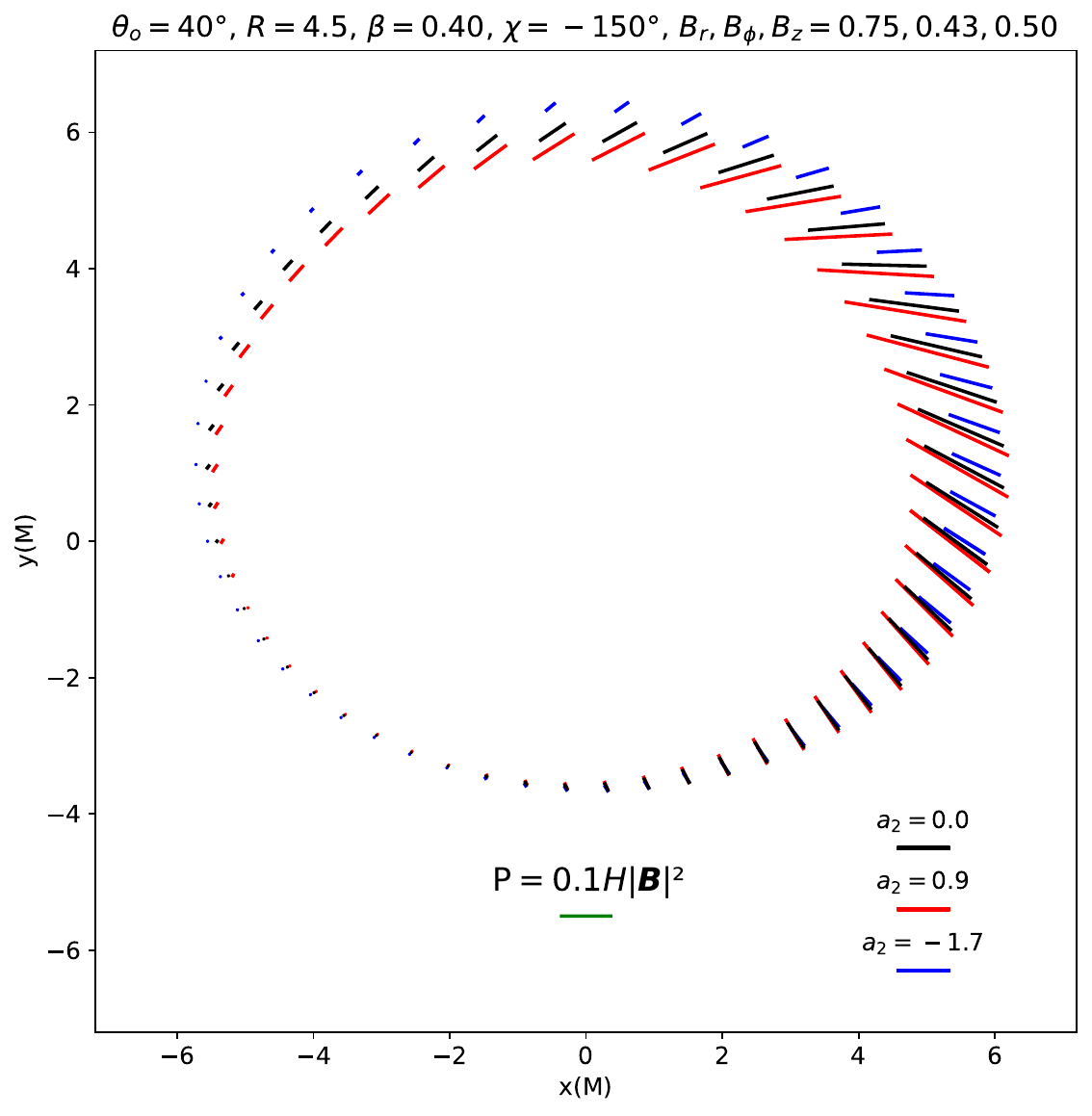}} &
            \hspace{2mm}
            {\includegraphics[scale=0.304,trim=0 0 0 0]{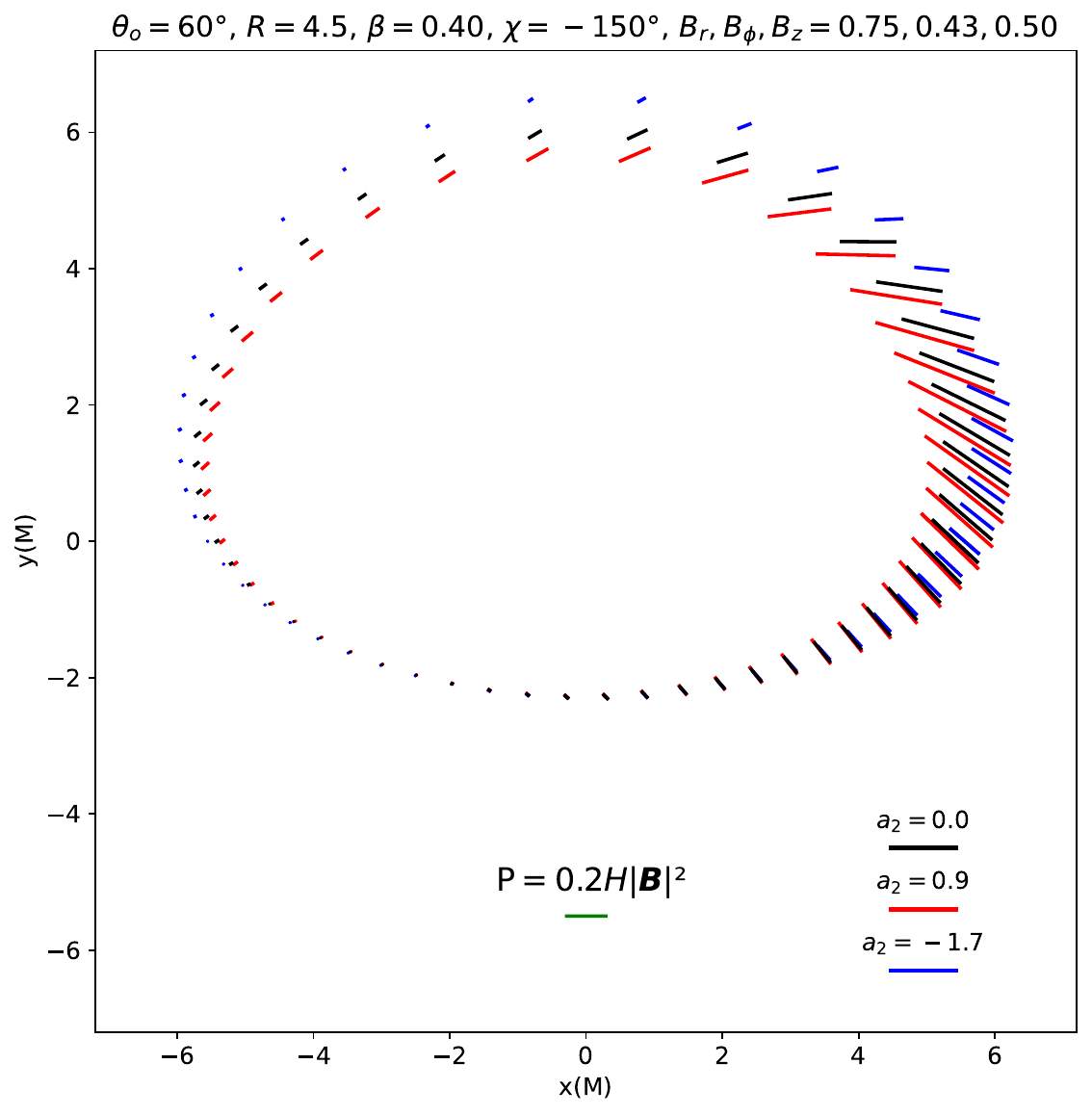}} &
            \hspace{2mm}
            {\includegraphics[scale=0.304,trim=0 0 0 0]{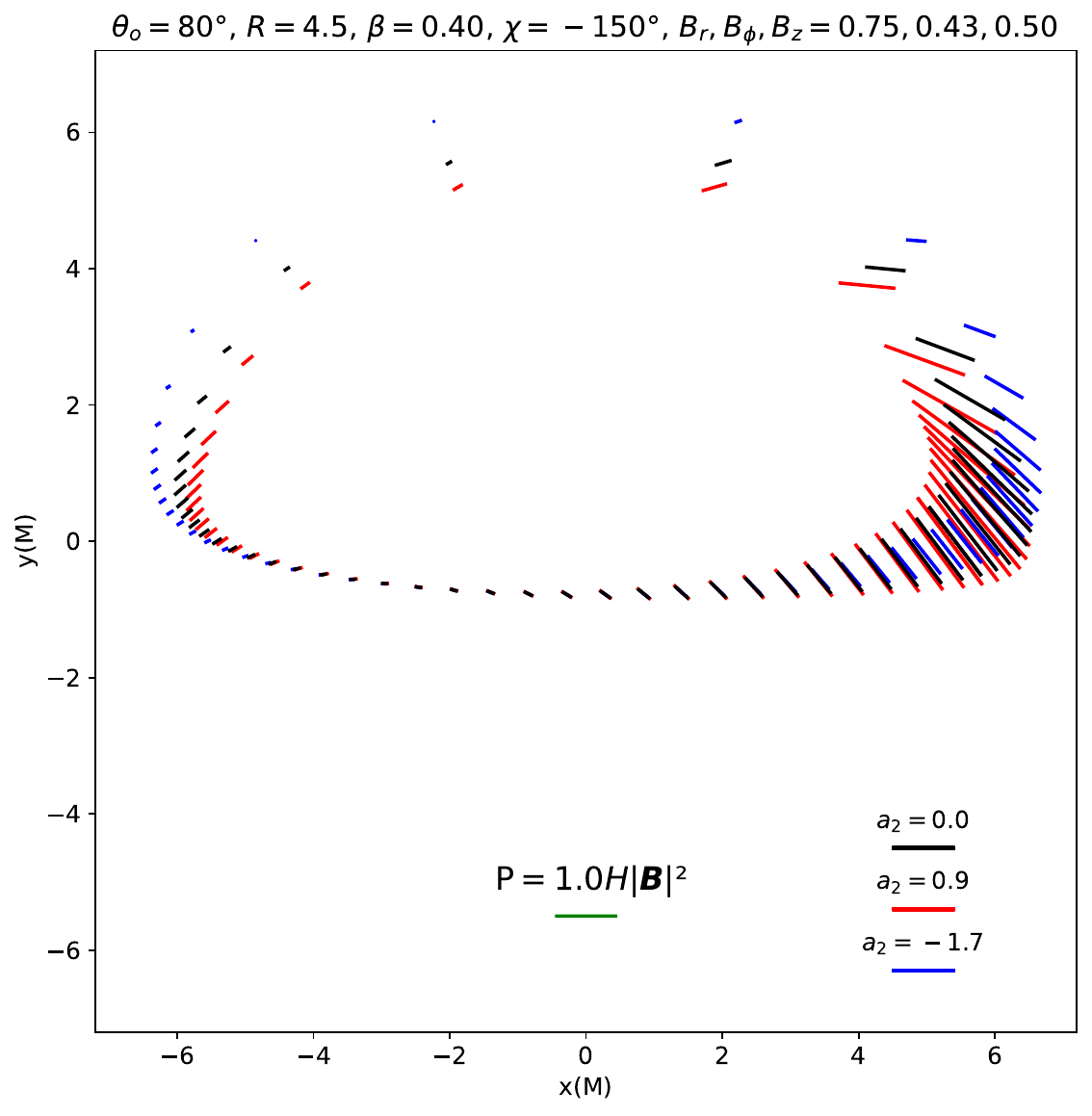}}
            \hspace{2mm}\\
        \end{tabular}
        \caption{Polarization patterns for different magnetic field configurations. In the top row, patterns are shown for a vertical $B_z=1$ (top left), radial $B_r=1$ (top middle), and azimuthal $B_\phi=1$ (top right) magnetic field, each at an emission radius $R=6$ and specifying the disk's inclination angle. The middle panel displays patterns for an equatorial magnetic field ($B_{\text{eq}}=1$) with varying $\theta_o$ values at $R=4.5$. In the bottom row, patterns are presented for $R=4.5$ with a magnetic field containing all field components. Different values of $a_2$ are considered in all cases, with $a_i=0$ for $i>2$.
        }
        \label{fig:pol_patterns}
    \end{figure*}

    \begin{figure*}[htbp]
        \centering
        \begin{tabular}{ccc}
            \hspace{-6mm}
            {\includegraphics[scale=0.29,trim=0 0 0 0]{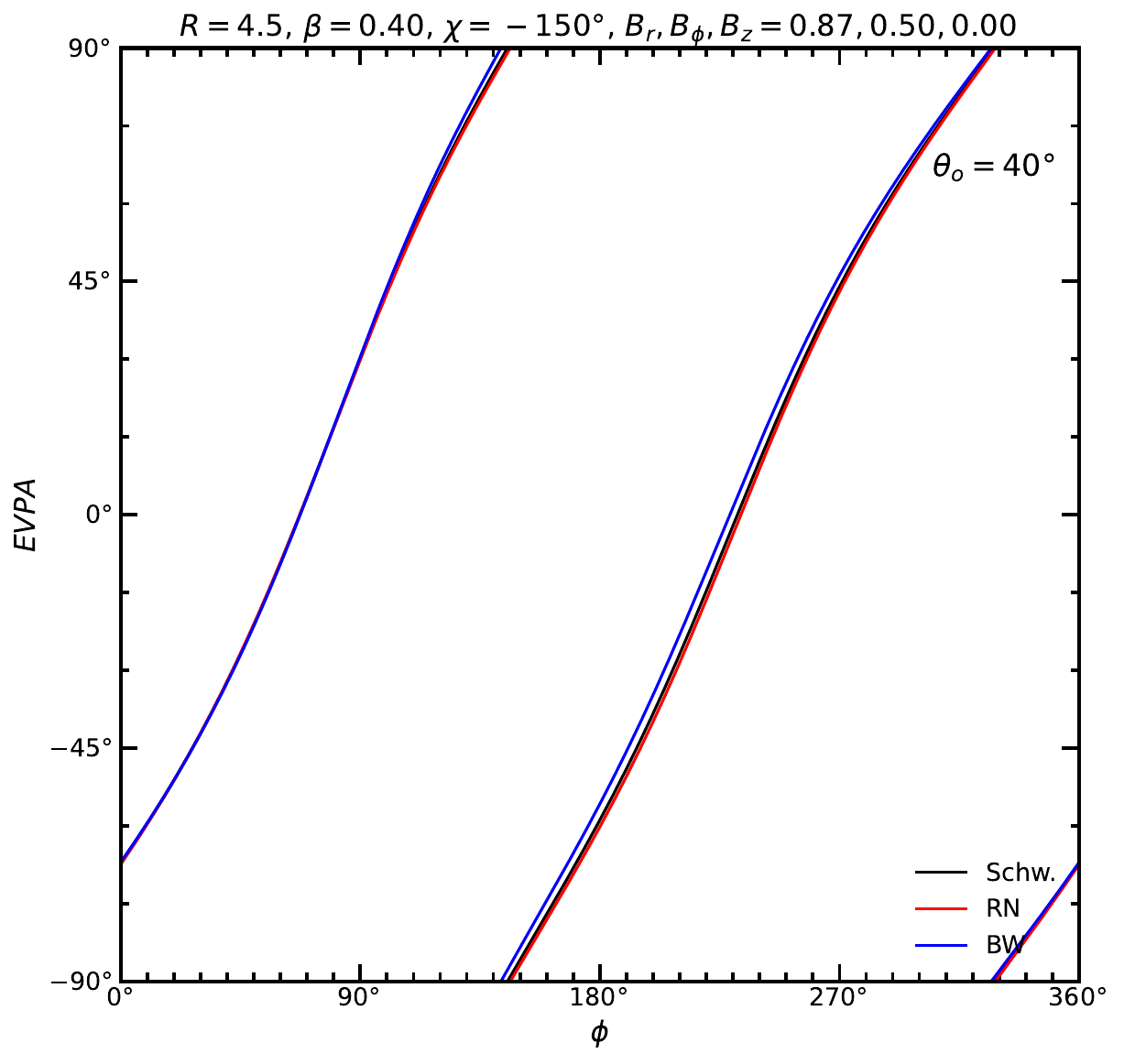}} &
            \hspace{2mm} 
            {\includegraphics[scale=0.29,trim=0 0 0 0]{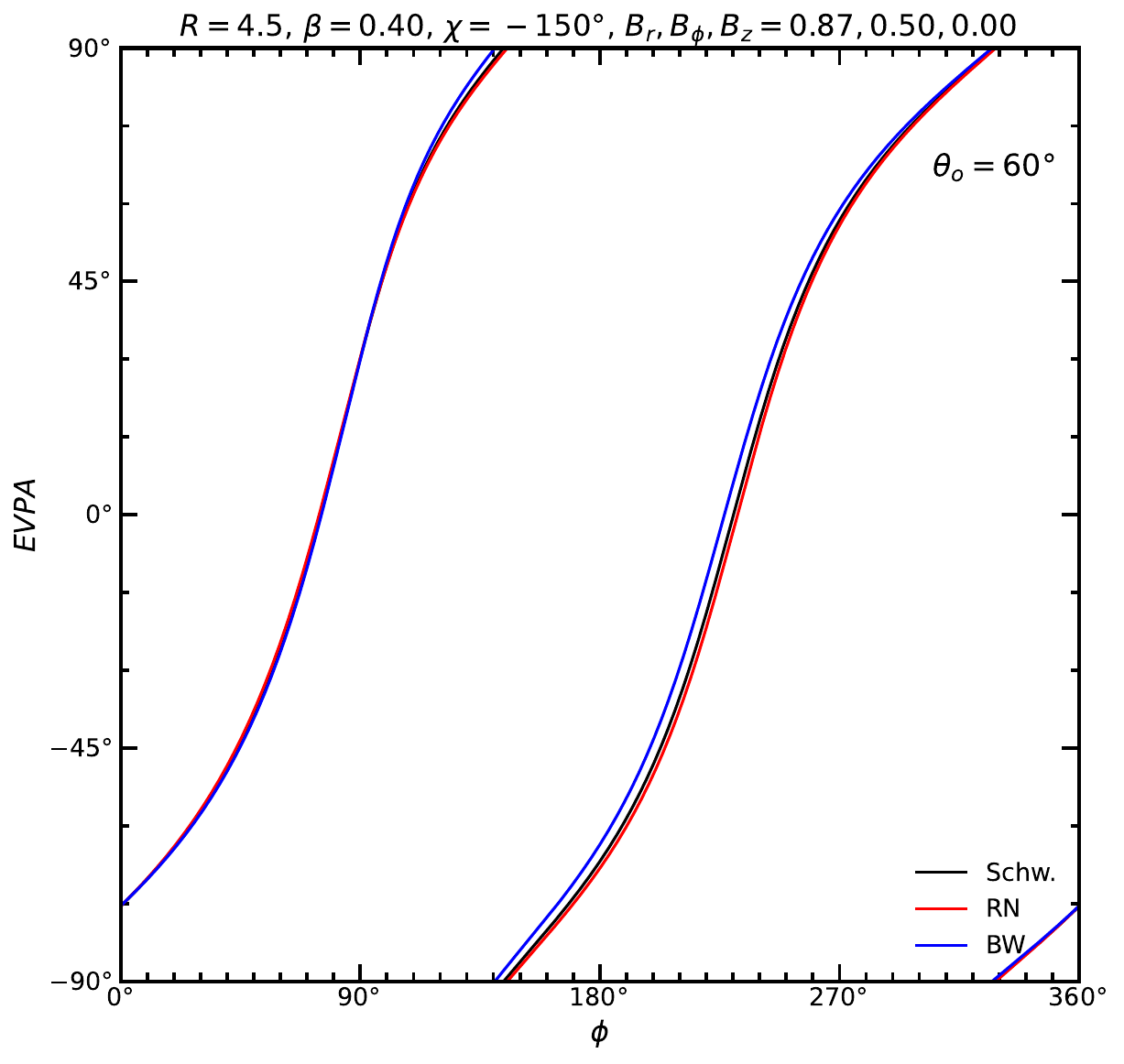}} &
            \hspace{2mm}  
            {\includegraphics[scale=0.29,trim=0 0 0 0]{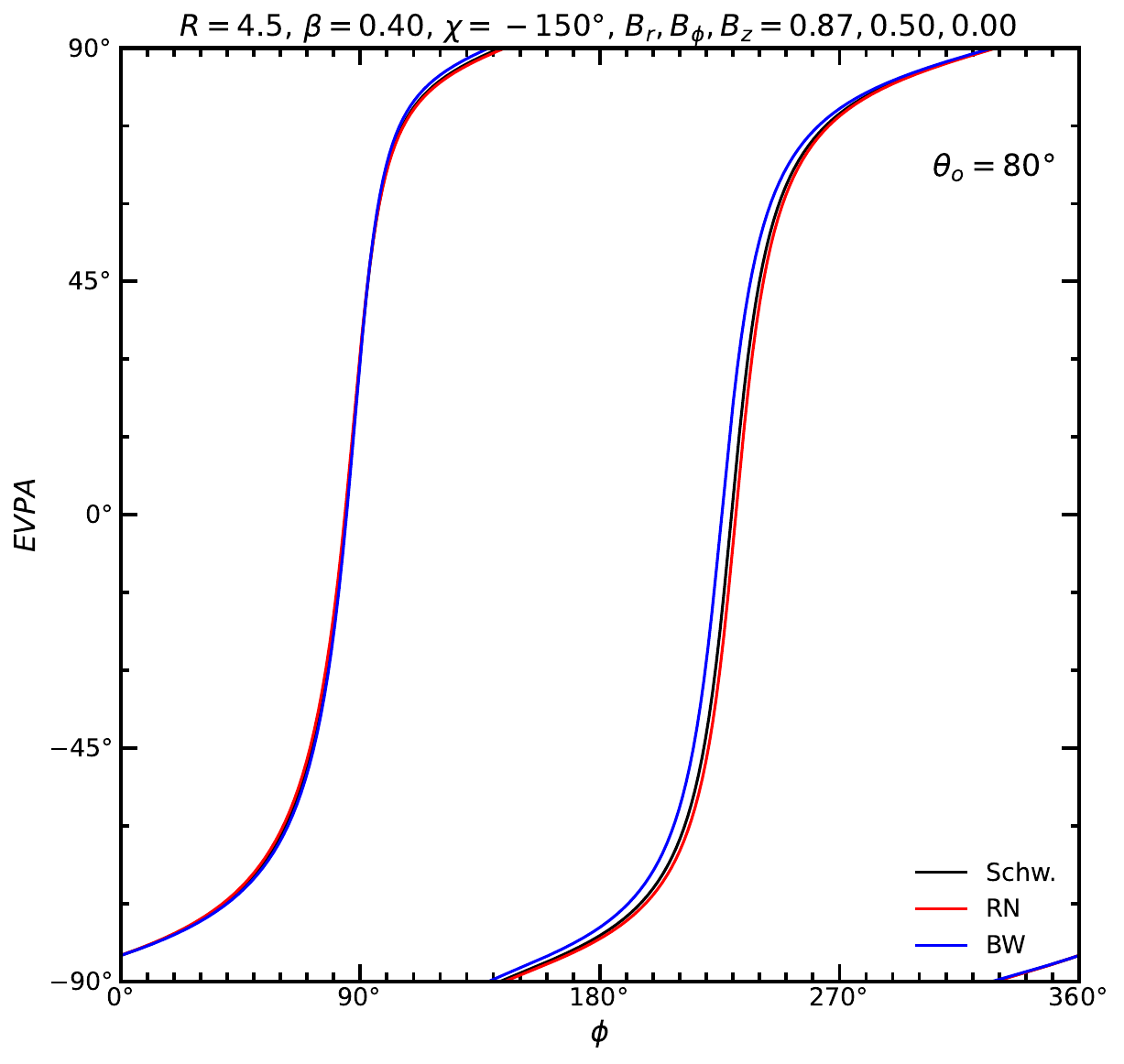}} 
            \hspace{2mm}\\
            \hspace{-6mm}
            {\includegraphics[scale=0.29,trim=0 0 0 0]{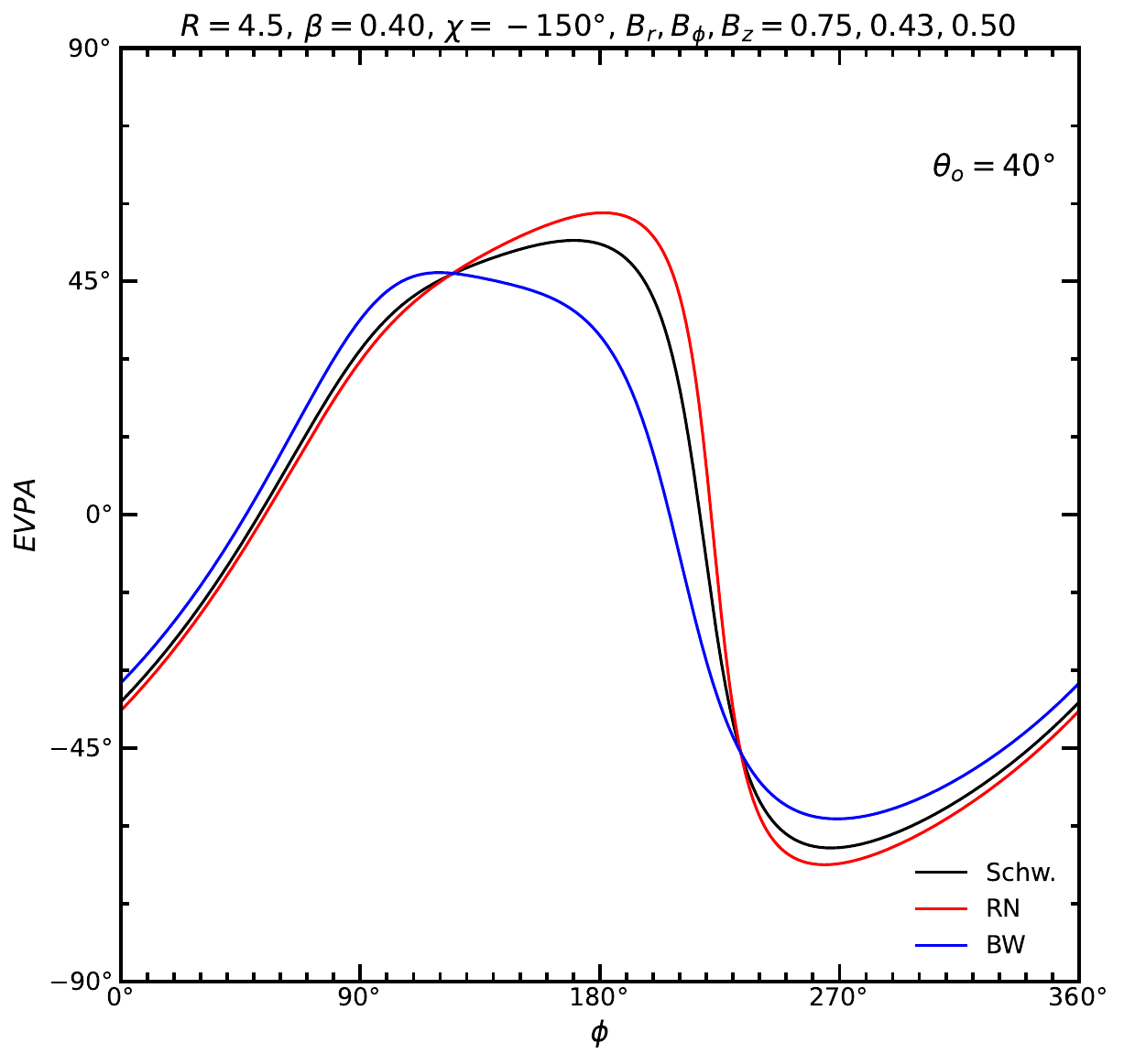}} &
            \hspace{2mm}
            {\includegraphics[scale=0.29,trim=0 0 0 0]{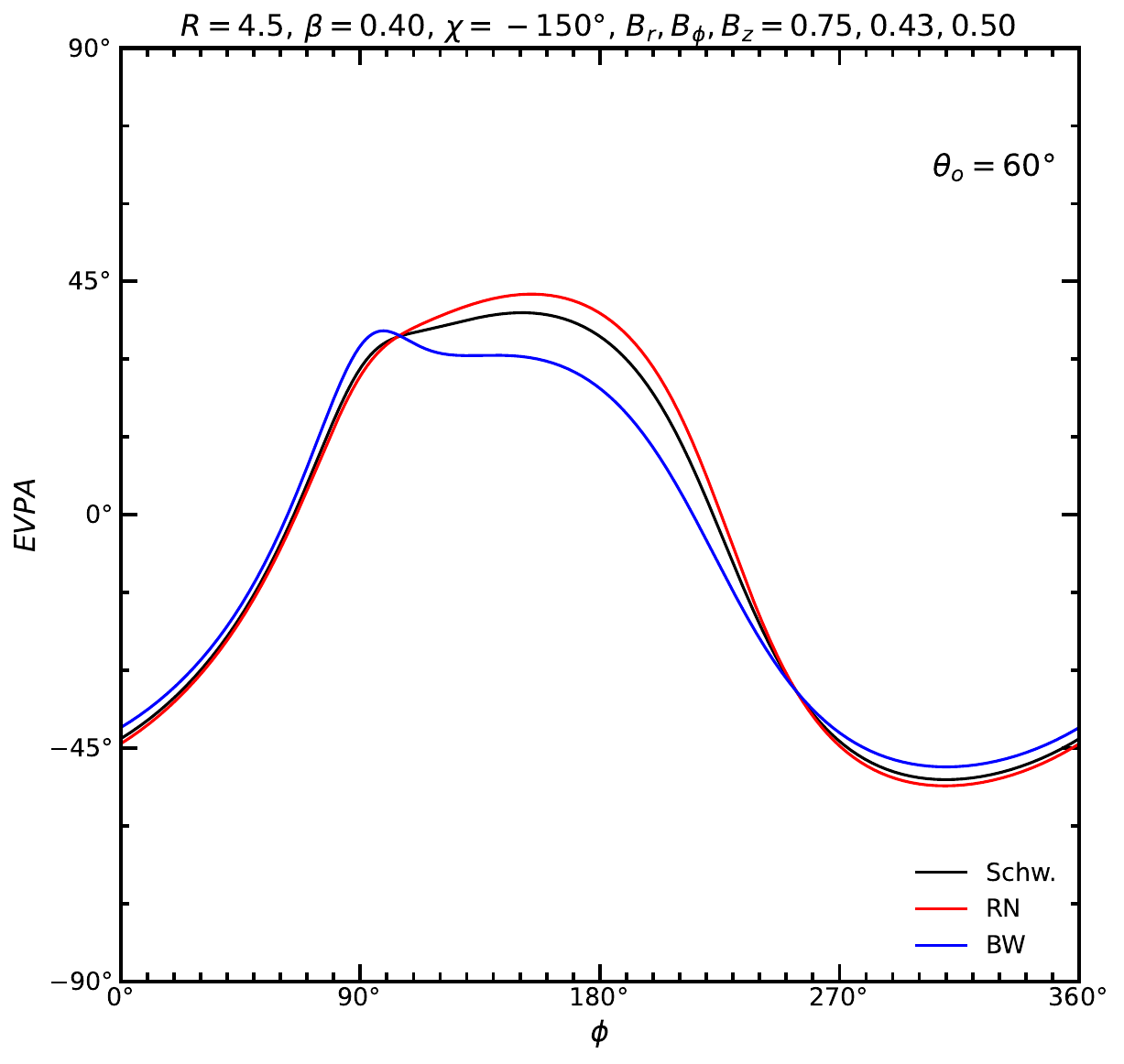}} &
            \hspace{2mm}
            {\includegraphics[scale=0.29,trim=0 0 0 0]{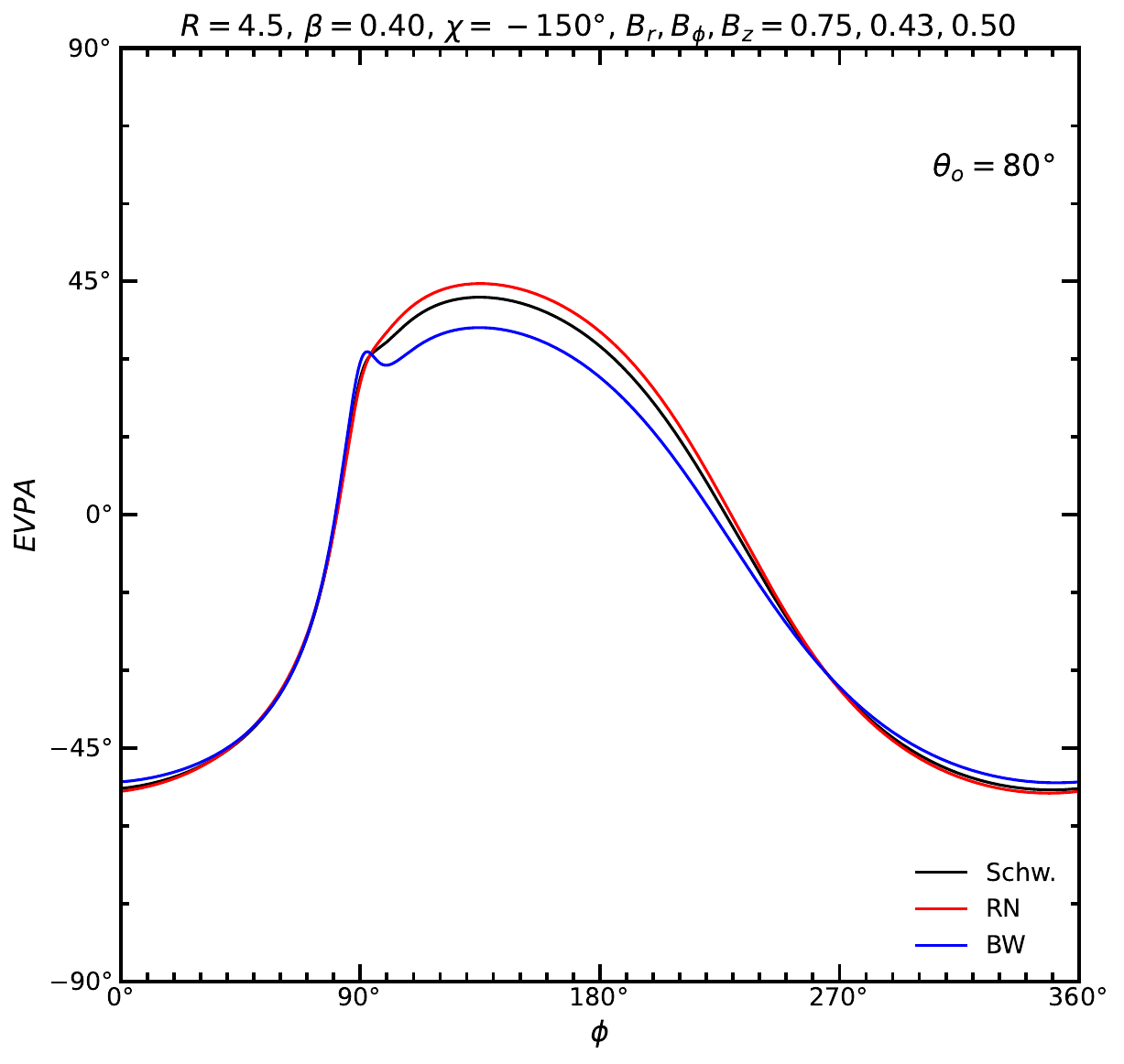}}
            \hspace{2mm}\\
        \end{tabular}
        \caption{The different EVPAs for a magnetic field $\vb*{B}=\vb*{B}_{\text{eq}}$ (top) and $\vb*{B}=\vb*{B}_{\text{eq}} +B_z \vb*{\tilde{z}}$ (bottom) at an emission radius $R=4.5$ for various disk inclinations. Schwarzschild ($a_2=0$), Reissner-Nordström ($a_2=0.9$), and braneworld ($a_2=-1.7$) black holes are considered.
        }
        \label{fig:pol_evpa}
    \end{figure*}
    
    \begin{figure*}[htbp]
        \centering
        \begin{tabular}{ccc}
            \hspace{-6mm}
            {\includegraphics[scale=0.305,trim=0 0 0 0]{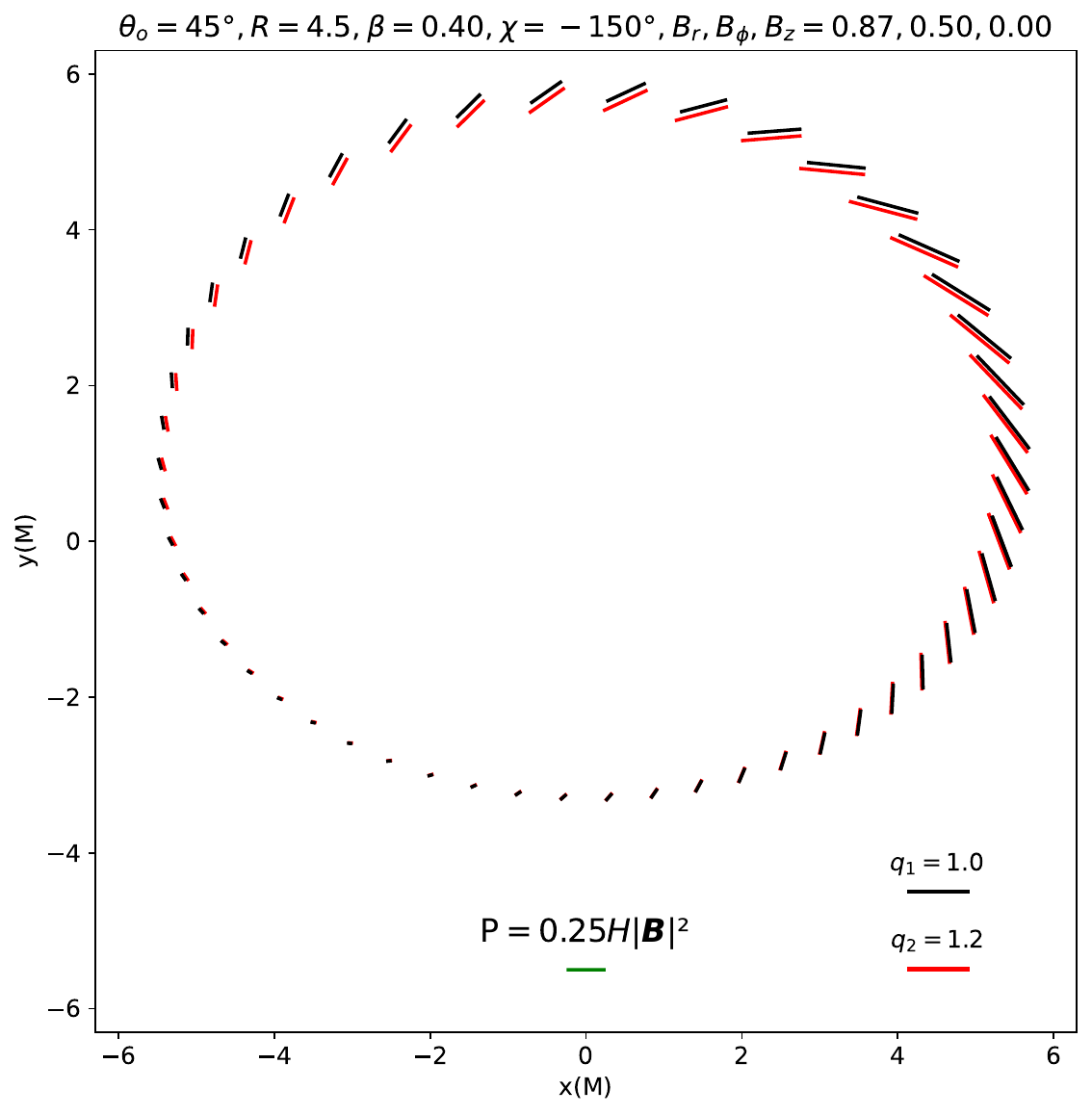}} &
            \hspace{2mm}  
            {\includegraphics[scale=0.305,trim=0 0 0 0]{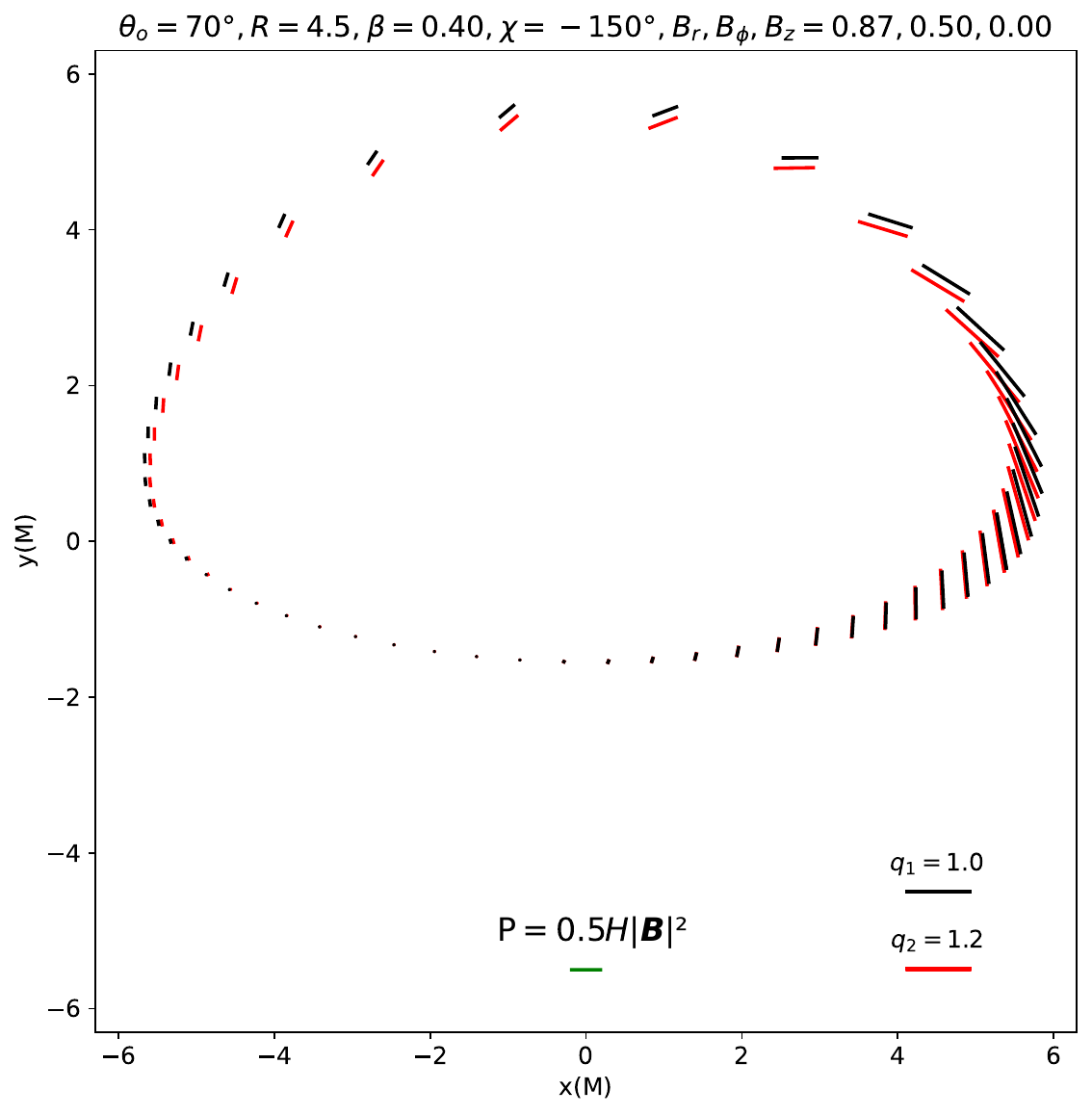}} &
            \hspace{2mm}
            {\includegraphics[scale=0.305,trim=0 0 0 0]{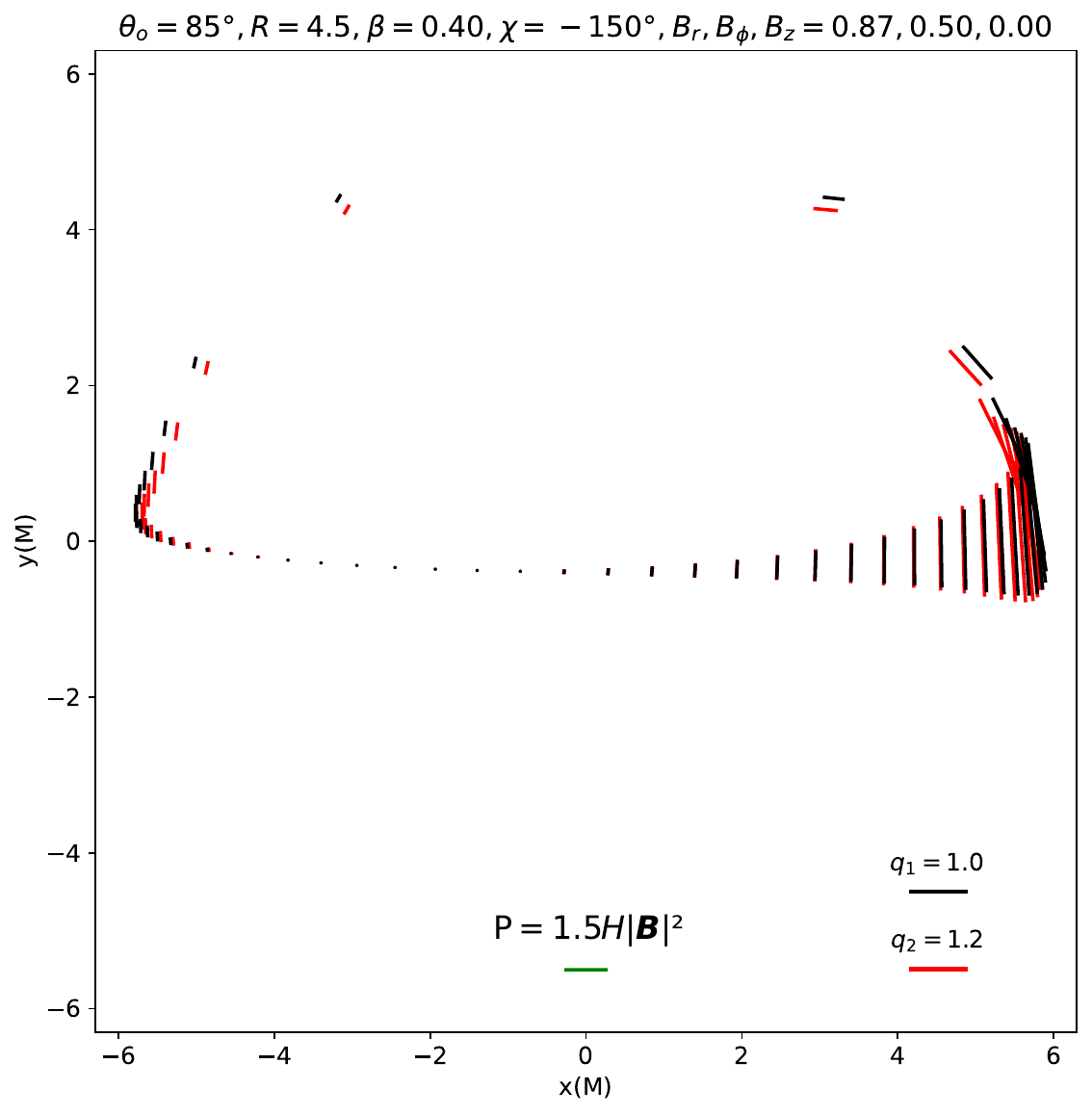}}
            \hspace{2mm}\\
        \end{tabular}
        \caption{Polarization patterns at the emission radius $R=4.5$ for an EMD black hole with $q=1$ (black ticks) and $q=1.2$ (red ticks) in an equatorial magnetic field ($B_{eq}=1$).}
        \label{fig:pol_emd}
    \end{figure*}   

    With this, we conclude the extension of the analytical formalism presented in \cite{EventHorizonTelescope:2021btj} to more general spherically symmetric metrics. Note that according to Eq. \eqref{eq:stokes_2}, the polarization vector angle is defined with respect to the $X'$ axis. However, in astronomical observations, it is more common to define it with respect to the $Y'$ axis. Between the Stokes parameters $Q$ and $U$ associated with both definitions, there is a simple global minus factor.

    To analyze the polarization patterns, we confine our study to the strong field regime. Initially, we restrict our considerations to metrics of the form \eqref{eq:ar6} with $a_1=-2$, $a_2\neq 0$, and the remaining $a_i$ set to zero. Subsequently, we conduct an analysis of the EMD black hole described by Eq. \eqref{eq:emd_bh} and approximated as outlined in Sec. \ref{subsec:isora}. Some polarization patterns with $a_2\neq 0$, and the remaining $a_i=0$  are depicted in Figure \ref{fig:pol_patterns}. They are calculated using 50 equally spaced ticks along $\phi$.
    
    In the first figure on the left in the top panel of Fig.\ref{fig:pol_patterns} we start by considering  an emission radius of $R=6$ and a vertical magnetic field such that $B_{z}=1$, $B_{\text{eq}}=0$. While the Doppler effect $(\beta=0)$ is not implicated, gravitational effects persist. A subtle increase in the EVPA is observed for the case of $a_2=-1.5$ in the counterclockwise direction on the right side of the image and in the clockwise direction on the left side of the image, i.e., toward the center of the image of the black hole. A subtle variation in the observed polarization intensities is also noticeable at the top of the image as $a_2$ decreases. This is explained by gravitational lensing effects on the original emission direction of light rays and the angle it forms with the direction of the magnetic field. For an inclination angle of $\theta_o=20^{\circ}$, it turns out that the value of $\sin\zeta$ increases as $a_2$ decreases. However, if we had plotted polarization patterns for purely vertical magnetic fields and higher angles $\theta_o$, we would have observed that generally, the intensity of polarized radiation decreases as the gravitational field increases (as is the case when $a_2$ decreases). In this scenario, due to the curvature effects on the trajectory of emitted light rays, they tend to align more with the vertical direction of the magnetic field, thus decreasing the value of $\sin\zeta$.

    Subsequently, we explore cases involving radial or toroidal magnetic fields, specifically $B_{z}=0$ and $B_{r}=1$ (radial magnetic field) or $B_{\phi}=1$ (toroidal magnetic field), corresponding to $\eta=0^{\circ}$ or $\eta=90^{\circ}$ respectively, with purely radial incoming velocities $(\chi=-180^{\circ})$. In both cases, the growth of polarization intensity correlates with an increase in the value of $a_2$. Additionally, the presence of the Doppler effect and the aberration of emitted light are observed. For further elucidation on these phenomena, refer to \cite{EventHorizonTelescope:2021btj}. 
    
    In the case of a radial magnetic field (top middle), it is observable that as $a_2$ increases, the ring appears slightly smaller than its corresponding value at $R=6$. This effect is more pronounced in the upper part of the image, corresponding to the region farther away from the disk relative to the observer's frame, where the gravitational lensing effect is most prominent. In turn, for the toroidal magnetic field case (top right), we can observe an almost negligible difference between the EVPA of the three studied cases.
    
    For the three patterns at the middle of Figure \ref{fig:pol_patterns}, we consider an equatorial magnetic field with both radial and azimuthal components. For simplicity, we assume the velocity $\vb*{\beta}$ is in the same direction as the magnetic field $\vb*{B}_{eq}$ but in the opposite sense, satisfying $\eta=\chi + \pi$, and ${B_{eq}}=1$. With a fixed value of $R=4.5$, we increase the disk inclination angle $\theta_{o}$ and observe how the strong gravitational lensing effect becomes increasingly significant. For the ring with $a_2=-1.5$, its polarization pattern is more affected due to the aforementioned conditions, resulting in an enlargement of the ring. Additionally, note the abrupt growth in intensity and polarization direction on the right side of the figure with $\theta_{o}=80^{\circ}$ (middle right). This is attributed to the Doppler effect and the projection of the fluid velocity being larger along the line of sight. 
    
    In the three polarization patterns of the lower part of the Figure \ref{fig:pol_patterns}, we consider a general magnetic field with the same parameters as in the middle panel, but in addition of a magnetic field on the disk plane, we also consider a non-zero vertical component $B_z \neq 0$. In particular, we take $B_z = 0.5$ such that $B_{eq}^2 + B_z^2=1$. Here, we assume that due to Faraday rotation effects, direct rays coming from the far side of the disk are depolarized. Therefore, we only consider the polarization of rays coming from the side closest to the observer ($Z>0$), and thus we take $\eta=\chi+\pi$ (for more details, refer to Sec. 3.5 of \cite{EventHorizonTelescope:2021btj}). It can be observed that the analysis for the cases where only $B_{eq} \neq 0$ is still valid, except for a change in the EVPA of the polarization due to the presence of the the vertical component of the magnetic field.

     \begin{figure*}[htbp]
        \centering
        \begin{tabular}{ccc}
            \hspace{-6mm}
            {\includegraphics[scale=0.27,trim=0 0 0 0]{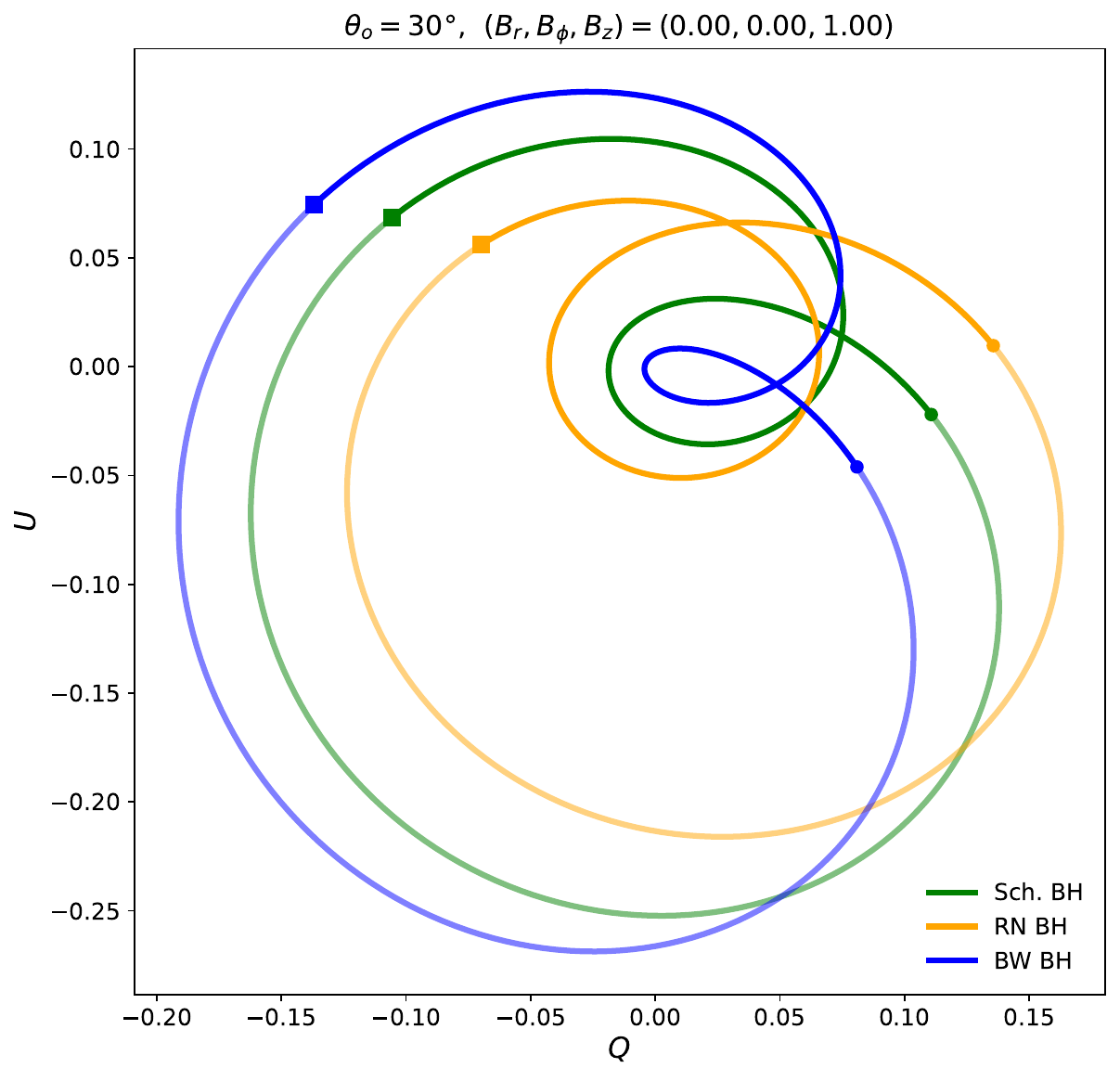}} &
            \hspace{2mm} 
            {\includegraphics[scale=0.27,trim=0 0 0 0]{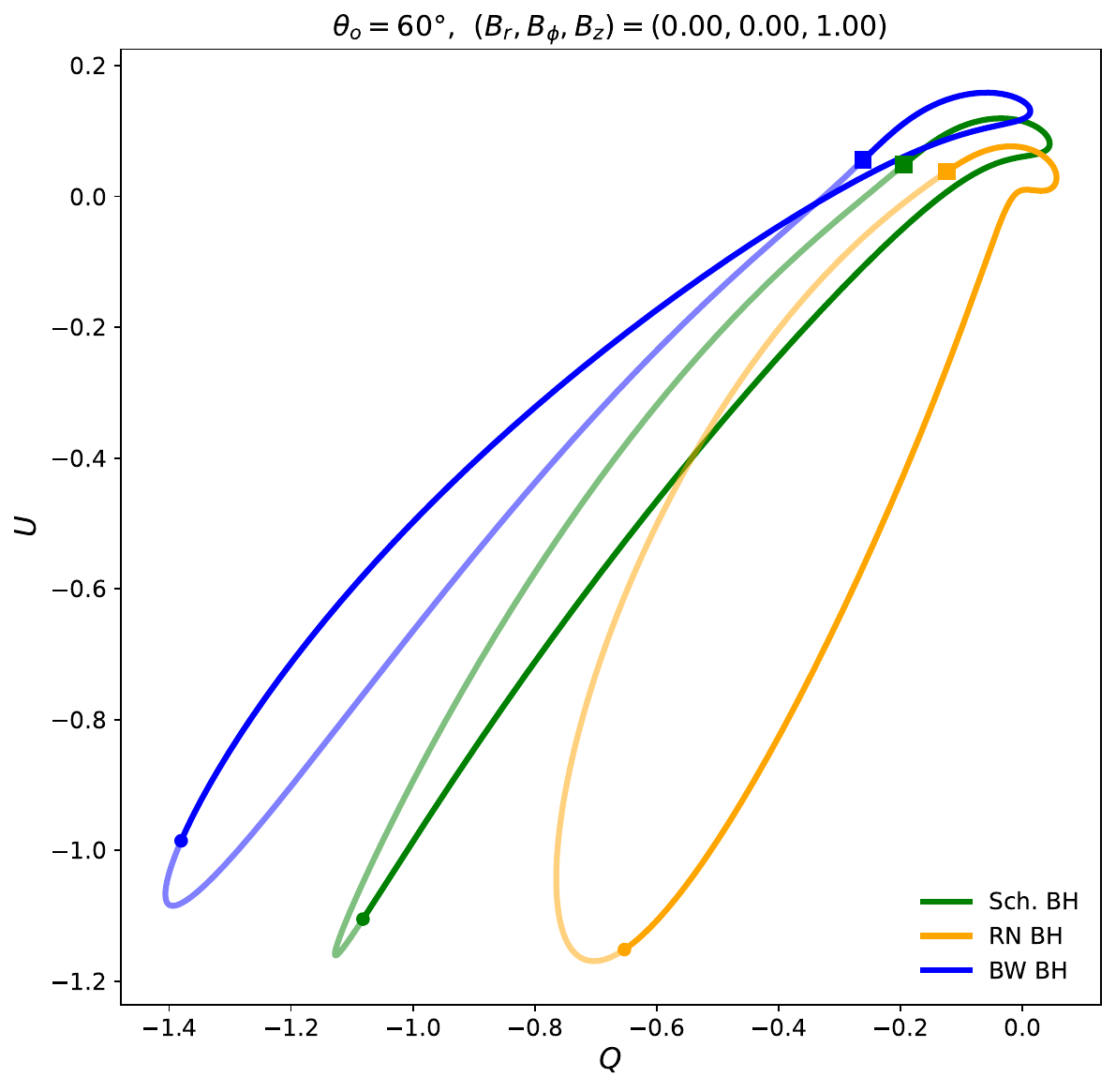}} &
            \hspace{2mm}  
            {\includegraphics[scale=0.27,trim=0 0 0 0]{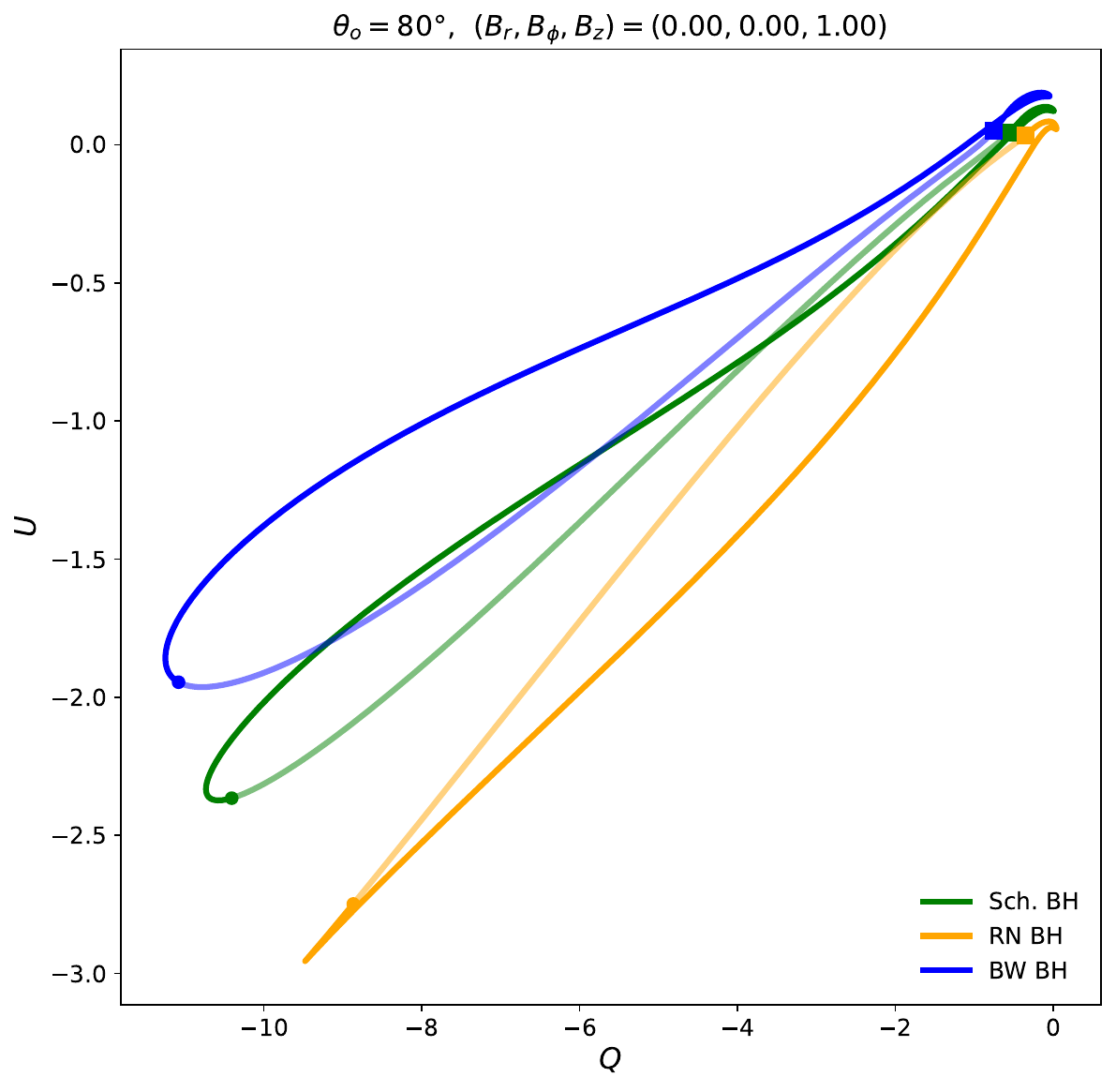}} 
            \hspace{2mm}\\
            \hspace{-6mm}
            {\includegraphics[scale=0.27,trim=0 0 0 0]{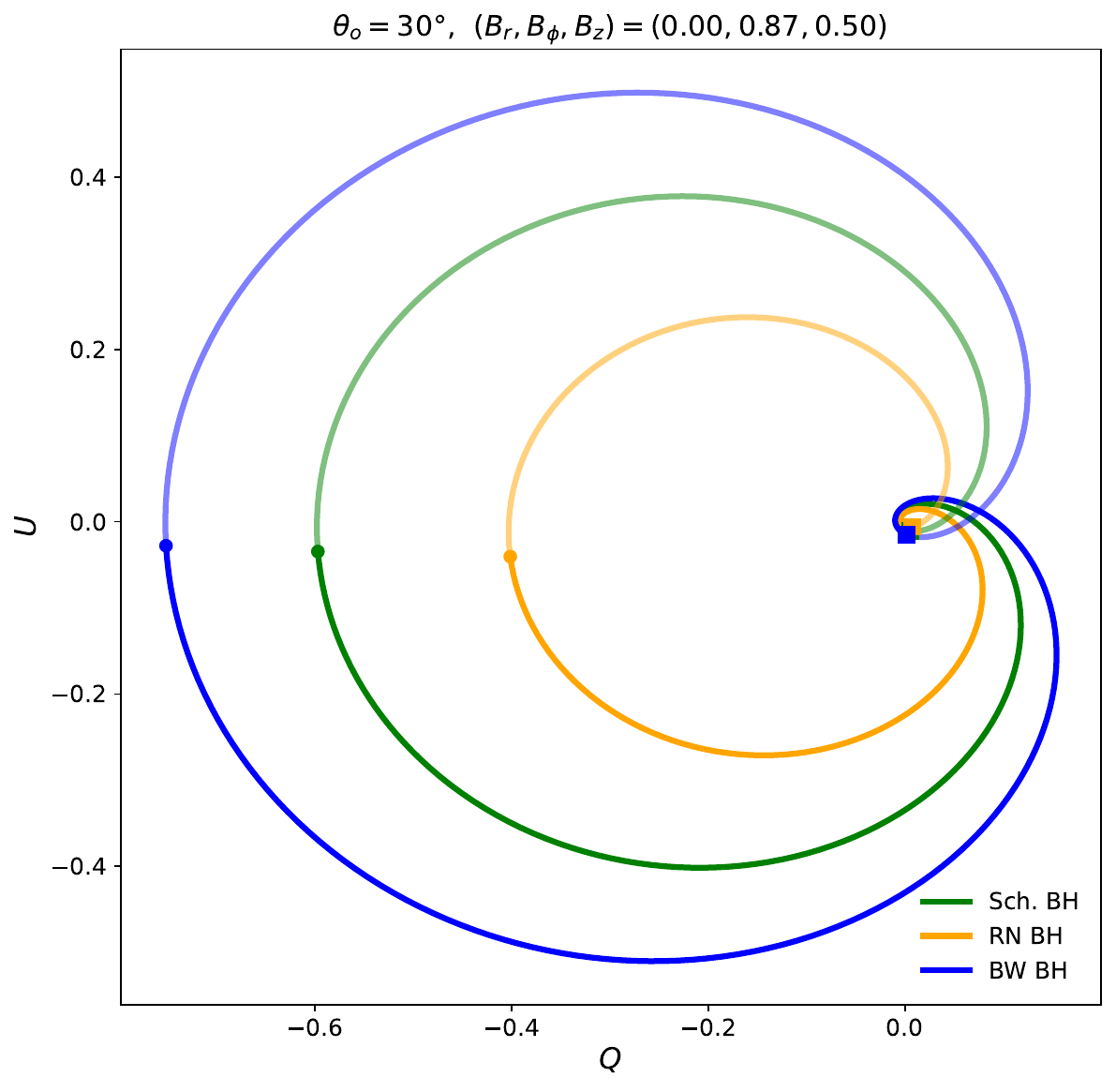}} &
            \hspace{2mm}
            {\includegraphics[scale=0.27,trim=0 0 0 0]{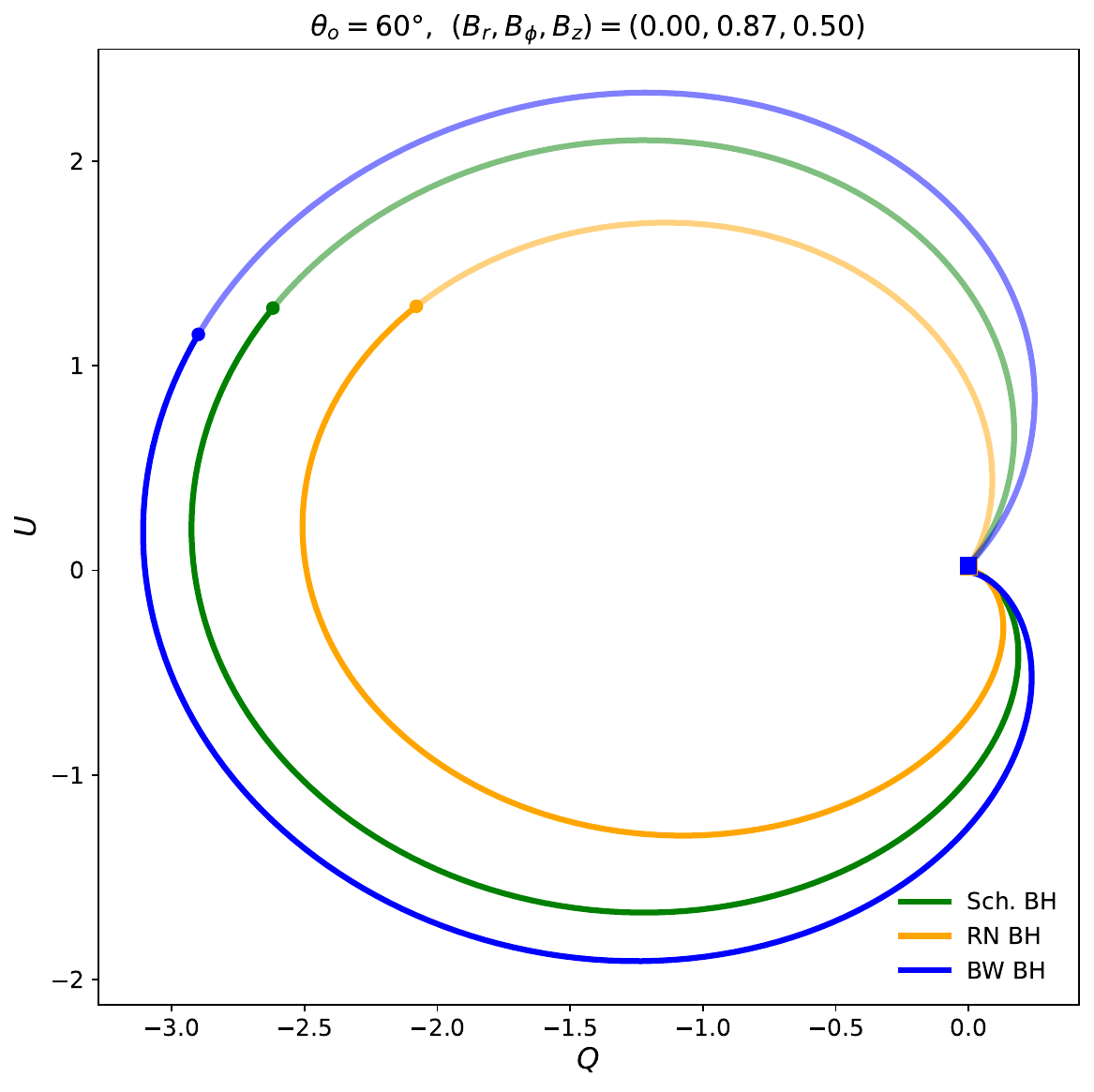}} &
            \hspace{2mm}
            {\includegraphics[scale=0.27,trim=0 0 0 0]{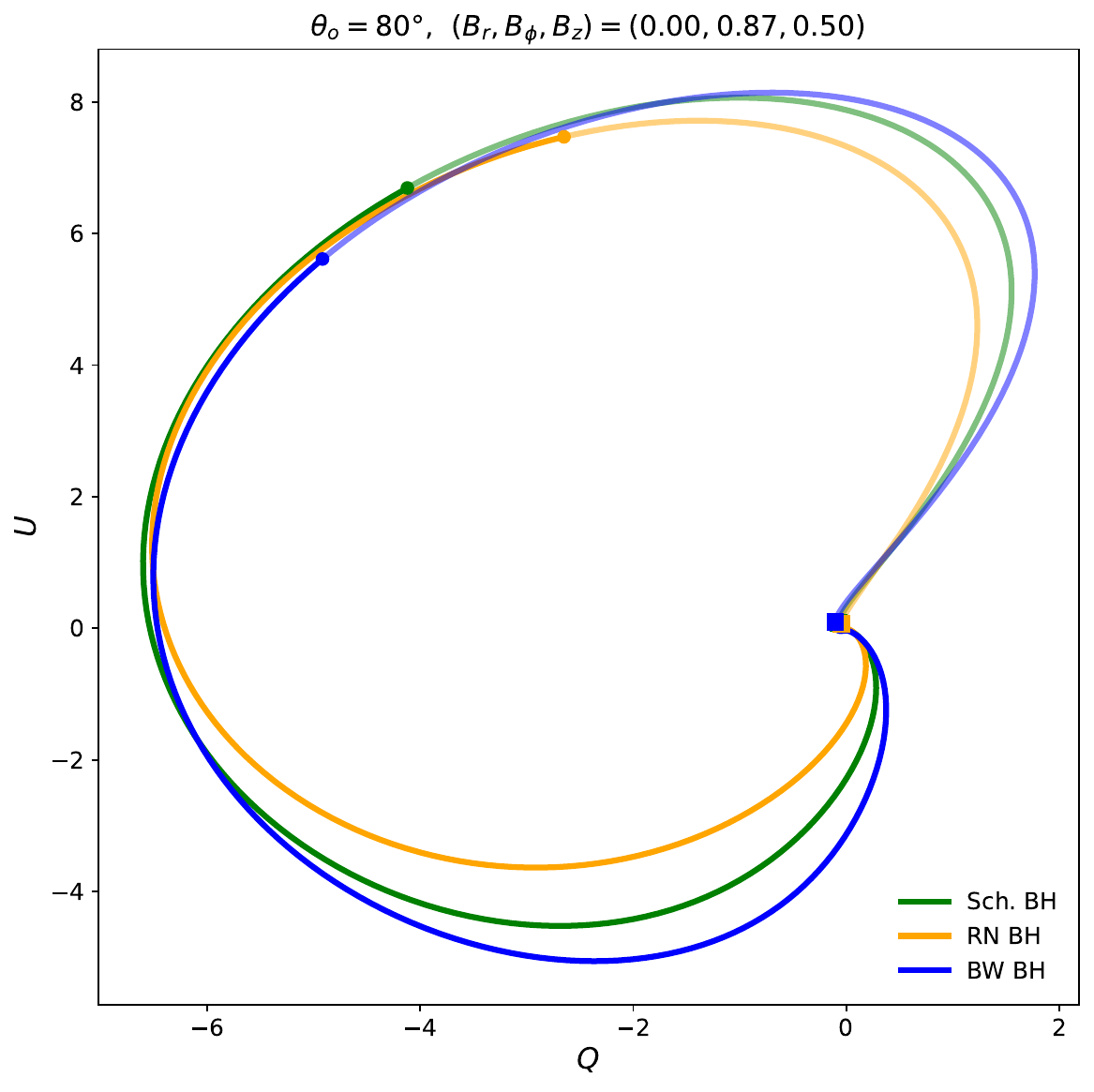}}
            \hspace{2mm}\\
            \hspace{-6mm}
            {\includegraphics[scale=0.27,trim=0 0 0 0]{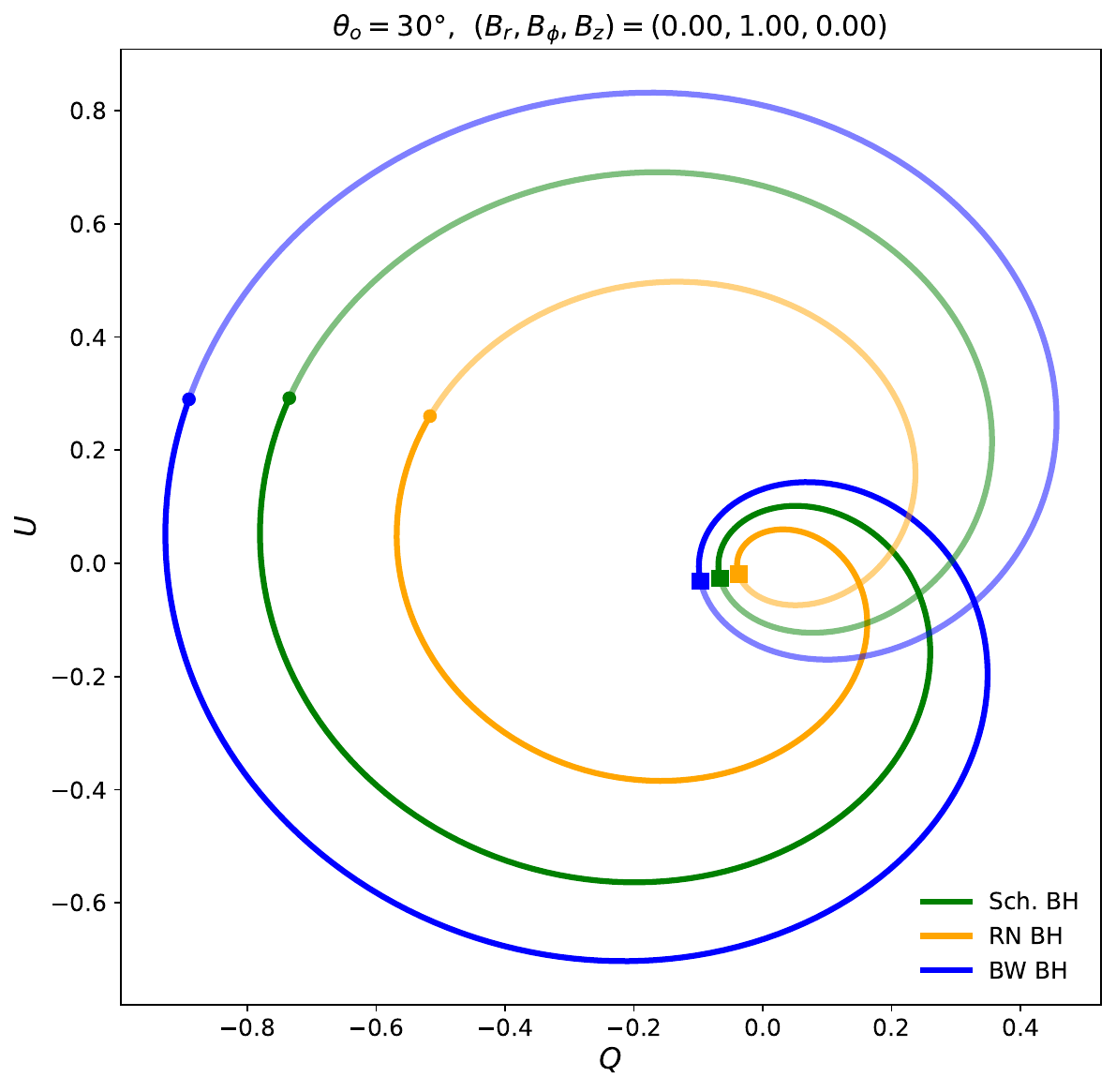}} &
            \hspace{2mm}
            {\includegraphics[scale=0.27,trim=0 0 0 0]{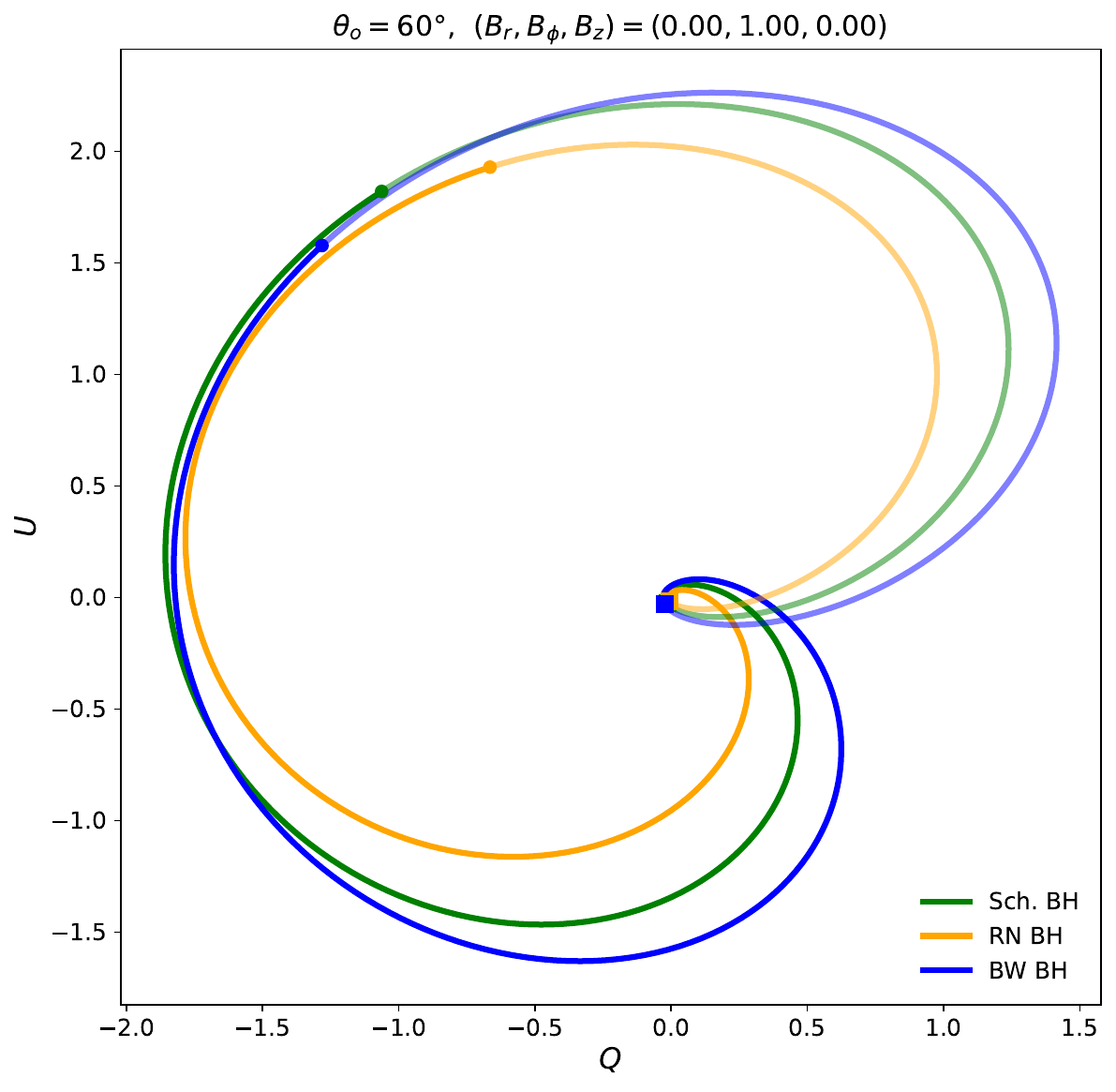}} &
            \hspace{2mm}
            {\includegraphics[scale=0.27,trim=0 0 0 0]{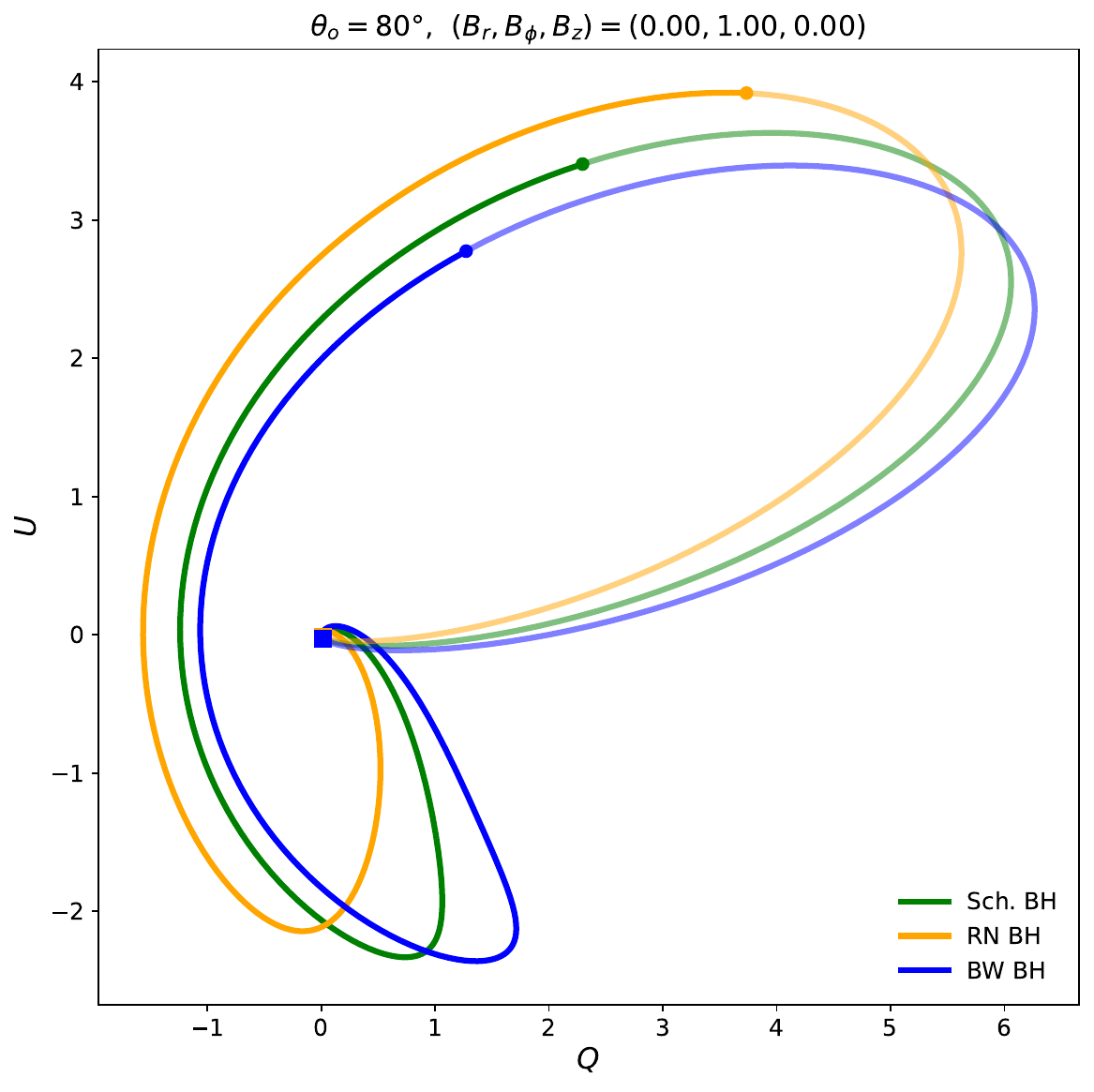}}
            \hspace{2mm}\\
        \end{tabular}
        \caption{$QU$-diagrams depict a hotspot modeled as a point source orbiting at the innermost stable circular orbits (ISCOs) for Schwarzschild, Reissner-Nordstr\"om, and braneworld black holes. The top panel illustrates a pure vertical magnetic field, the middle panel shows a combination of toroidal and vertical magnetic fields, and the bottom panel displays a pure toroidal field. In all cases, the hotspot orbits clockwise. For further details, refer to the main text.  
        }
    \label{fig:quloops}
    \end{figure*}

\begin{figure*}[htbp]
        \centering
        \begin{tabular}{ccc}
            \hspace{-6mm}
            {\includegraphics[scale=0.27,trim=0 0 0 0]{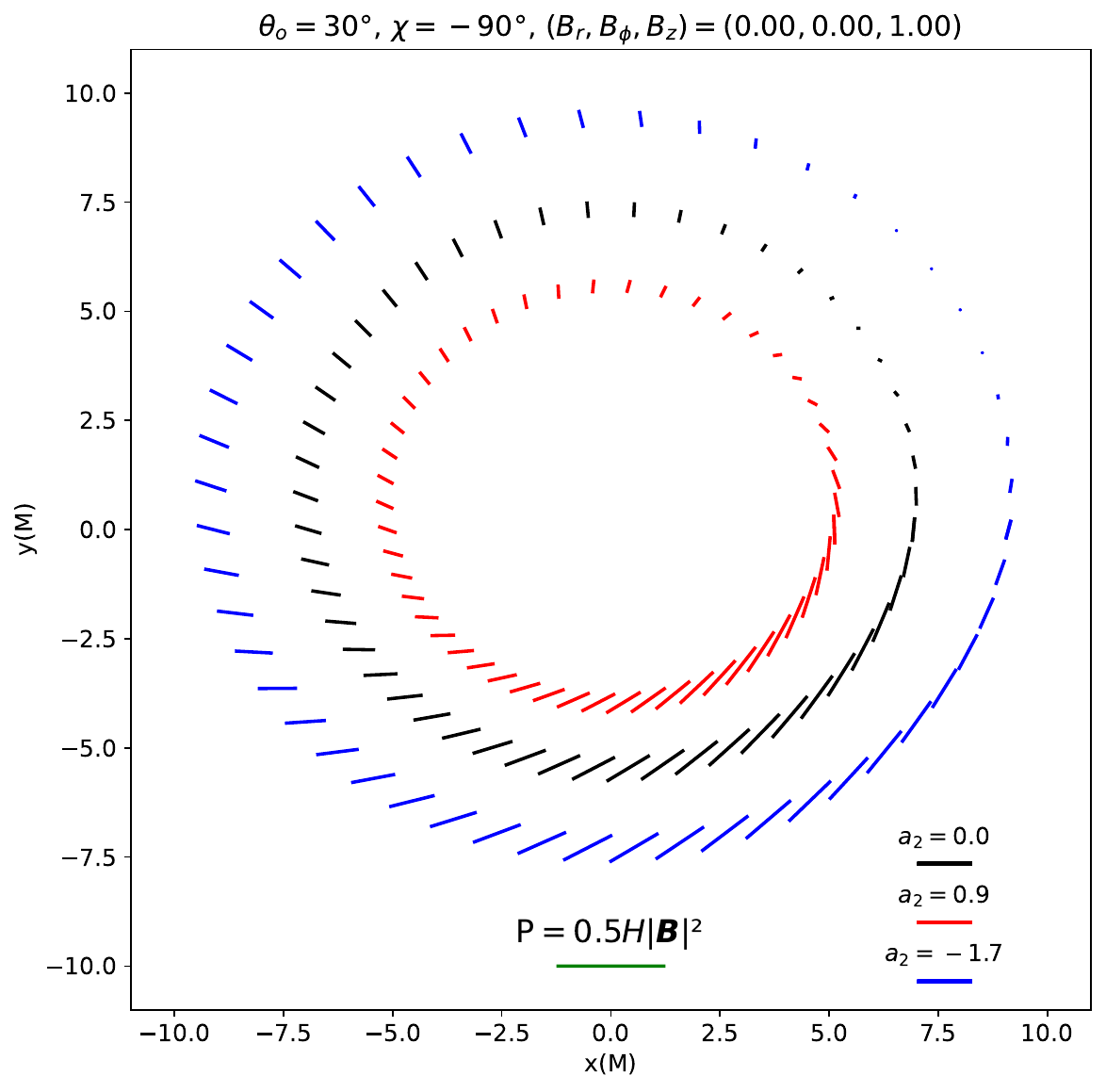}} &
            \hspace{2mm} 
            {\includegraphics[scale=0.27,trim=0 0 0 0]{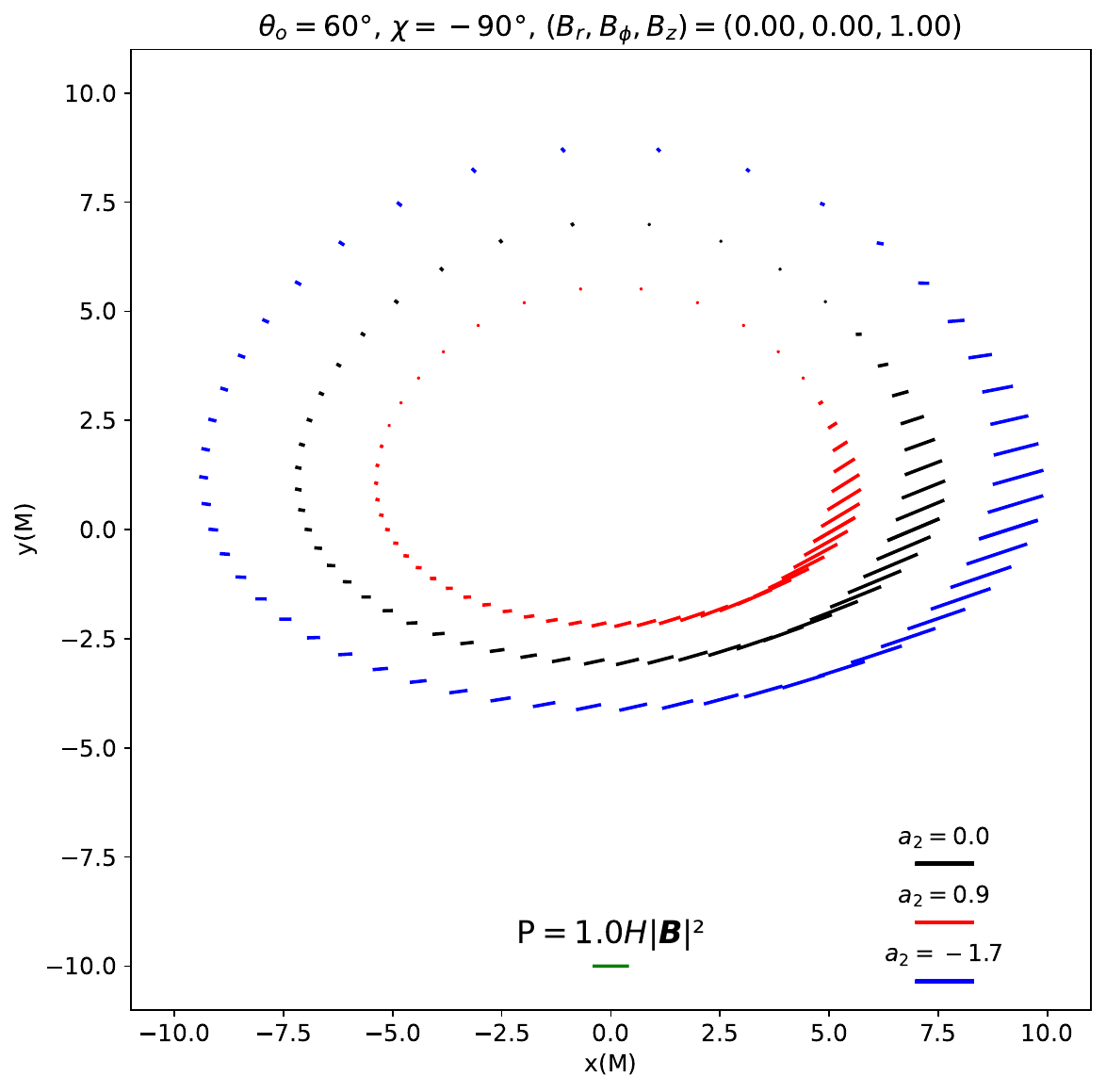}} &
            \hspace{2mm}  
            {\includegraphics[scale=0.27,trim=0 0 0 0]{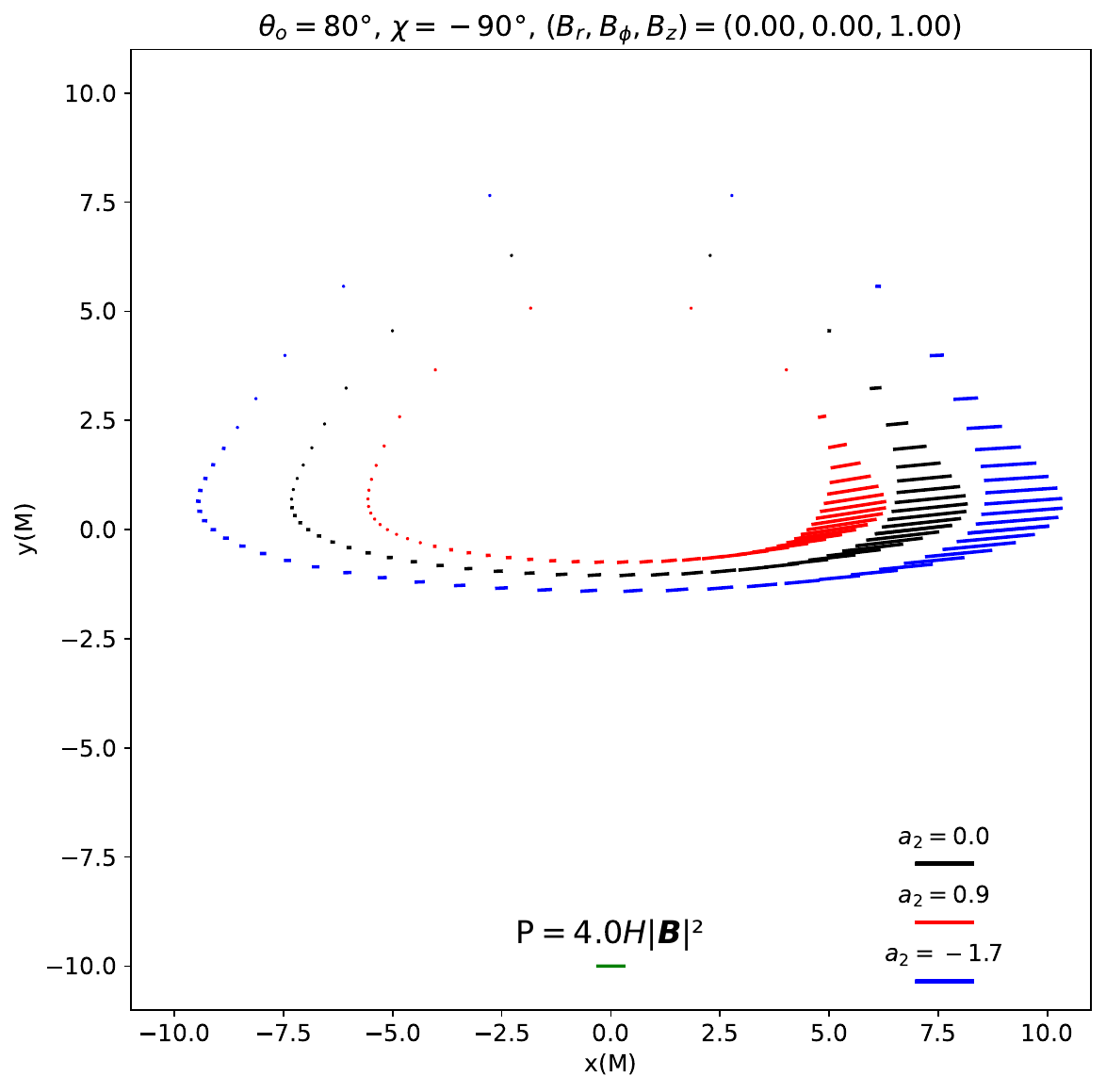}} 
            \hspace{2mm}\\
            \hspace{-6mm}
            {\includegraphics[scale=0.27,trim=0 0 0 0]{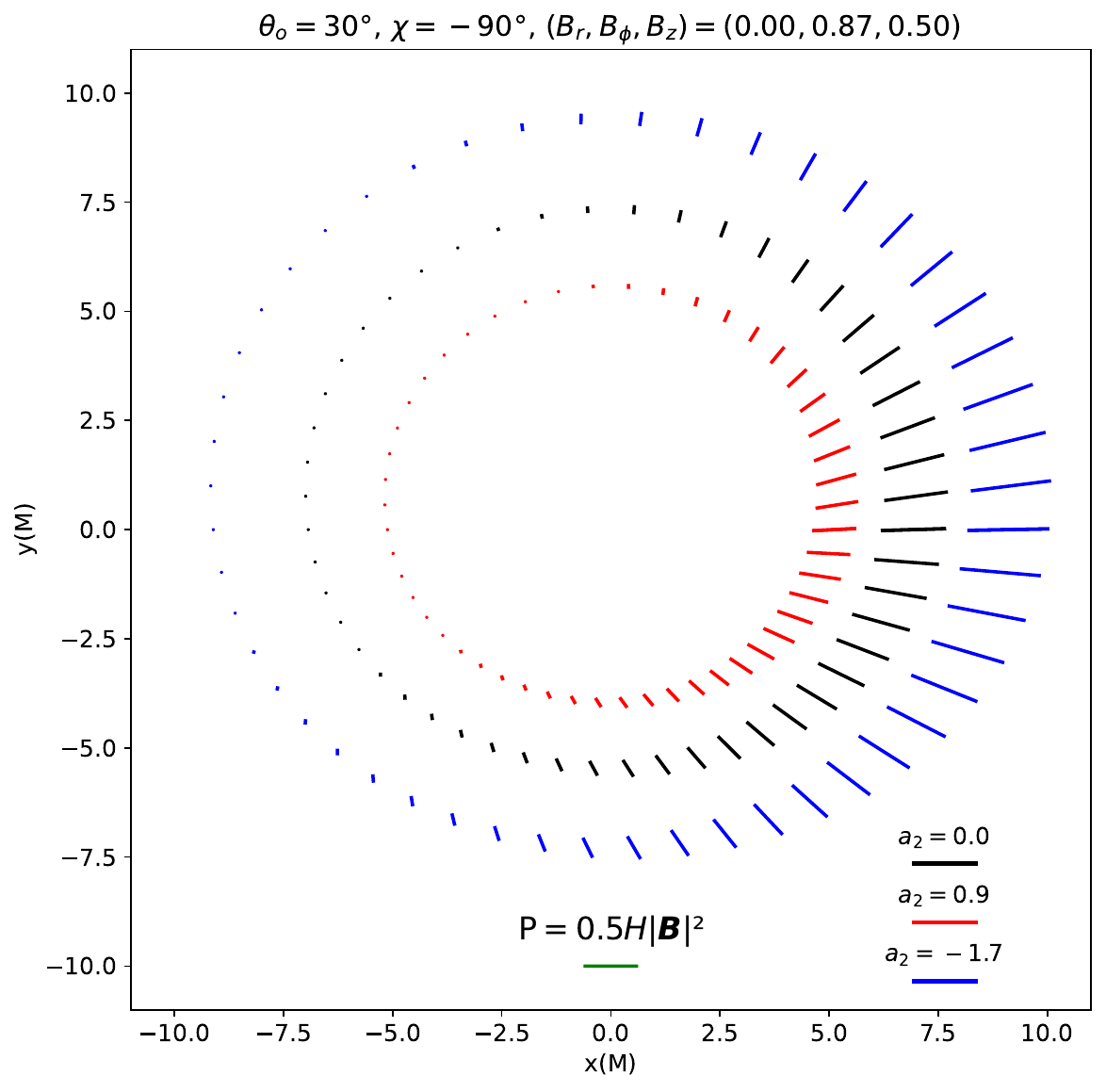}} &
            \hspace{2mm}
            {\includegraphics[scale=0.27,trim=0 0 0 0]{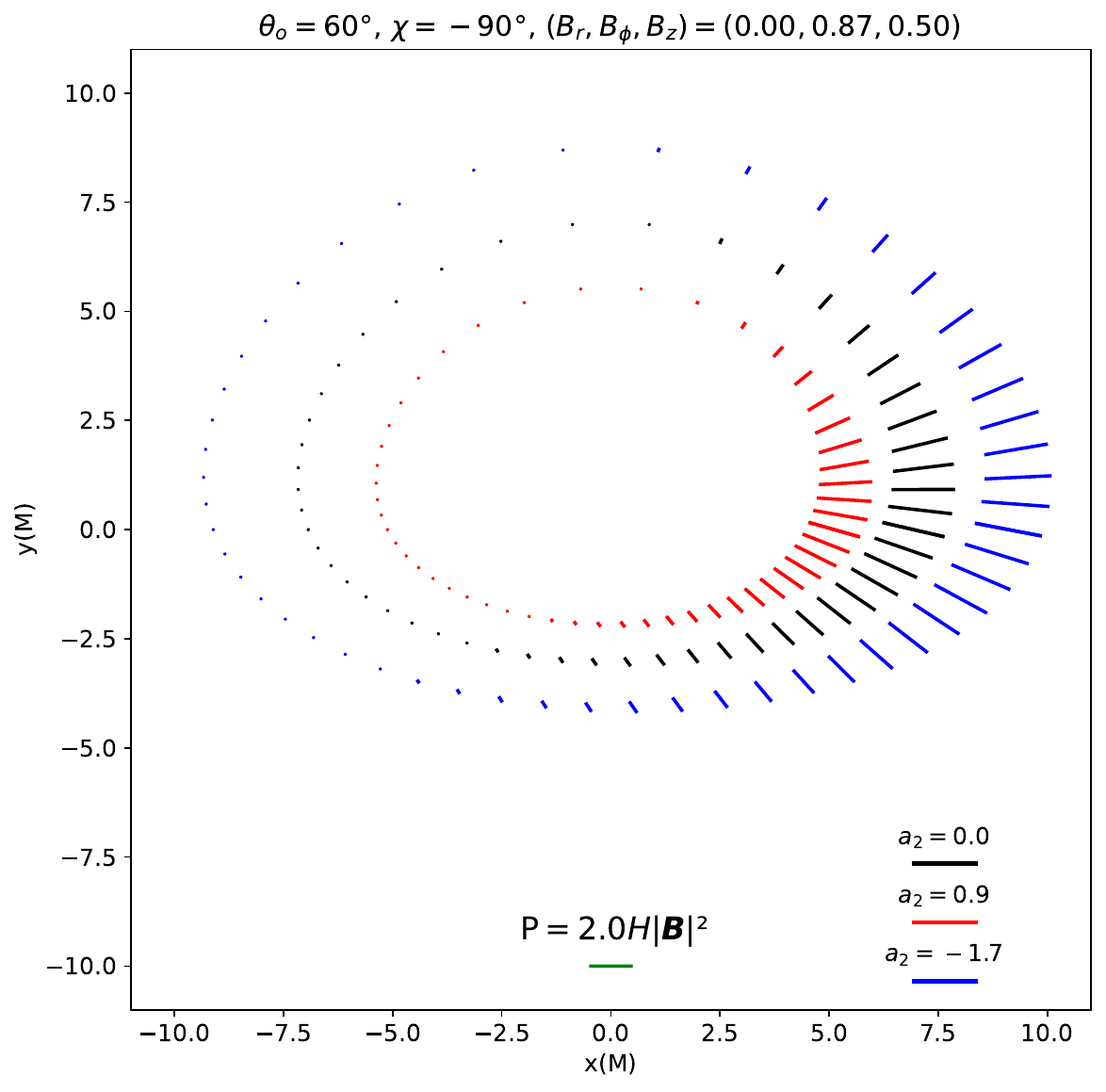}} &
            \hspace{2mm}
            {\includegraphics[scale=0.27,trim=0 0 0 0]{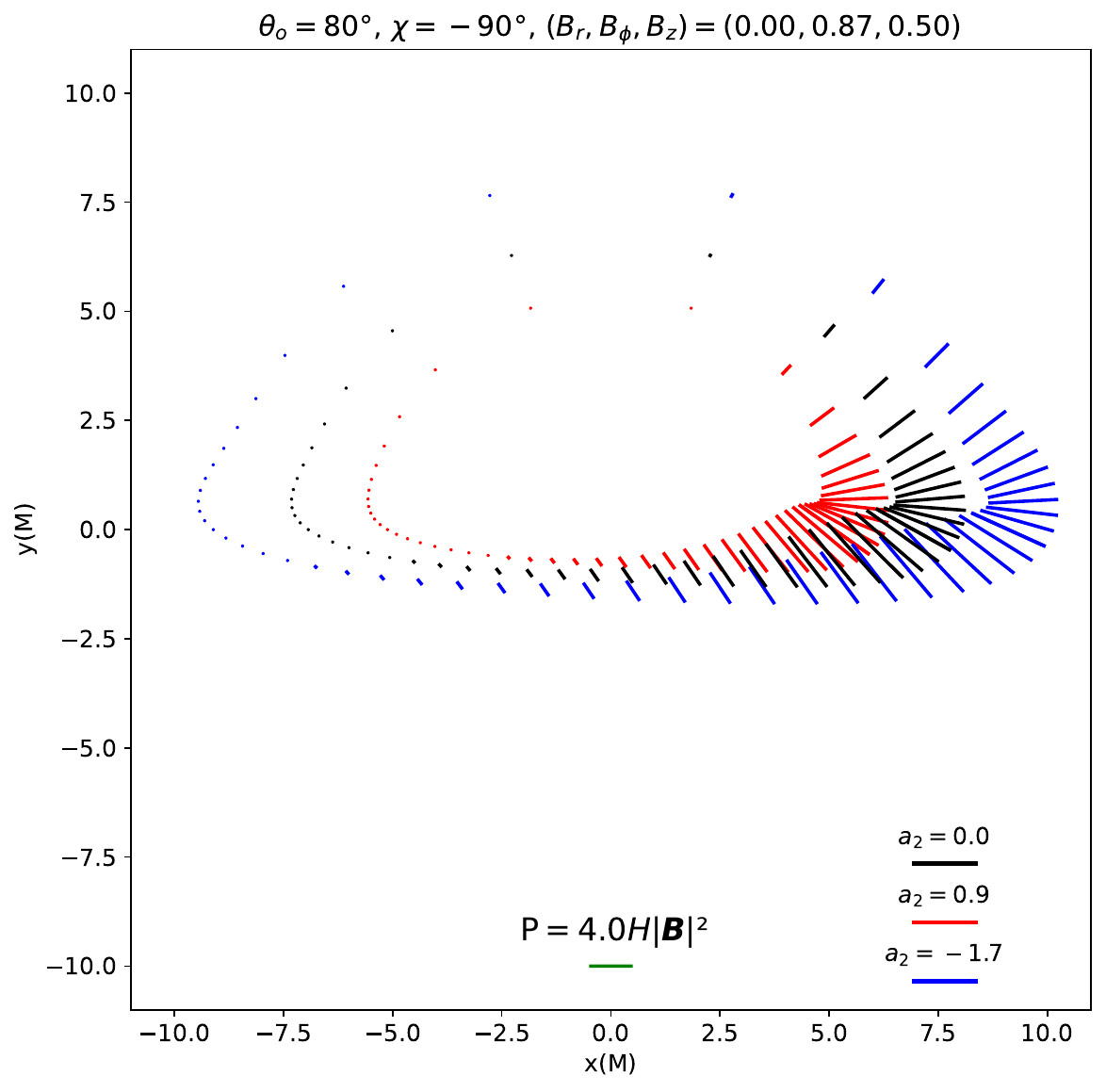}}
            \hspace{2mm}\\
            \hspace{-6mm}
            {\includegraphics[scale=0.27,trim=0 0 0 0]{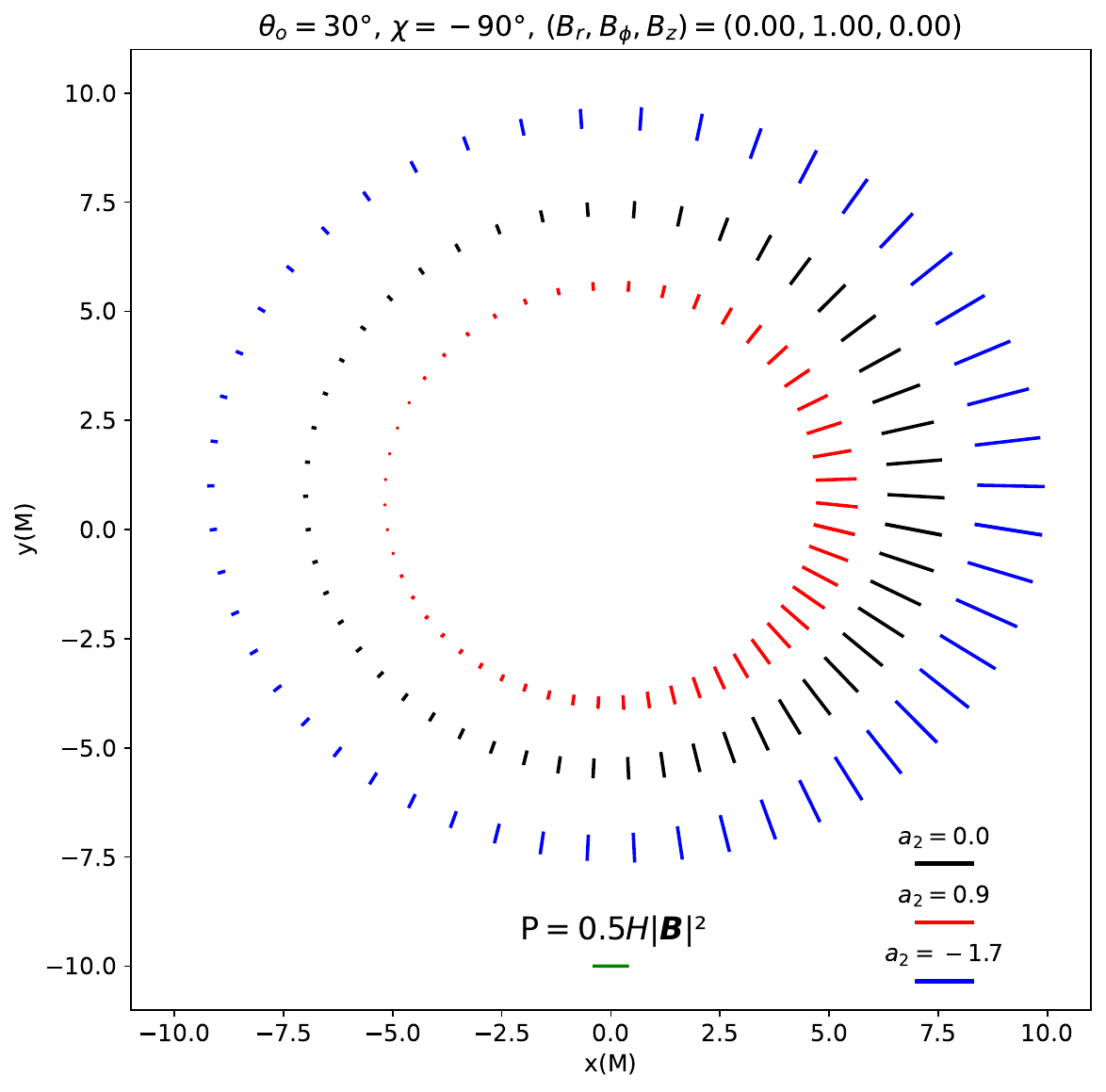}} &
            \hspace{2mm}
            {\includegraphics[scale=0.27,trim=0 0 0 0]{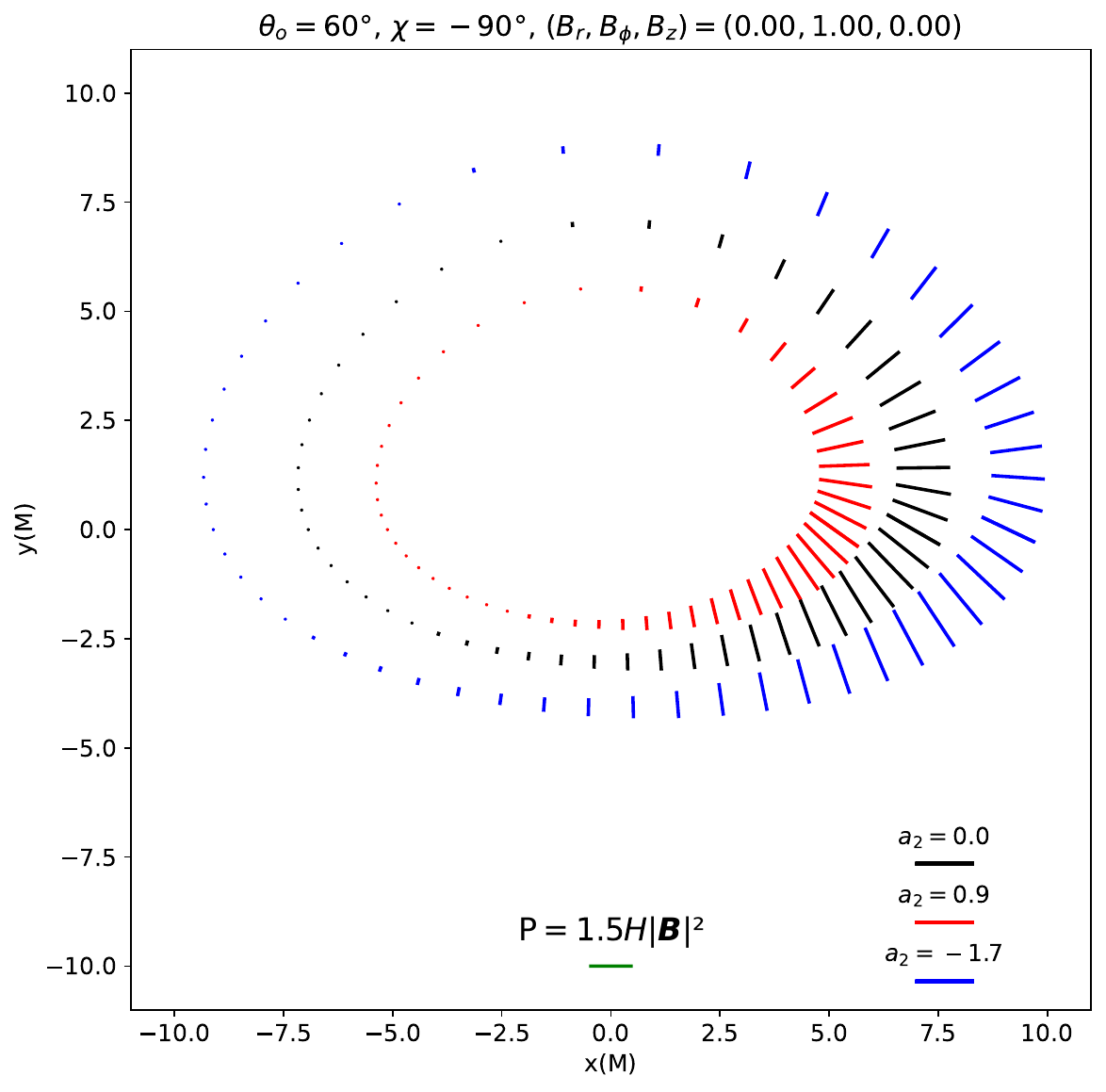}} &
            \hspace{2mm}
            {\includegraphics[scale=0.27,trim=0 0 0 0]{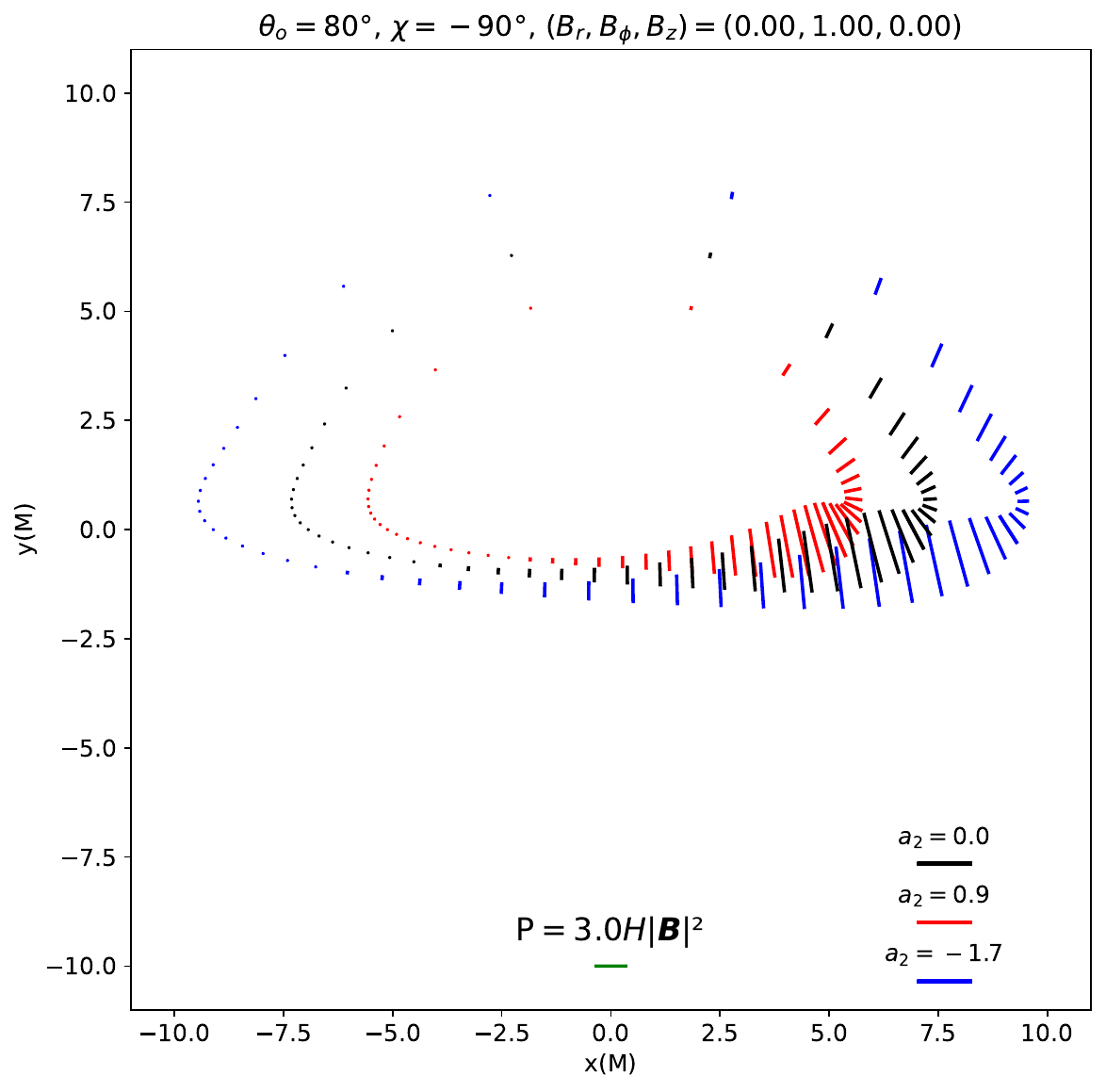}}
            \hspace{2mm}\\
        \end{tabular}
        \caption{Polarization patterns corresponding to the $QU$-diagrams of Fig.\ref{fig:quloops}.}
    \label{fig:quloops2}
    \end{figure*}

    In Figure \ref{fig:pol_evpa}, we show the corresponding variation of the polarization angle with $\phi$ associated to the cases of the middle and bottom of Fig.\ref{fig:pol_patterns}. For the case where only the equatorial magnetic field is considered, i.e. $\vb*{B}_{eq} \neq 0$, few variations are observed among the different black holes considered. A subtle difference in the range $\phi \in (180^{\circ}, 270^{\circ})$ is noticeable between the braneworld like black hole and the other two for a disk inclination of $\theta_o=80^{\circ}$. This is due to a combination  of the effects of the metric and the particular  magnetic field configuration we are considering.  On the other hand, if a general field is considered (i.e., a combination of equatorial and vertical components of the magnetic field), the variations in the polarization angle are more pronounced, with the largest differences found at less disk inclinations.

    Finally, we consider an EMD-black hole with the metric given by \eqref{eq:emd_bh} and doing a  Taylor expansion of $A(r)$ with respect to $q$ as in Sec.\ref{subsec:isora}. The polarization patterns are illustrated in Fig. \ref{fig:pol_emd}. We maintain an equatorial magnetic field of $B_{eq}=1$, preserving the same relationship with the fluid velocity $\beta=0.40$ (same direction, opposite sense), and the same emission radius $R=4.5$. A strong resemblance can be noticed between the polarization shown in Fig. \ref{fig:pol_emd} and the three patterns at the bottom of Fig. \ref{fig:pol_patterns}, except for the fact that in the EMD-black hole, a greater angular separation between the ticks can be seen at the top as the inclination of the disk increases, indicating the strong gravitational lensing effect on the light rays coming from the farthest part of the accretion disk. As in the Reissner-Nordstr\"on metric (where $a_2=q^2$), an increase in intensity is also observed as the value of $q$ grows. However, in the case of the EMD metric, this increase in intensity is more pronounced compared to the Reissner-Nordström case, as with the increase of $q$, $A(R)$ tends to $1$ more rapidly.
\subsection{QU-diagrams of orbiting hotspots}\label{Sec:QU_loops}
Consider now an electromagnetic radiation-emitting hotspot, modeled as a point source orbiting at the ISCO of the black hole. If such a hotspot emits polarized radiation, then both the magnitude and direction of polarization will depend on the relative position among the source, the black hole, and the observer. As mentioned in the introduction, various measurements through specialized instruments allow us to hypothesize that observed flares correspond to hotspots orbiting in the vicinity of the ISCO. In Figure \ref{fig:quloops}, we show different $QU$-diagrams for a hotspot following Keplerian orbits at the ISCO with its corresponding velocity $\beta$. The metrics used are the Schwarzschild metric, the Reissner-Nordstr\"om metric (with $a_2=0.9$), and a braneworld metric with $a_2=-1.7$. The corresponding ISCOs are located at $r_{\text{ISCO}}=6.0$ (Schwarzschild), $r_{\text{ISCO}}=4.286783377$ (RN), and $r_{\text{ISCO}}=8.072671086$ (BW). The corresponding velocity values $\beta$ are 0.5; 0.5625253047 and 0.4544352296, respectively.

In all cases, we use the usual astronomy convention, meaning here we take a global negative sign with respect to the definitions given in Eq.\eqref{eq:stokes_2}.  The colored circles correspond to $\phi=0$ ($\phi=2\pi$), and the squares to $\phi=\pi$. The direction of orbit circulation is assumed to be clockwise; that is, considering the parts of the loops plotted in darker colors, the temporal variation occurs in the direction from the square to the circle. 

In the top panel of Figure \ref{fig:quloops}, we show how these diagrams vary for a purely vertical magnetic field and three different inclinations of the observer.
We can see that the topology changes drastically depending on the observer's inclination (this was already observed in \cite{Gelles:2021kti}). We can also observe that in the case of the formation of two loops (top left), the enclosed area of the inner loop decreases as $a_2$ decreases, while the opposite occurs for the outer loop. On the other hand, for higher inclinations, we can see that not only does the topology change, but the inclination of the closed curves also strongly depends on the metric being considered. 

In the central row of the panel, we can see the shape of the diagrams for the case of a vertical magnetic field in combination with a toroidal field. In these cases, the topology does not vary as strongly with the observer's inclination, however, we can again observe that as $a_2$ decreases, the corresponding closed curves cover a larger area, with the inner loop decreasing significantly in size. Finally, in the bottom panel, we consider a purely toroidal magnetic field. In this case, we can see that the inner loop, in contrast to the case of the purely vertical field, now increases in size as $a_2$ decreases. On the other hand, the inner loop decreases in size for increasing observation inclination angles.

For completeness, in Fig.\ref{fig:quloops2}, we show the corresponding polarization patterns. More details regarding the topology of the different $QU$-diagrams and their analytical dependence on the various parameters characterizing both the metric and the magnetic field will be presented elsewhere. For a description of the geometry of these diagrams and their causes, both in terms of special relativity and the gravitational field, refer to \cite{Vincent2023}.
   
\section{Final remarks}\label{Sec:final}In this work, we have introduced approximate analytical formulas (given by Eqs.\eqref{eq:belo-altoorden}, \eqref{eq:our} and  \eqref{eq:Belo-general}) that establish a direct link between the emission points of light rays and observation points in the asymptotic region. We have demonstrated the broad applicability of these formulas by employing them across various spherically symmetric spacetimes. As a practical demonstration, we have conducted thorough analytical investigations into accretion disk imaging and synchrotron radiation polarimetry. In contrast to the computational complexity associated with images produced using numerical ray-tracing techniques, our analytical formulas offer a swift solution for synthesizing accretion disk images and polarization patterns. These images can be generated within fractions of seconds on a standard notebook, highlighting the efficiency and accessibility of our approach. In a remarkable way, even though very simple, Beloborodov's generalized formula Eq.\eqref{eq:Belo-general} provides very good accuracy for a large family of metrics, even for inclination angles of the accretion disk with respect to the observer as large as 70°.  Although our study has been limited to the investigation of images of thin accretion disks around black holes, the potential applications of our formulas are much broader. First, even when considering spinning black holes, it's important to note the insights from \cite{Cardenas-Avendano:2022csp,Loktev:2023cty}. These studies compare polarimetric observations of Schwarzschild with Kerr for direct rays and demonstrate the effectiveness of analytical formulas in approximating disk images and polarization patterns in Kerr, even for black holes with moderate spins. We hypothesize a similar scenario between the discussed more general black holes and their rotating counterparts. To address this, one should develop a formalism similar to that discussed in \cite{Loktev:2023cty} and compare the analytical approximations with those obtained numerically using ray-tracing codes such as Skylight \cite{Pelle:2022phf}. The outcomes of these comparative studies will be detailed in forthcoming research. 

Moreover, the formulas derived in this work have the potential to be generalized for more diverse spherically symmetric spacetimes where the $g_{rr}$ component of the metric differs from $1/g_{tt}$. This broader class of metrics encompasses black holes, wormholes, boson stars and other exotic compact objects. An analysis of these generalized formulas will be presented separately.

It is worth noting that while our discussion has been primarily focused on thin disks, the analytical formulas are versatile enough to analyze disks with structure and thickness, such as toroidal disks. Additionally, there is significant interest in examining luminosity curves and their polarization from pulsars. We believe that the formulas developed in this work can make a substantial contribution to their analysis, especially for metrics describing objects more general than Schwarzschild. 

Furthermore, the generalization of the formulas developed for the study of massive particles opens up other potential applications, particularly in describing luminosity curves of neutrino sources. Additionally, our formulas can be applied to investigate electromagnetic signals of binary systems where one of the components is a compact object.
These and other related issues will be addressed in future works.\\\\

\subsubsection*{Acknowledgments}

We acknowledge financial support from CONICET, SeCyT-UNC. J.Claros is supported by a doctoral fellowship from CONICET. E. Gallo expresses gratitude to Guillermo D\'{i}az, Tufi Hames and B. Irra for the insightful discussions that have inspired this work.

\appendix
\section{}\label{app:A}

\subsection{Explicit expressions for the integrals $I_1$ and $I_2$ associated to \eqref{eq:our_o6} }
The explicit forms of integrals $I_1$ and $I_2$, as defined by Eqs. \eqref{eq:i1} and \eqref{eq:i2} respectively, with $A(r)$ given by \eqref{eq:ar6}, are:
\begin{widetext}
\begin{eqnarray}
I_1&=&-\frac{1}{3} + \frac{a_2}{5R^2} + \frac{a_3}{3R^3} + \frac{3a_4}{7R^4} + \frac{a_5}{2R^5} + \frac{5a_6}{9R^6},\\
I_2&=&
\frac{23}{15}+a_1\left(\frac{9a_1}{7R^2} + \frac{8}{3R}\right) +  a_2 \left(\frac{53a_2^2}{45R^4} +\frac{12a_1}{5R^3} + \frac{242}{105R^2}\right) \nonumber\\ 
&&+  a_3 \left(\frac{37a_3^2}{33R^6} +\frac{34a_2}{15R^5} + \frac{20a_1}{9R^4} + \frac{2}{R^3}\right) +  a_4 \left(\frac{99a_4^2}{91R^8} +\frac{46a_3}{21R^7} + \frac{834a_2}{385R^6} + \frac{72a_1}{35R^5} + \frac{110}{63R^4}\right) \nonumber\\
&&+ a_5 \left(\frac{16a_5^2}{15R^{10}} + \frac{15a_4}{7R^9} + \frac{83a_3}{39R^8} + \frac{31a_2}{15R^7} + \frac{21a_1}{11R^6} + \frac{23}{15R^5}\right) \nonumber\\
&&+a_6 \left(\frac{161a_6^2}{153R^{12}}+\frac{19a_5}{9R^{11}} + \frac{662a_4}{315R^{10}} + \frac{130a_3}{63R^9} + \frac{1154a_2}{585R^8} + \frac{16a_1}{9R^7} + \frac{134}{99R^6}\right).
\end{eqnarray}
\end{widetext}

\subsection{Derivation of Eqs. \eqref{eq:phis_relation}, \eqref{cosvarp} and \eqref{sinvarp}}

Let us consider Figure \ref{fig:framework}. Let $\vb*{e_{X'}},\vb*{ e_{Y'}}$ be the two orthonormal basis vectors respectively aligned with the $X'$ and $Y'$ axes of the observer $O'$. Let $\vb*{ e_X}, \vb*{ e_Y}, \vb*{ e_Z}$ denote the unit vectors associated with the $\overline{OX}, \overline{OY},\overline{OZ}$ axes respectively and let $\vb*{e_{Z''}}$ be the unit vector associated with the $\overline{OZ''}$ axis. Then, we have the following relations between these unit basis vectors:
\begin{eqnarray}
    \vb*{ e_{X'}}&=&\vb*{ e_{X}},\\
\vb*{ e_{Y'}}&=&\vb*{ e_{Z''}}=\cos\theta_o \vb*{ e_Y}+\sin\theta_o \vb*{ e_Z}.
  \end{eqnarray}

In a similar manner, if $\vb*{ e_{X''}}$ denotes the unit vector along the $\overline{OX''}$ axis and noting that $\overline{OX''}||\overline{O'P'}$, we have:
\begin{equation}\label{eq:alt1x''}
\begin{split}
    \vb*{ e_{X''}}&=\cos\varphi\vb*{ e_{X'}}+\sin\varphi\vb*{ e_{Y'}}\\
    &=\cos\varphi \vb*{ e_{X}}+\sin\varphi\cos\theta_o \vb*{ e_Y}+\sin\varphi\sin\theta_o \vb*{ e_Z}.
    \end{split}
\end{equation}
The same vector expressed in terms of the unit vector $\vb*{ o}$ can be represented as $\vb*{ e_{X''}}=\frac{1}{\sin\psi}\vb*{ o}\times \frac{\vb*{R}}{R}\times \vb*{ o}$. Considering that
\begin{equation}
\frac{\vb*{R}}{R}=\cos\phi \vb*{ e_{X}}+\sin\phi\vb*{ e_{Y}},
\end{equation}
and $\vb*{ o}=-\sin\theta_o\vb*{ e_{Y}}+\cos\theta_o\vb*{ e_{Z}}$, we obtain:

\begin{equation}\label{eq:alt2x''}
    \begin{split}
        \vb*{ e_{X''}}=\frac{\cos\phi}{\sin\psi}\vb*{ e_{X}}+\frac{\sin\phi\cos^2\theta_o}{\sin\psi}\vb*{ e_{Y}}+\frac{\sin\phi\cos\theta_o\sin\theta_o}{\sin\psi}\vb*{ e_{Z}}.
    \end{split}
\end{equation}
Comparing Eqs. \eqref{eq:alt1x''} and \eqref{eq:alt2x''}, we arrive at:
\begin{eqnarray}
\cos\varphi &=& \frac{\cos\phi}{\sin\psi},\label{eq:covarphipsi} \\
\sin\varphi &=& \frac{\sin\phi\cos\theta_o}{\sin\psi}, \label{eq:sinvarphipsi} 
\end{eqnarray}
which agree with Eqs. \eqref{cosvarp} and \eqref{sinvarp} if we take into account the relation Eq. \eqref{eq:cospsi_eq}, leading to:
\begin{equation}
\sin\psi = \sqrt{1 - \sin^2\theta_o\sin^2\phi}.
\end{equation}

Eq.\eqref{eq:phis_relation} is a direct consequence of
\eqref{eq:covarphipsi} and \eqref{eq:sinvarphipsi}.

\subsection{Derivation of Eq.\eqref{eq:cospsilum}}

The expression \eqref{eq:cospsilum}, originally derived by Luminet using spherical trigonometry \cite{Lumi:1979}, can be straightforwardly obtained from Eqs. \eqref{eq:covarphipsi} and \eqref{eq:sinvarphipsi}, which imply:

\begin{eqnarray}
\cos^2\phi &=& \cos^2\varphi(1 - \cos^2\psi), \label{eq:cophipsi} \\
\sin^2\phi &=& \frac{\sin^2\varphi(1 - \cos^2\psi)}{\cos^2\theta_o}. \label{eq:sinphipsi}
\end{eqnarray}

By adding \eqref{eq:cophipsi} and \eqref{eq:sinphipsi} and solving for $\cos\psi$, we obtain Eq. \eqref{eq:cospsilum} (the global minus sign in Eq. \eqref{eq:cospsilum} arises because as $\varphi$ ranges from $0$ to $\pi$, $\psi$ takes values in the range $[\pi/2, \pi]$, and for $\varphi\in(\pi,2\pi)$, $\psi$ takes values in the interval $(0, \pi/2)$).

\section{}\label{app:B}
Let us consider a spherically symmetric metric described as in Eq.\eqref{eq:METRIC} with \begin{equation}
    A(r)=1+\frac{a_1}{r}+\frac{a_2}{r^2}.
\end{equation}
After substituting the expressions for the energy $\mathcal{E}$, orbital angular momentum $\mathcal{J}$, and angular velocity $\Omega$ of circular orbits described by \eqref{eq:energy_part}, \eqref{eq:momang_part}, and \eqref{eq:velang_part} respectively into Eq. \eqref{eq:flux_emit} for the emitted flux, we obtain:
    \begin{widetext}
    \begin{equation}
    \begin{aligned}
        F_e (r) =&-\frac{dM/dt}{4\pi} \frac{8 a_2 + 3 a_1 r}{4 r^2 \mu (4 a_2 + r(3 a_1 + 2 r))} \Bigg\lbrace 2 \left(\mu - \mu_o\right) - 4 \sqrt{2 a_2} \left[\arccot\left(\frac{\sqrt{2 a_2}}{\mu}\right) - \arccot\left(\frac{\sqrt{2 a_2}}{\mu_o}\right)\right] \\
        &+ \mu_+ \left[\arctan\left(\frac{2 \mu}{\mu_+}\right) - \arctan\left(\frac{2 \mu_o}{\mu_+}\right)\right] + \mu_- \left[\arctan\left(\frac{2 \mu}{\mu_-}\right) - \arctan\left(\frac{2 \mu_o}{\mu_-}\right)\right]\Bigg\rbrace, \\
        \text{where}\\
        \mu   =&\sqrt{-2 a_2 - a_1 r}, \\
        \mu_o =&\sqrt{-2 a_2 - a_1 r_{\text{isco}}}, \\
        \mu_+ =&\sqrt{-a_1 \left(3 a_1 + \sqrt{9 a_1^2 - 32 a_2}\right) + 8 a_2},\\
        \mu_- =&\sqrt{-a_1 \left(3 a_1 - \sqrt{9 a_1^2 - 32 a_2}\right) + 8 a_2}.
    \end{aligned}      
    \end{equation}    
    \end{widetext}
    

\end{document}